\documentclass[a4paper]{article}
\usepackage[english]{babel}
\usepackage{ulem}
\usepackage{algorithmicx, algpseudocode}
\usepackage{algorithm}
\usepackage[letterpaper,top=2cm,bottom=2cm,left=3cm,right=3cm,marginparwidth=1.75cm]{geometry}

\usepackage{amsmath}
\usepackage{subcaption}
\usepackage{amsthm}
\usepackage{amssymb}
\usepackage{comment}
\usepackage{mwe}
\usepackage{amsfonts}
\usepackage{xcolor}
\usepackage{graphicx}
\usepackage{verbatim}
\usepackage{float}
\usepackage[colorlinks=true, allcolors=black]{hyperref}
\usepackage[toc]{appendix}
\newtheorem{corollary}{Corollary}
\usepackage{multirow}
\newtheorem{theorem}{Theorem}
\newtheorem{remark}{Remark}

\newtheorem*{theorem*}{Picard's theorem}
\newcommand{\stkout}[1]{\ifmmode\text{\sout{\ensuremath{#1}}}\else\sout{#1}\fi}

\title{ Data-Driven Models for studying the Dynamics of the COVID-19 Pandemics}
\author{Rawan Madi \thanks{American University of Beirut (AUB), Beirut, Lebanon. (rhm26@mail.aub.edu)} \and  Sophie Moufawad \thanks{American University of Beirut (AUB), Beirut, Lebanon.  (sm101@aub.edu.lb) } \and Nabil Nassif \thanks{American University of Beirut (AUB), Beirut, Lebanon.  (nn12@aub.edu.lb) } \thanks{The authors would like to acknowledge: \\    $\qquad \qquad \bullet$ The programming contribution of  Miss Nadine Charaf and Miss Lara Tarakh during the Summer Research Camp 2022
    $\bullet$  AUB Center for Advanced Mathematical Sciences (CAMS) and Math department's support
}}
\begin{document}
\maketitle
\begin{abstract}     
     This paper seeks to study the evolution of the COVID-19 pandemic based on daily published data from Worldometer website, using a time-dependent SIR model. Our findings indicate that this model fits well such data, for different chosen periods and  different regions. 
     
      This well-known model, consisting of three disjoint compartments, susceptible , infected , and removed , depends in our case on two time dependent parameters, the infection rate $\beta(t)$ and the removal rate $\rho(t)$.
     After deriving the model, we prove the local exponential behavior of the number of infected people, be it growth or decay. Furthermore, we extract a time dependent replacement factor $\sigma_s(t) ={\beta(t)}s(t)/{\rho(t) }$, where $s(t)$ is the ratio of susceptible people at time $t$. In addition, $i(t)$ and $r(t)$ are respectively the ratios of infected and removed people, based on a population of size $N$, usually assumed to be constant.

      Besides these theoretical results, the report provides simulations on the daily data obtained for  Germany,  Italy, and the entire World, as collected from Worldometer over the period stretching from April 2020 to June 2022.
      The computational model consists of the estimation of $\beta(t)$, $\rho(t)$ and $s(t)$ based on the time-dependent SIR model. The validation of our approach is demonstrated by comparing the profiles of the collected $i(t), r(t)$ data and those obtained from the SIR model with the approximated parameters.
      
We also consider matching the data with a constant-coefficient SIR model, which seems to be working only for short periods. Thus, such model helps understanding and predicting the evolution of the pandemics for short periods of time where no radical change occurs.

\end{abstract}

\section{Introduction}

 Since 2019 and during its early stages, COVID-19 became a worldwide concern, as it was spreading fast and leading to death in certain cases \cite{ncbi2}. Not enough data was presented to comprehend its mechanisms and means of transmission \cite{RMV}. The degree of severity of the symptoms of this virus often ranged from none to extremes. Many people who were asymptomatic were spreading the virus because they were unaware they had it in the first place \cite{last}, \cite{time-dependent}. These unreported cases made it difficult to track the severity of the virus and study it smoothly. A study in Brazil showed that neglecting unreported cases has had ``contradictory effects in maintaining the disease" \cite{Brazil}.  With the progression of the virus, more data was collected and different mathematical and statistical models along with AI were 
 employed in attempting to address several scientific and healthcare imminent questions. Thousands of papers were written on this topic, some of these were compiled in the 2021 paper \cite{litrev1} and the 2022 paper \cite{litrev2}.
 
These models attempt at studying on a global or local scale, different parameters,  related to the pandemics, such as mortality, etc. Regardless of their objectives, those models used different types of approaches: some are strictly relying on statistical software, others use innovative data science combined to  artificial intelligence. Other approaches were simply of compartmental nature.

Since COVID is very similar in characteristics to SARS, the model frequently worked with is the constant coefficient SIR model, originally developed by Kermack and McKendrick \cite{heathcote}. This model helps in finding, or estimating, through numerical simulations, various important parameters such as, the maximum number of infected people \cite{bar}, the number of people that need to be vaccinated \cite{herd}, the expected duration of a pandemic. It also helps study the reproduction factor, a key parameter revealed by the constant coefficients SIR model, which sheds light on the progression,  sustainability, and decay of the pandemic. Thus, on the basis of theoretical results, added to numerical simulations run on available large amounts of data, one has powerful tools that provide better understanding of the phenomena and issues at hand \cite{bar}.\\
 However, sample of the surveyed results indicate that the constant coefficient SIR model is not well-fitted for analyzing the effects of vaccination, immunity, lock downs and social distancing in reducing the transmission and contact rates and in explaining the rapid rise or rapid decay in the number of infected people within a population, especially in what may be considered as the late-phase(s) of COVID-19 (\cite{RMV}, \cite{time-dependent}). 
 Therefore, implementing protection and actions in anticipation of any future new outbreak of the pandemics, requires more complex models to forecast values in the long run. Some variants of the constant coefficients SIR model could provide more accurate estimations of parameter values \cite{ncbi2}, \cite{kroger}, \cite{discrete}, \cite{RCMMAG}. However, the data needed to be able to use such variants is normally not available. 
 
In this work, our starting point is the data we have daily collected daily between April 2020 and June 2022, from the {\bf Worldometer Corona Virus Platform}. The main reported records on each day $t$, are as follows:  
\begin{enumerate}
    \item $I(t)$, the number of infected people,
     \item $C_N(t)$, the number of newly infected persons,
    \item  $C_I(t)$, the cumulative number of infected cases,
    \item $C_D(t)$, the cumulative number of deceased, out of the infected population,
    \item $C_R(t)$, the cumulative number of recovered, out of the infected population,
   \end{enumerate}
Based on these real time data, we attempt to develop models that would at least reproduce the data being reported, particularly in terms of the number of infected cases $I(t)$. Using the basic balance equation, that governs the dynamics of a pandemics, namely:
$$I'(t)=F_I(t)-F_R(t)$$
we derive successively a time-dependent IR \eqref{sys:basicIR} followed by   \eqref{eq:SIR}, a time-dependent SIR models. Such perception of the pandemics naturally excludes a constant coefficients SIR model, as was mentioned in the works cited   above. Such models depend on  time-dependent coefficients measuring the impact of infective and removal (death and recovery) forces. 
At that point of our research, independently of other types of data such as those people declared medically immune or reported as vaccinated or being in contact with others, the models derived allow to:
\begin{enumerate}
    \item reproduce the collected data, with at least 90 $\%$ accuracy.
    \item monitor the pandemics on short periods of time, revealing either outbreak or decay.
    \item provide in case of outbreak, the level of confinement that should be adopted by health authorities. 
\end{enumerate}
 
\noindent The remaining part of this paper is divided as follows.

In Section \ref{sec:Deriv}, we derive the models and then prove the exponential growth and decay of the pandemics,
particularly Theorems \ref{exponential-1} and \ref{exponential-2}.
  
In Section 3, we perform simulations on real data for  Germany,  Italy, and the world, as collected from Worldometer over a span of two years.  First, the appropriate compartments' data is extracted and used to approximate the time-dependent parameters of the model. Then, the time dependent SIR model is discretized using Runge-Kutta schemes to regenerate the worldometer data and test the accuracy of this model. Moreover, we graph and discuss the evolution of the pandemic's compartments and parameters during the outbreaks, on some random interval and on the entire period. The models did not reveal any pattern as to how the pandemics would generally outbreak, in spite of people being vaccinated, and/or confined.

Alternatively,  we explore in section 4, the possibility of developing a short-term constant-coefficient model to predict the pandemics evolution.

Finally, we conclude in section \ref{sec:conclusion} and suggest tracks for future research. 

\section{The Time-Dependent IR and SIR Models}\label{sec:Deriv}

\indent Both IR and SIR models are compartmental models based on categorizing the population into respectively two and three groups or compartments.
\subsection{Derivation of the Time-Dependent IR Model}
In the IR model one has only the infected and removed categories, this last category including those people who were originally infected, and have either died or recovered. There are some basic hypotheses one  considers when using a compartmental model. In the case of an IR model, one assumes that over the period under study:
\begin{enumerate} \label{assumptions}
\item An infected person is automatically active and remains in the infected category until removal through recovery or death. 
\item Once a person is in the removed compartment, he/she cannot go backward to the infected compartment (A recovered person has permanent immunity over the period under scrutiny).
\item Population birth and death rates are not taken into consideration.
\end{enumerate}

The derivation of the IR  model comes from the original idea in Kermack-McKendrick Model which considers the pandemics as the resultant of two forces: the infective $F_I(t)$ and the removal forces $F_R(t)$. 
\begin{equation}\label{sys:basicIR}
\begin{cases}
 I'(t) \, = \, F_I (t) \, - \, F_R(t)\\
 R'(t)  = \, F_R(t) 
\end{cases}
\end{equation}
At this point, one must make additional assumptions, the simplest being the linear dependence of $F_R (t)$ and $F_I(t)$ on $I(t)$, specifically:
    \begin{itemize}
        \item $F_R(t) = \rho(t)I(t)$, with:
        \begin{itemize}
            \item $\rho(t)=\mu(t)+\gamma(t)$ is the \textbf{total} removal rate per individual at time $t$. 
            \item $\mu(t)$ being the death rate at time $t$ per infected individual, and
            \item $\gamma(t)$  the recovery rate at time $t$ per infected individual. 
        \end{itemize}
\item  As for $F_I(t)$ we make the following assumption. It is also proportional to $I(t)$, with a proportionality factor  $\beta_s(t)$, where:
\begin{equation}\label{eq:infF} \displaystyle F_I(t) = \beta_s(t) I(t)\end{equation}
\end{itemize}
Thus in this case, \eqref{sys:basicIR} reduces to a linear initial value ordinary differential equation:
\begin{equation}\label{sys:basicIR-1}
\begin{cases} 
I'(t)=(\beta_s(t)-\rho(t))I(t),&\\
R'(t) = \rho(t)I(t).&
\end{cases}
\end{equation}
that can be solved as a decoupled system on any interval $[t_0,T]$, provided that $I(t_0)$ and $R(t_0)$ are given, and $\beta_s(t)$ and $\rho(t)$ approximated using \eqref{eq:betaApp} and \eqref{eq:rhoApp}.\\
Solving first for $I(t)$ on the interval $[t_i,t_{i+1}]$  and then for $R(t)$, one gets :
\begin{eqnarray}
I(t_{i+1})&=&I(t_i)e^{\int_{t_i}^{t_{i+1}}{(\beta_s(t)-\rho(t))dt}}, \label{eq:IRi}\\
R(t_{i+1}) &=& R(t_{i}) + \int_{t_i}^{t_{i+1}} \rho(t) \, I(t) dt \vspace{-3mm}
\end{eqnarray}
which numerically can be solved using any approximation formula for the integrals $\int_{t_i}^{t_{i+1}}{(\beta_s(t)-\rho(t))dt}$ and $\int_{t_i}^{t_{i+1}} \rho(t) \, I(t) dt$.

\noindent Note from \eqref{eq:IRi} the exponential behavior of $I(t)$, be it growth or decay, depending if the ratio $\dfrac{\beta_s}{\rho}$ is greater than or less than 1 respectively. As for $R(t)$, it is always increasing given the positivity of $\rho(t)I(t)$.

Naturally, this IR model is a flat two-compartments model. It does not consider, the stage(s) before infection, and after recovery. For example, after recovery, a person may become immune or may get infected another time, i.e. be part of the infective forces.

\noindent Moving forward in modeling the pandemics, one must look into the pre-infection stages, considering the  main factors for its spreading are the following:
\begin{enumerate}
    \item  Power of the virus to be transmitted,  $\beta_v(t)$ 
\item Immunity of the non-infected people (vaccine, individual immunity),  $\beta_m(t)$
\item The probability to get infected (mask-wearing, getting in contact with an infected individual,...),  $p(t)$
\end{enumerate}
where assuming these factors are multiplicative, i.e.  $\beta_s(t) = \beta_v(t) \beta_m(t) p(t)$, then \eqref{eq:infF}  becomes
\begin{equation}\label{eq:infF2} F_I(t) = \beta_v(t)\, \beta_m(t) \,p(t) \,I(t).
\end{equation}

As we are considering Data-Driven modeling, at this stage we do not have any available data that could measure $\beta_v(t)$ and $\beta_m(t)$ independently. 

On the other hand, one could attempt to categorize the non-infected and non-removed in subcategories, such as
\begin{itemize}
    \item the category of people with a high-level of immunity (strongly positive antibody test), that will protect them from the virus. 
    \item the category of people who were tested negative but yet were infected.
    \item the category of isolated people.
\end{itemize}
However, since we do not have enough data  on these subcategories, we lump them all in one ``susceptible" compartment, leading to an SIR model.

\subsection{Derivation of the Time-Dependent SIR Model}
At this point, the additional assumptions to move ahead from the IR model is that all people who are not infected or removed, have equal chance $p(t) = \dfrac{S(t)}{N(t)}$ of becoming infected. Thus, these belong to a third susceptible compartment, $S(t)$, such that     
\begin{equation} \label{eq:SIRN}
    S(t) + I(t) + R(t) = N(t)\vspace{-2mm}
\end{equation}
where $N(t)$ is the total size of the population. Since in the literature $\dfrac{S(t)}{N(t)}$ is denoted by $s(t)$, thus we will denote $p(t)$ by $s(t)$. Therefore, \eqref{eq:infF2} becomes
\begin{equation}
    F_I(t) = \beta(t) s(t) I(t),
\end{equation} 
   where $\beta(t) = \beta_v(t)*\beta_m(t)$.\vspace{1mm}\\
   \noindent We derive \eqref{eq:SIRN} with respect to $t$ and use \eqref{sys:basicIR-1}, to get 
\begin{equation} S'(t) = N'(t) - R'(t) - I'(t)  = N'(t) - F_I(t)
\end{equation}

To get the well-known Kermack–McKendrick SIR model,
we assume a closed population, i.e.  $N(t)=N$ is fixed. Accordingly, $N'(t) = 0$ and we get system \eqref{eq:SIR}
\begin{equation}\label{eq:SIR}
\begin{cases}
S'(t) \, = \, -\beta(t) \cdot \dfrac{S(t)}{N}  \cdot I(t) \\
I'(t) \, = \, \beta(t) \cdot \dfrac{S(t)}{N}  \cdot I(t) - \rho(t) \cdot I(t)  \\
R'(t) = \, \rho(t) \cdot I(t)\\
S(t_0) = S_0 ; \; I(t_0)=I_0;\; R(t_0)= R_0
\end{cases}
\end{equation}

Normalizing the equations, i.e. dividing by the assumed constant size $N$ of the population on both sides of each equation, one gets:
\begin{equation} \label{sir}
\begin{cases}
s'(t) \, = \, -\beta(t) \cdot s(t) \cdot i(t)&\\
i'(t) \, = \, \beta(t) \cdot s(t) \cdot i(t) - \rho(t) \cdot i(t)& \\
r'(t) \,= \, \rho(t) \cdot i(t)&\\
s(t_0) = s_0 ; \; i(t_0)=i_0;\; r(t_0)= r_0
\end{cases}
\end{equation}
where
\begin{itemize}
    \item  $ s(t) \, = \, \dfrac{S(t)}{N}, \, i(t) \, = \, \dfrac{I(t)}{N},$ and $ r(t) \, = \, \dfrac{R(t)}{N}$.
    \item $ 0 \leq s(t),i(t),r(t) \leq 1$, and $s(t)+i(t)+r(t) =1 $ 
    \item $S(t)$ is at least $1$ for $ t\,< \,\infty $ since there must be at least one susceptible person during a given period of time.
\end{itemize}

\vspace{1mm}

\subsection{Theoretical Study of the SIR Model}\label{sec:ana}
In this section, we will be focusing on the theoretical aspects of the model. A theoretical analysis is significant prior to computational work, as it presents a basis for our work and is means to find out whether the work is correct or not. We analyze the time-dependent SIR model by listing three theorems along with their proofs which describe the changes in the three compartments and especially the exponential growth and decay in the infected compartment. 

\begin{theorem} \label{thm:sandr}
 $s(t)$ is strictly decreasing if $i(t) > 0$ and stagnates if $i(t) = 0$. Whereas $r(t)$ is strictly increasing if $i(t) > 0$ and stagnates if $i(t) = 0$.
\end{theorem}
\begin{proof}\label{firstproof}
\begin{enumerate}
    \item Since $\beta(t) > 0$, $s(t) > 0 \; \forall t\,>\,0$, and $i(t) \geq 0 \; \forall t\,>\,0$, then by (\ref{sir})-(a) $$s'(t) = -\beta(t) \cdot s(t) \cdot i(t) \leq 0 \qquad \forall t > 0$$ which implies that $s(t)$ is strictly decreasing when $i(t) >0$, and stagnates when $i(t) = 0$.
    \item Since $\rho(t) > 0$ and $i(t) \geq 0 \; \forall t >0$, then by (\ref{sir})-(c) $$r'(t) = \rho(t) \cdot i(t) \geq 0 \qquad \forall t > 0$$ which implies that R is strictly increasing when $i(t) >0$, and stagnates when $i(t) = 0$.
\end{enumerate}
\end{proof}
\noindent Now, we write equation \eqref{sir}-(b) as follows:
\begin{equation}\label{eq:iprim0}
i'(t) \, = \, (\beta(t) s(t) - \rho(t)) \cdot i(t),    
\end{equation}

Thus the variation of the ratio of number of infected $i(t)$ depends on the sign of $\beta(t) s(t) - \rho(t)$. Hence if:
$$\rho(t)(s(t) \dfrac{\beta(t)}{\rho(t)} - 1)\,
\begin{cases} 
>0,& \mbox{ there is an outbreak of the pandemics}\\
<0,& \mbox{ then the pandemic is in an decreasing mode.}
\end{cases} $$
We introduce now:
\begin{enumerate}
    \item The basic reproduction factor at time $t$:  $\sigma(t)  := \dfrac{\beta(t)}{\rho(t)}$, is the maximum number of secondary infections introduced when one infective is introduced in a totally susceptible population.
    \item The replacement number at time $t$: $\sigma_s(t)=\sigma(t)s(t) = \dfrac{\beta_s(t)}{\rho(t)}$ which is the number of secondary infections produced by a typical infective in a population in which the ratio of susceptible population is $s(t)$. 
\end{enumerate}
and therefore equation (\ref{eq:iprim0}) can be rewritten as:
\begin{equation}\label{eq:iprime}
i'(t)=(\sigma_s(t)-1)\rho(t)i(t)
\end{equation}

\subsubsection*{Criteria for Growth and Decay of the Pandemics}
Then, one can state:
\begin{theorem}\label{theo:pandbehav} The evolution of the disease obeys the following facts:
\begin{enumerate}
    \item If $\sigma_s(t)<1$, then $i'(t)<0$, and the pandemics is in decreasing mode.
    \item If $\sigma_s(t)>1$, then $i'(t)>0$ and there is an outbreak of the pandemics.
    \item If $\sigma_s(t)=1$, then $i'(t)=0$ and the pandemics stagnates.
\end{enumerate}
\end{theorem}
\begin{proof}
This result is simply obtained from \eqref{eq:iprime}.
\end{proof}
\noindent Thus, to monitor the evolution of the pandemics, one can simply follow-up the variation of $\sigma_s(t)$.\\ 
As a consequence, we have the following result in case the pandemics is dying out.
\begin{corollary}\label{cor:1}
If $\sigma(t)<1$, then $\sigma_s(t)<1$ and the pandemics is in decreasing mode.
\end{corollary}
\begin{proof}
Since $s(t)\le 1$ then $\sigma_s(t)=s(t)\sigma(t)\le\sigma(t)<1$ and therefore the first case of Theorem \ref{theo:pandbehav} applies.
\end{proof}
\noindent On the other hand if $\sigma(t)>1$, the evolution of the pandemics is a function of the susceptible fraction $s(t)$. Specifically, one has:
\begin{corollary}\label{cor:2} In case $\sigma(t)>1$, then: 
\begin{enumerate}
    \item If $s(t)>\dfrac{1}{\sigma(t)}$, there is an outbreak of the pandemics.
    \item If $s(t)<\dfrac{1}{\sigma(t)}$, the pandemics decreases.
    \item If $s(t)=\dfrac{1}{\sigma(t)}$, the pandemics stagnates.
\end{enumerate}
\end{corollary}
\begin{proof}
This result is in direct relation with the identity $\sigma_s(t)=s(t)\sigma(t)$ followed with the application of Theorem \ref{theo:pandbehav}. 
\end{proof}
\noindent We can now make the following remarks.
\begin{remark} On the basis of Theorem \ref{theo:pandbehav}, monitoring the pandemics can be done by following the replacement number $\sigma_s(t)$.
\end{remark}
\begin{remark} Thus, when $\sigma(t)>1$, $\dfrac{1}{\sigma(t)}=\dfrac{\rho(t)}{\beta(t)}$ provides a level under which the susceptible fraction of the population should be kept if one is to avoid outbreak of the pandemics.\\
To increase $\dfrac{1}{\sigma(t)}=\dfrac{\rho(t)}{\beta(t) }$, one should either reduce the contact function $\beta(t)$ or increase the removal function $\rho(t)$.
\end{remark}
\noindent The following result provides  further clarification on the growth of infected $I(t)$ when $\sigma(t)<1$.\\
\begin{theorem}\label{exponential-1} Assume the rate function $\beta(.),\,\mu(.),\,\gamma(.)$ are continuous, with the existence  of $t_0<t_1$ such that $\sigma(t)<1$ for $t\in[t_0,t_1]$. Let also, $\rho_{\min}=\min\limits_{[t_0,t_1]}{\{\rho(t)\}}$ and $\sigma_{\max}=\max\limits_{[t_0,t_1]}{\{\sigma_s(t)\}}<1$.  Then,
$$i(t)\le i(t_0)e^{-\kappa_0(t-t_0)}, \quad \forall t\in[t_0,t_1], $$ with $\kappa_0=\rho_{\min}(1-\sigma_{\max})$.
\end{theorem}
\begin{proof} Dividing \eqref{eq:iprime}  by $i(t)$ yields:
\begin{equation}\label{ode}
    \dfrac{1}{i(t)}\dfrac{di}{dt}(t)=\rho(t)(\sigma_s(t)-1)\Longleftrightarrow \frac{d}{dt}\ln(i(t))=\rho(t)(\sigma_s(t)-1).
\end{equation}
Integration from $t_0$ to $t$ gives:
\begin{eqnarray}
  \ln\left(\dfrac{i(t)}{i(t_0)}\right)&=&\int_{t_0}^t{\rho(t')(\sigma_s(t')-1)\,dt'},   \nonumber\\   i(t)&=&i(t_0)\exp{\left(\int_{t_0}^t{\rho(t')(\sigma_s(t')-1)\,dt'}\right)}.\label{formula_I}
\end{eqnarray}
Now $\sigma_s(t')-1\le\sigma_{\max}-1<0$ and therefore: 
$$\rho(t')(\sigma_s(t')-1)\le \rho(t')(\sigma_{\max}-1)<0, $$
which leads to:
$$ \rho(t')(\sigma_s(t')-1)\le\rho(t')(\sigma_{\max}-1)\le(\sigma_{\max}-1)\rho_{min}=-\kappa_0<0.$$
Combining this inequality with (\ref{formula_I}), yields to the result of theorem.
\end{proof}
\noindent Turning now to the case when $\sigma_s(t)\ge 1$, one notes that $\dfrac{di(t)}{dt}$ becomes non-negative, leading to the following ``exponential blow-up'' result.
\begin{theorem}\label{exponential-2} Assume $\beta(.),\,\mu(.),\,\gamma(.)\in C[t_0,t_1]$ with  $\sigma_s(t)\ge \sigma_0>1$ for $t\in[t_0,t_1]$, with \\$\rho_{\min}=\min\limits_{[t_0,t_1]}\{\rho(t)\}$, then:
$$i(t)\ge i(t_0)e^{\kappa_1(t-t_0)},\quad \forall t\in[t_0,t_1],$$ with $\kappa_1=\rho_{\min}(\sigma_0-1)$.
\end{theorem}
\begin{proof} We proceed as in the previous theorem. Based on  \eqref{ode}, one writes:
$$i(t)=i(t_0)\exp{\left(\int_{t_0}^t{\rho(t')(\sigma_s(t')-1)\,dt'}\right)}\ge i(t_0)\exp\left(\rho_{\min}(\sigma_0-1)(t-t_0)\right),$$
which leads to the result of the theorem.
\end{proof}

\section{Simulations on Real Data}
 The data we get is from {Worldometers} \cite{worldometer}, a reference website that collects data and statistics on different issues, related to health, demography, economy, food, environment, and media.
 COVID 19 data has been regularly collected from  \href{http://www.worldometers.info/coronavirus/}{Worldometers Corona Virus} on a set $\mathcal{T}_n:=\{t_0,t_1,...,t_n\}$ of days over a global period stretching from April 4, 2020 until June 30, 2022 for Italy, Germany and the world.

 The data, collected over the past two years, provides information on the total tests done in a day, total deaths and recovered, number of infected people in a day, the critical cases, the population size, and the total cases in some countries or in the whole world. 
The data fields being sought are indicated in Table \ref{tab:data}. 
\begin{table}[H]
\begin{center}
\begin{tabular}{|c|c|c|c|c|}
\hline
  {\bf{Date}} &{\bf  Cumulative}  &{\bf Cumulative}  &{\bf Cumulative}&{\bf Active}  \\
 $t_i$& {\bf  Cases $C_I(t_i)$} & {\bf Deaths $C_D(t_i)$} &{\bf Recovered $C_R(t_i)$}& {\bf Cases ($I(t_i)$)}\\
  \hline
{\bf t$_0$}& ...... & .......&......&......\\
\vdots &...... &...... &......& ...... \\
{\bf t$_n$ }&...... &...... &......&......\\
\hline
\end{tabular}
\caption{Source Data Fields}\label{tab:data}
\end{center}
\end{table}
\noindent The collected data allows the extraction of time-dependent parameters $N(t)$, $I(t)$ and $R(t)$, leading to $S(t)$. as detailed in section \ref{sec:compart}. Then, in section \ref{sec:Param} we approximate the time-dependent rates  $\beta(t)$ and $\rho(t)$, and compute $\sigma(t)$ and $\sigma_s(t)$ which are key indicators of the pandemics' evolution. 
To validate the obtained parameters, we run the SIR model using the approximated $\beta(t)$ and $\rho(t)$ and compare the estimated results to the actual ones as discussed in section \ref{sec:valid}.

In section \ref{sec:worldtest} we test this approach on 
 real data from Italy, Germany, and the world over several periods that include two outbreaks, a random period, and the entire period. Moreover, an analysis of the results and the statistical properties of the parameters is provided.

\subsection{Extraction of Compartments }\label{sec:compart}%

First, in section \ref{sec:N} we define the total population as an average. In section \ref{sec:I} we extract the  number of infected people $I(t)$ and the number of removed people $R(t)$ from collected data.
Then, the number of susceptible $S(t)$ is deduced in section \ref{sec:S}. Given $S(t),I(t),R(t)$ and $N$, we get the normalized fractions $s(t), i(t)$, and $r(t)$.
\subsubsection*{Population size N}\label{sec:N}
We will consider the population to be constant and not dynamic.  Thus, we take the average of the given total population $N(t)$ over some given period $[t_0,t_j]$, i.e. 
$$N = \dfrac{1}{j+1}\sum\limits_{i=0}^j N(t_i)$$
For example, the average worldwide total population  over the period of two years is $N = 7,305,100,000$.

\subsubsection*{Extracting I(t) and R(t)}\label{sec:I}
The collected Woldometer data provides the number of:
\begin{enumerate}
\item cumulative infected till a certain day $t_i$, $C_I(t_i)$
\item cumulative recovered till a certain day $t_i$, $C_R(t_i)$
    \item cumulative dead  till a certain day $t_i$, $C_D(t_i)$
    \item active cases on a certain day $t_i$, $I(t_i)$
\end{enumerate}

\noindent Thus, the total number of infected people at day $t_i$, $I(t_i)$ is extracted from data directly. Then, the fractions of infected people is ${i}(t) = {I}(t)/N$.

\noindent The total number of removed people is the sum of total recovered and total dead, i.e. $$R(t_i) = C_R(t_i) + C_D(t_i).$$ 
Thus, $R(t)$ is an increasing function as proven in Theorem \ref{theo:pandbehav} and shown in the simulation plots. Then, the fractions  of removed people is ${r(t)} = {R(t)}/N$.
\subsubsection*{Deducing S(t)}\label{sec:S}
One cannot directly measure the number of susceptible people. So, to find this value, we  use the assumption from our basic SIR model: \\ At any instant of time,
$$S(t)+I(t)+R(t)=N(t)$$
In our case, $N(t):=N$ is constant as discussed in section \ref{sec:N}. Thus, $S(t) = N - I(t) - R(t)$, $s(t) = S(t)/N$, where
\begin{equation}\label{eq:s(t)}
    s(t)=1-i(t)-r(t).
\end{equation}

Once the values $s(t), i(t)$ and $r(t)$ are extracted, we create the vectors ${I}_{ex}$, ${R_{ex}}$ and  $S_{ex}$ 
 of the daily infected fraction ${i}(t)$,  the total cumulative removed fraction $r(t)$, and susceptible fraction $s(t)$ for the chosen period of time.
\subsection{Extraction of Parameters }\label{sec:Param}
Given the extracted $s(t), i(t)$ and $r(t)$ as discussed in section \ref{sec:compart}, we consider two methods for approximating  the parameters $\beta(t)$, and $\rho(t)$. 
The first relies exclusively on the data as detailed in section \ref{sec:ParamM1}.
The second uses the SIR model along with the data to approximate $\beta(t)$, and $\rho(t)$ as detailed in section \ref{sec:ParamM2}.

Finally, $\sigma(t) = {\beta(t)}/{\rho(t)}$ and $\sigma_s(t) = s(t)\cdot \sigma(t)$ can be estimated. 
\subsubsection{Estimating the parameters $\boldsymbol{ \beta(t)}$, $\boldsymbol{\rho(t)}$ using the Real Data - Method 1}\label{sec:ParamM1}
The infective and removal forces $F_I(t), F_R(t)$ are proportional to $I(t)$.\\
Thus, given the raw data provided by woldomneter, we can extract $\beta_s(t)$ and $\rho(t)$ which are independent of any adopted model as follows:
\begin{equation}
\beta_s(t) = \dfrac{C_I'(t)}{I(t)}, \qquad  
\rho(t) = \dfrac{R'(t)}{I(t)}  \end{equation}
To compute $\beta_s(t_i)$ and $\rho(t_i)$, we use the following backward difference approximations:
\begin{eqnarray}
\beta_s(t_i)&\approx& \dfrac{C_I(t_i)-C_I(t_{i-1})}{I(t_i)(t_i-t_{i-1})}, \label{eq:betaApp}\\
\rho(t_i)&\approx& \dfrac{R(t_i)-R(t_{i-1})}{I(t_i)(t_i-t_{i-1})}.\label{eq:rhoApp}
\end{eqnarray}
If we use the SIR model, then  $\beta(t) = \dfrac{\beta_s(t)}{s(t)} = \dfrac{C_I'(t)}{I(t)s(t)}  $.\\
\subsubsection{Approximating the parameters $\boldsymbol{ \beta(t)}$, $\boldsymbol{\rho(t)}$ - Method 2}\label{sec:ParamM2}
 We approximate $\beta(t)$ and $\rho(t)$, using the data and SIR model by simple algebra. \\Using the following equations in \eqref{sir}: 
\begin{eqnarray}  \label{parameterequations}
s'(t) &=& - \beta(t) s(t)i(t)\nonumber \\
r'(t) &=&  \rho(t) i(t) \nonumber
\end{eqnarray}
we get 
\begin{equation}
    \beta(t) = - \frac{s'(t)}{s(t)i(t)}\qquad and \qquad 
 \rho(t)  =  \frac{r'(t)}{i(t)}.
\end{equation}

Hence, we need the values of $s'(t)$ and $r'(t)$. To find $s'
(t)$ and $r'(t)$,  we use Finite Differences  (forward, backward, and central differences). It is worth noting that the first derivative (whether $s'(0), \; i'(0), \; r'(0)$) can only be calculated using forward difference, since there are no values that precede. Similarly, the last derivative ($s'(k), \; r'(k), \; i'(k)$ where k is the last point/day in the collected data) can only be obtained using backward difference. For the derivatives in between, we use the most accurate among the three which is the central difference.

After doing so, we find the parameters $\beta(t)$ and  $\rho(t)$ using \eqref{parameterequations}. 

\subsection{Validation of Obtained Results}\label{sec:valid}
Given the daily Worldometer Data described in Table \ref{tab:data}, 
 we have the values of $I(t_n)$ and $R(t_n)$ for $t_n \in [t_0, T]$ with an interval of 1 day, i.e. $t_{n+1}-t_n = 1$.  Then, we can extract $i(t_n), r(t_n),s(t_n),$ and approximate $\beta(t_n)$ and $\rho(t_n)$ for the given/chosen data set as described in sections \ref{sec:compart} and \ref{sec:Param} . Thus, we can get the reproduction factor $\sigma(t_n)$ and the replacement number $\sigma_s(t_n)$ which can be used to analyse the situation.

 But before that, we need to validate the obtained results. We do so by solving the Direct SIR Initial Value Problem  \eqref{sir} using the obtained $\beta(t_n)$ and $\rho(t_n)$ values using Method1 and Method2, along with $s(t_0),i(t_0),r(t_0)$. We solve IVP \eqref{sir} over the considered time interval $[t_0, T]$ as discussed in the next paragraph, to obtain the approximate computed solutions $s_c(t_n),i_c(t_n),r_c(t_n)$ for $n\geq 1$ and $T = t_0 + k$. Then, we evaluate the $L_2$ and $L_\infty$ relative error norms as shown in Table \ref{tableRef} where $S_{ex}, I_{ex}, R_{ex}$ are the vectors containing the extracted values from the Worldometer data, whereas $S_{c}, I_{c}, R_{c}$ are the vectors of the approximate computed Direct problem solutions $s_c(t_n),i_c(t_n),r_c(t_n)$.
 \begin{table}[H] 
\centering
{\renewcommand{\arraystretch}{1.1}
\begin{tabular}{||c |c c c||} 
 \hline 
 &&& \vspace{-3mm}\\
 Norm & S & I & R \\ [0.5ex] 
 \hline\hline
 &&& \vspace{-3mm}\\
 $L_2$ & $\dfrac{||S_{ex} - S_{c}||_2}{||S_{ex}||_2}$&$\dfrac{||I_{ex} - I_{c}||_2}{||I_{ex}||_2}$ &  $\dfrac{||R_{ex} - R_{c}||_2}{||R_{ex}||_2}$ \\ &&& \vspace{-3mm}\\
 \hline
 &&& \vspace{-3mm}\\
 $L_\infty$ & $\dfrac{||S_{ex} - S_{c}||_\infty}{||S_{ex}||_\infty}$&$\dfrac{||I_{ex} - I_{c}||_\infty}{||I_{ex}||_\infty}$ &  $\dfrac{||R_{ex} - R_{c}||_\infty}{||R_{ex}||_\infty}$ \\ 
 \hline
\end{tabular}}
\caption{The $L_2$ and $L_\infty$ relative errors of the computed S,I,R from the time-dependent model with respect to the S,I,R collected from data}
\label{tableRef}
\end{table}
\paragraph{Direct Problem Discretization}\label{sec:directpb}
It is possible to discretize the initial value problem $y'(t) = f(t,y(t))$ with $y(t_0) = y_0$ for $t\in [t_0,T]$
using Runge-Kutta schemes.
The most known and used RK-scheme is the $4^{th}$ order RK-scheme defined  for the time step $h = t_{n+1}-t_n$  as: \\
\begin{equation*}
    \begin{cases}
    \displaystyle
    k_1 = f(t_n, y_n)\\
    k_2 = f(t_n + \frac{h}{2}, y_n + \frac{k_1}{2})\\
    k_3 = f(t_n + \frac{h}{2}, y_n + \frac{k_2}{2})\\
    k_4 = f(t_n + h, y_n + k_3)\\
    y_{n+1} = y_n + \dfrac{h}{6}(k_1 + 2k_2 + 2k_3 + k_4) \\
    \end{cases}
\end{equation*}

The function handle $f(t,y(t))$, $y(t)$, $y_0$, and $h$ have to be defined. For the SIR problem \eqref{sir}, $y(t)$ and $y_0$ are vectors, and $f(t,y(t))$ is actually a vectorized function depending also on $\beta(t)$ and $\rho(t)$ defined by 
$$y(t) = \begin{bmatrix}
 s(t)\\ 
 i(t)\\
 r(t)   
\end{bmatrix}, \;\;\; y_0 = \begin{bmatrix}
    s(t_0)\\
    i(t_0)\\
    r(t_0)
\end{bmatrix}, \;\;\;  f(t,y(t), \beta(t), \rho(t)) = \begin{bmatrix}
    -\beta(t)\cdot s(t)\cdot i(t)\\ \beta(t)\cdot s(t) \cdot i(t) - \rho(t)\cdot i(t)\\ \rho(t)\cdot i(t) 
\end{bmatrix}.
$$

Given that $\beta(t_n)$ and $\rho(t_n)$ are approximated on a daily basis for $t_0 \leq t_n \leq T = t_k = t_0+k$ and $n=0,1,\cdots, k$, we set $h=2$ days. 
Note that it is possible to set $h=1$ day, but then evaluating $$f(t_n+h/2,y_n+k/2) := f \big{(}t_n+h/2,\,y_n+k/2,\, \beta(t_n+h/2),\, \rho(t_n+h/2)\big{)} $$
requires interpolating the data sets $\beta(t_n)$ and $\rho(t_n)$ to get the values $\beta(t_n + 0.5)$ and $\rho(t_n +0.5)$.

 Throughout the paper, we   use the fourth order Runge-Kutta scheme for our simulations. 

\subsection{Testings and Simulations}\label{sec:worldtest}
The data collected extends over two years, mainly from April 2020 till June 2022, and highlights two strong changes in the number of infected people.
The duration of these outbreaks differ between Italy, Germany, and the world, and are summarized in Table \ref{tab:outbreaks}. 
 
\begin{table}[H]
    \centering
    \setlength{\tabcolsep}{10pt}
{\renewcommand{\arraystretch}{1.4}
    \begin{tabular}{
    |c|c|c|c|}\cline{2-4}
     \multicolumn{1}{c|}{}   &  First Outbreak & Second Outbreak& Other Periods\\ \hline
  Italy   &   20/09/2020 - 21/12/2020  &  26/10/2021 - 25/02/2022 & 03/02/2021 - 03/07/2021\\ \hline
  Germany  &   6/10/2020 - 21/02/2021    & 02/01/2022 - 22/05/2022 & 09/03/2021 - 17/04/2021\\ \hline
  World  & 04/04/2020 - 21/06/2021& 22/06/2021 - 24/06/2022& 24/10/2021 - 04/02/2022\\
  \hline
    \end{tabular}\vspace{-3mm}
    \caption{The considered periods for testing}
    \label{tab:outbreaks}}
\end{table}

In this section, we use Italy's, Germany's, and the world's data to approximate the parameters as discussed in section \ref{sec:compart} for the first outbreak (section \ref{sec:firstoutb}), the second outbreak (section \ref{sec:secondoutb}), random periods (section \ref{sec:random}), and the whole two years period (section \ref{sec:twoywar}) . The obtained parameters $\beta_s(t), \rho(t)$ using method 1 are then used in the SIR model to obtain the simulated values of $s(t), i(t), r(t)$ as discussed in section \ref{sec:directpb}. We compare the real data to the simulated results to validate the efficiency of the SIR model in fitting the real-life data and the accuracy of the computed parameters. Note that we have also tested the IR model where the parameters $\beta(t), \rho(t)$ are obtained using methods 1 or 2. However, we do not show the results in this report as they are identical to that of the SIR model. 

\subsubsection*{First Outbreak}\label{sec:firstoutb}
\begin{figure}[H]
\begin{subfigure}{0.5\textwidth}
\centering
\includegraphics[scale=0.28]{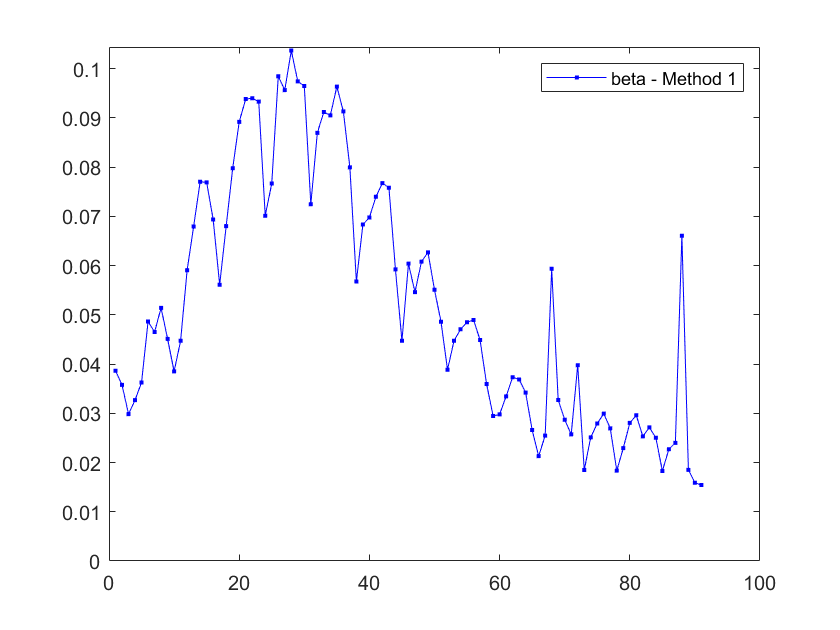} \vspace{-2mm}
\caption{Infection Rate $\beta(t)$ - Method 1}
\label{fig:beta1stoutbreakitalyM1}
\end{subfigure}
\hfill
\begin{subfigure}{0.5\textwidth}
\centering
\includegraphics[scale=0.28]{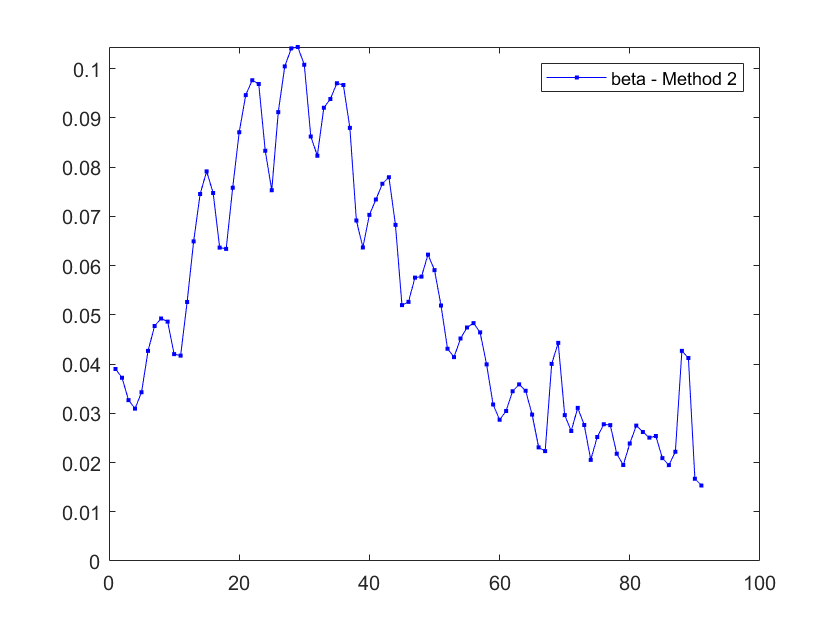} \vspace{-2mm}
\caption{Infection Rate $\beta(t)$ - Method 2}
\label{fig:beta1stoutbreakitalyM2}
\end{subfigure}
\newline
\begin{subfigure}{0.5\textwidth}
\centering
\includegraphics[scale=0.28]{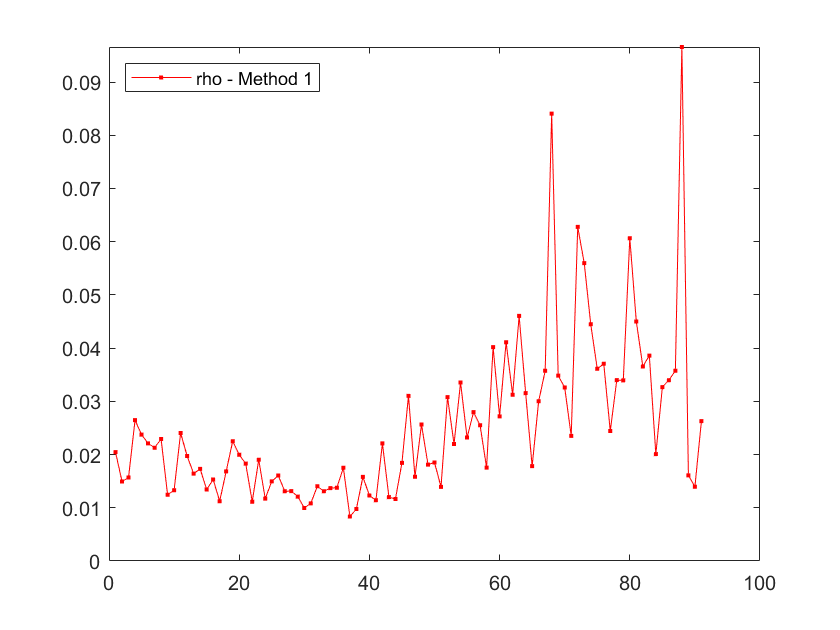} \vspace{-2mm}
\caption{Removal Rate $\rho(t)$ - Method 1 }
\label{fig:rho1stoutbreakitalyM1}
\end{subfigure}
\hfill
\begin{subfigure}{0.5\textwidth}
\centering
\includegraphics[scale=0.28]{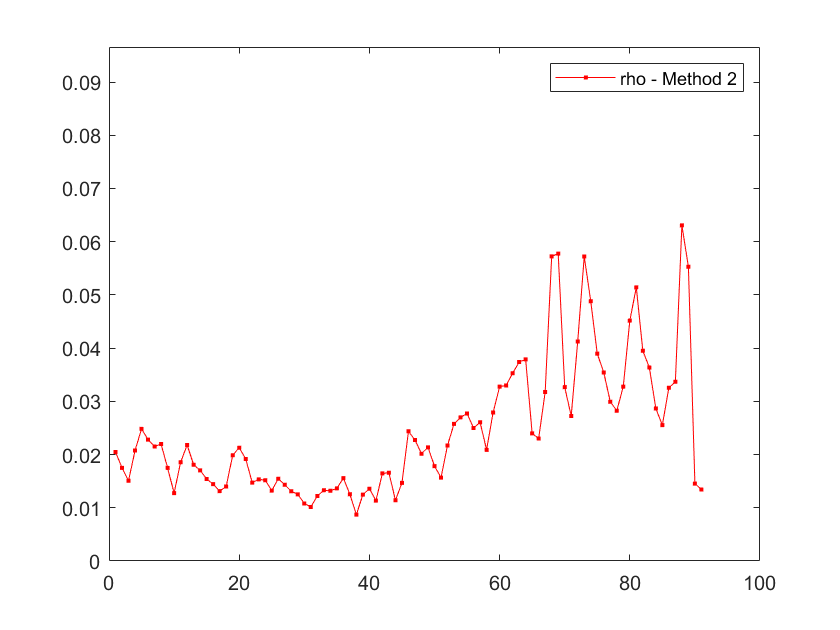} \vspace{-2mm}
\caption{Removal Rate $\rho(t)$ - Method 2 }
\label{fig:rho1stoutbreakitalyM2}
\end{subfigure}
\newline
\begin{subfigure}{0.5\textwidth}
\centering
\includegraphics[scale=0.28]{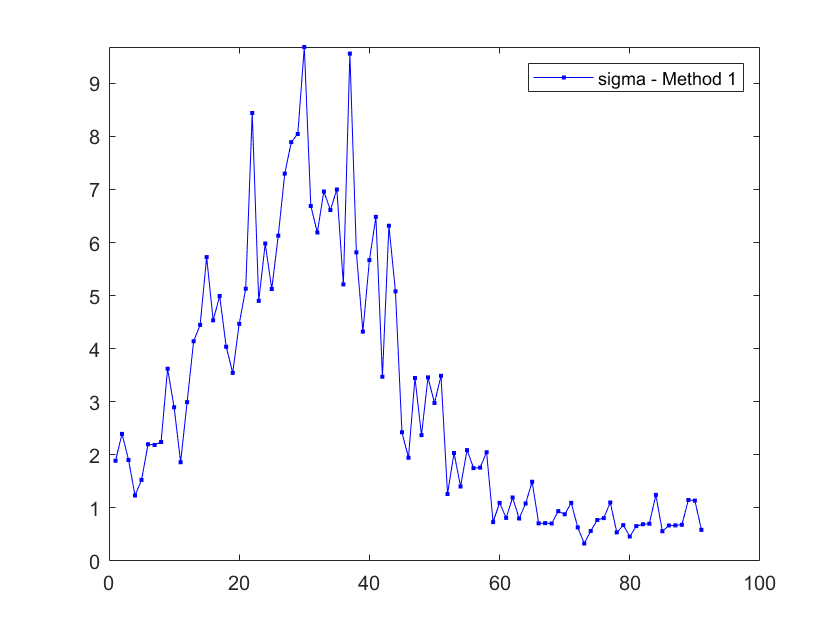} \vspace{-2mm}
\caption{Reproduction Factor $\sigma(t)$ - Method 1}
\label{fig:sigma1stoutbreakitalyM1}
\end{subfigure}
\hfill
\begin{subfigure}{0.5\textwidth}
\centering
\includegraphics[scale=0.28]{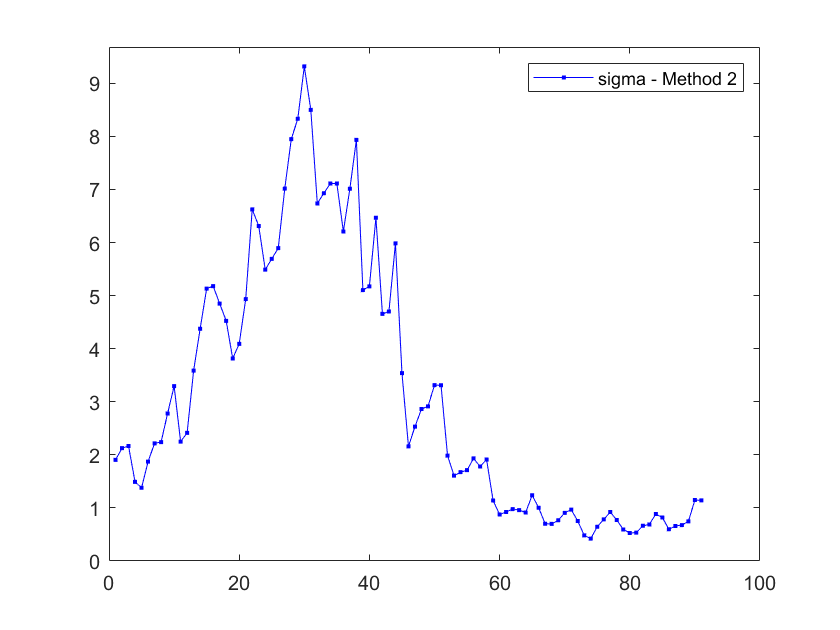} \vspace{-2mm}
\caption{Reproduction Factor $\sigma(t)$ - Method 2}
\label{fig:sigma1stoutbreakitalyM2}
\end{subfigure}
\newline
\begin{subfigure}{0.5\textwidth}
\centering
\includegraphics[scale=0.28]{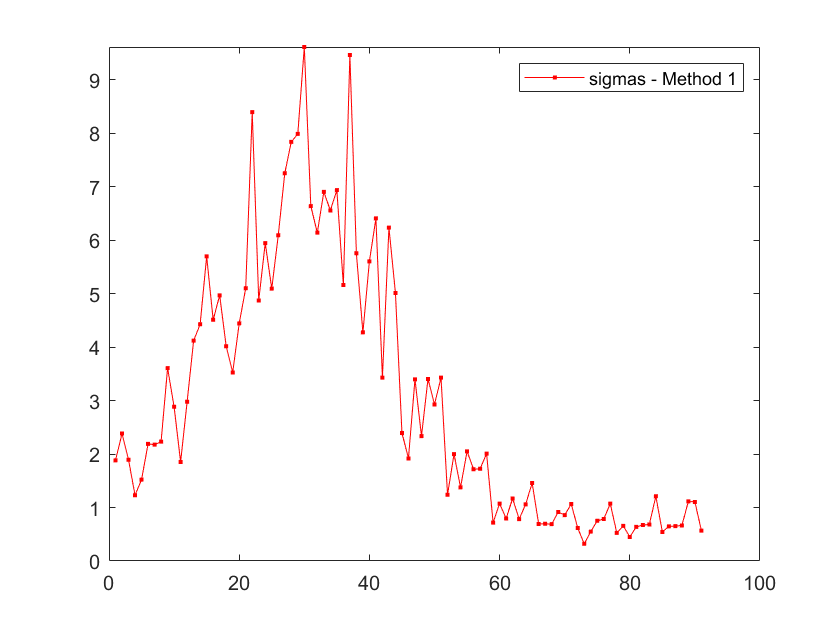}\vspace{-2mm} 
\caption{Replacement Number $\sigma_s(t)$ - Method 1}
\label{fig:sigmas1stoutbreakitalyM1}
\end{subfigure}\hfill
\begin{subfigure}{0.5\textwidth}
\centering
\includegraphics[scale=0.28]{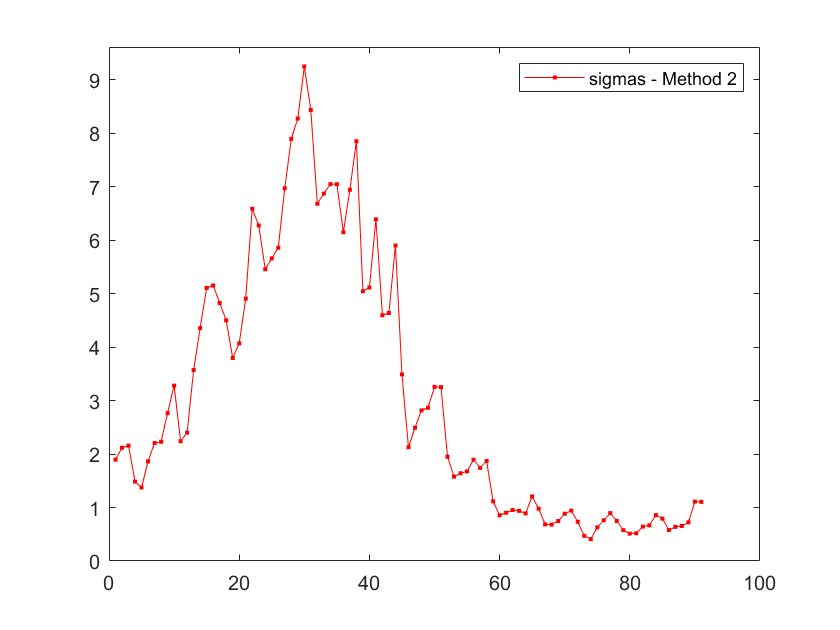}\vspace{-2mm} 
\caption{Replacement Number $\sigma_s(t)$ - Method 2}
\label{fig:sigmas1stoutbreakitalyM2}
\end{subfigure}
\caption{Parameters of SIR Model during First Outbreak in Italy }
\label{fig:parameters1stoutbreakitaly}\vspace{0mm}
\end{figure}

\begin{figure}[H]
\begin{subfigure}{0.5\textwidth}
\centering
\includegraphics[scale=0.28]{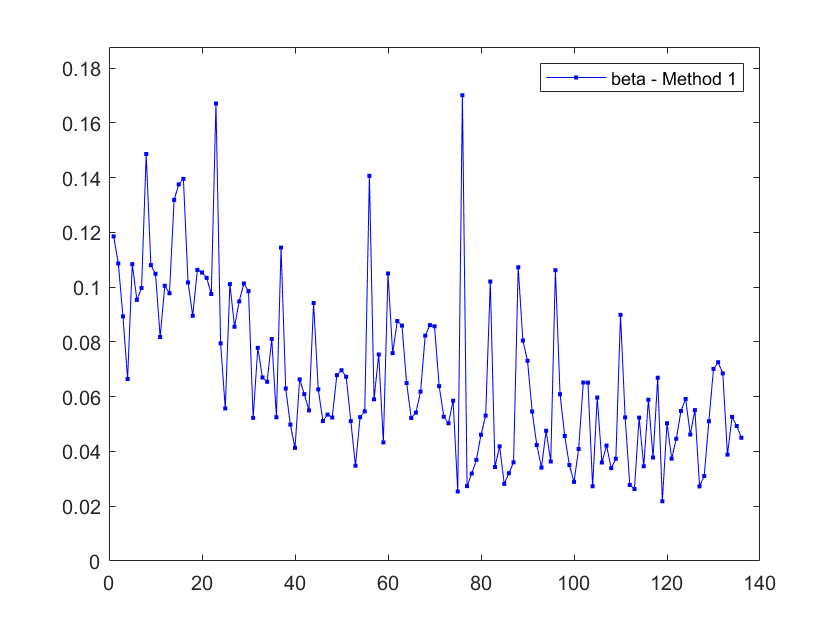} \vspace{-2mm}
\caption{Infection Rate $\beta(t)$ - Method 1}
\label{fig:beta1stoutbreakgermanyM1}
\end{subfigure}
\hfill
\begin{subfigure}{0.5\textwidth}
\centering
\includegraphics[scale=0.28]{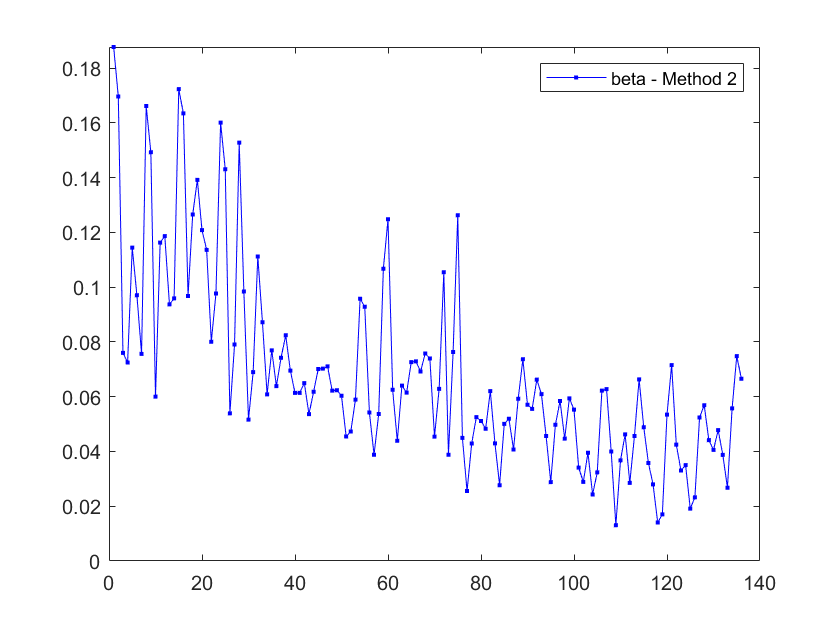} \vspace{-2mm}
\caption{Infection Rate $\beta(t)$ - Method 2}
\label{fig:beta1stoutbreakgermanyM2}
\end{subfigure}
\newline
\begin{subfigure}{0.5\textwidth}
\centering
\includegraphics[scale=0.28]{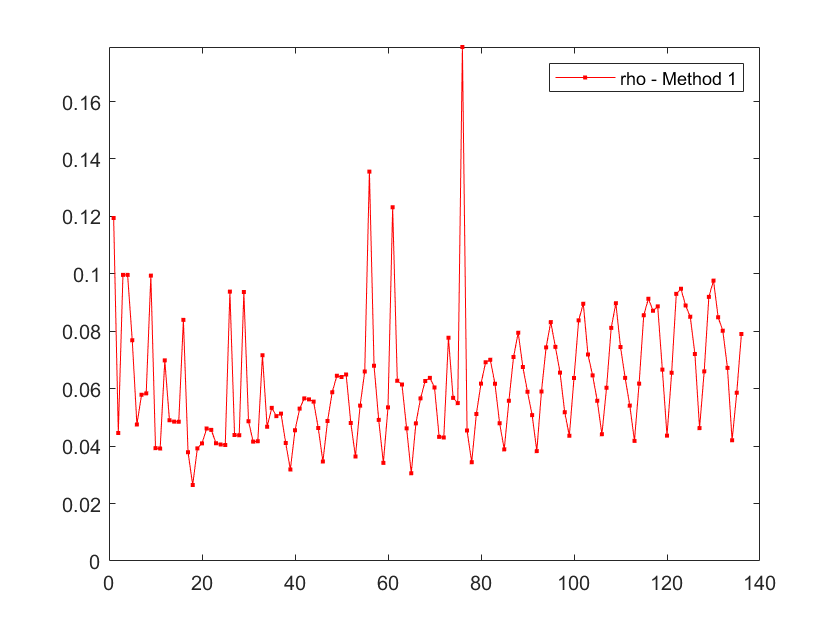} \vspace{-2mm}
\caption{Removal Rate $\rho(t)$ - Method 1 }
\label{fig:rho1stoutbreakgermanyM1}
\end{subfigure}
\hfill
\begin{subfigure}{0.5\textwidth}
\centering
\includegraphics[scale=0.28]{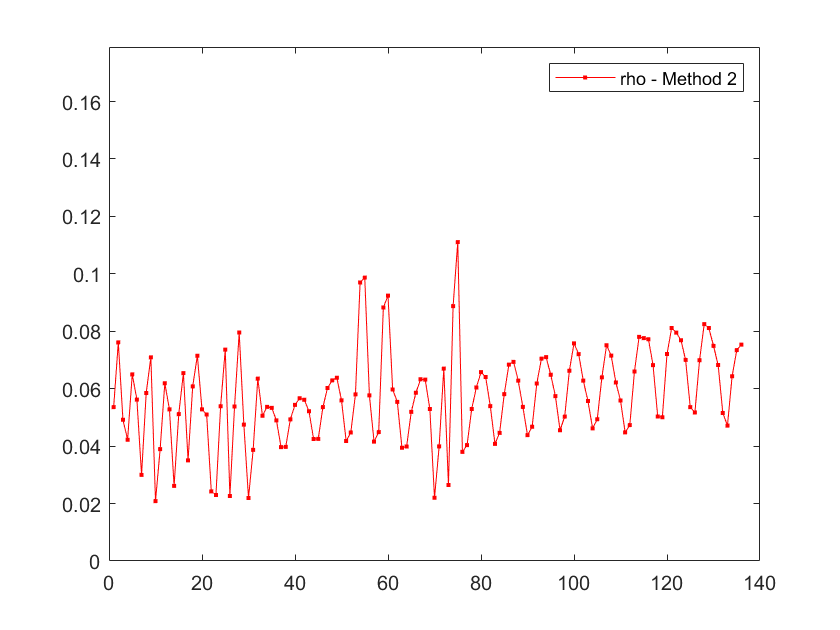} \vspace{-2mm}
\caption{Removal Rate $\rho(t)$ - Method 2 }
\label{fig:rho1stoutbreakgermanyM2}
\end{subfigure}
\newline
\begin{subfigure}{0.5\textwidth}
\centering
\includegraphics[scale=0.28]{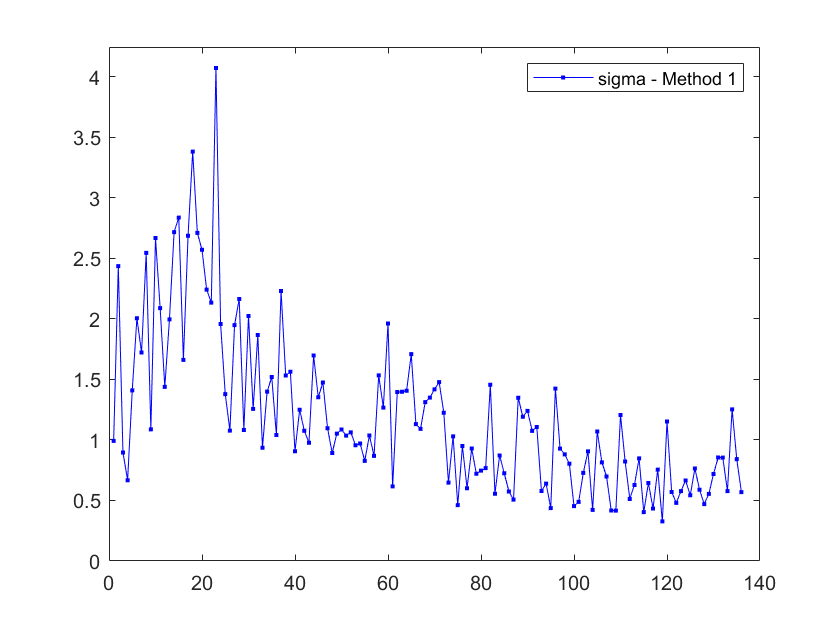} \vspace{-2mm}
\caption{Reproduction Factor $\sigma(t)$ - Method 1}
\label{fig:sigma1stoutbreakgermanyM1}
\end{subfigure}
\hfill
\begin{subfigure}{0.5\textwidth}
\centering
\includegraphics[scale=0.28]{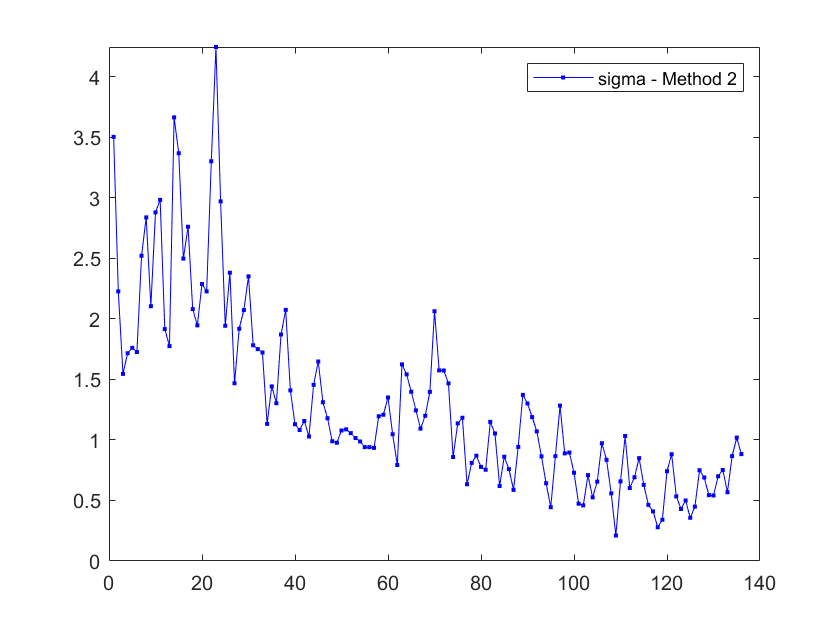} \vspace{-2mm}
\caption{Reproduction Factor $\sigma(t)$ - Method 2}
\label{fig:sigma1stoutbreakgermanyM2}
\end{subfigure}
\newline
\begin{subfigure}{0.5\textwidth}
\centering
\includegraphics[scale=0.28]{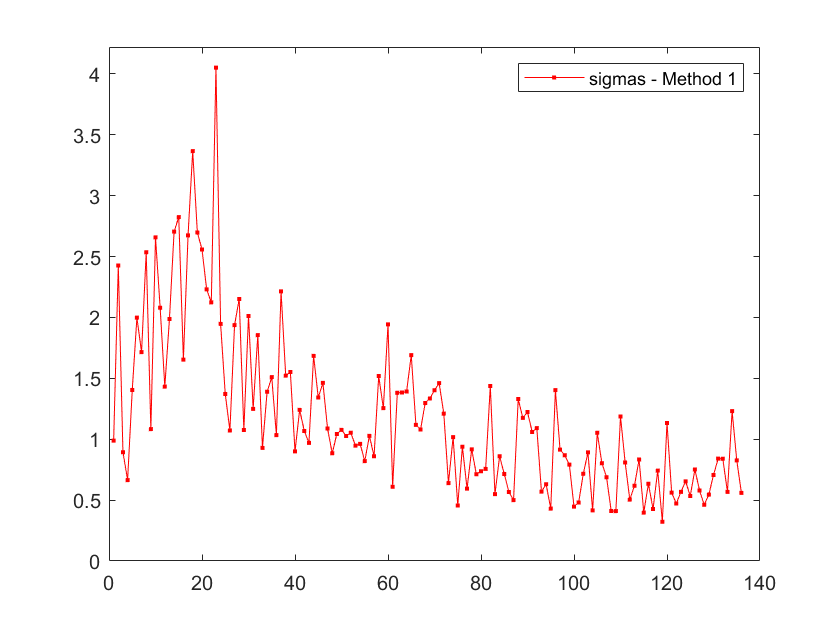}\vspace{-2mm} 
\caption{Replacement Number $\sigma_s(t)$ - Method 1}
\label{fig:sigmas1stoutbreakgermanyM1}
\end{subfigure}\hfill
\begin{subfigure}{0.5\textwidth}
\centering
\includegraphics[scale=0.28]{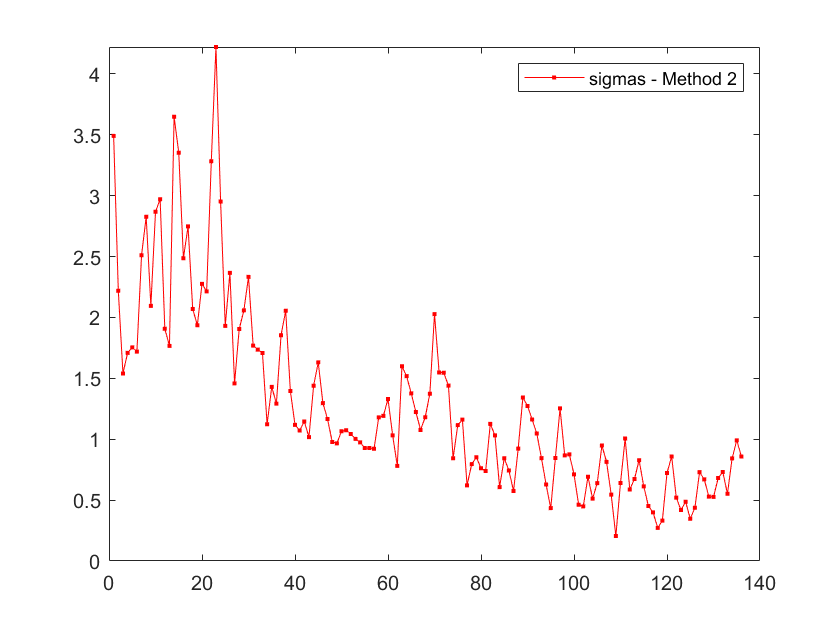}\vspace{-2mm} 
\caption{Replacement Number $\sigma_s(t)$ - Method 2}
\label{fig:sigmas1stoutbreakgermanyM2}
\end{subfigure}
\caption{Parameters of SIR Model during First Outbreak in Germany }
\label{fig:parameters1stoutbreakgermany}\vspace{0mm}
\end{figure}

\begin{figure}[H]
\begin{subfigure}{0.5\textwidth}
\centering
\includegraphics[scale=0.28]{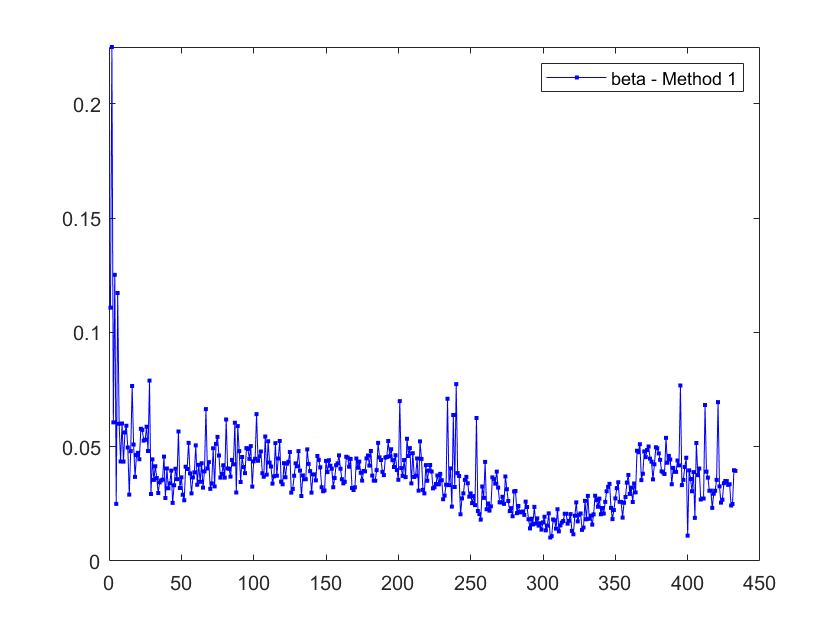} \vspace{-2mm}
\caption{Infection Rate $\beta(t)$ - Method 1}
\label{fig:beta1stoutbreakworldM1}
\end{subfigure}
\hfill
\begin{subfigure}{0.5\textwidth}
\centering
\includegraphics[scale=0.28]{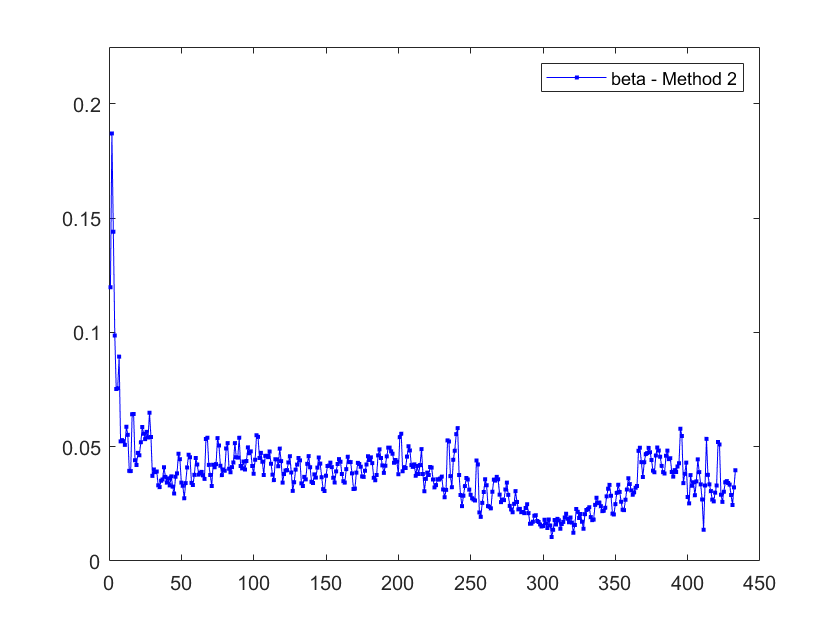} \vspace{-2mm}
\caption{Infection Rate $\beta(t)$ - Method 2}
\label{fig:beta1stoutbreakworldM2}
\end{subfigure}
\newline
\begin{subfigure}{0.5\textwidth}
\centering
\includegraphics[scale=0.28]{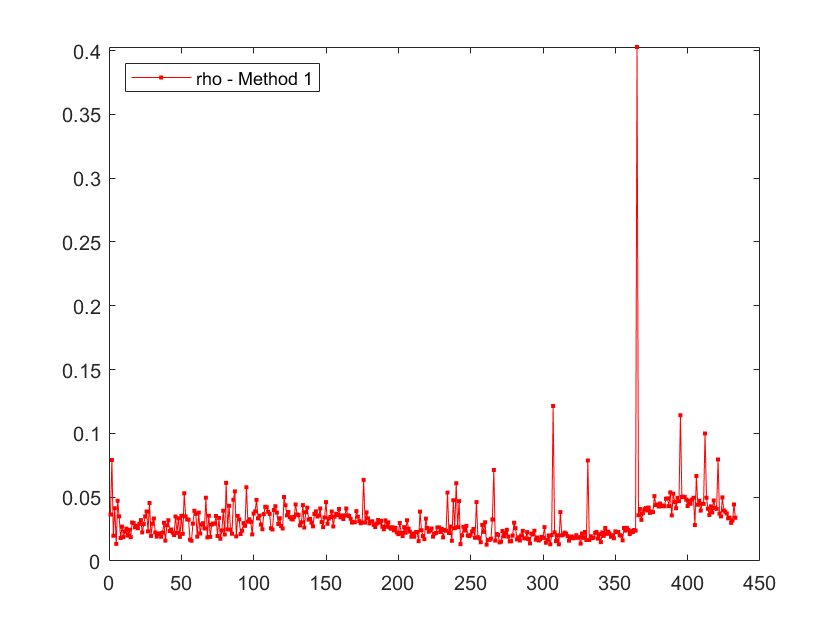} \vspace{-2mm}
\caption{Removal Rate $\rho(t)$ - Method 1 }
\label{fig:rho1stoutbreakworldM1}
\end{subfigure}
\hfill
\begin{subfigure}{0.5\textwidth}
\centering
\includegraphics[scale=0.28]{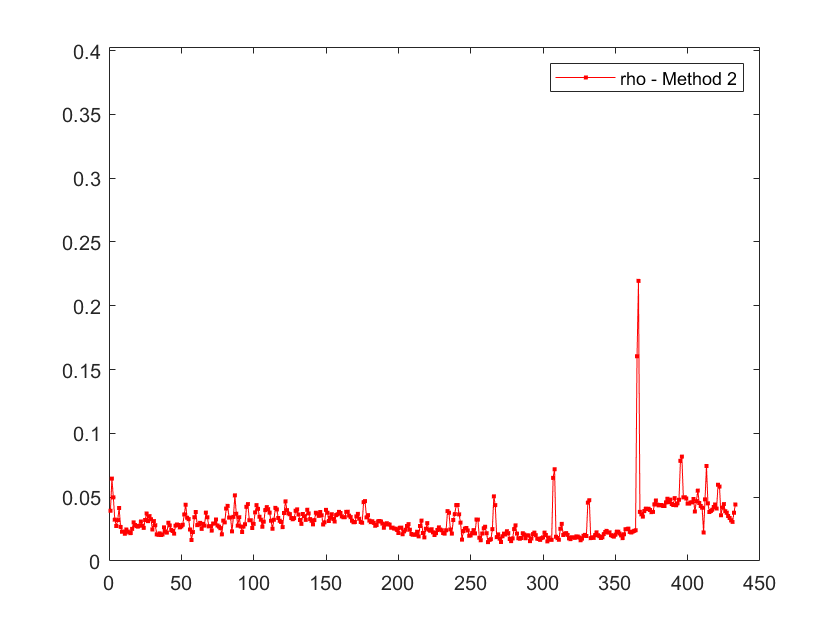} \vspace{-2mm}
\caption{Removal Rate $\rho(t)$ - Method 2 }
\label{fig:rho1stoutbreakworldM2}
\end{subfigure}
\newline
\begin{subfigure}{0.5\textwidth}
\centering
\includegraphics[scale=0.28]{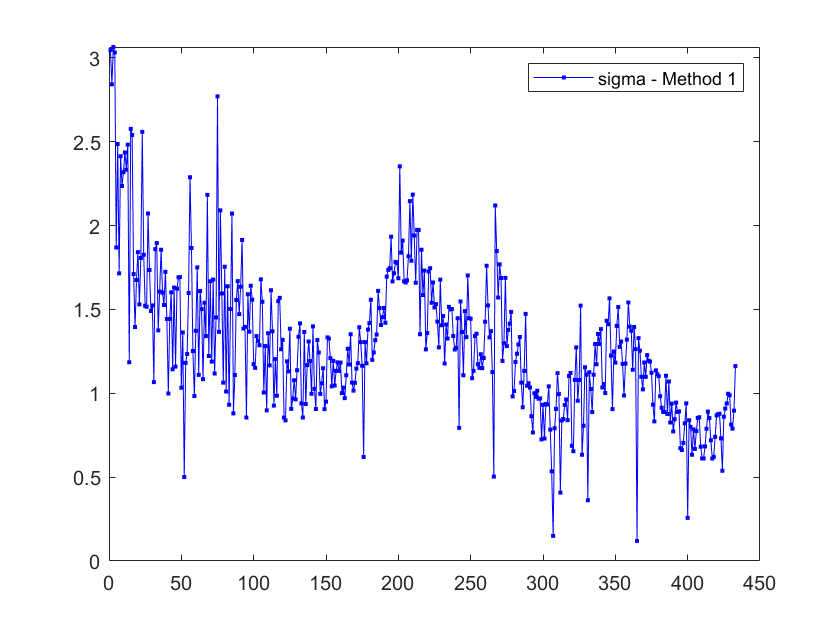} \vspace{-2mm}
\caption{Reproduction Factor $\sigma(t)$ - Method 1}
\label{fig:sigma1stoutbreakworldM1}
\end{subfigure}
\hfill
\begin{subfigure}{0.5\textwidth}
\centering
\includegraphics[scale=0.28]{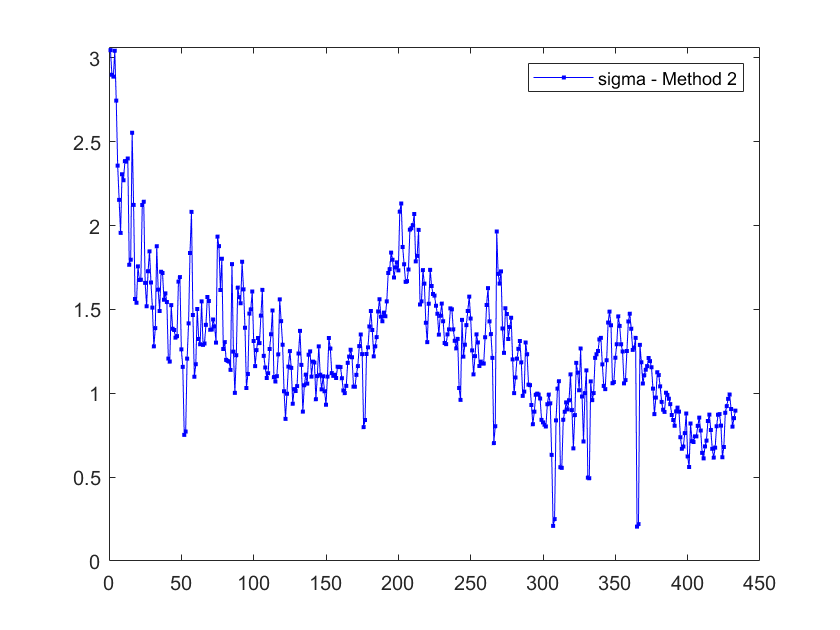} \vspace{-2mm}
\caption{Reproduction Factor $\sigma(t)$ - Method 2}
\label{fig:sigma1stoutbreakworldM2}
\end{subfigure}
\newline
\begin{subfigure}{0.5\textwidth}
\centering
\includegraphics[scale=0.28]{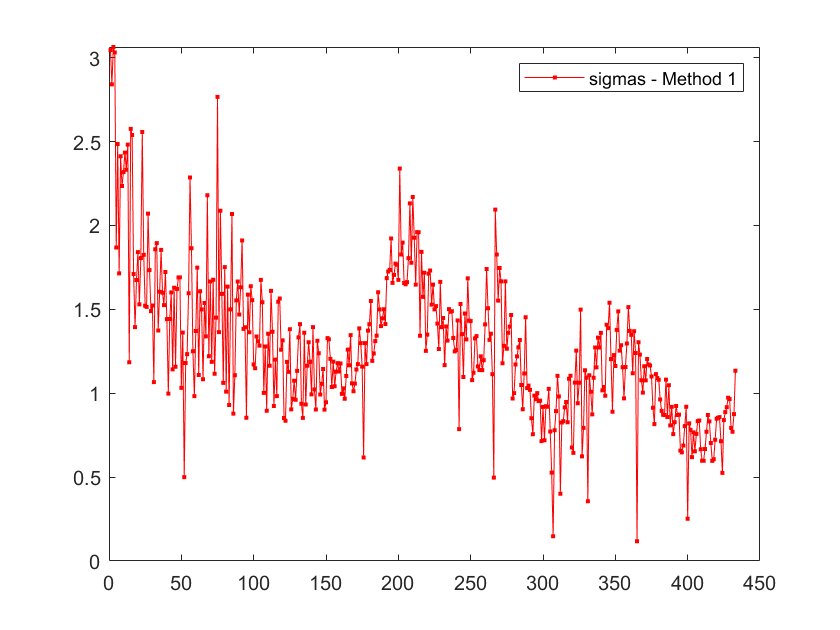}\vspace{-2mm} 
\caption{Replacement Number $\sigma_s(t)$ - Method 1}
\label{fig:sigmas1stoutbreakworldM1}
\end{subfigure}\hfill
\begin{subfigure}{0.5\textwidth}
\centering
\includegraphics[scale=0.28]{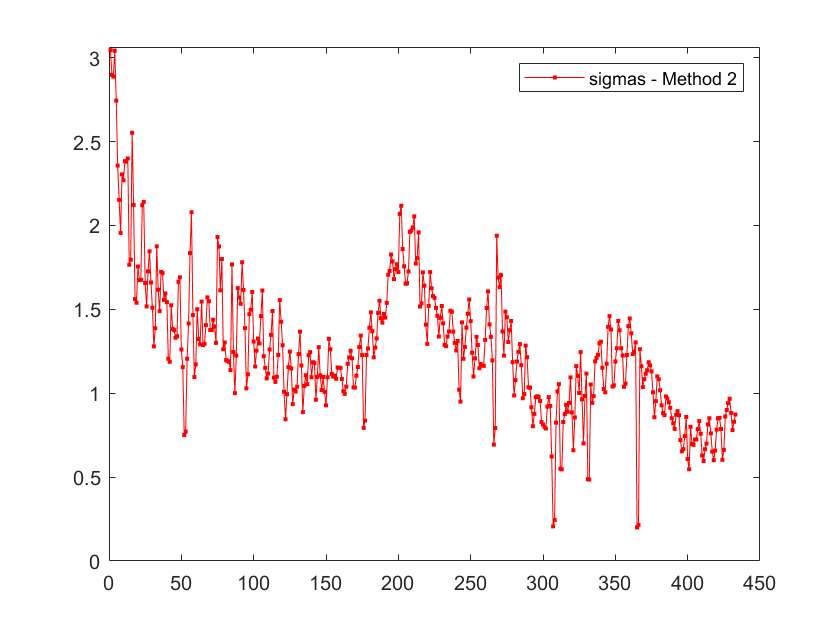}\vspace{-2mm} 
\caption{Replacement Number $\sigma_s(t)$ - Method 2}
\label{fig:sigmas1stoutbreakworldM2}
\end{subfigure}
\caption{Parameters of SIR Model during First Outbreak in the World }
\label{fig:parameters1stoutbreakworld}\vspace{0mm}
\end{figure}

Figures \ref{fig:parameters1stoutbreakitaly},  \ref{fig:parameters1stoutbreakgermany}, and \ref{fig:parameters1stoutbreakworld} plot the approximated parameters $\beta(t), \rho(t), \sigma(t)$, and $\sigma_s(t)$ using Method 1 (section \ref{sec:ParamM1}) and Method 2 (section \ref{sec:ParamM2}) during the first outbreak in Italy, Germany, and the world respectively. Clearly the results of both methods are similar in global behavior, with some different variations. These differences are detailed in  
 Tables \ref{table:statbetaFirst}, \ref{table:statrhoFirst} \ref{table:statsigmaFirst}, and \ref{table:statsigmasFirst} that summarize the statistical properties (mean, median, standard deviation) of the approximated parameters $\beta(t)$, $\rho(t)$, $\sigma(t)$, and  $\sigma_s(t)$ respectively. Note that the mean of the 4 parameters is larger than or equal to the median.

Moreover, Table \ref{table:table1strelparam} computes the relative L2 and Linfinity errors between the parameters computed using both methods. These errors are of order $10^{-1}$ for all parameters and all three countries. As for the absolute errors of the parameters at some time $t_i$, then by comparing the means, it is clear that it is of order $10^{-3}$ for $\beta$ and $\rho$ (except for germany, $10^{-2}$), and $10^{-1}$ for $\sigma$ and $\sigma_s$ (except for the world, $10^{-2}$).

\begin{table}[H]
\centering
\setlength{\tabcolsep}{10pt}
{\renewcommand{\arraystretch}{1.2}
\begin{tabular}{||c||c| c| c |c| c|c| c |c| c| c||} 
\cline{2-11}
  \multicolumn{1}{c||}{}& \multicolumn{10}{c||}{$\beta(t)$}\\
 \cline{2-11}
 \multicolumn{1}{c||}{}& \multicolumn{5}{c|}{Method 1} &  \multicolumn{5}{|c||}{Method 2}\\
 \cline{2-11}
  \multicolumn{1}{c||}{}& Me&Md&SD&Min&Max& Me&Md&SD &Min&Max\\
  \hline\hline
 Italy&0.05 &0.05 &0.03  &0.01 &0.10&0.05& 0.05& 0.03 &0.02&0.10 \\ 
\hline
  Germany&0.07&0.06&0.03&0.02&0.17& 0.07& 0.06&  0.04 &0.01&0.19  \\ \hline
  World &0.04&0.04&0.02&0.01&0.22& 0.04 &0.04 & 0.02&0.01&0.19 \\ 
\hline \hline
\end{tabular}\vspace{-3mm}
\caption{The Mean (Me), Median (Md) Standard Deviation (SD), Minimum (Min) and Maximum (Max) of the computed $\beta(t)$ using the two Methods during the first outbreak.}\label{table:statbetaFirst}}\vspace{-3mm}
\end{table}

\begin{table}[H]
\centering
\setlength{\tabcolsep}{10pt}
{\renewcommand{\arraystretch}{1.2}
\begin{tabular}{||c||c| c|c| c|c|c| c |c| c| c||} 
 \cline{2-11}
  \multicolumn{1}{c||}{}& \multicolumn{10}{c||}{$\rho(t)$}\\
 \cline{2-11}
 \multicolumn{1}{c||}{}& \multicolumn{5}{c|}{Method 1} &  \multicolumn{5}{|c||}{Method 2}\\
 \cline{2-11}
  \multicolumn{1}{c||}{}& Me&Md&SD&Min&Max& Me&Md&SD &Min&Max\\
  \hline\hline
 Italy&0.03&0.02&0.02&0.01&0.09 & 0.02&0.02& 0.01 &0.01& 0.06\\ 
\hline
  Germany&0.06&0.06&0.02&0.03&0.18& 0.06& 0.05& 0.02&0.02& 0.11
  \\ \hline
  World &0.03&0.03&0.02&0.01&0.40& 0.03 & 0.03 & 0.02&0.01&0.22 \\ 
\hline \hline
\end{tabular}\vspace{-3mm}
\caption{The Mean (Me), Median (Md), Standard Deviation (SD), Minimum (Min) and Maximum (Max) of the computed $\rho(t)$ using the two Methods during the first outbreak.}\label{table:statrhoFirst}}\vspace{-3mm}
\end{table}

\begin{table}[H]
\centering
\setlength{\tabcolsep}{10pt}
{\renewcommand{\arraystretch}{1.2}
\begin{tabular}{||c||c| c|c| c|c|c| c |c| c| c||} 
 \cline{2-11}
  \multicolumn{1}{c||}{}& \multicolumn{10}{c||}{$\sigma(t)$}\\
 \cline{2-11}
 \multicolumn{1}{c||}{}& \multicolumn{5}{c|}{Method 1} &  \multicolumn{5}{|c||}{Method 2}\\
 \cline{2-11}
  \multicolumn{1}{c||}{}& Me&Md&SD&Min&Max& Me&Md&SD &Min&Max\\
  \hline\hline
 Italy&3.04&2.09&2.45&0.30&9.68 & 3.13& 2.17 & 2.41 &0.42&9.32 \\ 
\hline
  Germany&1.19&1.05&0.67&0.33&4.07& 1.41& 1.09& 1.07
  &0.21&4.24\\ \hline
  World &1.29&1.24&0.44&0.12&3.06& 1.28 &1.23&  0.41&0.20&3.05 \\ 
\hline \hline
\end{tabular}\vspace{-3mm}
\caption{The Mean (Me), Median (Md), Standard Deviation (SD), Minimum (Min) and Maximum (Max) of the computed $\sigma(t)$ using the two Methods during the first outbreak.}\label{table:statsigmaFirst}}\vspace{-3mm}
\end{table}

\begin{table}[H]
\centering
\setlength{\tabcolsep}{10pt}
{\renewcommand{\arraystretch}{1.2}
\begin{tabular}{||c||c| c|c| c|c|c| c |c| c| c||} 
 \cline{2-11}
  \multicolumn{1}{c||}{}& \multicolumn{10}{c||}{$\sigma_s(t)$}\\
 \cline{2-11}
 \multicolumn{1}{c||}{}& \multicolumn{5}{c|}{Method 1} &  \multicolumn{5}{|c||}{Method 2}\\
 \cline{2-11}
  \multicolumn{1}{c||}{}& Me&Md&SD&Min&Max& Me&Md&SD &Min&Max\\
  \hline\hline
 Italy&3.00&2.05&2.44&0.32&9.61 & 3.10 &2.16 & 2.39&0.41&9.24 \\ 
\hline
  Germany&1.18&1.04&0.66&0.32&4.05& 1.28&
1.07& 0.77
  &0.21&4.22\\ \hline
  World &1.29&1.23&0.40&0.12&3.06& 1.26& 1.22&0.42&0.20& 3.05\\ 
\hline \hline
\end{tabular}\vspace{-3mm}
\caption{The Mean (Me), Median (Md), Standard Deviation (SD), Minimum (Min) and Maximum (Max) of the computed $\sigma_s(t)$ using the two Methods during the first outbreak.}\label{table:statsigmasFirst}}\vspace{-3mm}
\end{table}

 \begin{table}[H]
\centering
\setlength{\tabcolsep}{10pt}
{\renewcommand{\arraystretch}{1.4}
\begin{tabular}{||c| c| c| c |c||} 
 \hline
  &Norm & Italy  & Germany  & World  \\ 
 \hline\hline
\multirow{2}{*}{$\beta$} &$L_2$ & $  1.035* 10^{-1}$ &$ 4.199* 10^{-1}$ &  $  2.055 * 10^{-1}$  \\ 
 \cline{2-5}
  & $L_\infty$ & $2.247 * 10^{-1}$  & $ 6.993* 10^{-1}$ & $  4.049 * 10^{-1}$ \\ 
\hline
\multirow{2}{*}{$\rho$} &$L_2$ & $2.907* 10^{-1}$ &$ 4.267 * 10^{-1}$ &  $ 4.572 * 10^{-1}$ \\ 
\cline{2-5}
& $L_\infty$ & $ 4.910* 10^{-1}$ & $  9.722* 10^{-1}$ & $7.788 * 10^{-1}$ \\ 
\hline
 \multirow{2}{*}{$\sigma$} & $L_2$ & $1.615 * 10^{-1}$ &$3.222* 10^{-1}$ &  $  1.446 * 10^{-1}$ \\
\cline{2-5}
 & $L_\infty$ & $ 2.679* 10^{-1}$ & $6.036* 10^{-1}$ & $4.309 * 10^{-1}$ \\  
 \hline
\multirow{2}{*}{$\sigma_s$} &$L_2$ & $1.612* 10^{-1}$ &$3.224* 10^{-1}$ &  $  1.447 * 10^{-1}$ \\ 
 \cline{2-5}
  & $L_\infty$ & $2.673* 10^{-1}$  & $ 6.048* 10^{-1}$ & $  4.261 * 10^{-1}$ \\ 
 \hline
\end{tabular}
\caption{The $L_2$ and $L_\infty$ relative errors between the computed parameters based on Methods 1 and 2, for the first outbreak.}
\label{table:table1strelparam}}
\end{table}

\begin{figure}[H]
\begin{subfigure} {0.5\textwidth}
\centering
\includegraphics[scale=0.3]{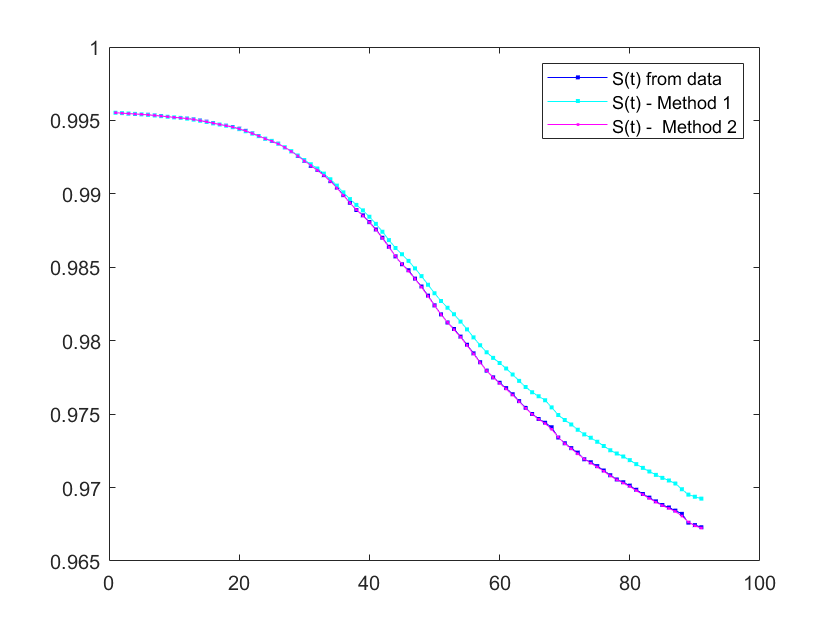}\vspace{-2mm}
\caption{Ratio of Susceptible s(t)}
\label{fig:SusceptiblefirstoutbreakItaly}
\end{subfigure}%
\begin{subfigure}{.5\linewidth}
\centering
\includegraphics[scale=.3]{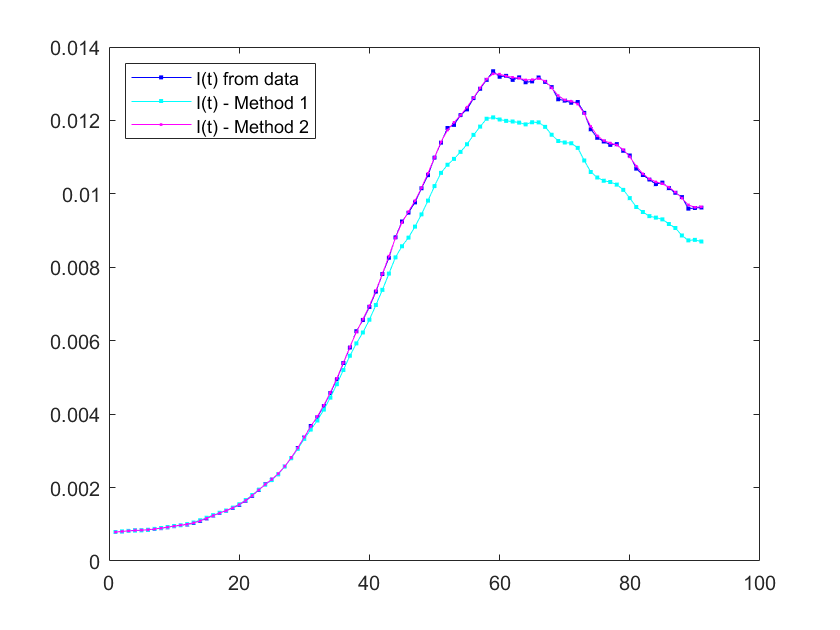}\vspace{-2mm}
\caption{Ratio of Infected i(t)}\label{fig:InfectedfirstoutbreakItaly}
\end{subfigure}
\begin{subfigure}{1.0\linewidth}
\centering
\includegraphics[scale=.3]{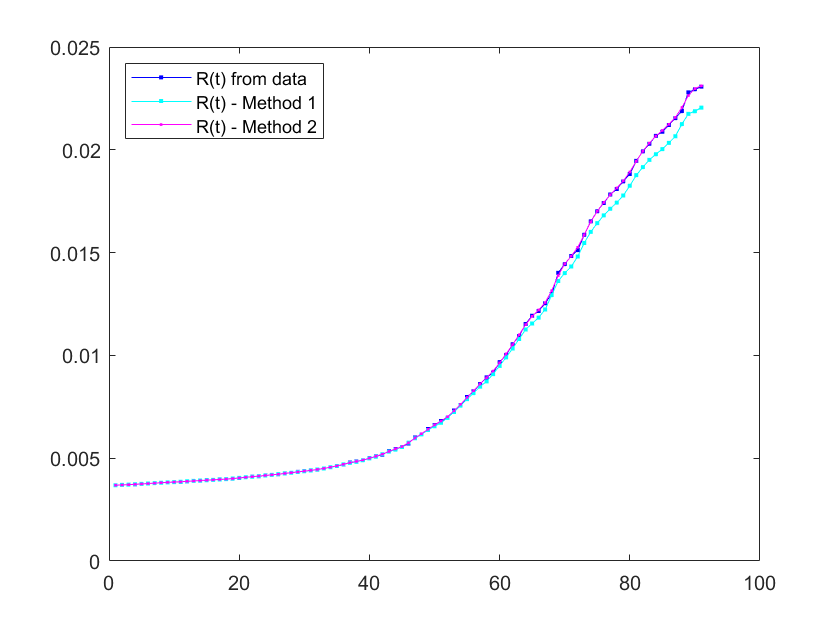}\vspace{-2mm}
\caption{Ratio of Removed r(t)}\label{fig:RemovedfirstoutbreakItaly}
\end{subfigure}

\caption{Comparison of Compartments' ratios from real data to those obtained using SIR model with approximated parameters, during the first Outbreak in Italy }\vspace{-0.5cm}
\label{fig:SIRfirstoutbreakitaly}
\end{figure}

\begin{figure}[H]
\begin{subfigure} {0.5\textwidth}
\centering
\includegraphics[scale=0.3]{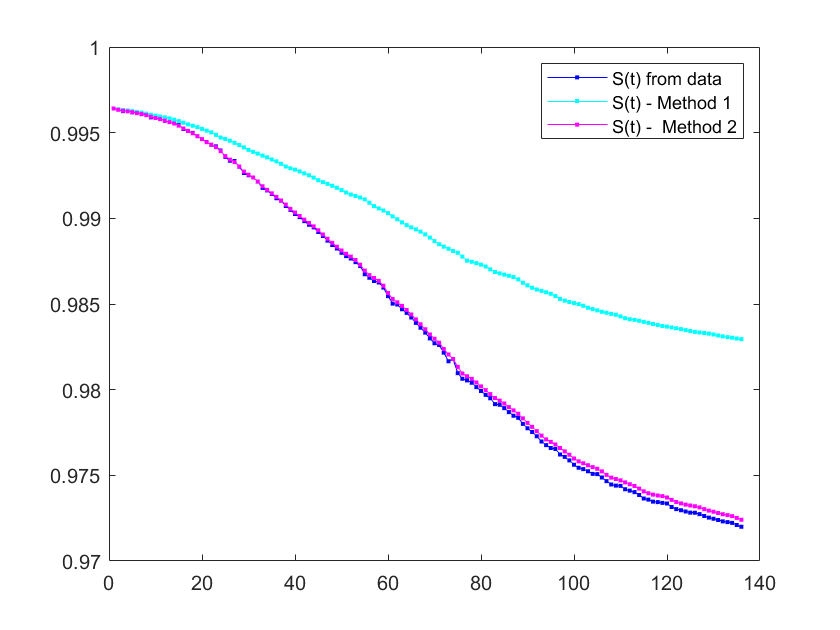}\vspace{-2mm}
\caption{Ratio of Susceptible s(t)}
\label{fig:Susceptiblefirstoutbreakgermany}
\end{subfigure}%
\begin{subfigure}{.5\linewidth}
\centering
\includegraphics[scale=.3]{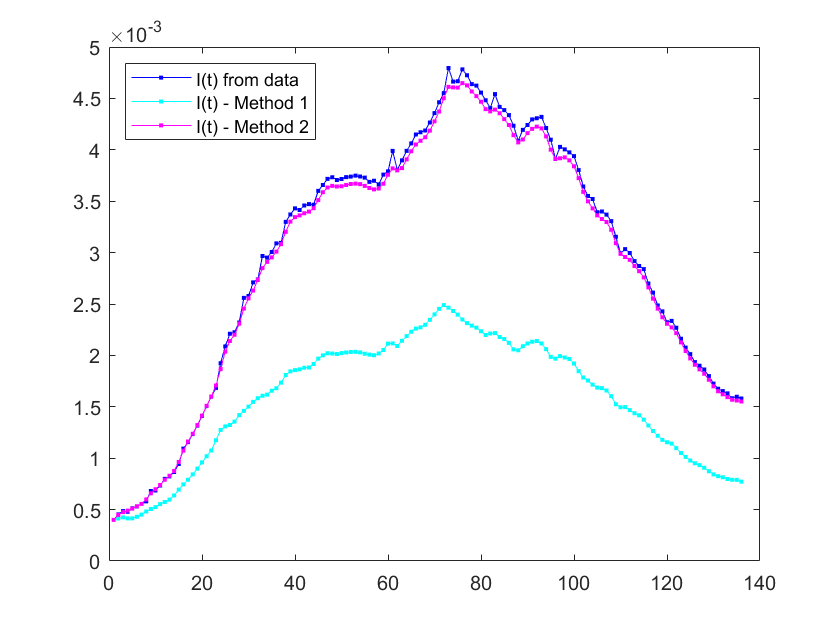}\vspace{-2mm}
\caption{Ratio of Infected i(t)}\label{fig:Infectedfirstoutbreakgermany}
\end{subfigure}
\begin{subfigure}{1.0\linewidth}
\centering
\includegraphics[scale=.3]{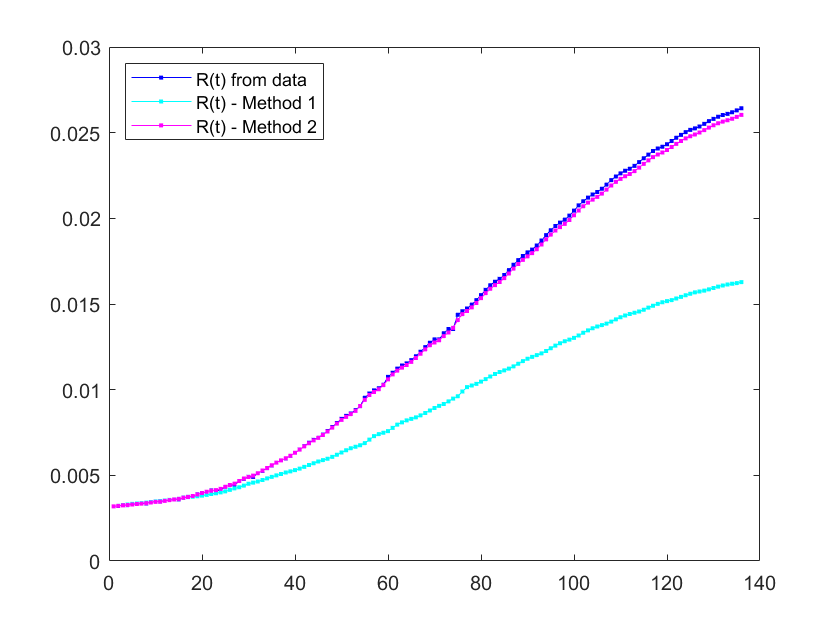}\vspace{-2mm}
\caption{Ratio of Removed r(t)}\label{fig:Removedfirstoutbreakgermany}
\end{subfigure}

\caption{Comparison of Compartments' ratios from real data to those obtained using SIR model with approximated parameters, during the first Outbreak in Germany}
\label{fig:SIRfirstoutbreakgermany}\vspace{-1cm}
\end{figure}

\begin{figure}[H]
\begin{subfigure} {0.5\textwidth}
\centering
\includegraphics[scale=0.3]{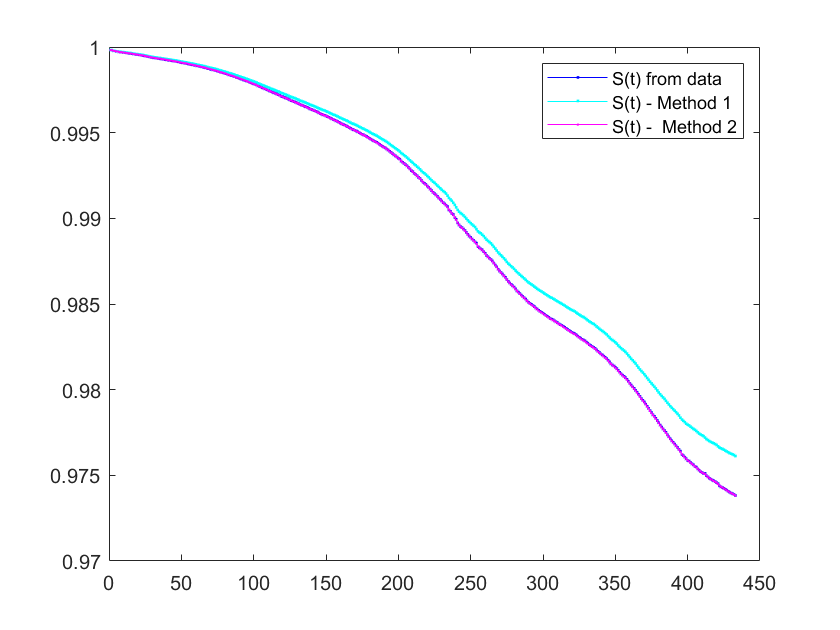}\vspace{-2mm}
\caption{Ratio of Susceptible s(t)}
\label{fig:Susceptiblefirstoutbreakworld}
\end{subfigure}%
\begin{subfigure}{.5\linewidth}
\centering
\includegraphics[scale=.3]{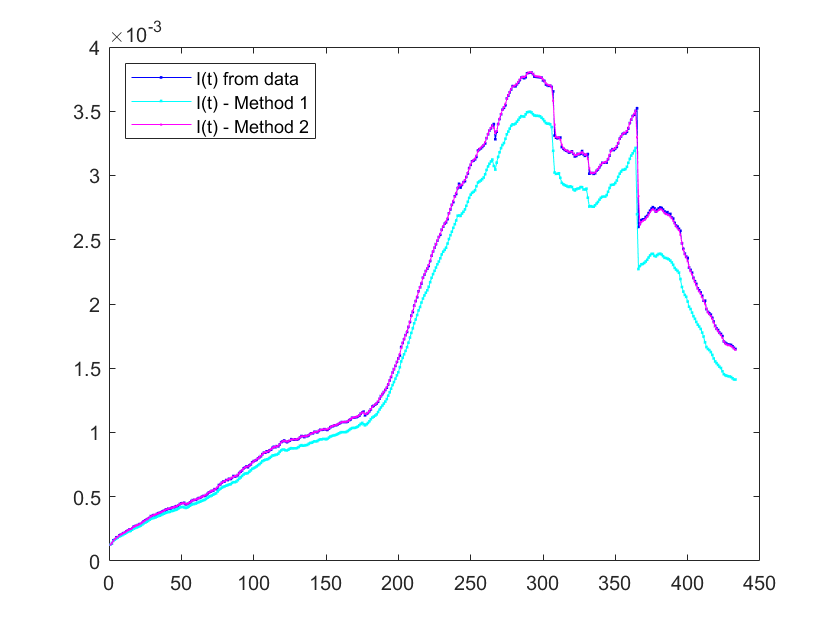}\vspace{-2mm}
\caption{Ratio of Infected i(t)}\label{fig:Infectedfirstoutbreakworld}
\end{subfigure}
\begin{subfigure}{1.0\linewidth}
\centering
\includegraphics[scale=.3]{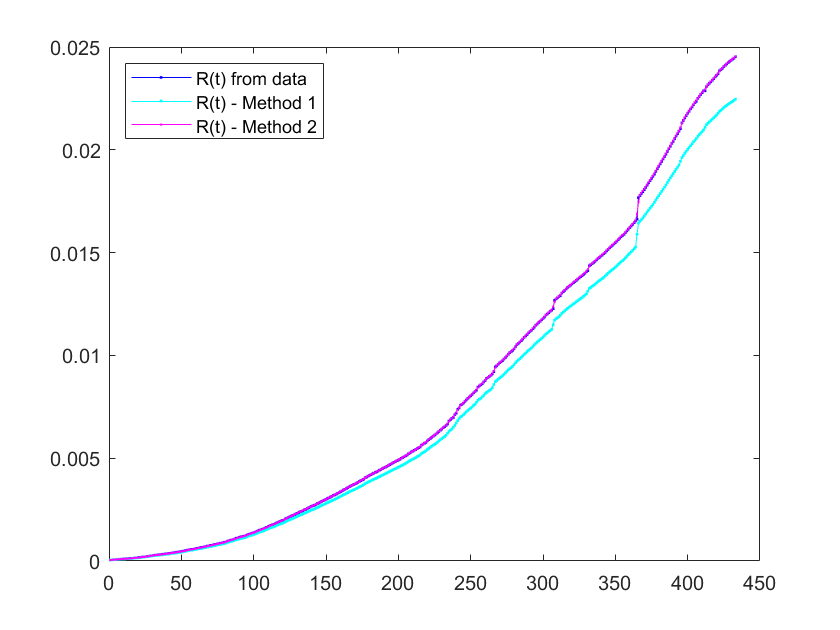}\vspace{-2mm}
\caption{Ratio of Removed r(t)}\label{fig:Removed1stoutbreakworld}
\end{subfigure}

\caption{Comparison of Compartments' ratios from real data to those obtained using SIR model with approximated parameters, during the first Outbreak in the World }
\label{fig:SIRfirstoutbreakworld}
\end{figure}

To validate these results, we run the SIR model with the obtained parameters $\beta(t), \rho(t)$ using method 1 or method 2, and the corresponding initial values as discussed in section \ref{sec:valid}. Figures \ref{fig:SIRfirstoutbreakitaly}, \ref{fig:SIRfirstoutbreakgermany},  and \ref{fig:SIRfirstoutbreakworld}  compare  the behavior of the ratio of compartments $s(t), i(t), r(t)$ of the real data to the simulated values. Moreover, Table \ref{table:table1stSIRrel} 
shows the relative errors between the real data and simulated values.
Clearly, the time-dependent SIR model simulates efficiently the real situation. The trends of the three compartments in all three simulations are as expected. The number of infected increased then decreased (somewhat exponential),  and naturally, the number of susceptible decreased whereas that 
 of the removed increased. The number of infected, even at the maximum, is much lower than that of  the susceptible.  Thus, even  in periods of outbreaks, most of the population is still not infected.

 However, it is clear that the simulated results using method 2 are more accurate than those obtained using method 1, with L2 relative errors are of order $10^{-3}$ versus $10^{-2}$ for $i(t)$ and $r(t)$, and $10^{-5}$ versus $10^{-3}$ for $s(t)$. This is expected since method 2 uses the SIR model to approximate $\beta$ and $\rho$, whereas method 1  relies only on data. 
 
 Moreover, note that there are some inconsistencies in the reported worldometer data during the first outbreak, specifically in the Case of Germany, which shows in the larger relative errors for method 1 (Table \ref{table:table1stSIRrel}) and in the difference in the simulated s,i,r in figure \ref{fig:SIRfirstoutbreakgermany}. When investigating about such inconsistency, we discovered that during the first outbreak the reported cumulative infected cases $C_I(t+\Delta t) - C_I(t)$ is not equal to the reported number of new cases at time $t+\Delta t$.
 
 For Italy and the whole world the simulated s,i,r using parameters obtained via method 1 or method 2 have the same shapes and variations with a small margine of error. This validates the time-dependent SIR model, and the methods used for computing the parameters.\\

\begin{table}[H]
\centering
\setlength{\tabcolsep}{10pt}
{\renewcommand{\arraystretch}{1.4}
\begin{tabular}{||c|c| c| c| c |c||} 
 \hline
 Method& &Norm & Italy  & Germany  & World  \\ 
 \hline\hline
\multirow{6}{*}{1}&\multirow{2}{*}{S} &$L_2$ & $  1.083* 10^{-3}$ &$6.938 * 10^{-3}$ &  $  1.066  * 10^{-3}$  \\ 
 \cline{3-6}
 & & $L_\infty$ & $ 1.952* 10^{-3}$  & $ 1.099 * 10^{-2}$ & $ 2.308  * 10^{-3}$ \\ 
 \cline{2-6}
&\multirow{2}{*}{I} &$L_2$ & $ 8.644* 10^{-2}$ &$ 4.799* 10^{-1}$ &  $  9.345 * 10^{-2}$ \\ 
 \cline{3-6}
&& $L_\infty$ & $ 9.745* 10^{-2}$ & $ 5.075 * 10^{-1}$ & $2.169 * 10^{-1}$ \\ 
  \cline{2-6}
 &\multirow{2}{*}{R} & $L_2$ & $ 3.362* 10^{-2}$ &$3.550* 10^{-1}$ &  $ 7.774 * 10^{-2}$ \\\cline{3-6}
 && $L_\infty$ & $ 4.615 * 10^{-2}$ & $ 3.840* 10^{-1}$ & $8.437 * 10^{-2}$ \\  
 \hline
 \multirow{6}{*}{2}&\multirow{2}{*}{S} &$L_2$ & $3.909 * 10^{-5}$ &$2.587* 10^{-4}$ &  $1.301 * 10^{-5}$ \\ 
 \cline{3-6}
 & & $L_\infty$ & $1.484 * 10^{-4}$  & $4.202 * 10^{-4}$ & $4.382 * 10^{-5}$ \\ 
 \cline{2-6}
&\multirow{2}{*}{I} &$L_2$ & $3.876 * 10^{-3}$  &$ 2.083 * 10^{-2}$ &  $8.112 * 10^{-3}$ \\ 
 \cline{3-6}
&& $L_\infty$ & $7.603 * 10^{-3}$ & $3.836* 10^{-2}$ &$6.270 * 10^{-2}$ \\ 
  \cline{2-6}
 &\multirow{2}{*}{R} & $L_2$ & $3.983 * 10^{-3}$ &$1.323 * 10^{-2}$ &  $1.984 *10^{-3}$ \\\cline{3-6}
 && $L_\infty$ & $7.699 * 10^{-3}$  & $1.465 * 10^{-2}$ & $1.020 * 10^{-2}$ \\  
 \hline
\end{tabular}
\caption{The $L_2$ and $L_\infty$ relative errors of the computed S,I,R from the time-dependent model with the S,I,R collected from data during the first outbreak, where $\beta,\rho$ are computed using Method 1 or 2. }\label{table:table1stSIRrel}}
\end{table}

Analyzing the approximated values of the parameters, we observe the following.     In Italy the infection rate has an increasing trend over the first 40 days followed by a decreasing one as seen in figures \ref{fig:beta1stoutbreakitalyM1} and \ref{fig:beta1stoutbreakitalyM2}. Whereas, in Germany  the infection rate is in a decreasing mode (figures \ref{fig:beta1stoutbreakgermanyM1} and \ref{fig:beta1stoutbreakgermanyM2}) and it is somehow stable worldwide (figures \ref{fig:beta1stoutbreakworldM1} and \ref{fig:beta1stoutbreakworldM2}). Yet, the maximum infection rate in this first outbreak occurs in the world, on the second day with a value of  $\approx 0.2$ 
as seen in figures \ref{fig:beta1stoutbreakworldM1} and \ref{fig:beta1stoutbreakworldM2}.
    On the other hand, the mean, median, and standard deviation of the infection rate are highest for Germany and lowest for the world.

     On the other hand, the removal rate for Italy (figures \ref{fig:rho1stoutbreakitalyM1} and \ref{fig:rho1stoutbreakitalyM2}) and Germany (figures \ref{fig:rho1stoutbreakgermanyM1} and  \ref{fig:rho1stoutbreakgermanyM2}) seem to have an increasing trend during the first outbreak, which implies a larger number of people are getting recovered in a smaller amount of time. As for the world (figures \ref{fig:rho1stoutbreakworldM1} and  \ref{fig:rho1stoutbreakworldM2}), the removal rate fluctuates within a certain range ([0.2, 0.5]) with the exception of a few peaks. The highest peak occurs around day 370 which corresponds to a sudden drop in the infected (figure \ref{fig:Infectedfirstoutbreakworld}).  

   The reproduction factor and replacement number have the same behavior, which is mainly proportional to that of $\beta(t)$. Thus, for Italy (figures \ref{fig:sigma1stoutbreakitalyM1}, \ref{fig:sigma1stoutbreakitalyM2}, \ref{fig:sigmas1stoutbreakitalyM1}  and \ref{fig:sigmas1stoutbreakitalyM2}), both $\sigma(t)$ and $\sigma_s(t)$ increase first to a maximum value of $\approx 9.6$ (Method 1) or $\approx 9.3$ (Method 2). 
   As for the world  (figures \ref{fig:sigma1stoutbreakworldM1}, \ref{fig:sigma1stoutbreakworldM2}, \ref{fig:sigmas1stoutbreakworldM1}  and \ref{fig:sigmas1stoutbreakworldM2}),   
$\sigma(t)$ and $\sigma_s(t)$ show a decreasing trend from a value of 3 to a value close to 1. Since the replacement number is on average greater than one, this reaffirms the fact that the period considered is indeed an outbreak period.

  Note that the mean, median, and maximum $\sigma_s(t)$  values  are the highest for Italy. This means that there are several countries that did not go through a strong outbreak during that period as Italy, which was one of the first countries to have a rapid spread of COVID-19.

\section{Do Constant Coefficient SIR's Predict  Covid-19 Dynamics?}
So far, all the discussions that were performed in the previous section, were merely analysing situations that already occurred to better understand, assess and improve performance in the future. However, they do not allow to extrapolate and predict how the pandemic will evolve in the coming days, weeks or months, since the time-dependent parameters $\beta(t)$ and $\rho(t)$ cannot be known beforehand. Their approximations in section \ref{sec:Param} were based on given data, which are not available in case of prediction. Unless these parameters can be determined based on some medical assumptions or observations, it is not possible to use a time-dependent SIR model to predict accurately the evolution of the pandemic.
Thus, we consider a constant-coefficient model where $\beta$ and $\rho$ are constants.

Unlike the time-dependent model, a constant-coefficient cannot depict the fluctuations in the compartments for large periods, and may also give totally incorrect results as shown in section \ref{sec:incomp}.  Hence, we consider a constant-coefficient model for predicting the evolution of the pandemic for a short period of time in section \ref{sec:predic}.

\subsection{Incompatibility of Constant-Coefficient Models for  Covid-19}\label{sec:incomp}
\begin{figure}[H]
\begin{subfigure} {0.5\textwidth}
\centering
\includegraphics[scale=0.3]{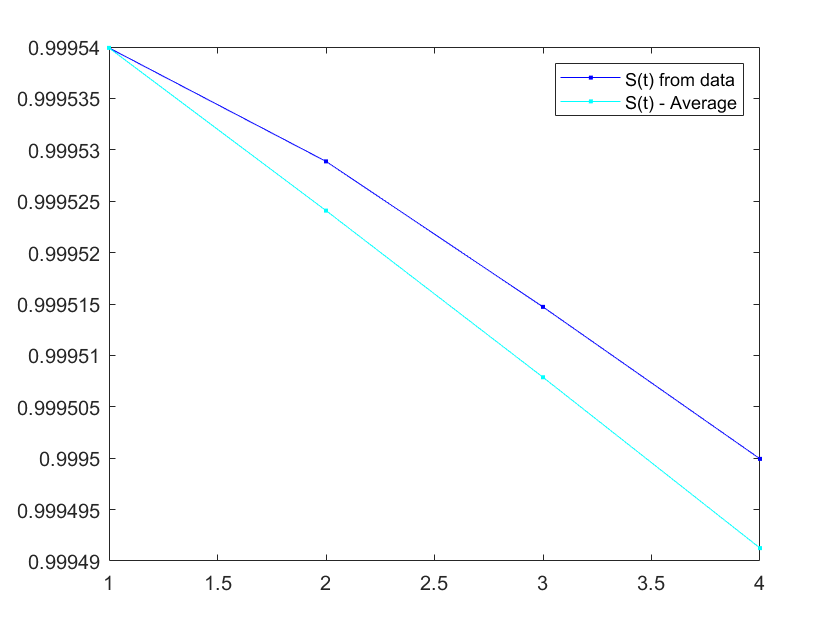}\vspace{-2mm}
\caption{Ratio of Susceptible s(t)}\label{fig:susceptibleavfour}
\end{subfigure}%
\begin{subfigure}{.5\linewidth}
\centering
\includegraphics[scale=.3]{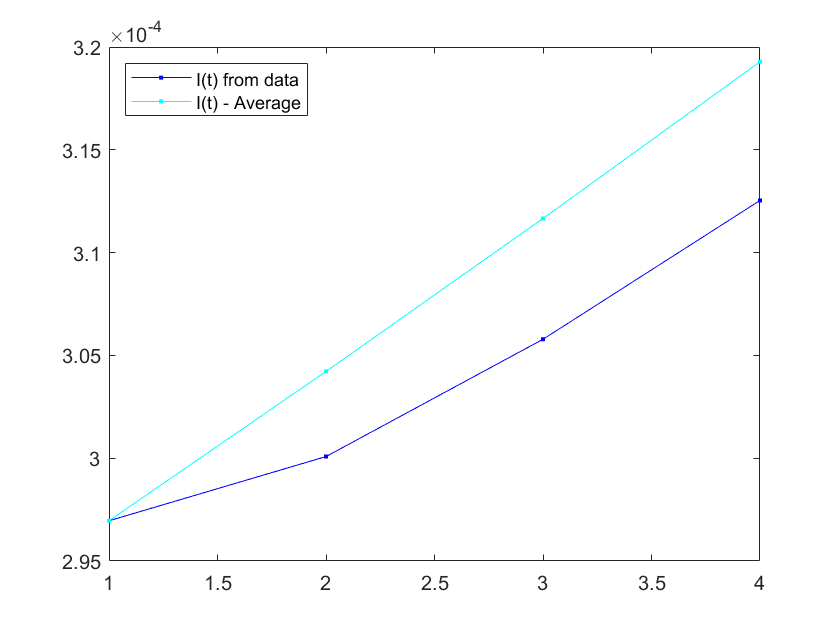}\vspace{-2mm}
\caption{Ratio of Infected i(t)}\label{fig:infectedavfour}
\end{subfigure}
\begin{subfigure}{1.0\linewidth}
\centering
\includegraphics[scale=.3]{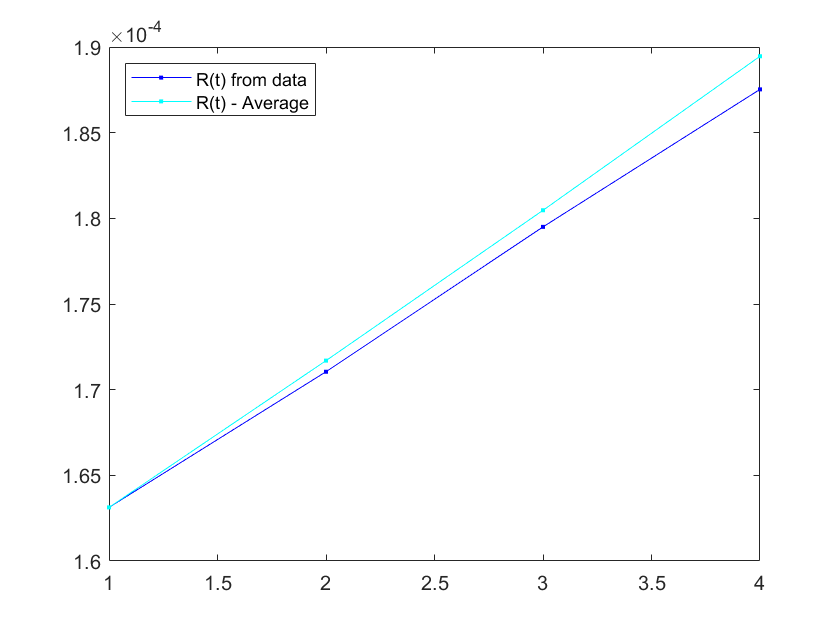}\vspace{-2mm}
\caption{Ratio of Removed r(t)}\label{fig:rav4}
\end{subfigure}
\caption{Time-Independent SIR Model Using the Average: First Interval}\vspace{-.5cm}
\label{fig:first}
\end{figure}

\begin{figure}[H]
\begin{subfigure}{.5\textwidth}
\centering
\includegraphics[scale=.3]{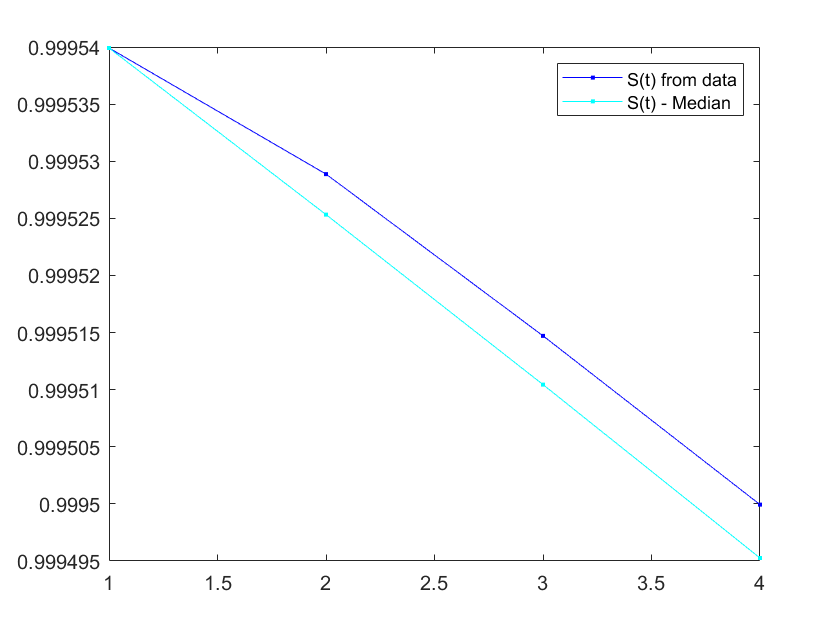}
\caption{Ratio of Susceptible s(t)}\label{fig:smedian1}
\end{subfigure}%
\begin{subfigure}{.5\textwidth}
\centering
\includegraphics[scale=.3]{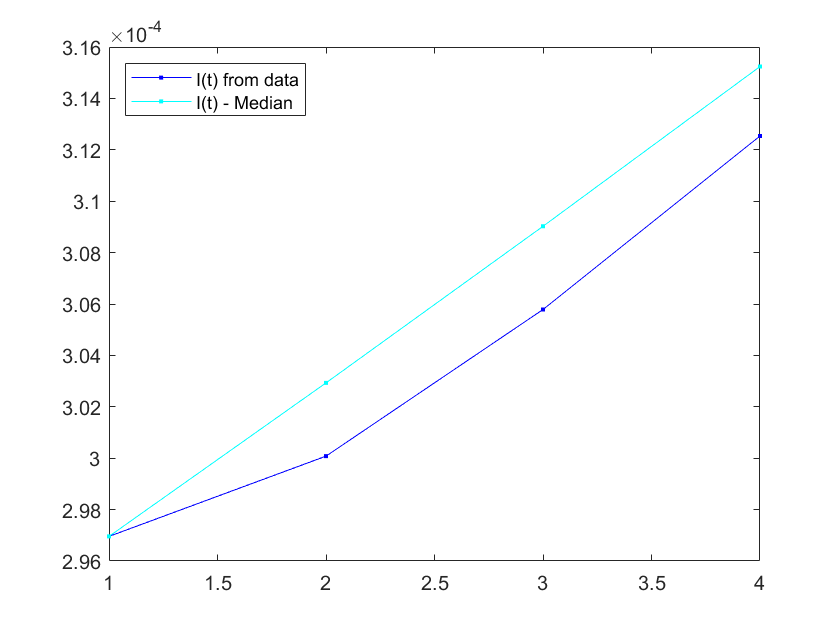}\vspace{-2mm}
\caption{Ratio of Infected i(t)}\label{fig:imedian1}
\end{subfigure}
\begin{subfigure}{1.0\linewidth}
\centering
\includegraphics[scale=.3]{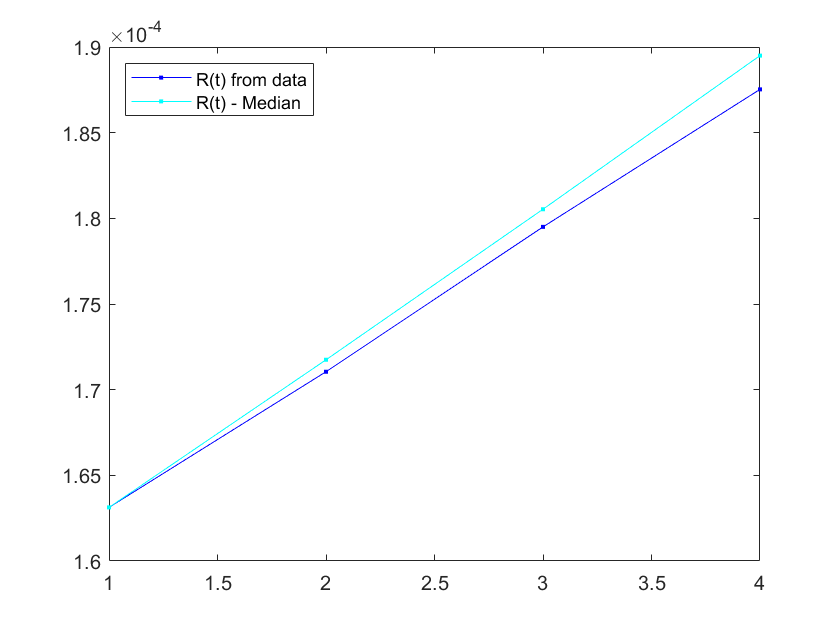}\vspace{-2mm}
\caption{Ratio of Removed r(t)}\label{fig:rmedian1}
\end{subfigure}
\caption{Time-Independent SIR Model Using the Median: First Interval}\vspace{-.5cm}
\label{fig:sirmedian1}
\end{figure}

\begin{figure}[H]
\begin{subfigure} {0.5\textwidth}
\centering
\includegraphics[scale=0.3]{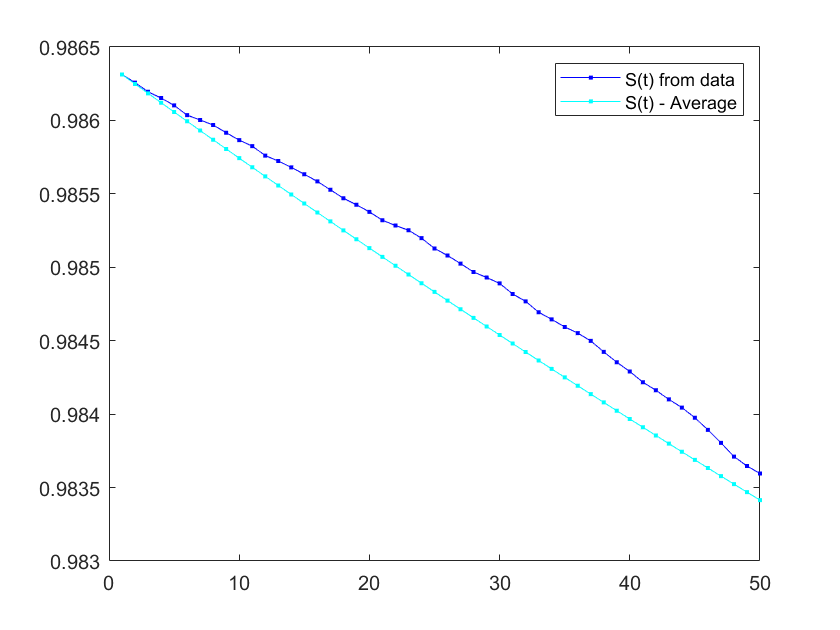}\vspace{-2mm}
\caption{Ratio of Susceptible s(t)}\label{fig:susceptibleav2}
\end{subfigure}%
\begin{subfigure}{.5\linewidth}
\centering
\includegraphics[scale=.3]{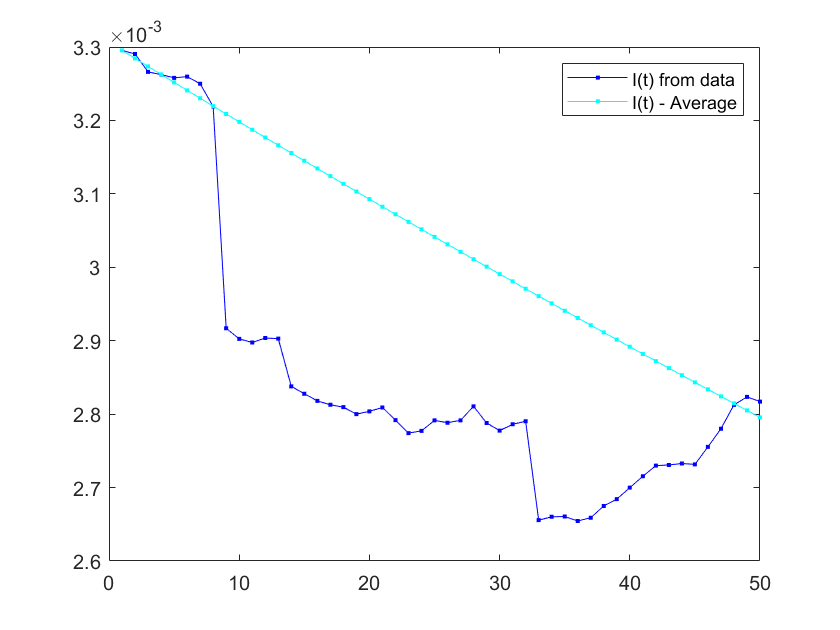}\vspace{-2mm}
\caption{Ratio of Infected i(t)}\label{fig:infectedav2}
\end{subfigure}
\begin{subfigure}{1.0\linewidth}
\centering
\includegraphics[scale=.3]{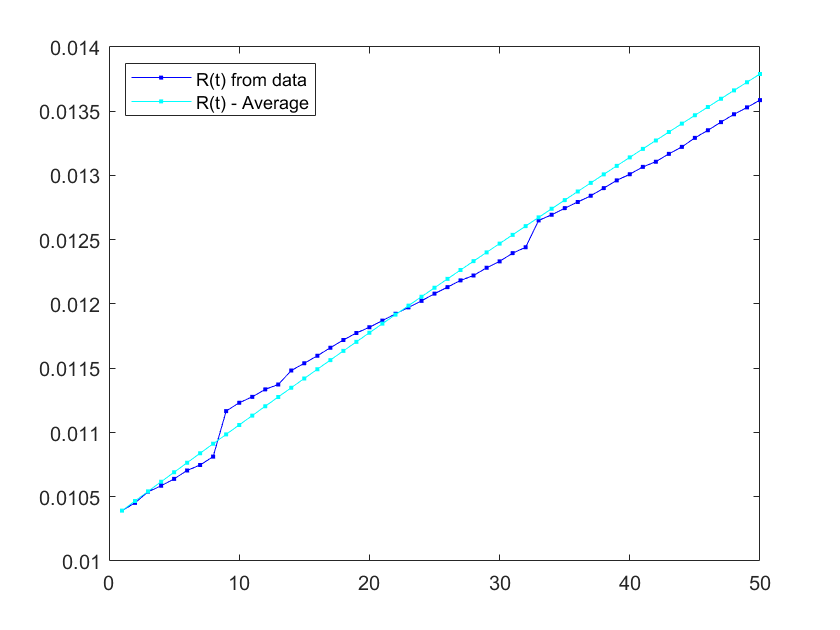}\vspace{-2mm}
\caption{Ratio of Removed r(t)}\label{fig:rav2}
\end{subfigure}
\caption{Time-Independent SIR Model Using the Average: Second Interval}\vspace{-.5cm}
\label{fig:siraverage2}
\end{figure}

\begin{figure}[H]
\begin{subfigure} {0.5\textwidth}
\centering
\includegraphics[scale=0.3]{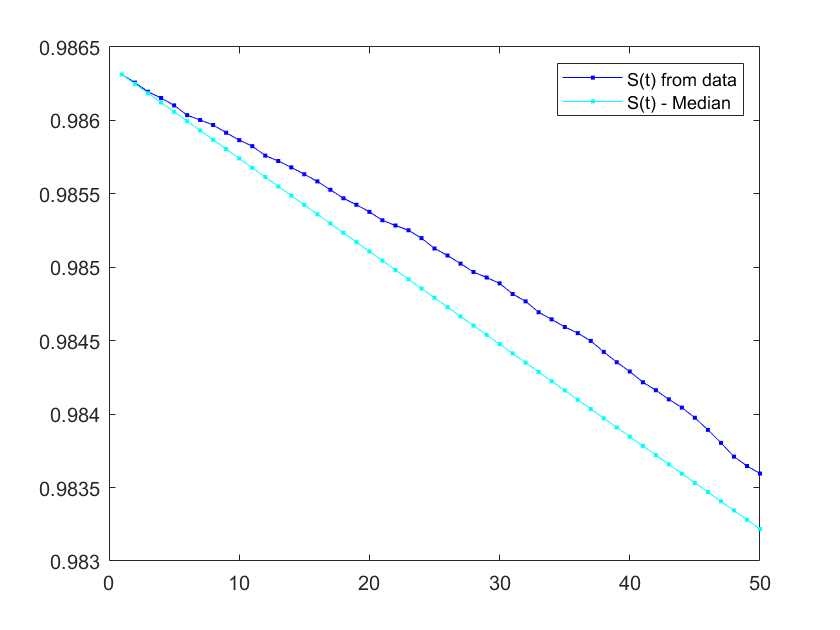}\vspace{-2mm}
\caption{Ratio of Susceptible s(t)}\label{fig:smedian2}
\end{subfigure}%
\begin{subfigure}{.5\linewidth}
\centering
\includegraphics[scale=.3]{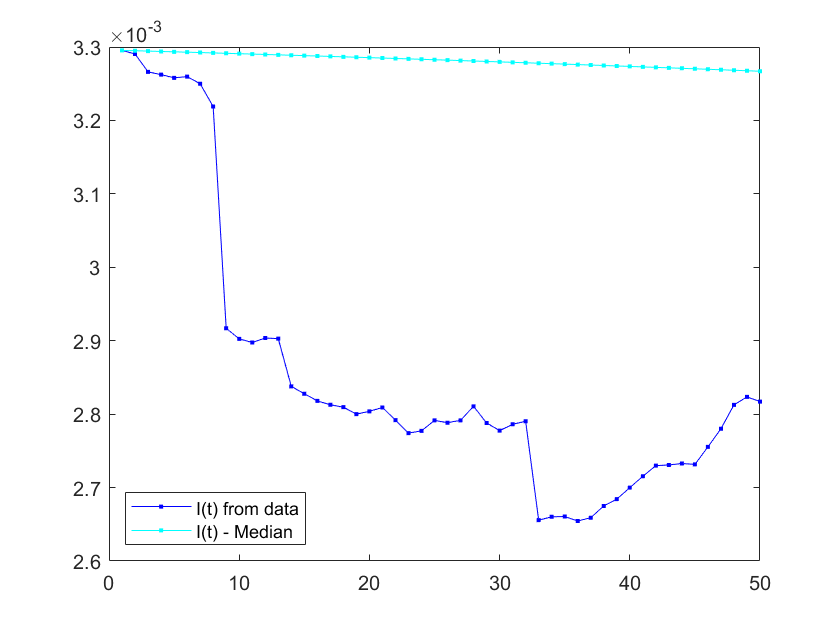}\vspace{-2mm}
\caption{Ratio of Infected i(t)}\label{fig:imedian2}
\end{subfigure}
\begin{subfigure}{1.0\linewidth}
\centering
\includegraphics[scale=.3]{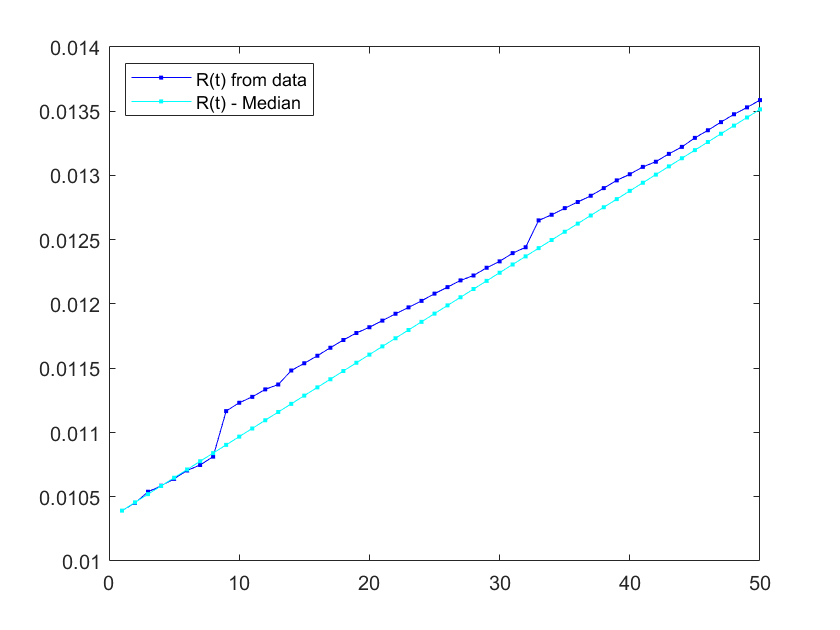}\vspace{-2mm}
\caption{Ratio of Removed r(t)}\label{R-world-median-second.png}
\end{subfigure}
\caption{Time-Independent SIR Model Using the Median: Second Interval}\vspace{-.5cm}
\label{fig:sirmedian2}
\end{figure}

\begin{figure}[H]
\begin{subfigure} {0.5\textwidth}
\centering
\includegraphics[scale=0.3]{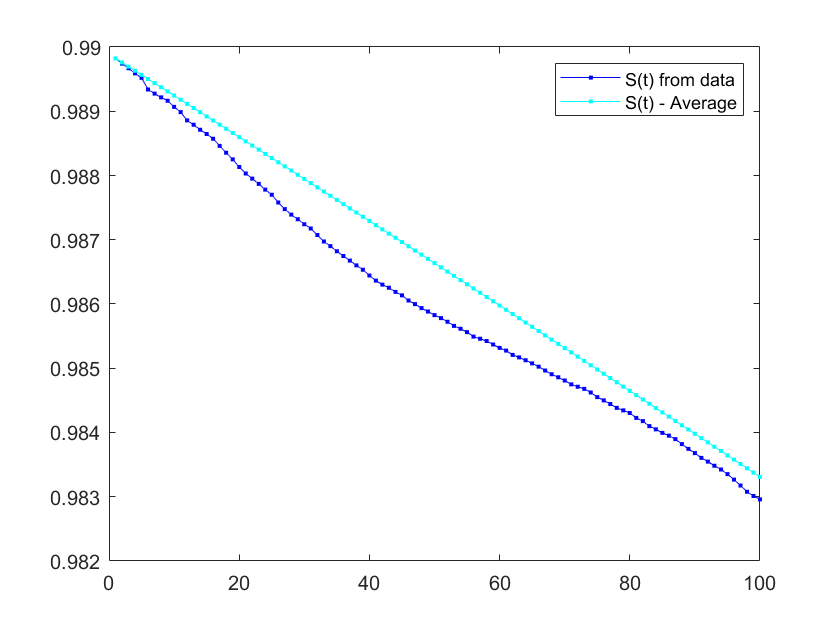}\vspace{-2mm}
\caption{Ratio of Susceptible s(t)}\label{fig:susceptibleav3}
\end{subfigure}%
\begin{subfigure}{.5\linewidth}
\centering
\includegraphics[scale=.3]{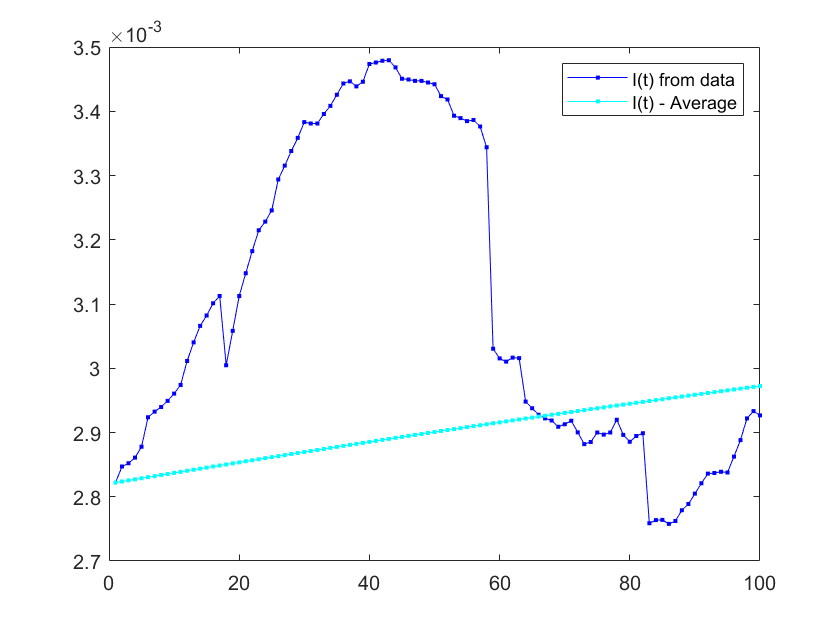}\vspace{-2mm}
\caption{Ratio of Infected i(t)}\label{fig:infectedav3}
\end{subfigure}
\begin{subfigure}{1.0\linewidth}
\centering
\includegraphics[scale=.3]{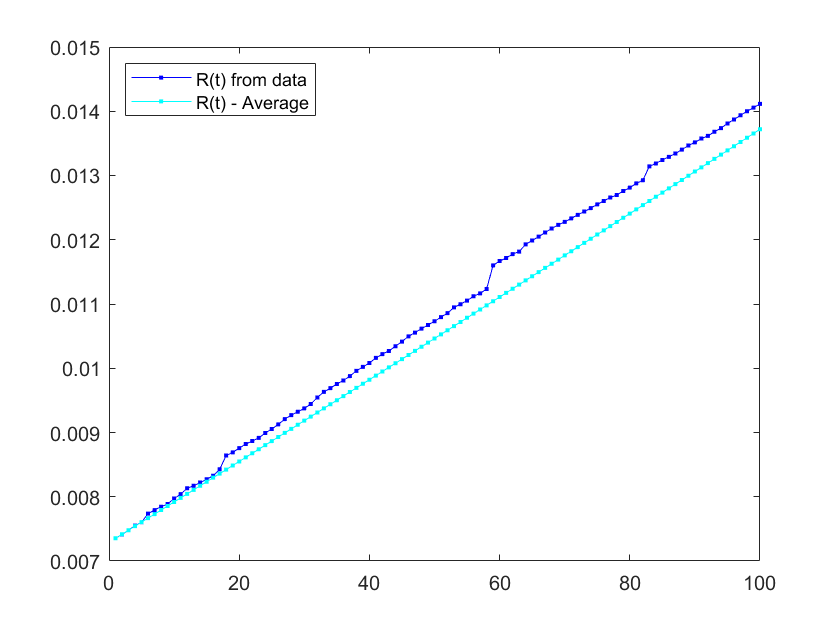}\vspace{-2mm}
\caption{Ratio of Removed r(t)}\label{fig:rav3}
\end{subfigure}
\caption{Time-Independent SIR Model Using the Average: Third Interval}\vspace{-.5cm}
\label{fig:sirav3}
\end{figure}

\begin{figure}[H]
\begin{subfigure} {0.5\textwidth}
\centering
\includegraphics[scale=0.3]{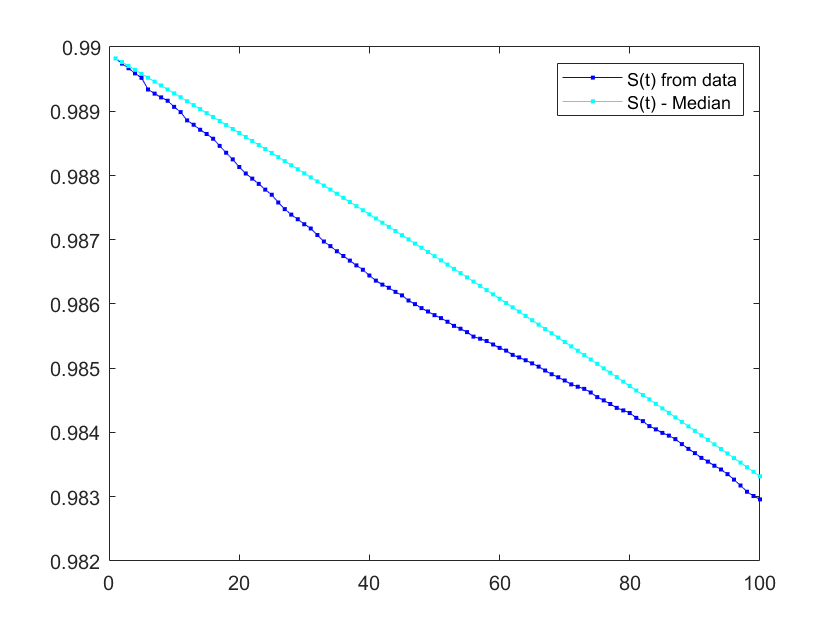}\vspace{-2mm}
\caption{Ratio of Susceptible s(t)}\label{fig:smedian3}
\end{subfigure}%
\begin{subfigure}{.5\linewidth}
\centering
\includegraphics[scale=.3]{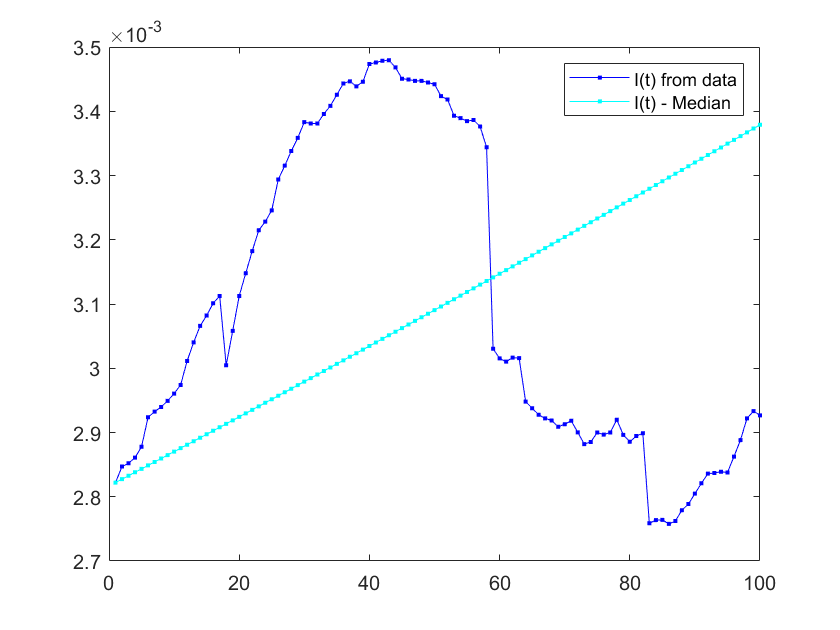}\vspace{-2mm}
\caption{Ratio of Infected i(t)}\label{fig:imedian3}
\end{subfigure}
\begin{subfigure}{1.0\linewidth}
\centering
\includegraphics[scale=.3]{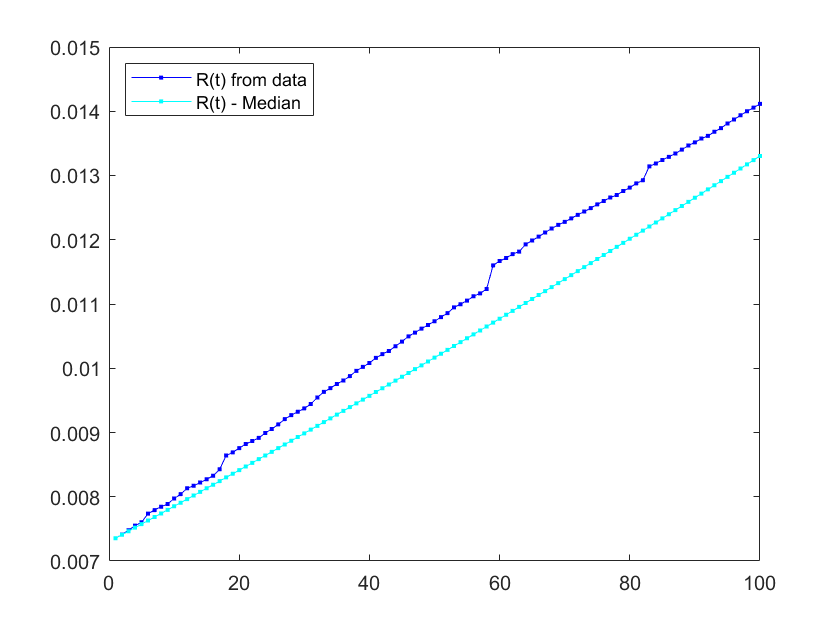}\vspace{-2mm}
\caption{Ratio of Removed r(t)}\label{R-world-median-third.png}
\end{subfigure}
\caption{Time-Independent SIR Model Using the Median: Third Interval}\vspace{-.5cm}
\label{fig:sirmedian3}
\end{figure}

In this section we consider the Constant-Coefficient SIR model, where $\beta, \rho, $ and $\sigma$ are constants, and test its efficiency for modeling the Covid-19 Pandemic using the worldometer data for the world. 
We consider three different increasing time intervals (Table \ref{tab:independentoutbreaks}), extract the compartments and parameters as discussed in sections \ref{sec:compart} and \ref{sec:ParamM1} (Method 1) respectively. Then, we transform each of the vectors with the $\beta(t)$ and $\rho(t)$ values for the chosen period of time into a constant, by taking the average,
or the median.
\begin{table}[H]
    \centering
    \setlength{\tabcolsep}{10pt}
{\renewcommand{\arraystretch}{1.4}
    \begin{tabular}{
    |c|c|c|}\hline
          First Interval & Second Interval& Third Interval \\ \hline
25/04/2020 - 29/04/2020 & 8/02/2021 - 29/03/2021& 19/12/2020 - 29/03/2021 \\
  \hline
    \end{tabular}\vspace{-3mm}
    \caption{The considered periods for testing}\vspace{-3mm}
    \label{tab:independentoutbreaks}}
\end{table}
\noindent Figures \ref{fig:first},  \ref{fig:siraverage2}, and \ref{fig:sirav3} show the difference between the simulated compartments when computing the averages of $\beta(t)$ and $\rho(t)$, and the real data. Whereas,  Figures \ref{fig:sirmedian1}, \ref{fig:sirmedian2} and \ref{fig:sirmedian3} show the difference between the simulated compartments when computing the medians of $\beta(t)$ and $\rho(t)$, and the real data. Moreover, the relative errors of the simulated S,I,R using the constant-coefficient model with the 2 parameters' calculations are summarized in Table \ref{tab:averageWorld} and compared to that of the time-dependent model.

\begin{table}[H]
\centering
\setlength{\tabcolsep}{10pt}
{\renewcommand{\arraystretch}{1.1}
\begin{tabular}{||c|c|c|c|c|c||} 
 \cline{2-6}
\multicolumn{1}{c|}{} &Period&Norm & S & I & R \\ 
 \hline\hline
\multirow{6}{*}{Average} &\multirow{2}{*}{4\; days} & $L_2$ &  $6.025*10^{-6}$ & $1.621* 10^{-2}$ &  $ 6.428*10^{-3}$  \\ 
\cline{3-6}
& &$L_\infty$ &   $8.682* 10^{-6}$ &  $2.158*10^{-2}$ & $1.031*10^{-2}$ \\ 
 \cline{2-6}
&\multirow{2}{*}{50\; days} &$L_2$ &  $ 2.573* 10^{-4}$& $   7.651*10^{-2}$ &  $ 9.725*10^{-3}$ \\ 
\cline{3-6}
& &$L_\infty$ &   $3.672 * 10^{-4}$ &  $ 9.645*10^{-2}$ & $  1.496*10^{-2}$\\
 \cline{2-6}
 &\multirow{2}{*}{100\; days} &$L_2$ &  $ 5.619*10^{-4}$& $    1.055*10^{-1}$ &  $  3.172*10^{-2}$ \\ 
 \cline{3-6}
& &$L_\infty$ &   $ 8.726 * 10^{-4}$ &  $ 1.695*10^{-1}$ & $3.979*10^{-2}$ \\ 
 \hline

\multirow{6}{*}{Median} &\multirow{2}{*}{4\; days} & $L_2$ &  $3.6301*10^{-6}$ & $8.377* 10^{-3}$ &  $  6.639*10^{-3}$  \\ 
\cline{3-6}
 &&$L_\infty$ &   $4.676* 10^{-6}$ &  $ 1.036*10^{-2}$ & $1.051*10^{-2}$ \\ 
\cline{2-6}
&\multirow{2}{*}{50\; days} &$L_2$ &  $3.314* 10^{-4}$& $   1.621*10^{-1}$ &  $ 1.313*10^{-2}$ \\ 
\cline{3-6}
 &&$L_\infty$ &   $ 4.698 * 10^{-4}$ &  $ 1.888*10^{-1}$ & $  1.944*10^{-2}$\\
 \cline{2-6}
 &\multirow{2}{*}{100\; days} &$L_2$ &  $  6.411*10^{-4}$& $    1.105*10^{-1}$ &  $   5.837*10^{-2}$ \\ 
 \cline{3-6}
 &&$L_\infty$ &   $9.766 * 10^{-4}$ &  $  1.553*10^{-1}$ & $6.636*10^{-2}$ \\ 
 \hline

\multirow{6}{*}{Time-dependent} &\multirow{2}{*}{4\; days} & $L_2$ &  $6.371*10^{-6}$ & $1.695* 10^{-2}$ &  $ 7.426*10^{-3}$  \\ 
\cline{3-6}
& &$L_\infty$ &   $8.242* 10^{-6}$ &  $2.266*10^{-2}$ & $ 9.955*10^{-3}$ \\ 
\cline{2-6}
&\multirow{2}{*}{50\; days} &$L_2$ &  $3.109* 10^{-5}$& $  2.103*10^{-2}$ &  $4.419*10^{-3}$ \\ 
\cline{3-6}
& &$L_\infty$ &   $8.989 * 10^{-5}$ &  $  9.060*10^{-2}$ & $ 2.194*10^{-2}$\\
 \cline{2-6}
& \multirow{2}{*}{100\; days} &$L_2$ &  $  8.177*10^{-5}$& $     1.914*10^{-2}$ &  $  6.048*10^{-3}$ \\ 
\cline{3-6}
 &&$L_\infty$ &   $1.389 * 10^{-4}$ &  $  1.076*10^{-1}$ & $1.892*10^{-2}$ \\ 
 \hline
\end{tabular}}
\caption{Relative errors between the S,I,R values computed from the constant-coefficient model with average or median $\beta, \rho$ values or time-dependent model; and the S,I,R values collected from the world data for different periods. }\label{tab:averageWorld}
\end{table}

Given that $s(t)$ is decreasing and $r(t)$ is increasing for any time $t$, then the simulated values approximate these trends. However, the constant-coefficient model using the average or the median, doesn't capture the variations in $i(t)$ for larger time intervals.
Thus, the constant-coefficient model shows its inefficiency and inaccuracy, since the larger the time interval used, the larger the relative errors as shown in Table  \ref{tab:averageWorld}, where the relative errors are the largest for $i(t)$. 

However, it is worth noting that for the 4-days period, the relative errors of the constant-coefficient with average are of the same order as the time-dependent model. Since the parameters should not vary much during a small interval of time, taking the average is a good representation of their actual values and thus does not alter the real data by a huge factor. Moreover, the small magnitude of the errors suggest the possibility of predicting the 
 the evolution of the pandemic for a few days using the constant-coefficient model.
 Note that the same tests were performed with $\beta(t)$ and $\rho(t)$ computed using Method 2 (section \ref{sec:ParamM2}) instead of Method 1, and the similar results are shown in Appendix \ref{sec:append}.

\subsection{Predicting}\label{sec:predic}
Given that the $\beta(t)$ and $\rho(t)$ are not know beforehand, and unless they can be approximated using some medical or statistical assumptions, one cannot use the time-dependent model for prediction.
Moreover, due to the inefficiency of the constant-coefficient model for simulating accurately the COVID-19 pandemic, which is highly time dependent, then it cannot be used for prediction purposes over large periods of time. We consider small periods of times, not exceeding one month and check if the   constant-coefficient SIR model can predict the dynamics and evolution of the COVID-19 pandemics.

Many studies consider some fixed value for $\beta$ and $\rho$ when predicting the evolution of the pandemics. However, based on the testings in section \ref{sec:worldtest} the average of $\beta(t)$ and $\rho(t)$ vary over different periods of time and for different countries. Thus, we consider computing the constant $\beta$ and $\rho$ based on the data.

Given the Worldometer data up till some day $t_i$, and assuming we want to predict the dynamics of COVID-19 for the next $k\leq 30$ days, i.e $t = [t_i+1, t_i+k]$, we compute the values of the parameters $\beta(t)$ and $\rho(t)$ for the previous $k$ days, i.e.  $t = [t_i - k+1, t_i]$, using Method 1 (section \ref{sec:ParamM1}). Then, we take the average of the obtained vectors to define the constants $\beta$ and $\rho$, run the constant-coefficient SIR model starting with $t_0 = t_i$ with the obtained constant parameters, and compare the obtained compartments to the exact data. 

Note that it is possible to approximate the constant parameters $\beta$ and $\rho$ in different ways, such as taking the average $\beta(t)$ and $\rho(t)$ computed over some fixed interval independent of the duration $k$. However, these options are not considered in this paper. 

\begin{figure}[H]
\begin{subfigure} {0.5\textwidth}
\centering
\includegraphics[scale=0.32]{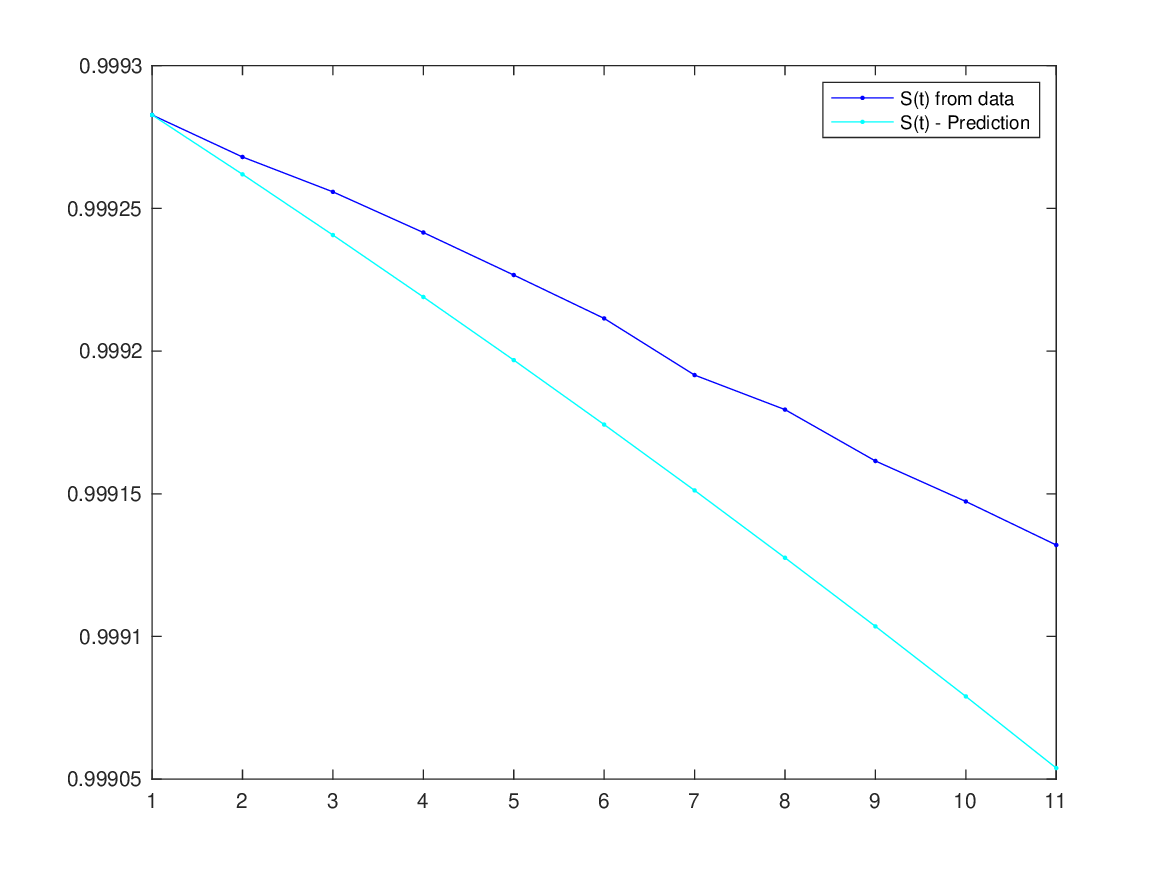}\vspace{-2mm}
\caption{t1}
\end{subfigure}%
\begin{subfigure}{.5\linewidth}
\centering
\includegraphics[scale=.32]{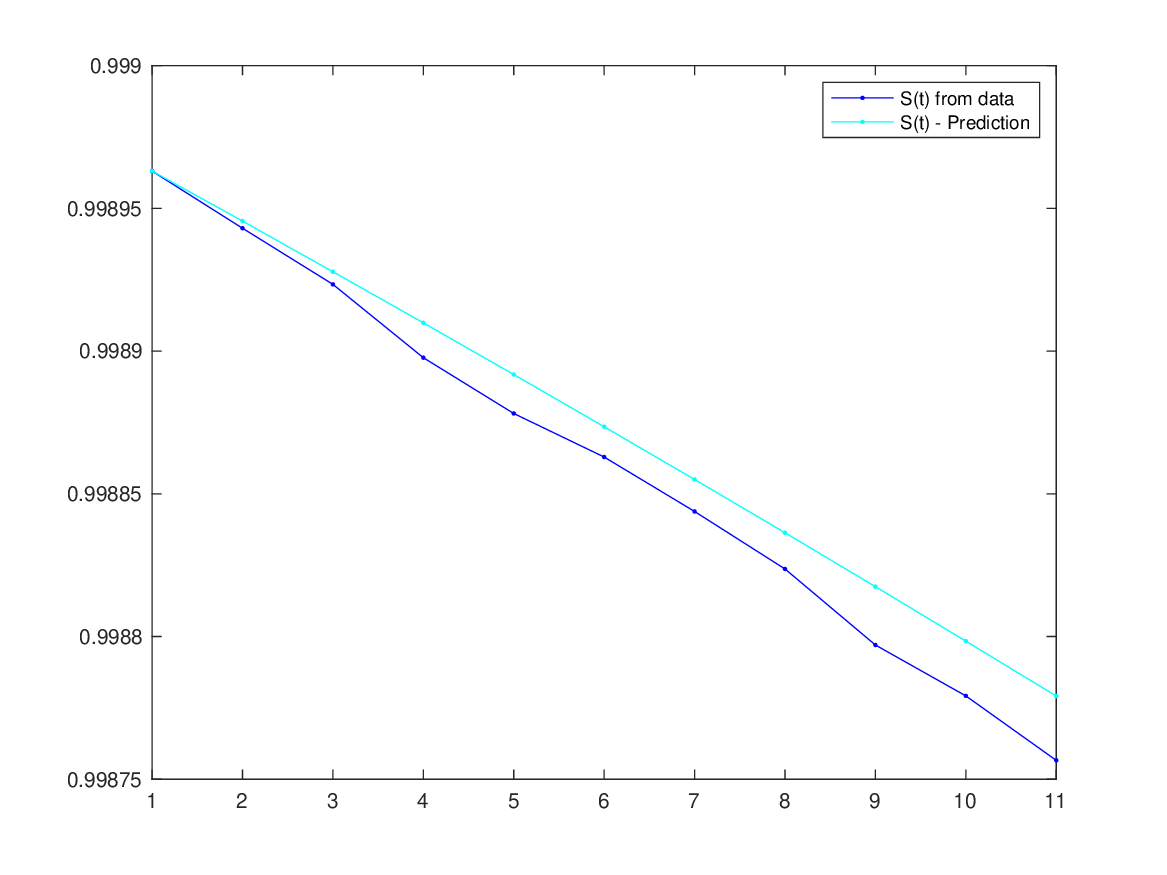}\vspace{-2mm}
\caption{t2}
\end{subfigure}
\begin{subfigure}{0.5\linewidth}
\centering
\includegraphics[scale=.32]{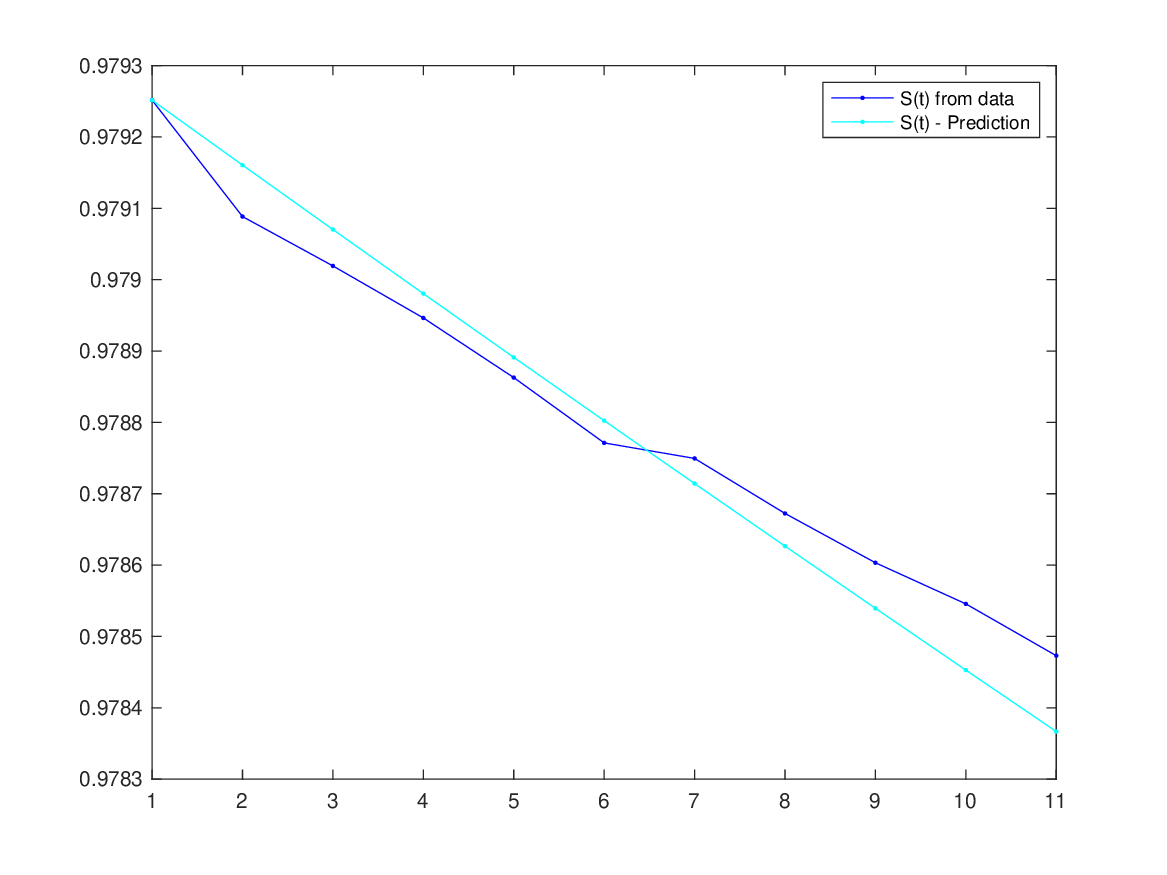}\vspace{-2mm}
\caption{t3}
\end{subfigure}
\begin{subfigure}{0.5\linewidth}
\centering
\includegraphics[scale=.32]{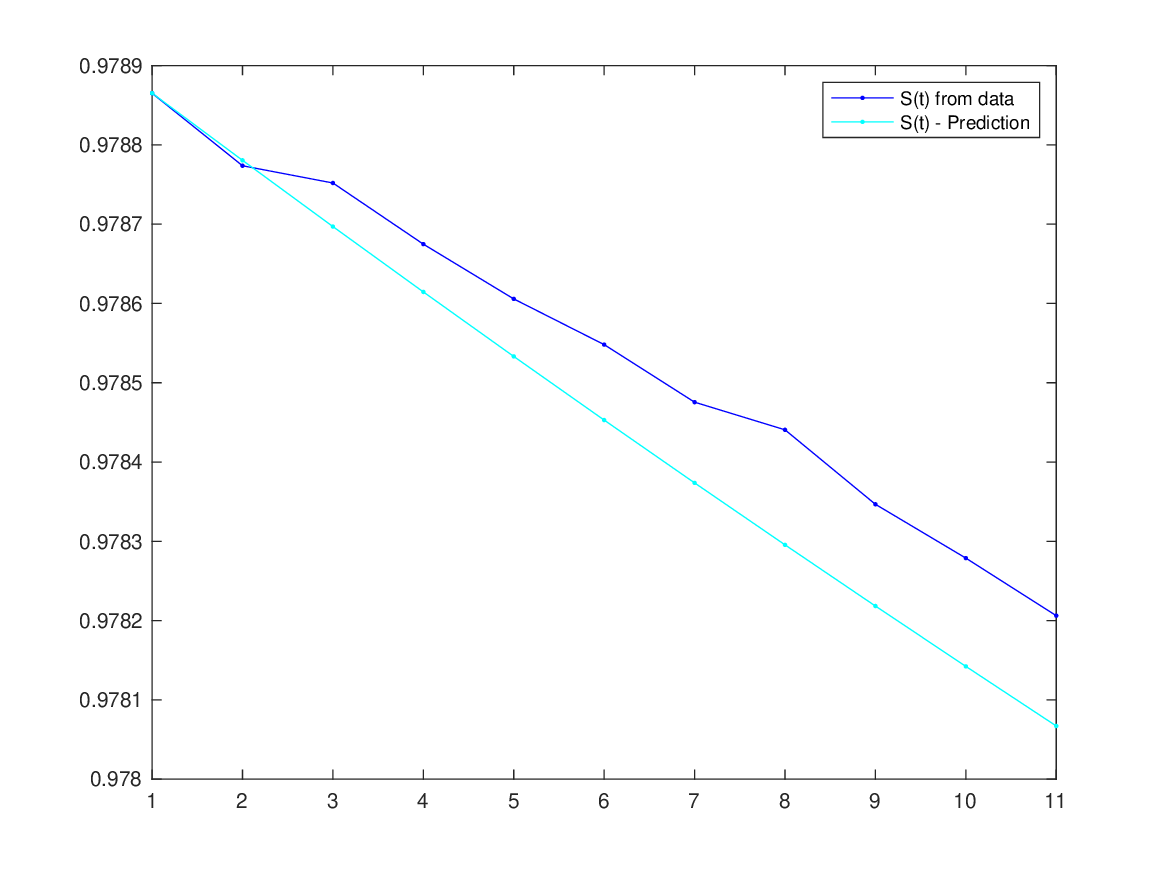}\vspace{-2mm}
\caption{t4}
\end{subfigure}
\caption{Prediction of $s(t)$ for the 4 starting dates and duration of 10 days
}\vspace{-.5cm}
\label{fig:predictionS10}
\end{figure}

\begin{figure}[H]
\begin{subfigure} {0.5\textwidth}
\centering
\includegraphics[scale=0.32]{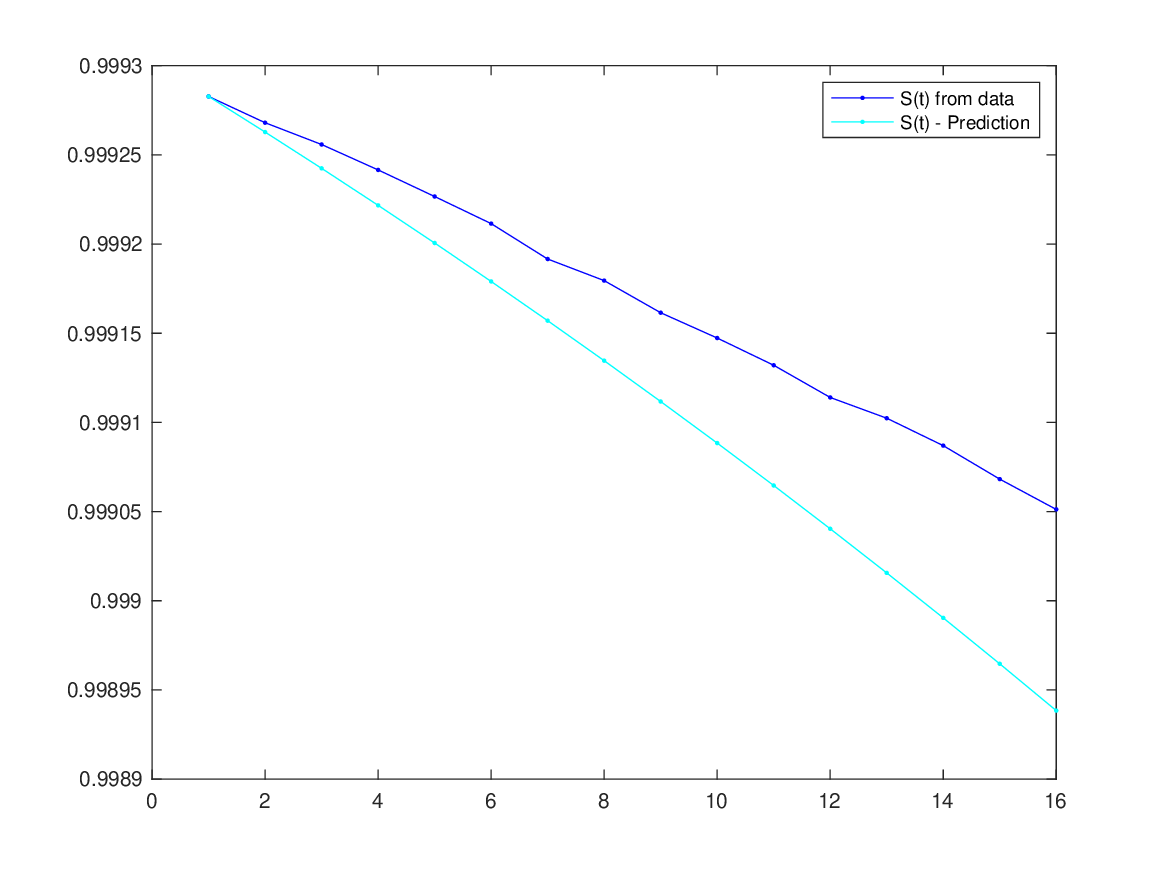}\vspace{-2mm}
\caption{t1}
\end{subfigure}%
\begin{subfigure}{.5\linewidth}
\centering
\includegraphics[scale=.32]{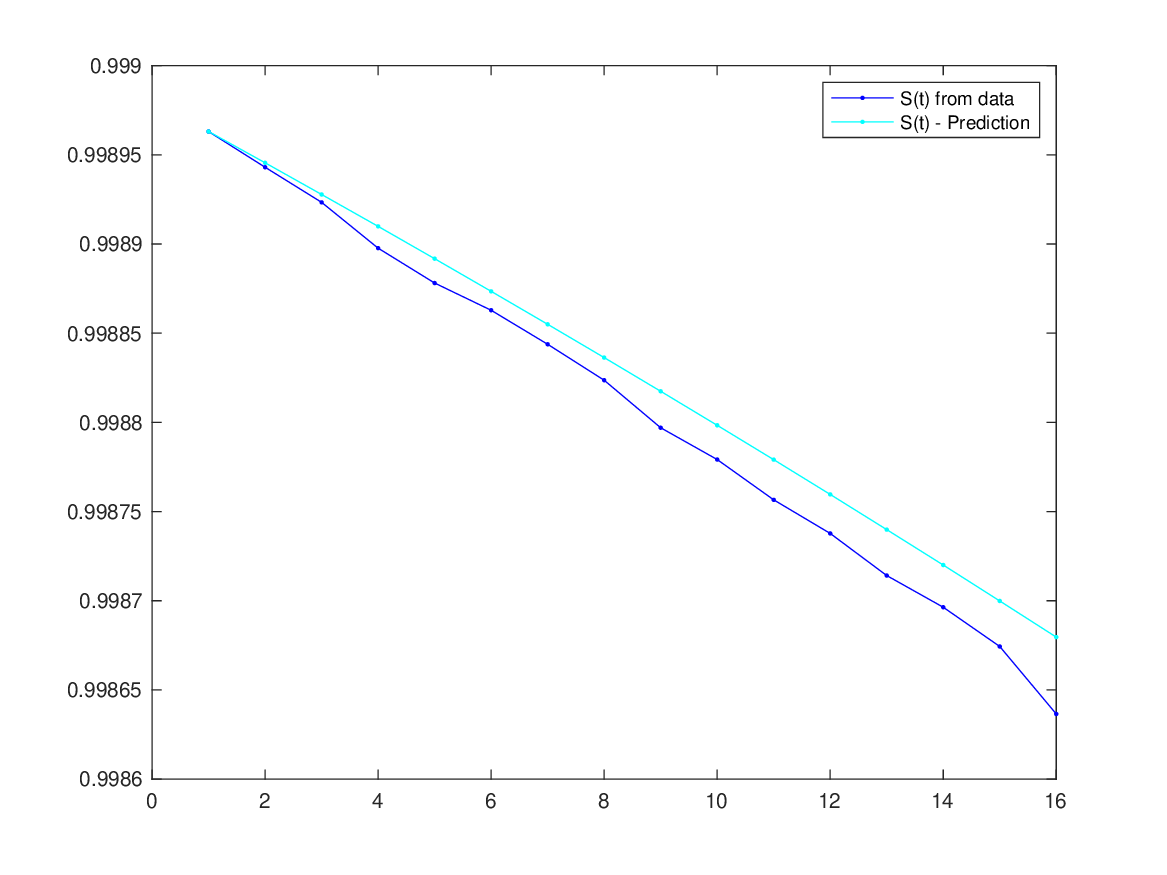}\vspace{-2mm}
\caption{t2}
\end{subfigure}
\begin{subfigure}{0.5\linewidth}
\centering
\includegraphics[scale=.32]{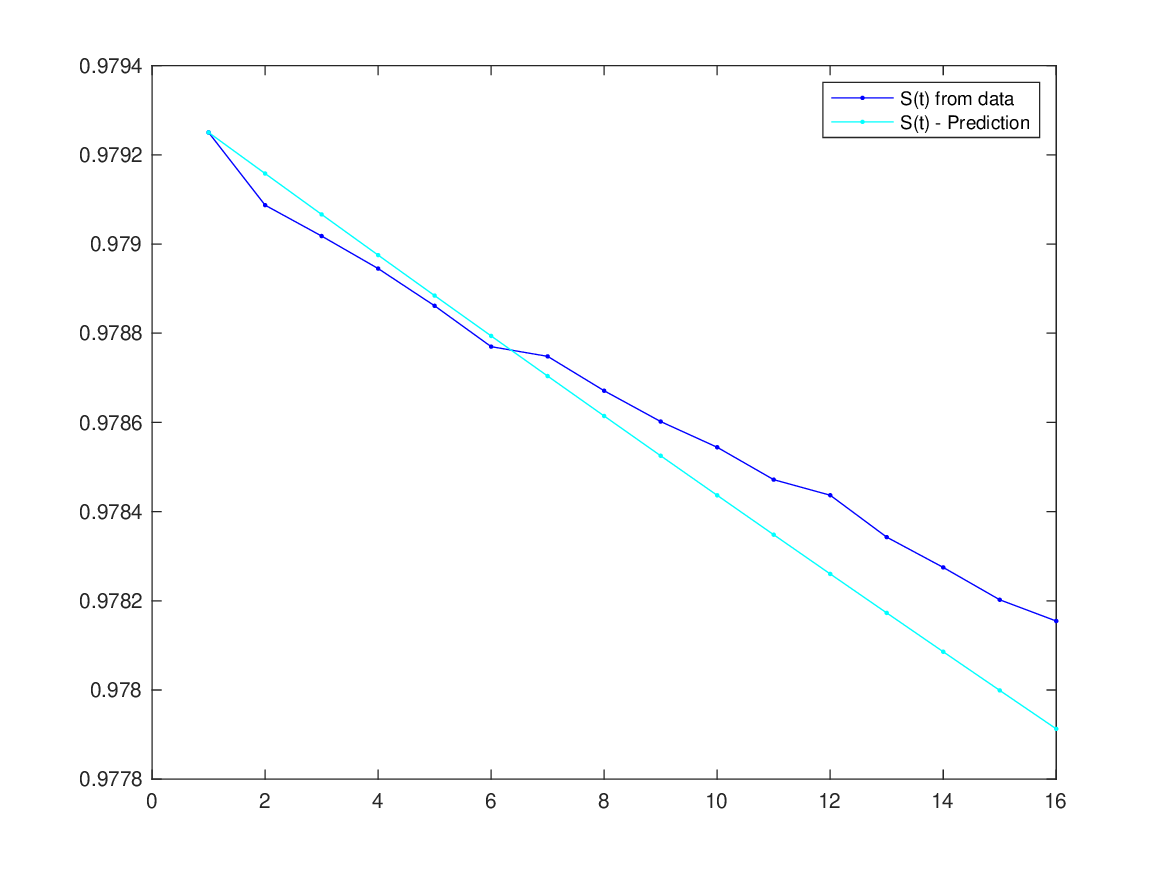}\vspace{-2mm}
\caption{t3}
\end{subfigure}
\begin{subfigure}{0.5\linewidth}
\centering
\includegraphics[scale=.32]{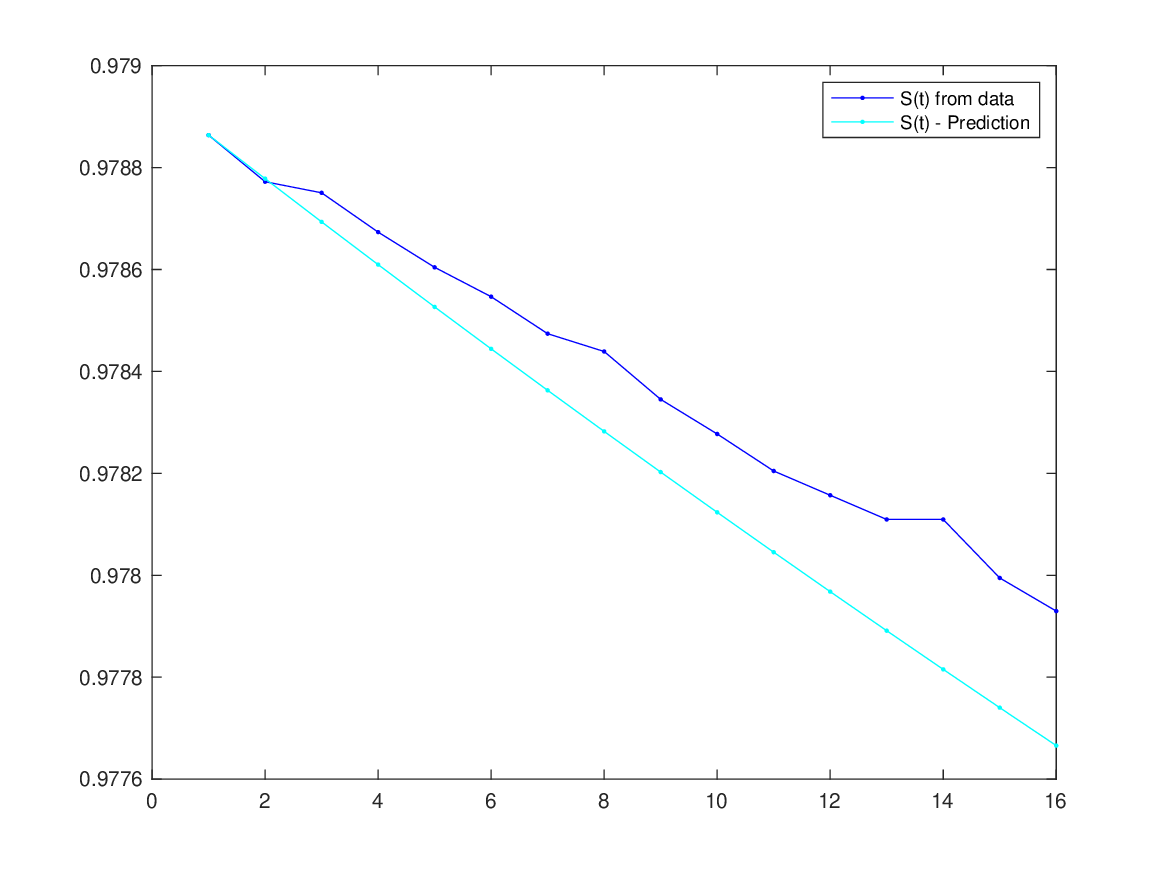}\vspace{-2mm}
\caption{t4}
\end{subfigure}
\caption{Prediction of $s(t)$ for the 4 starting dates and duration of 15 days
}\vspace{-.5cm}
\label{fig:predictionS15}
\end{figure}

\begin{figure}[H]
\begin{subfigure} {0.5\textwidth}
\centering
\includegraphics[scale=0.32]{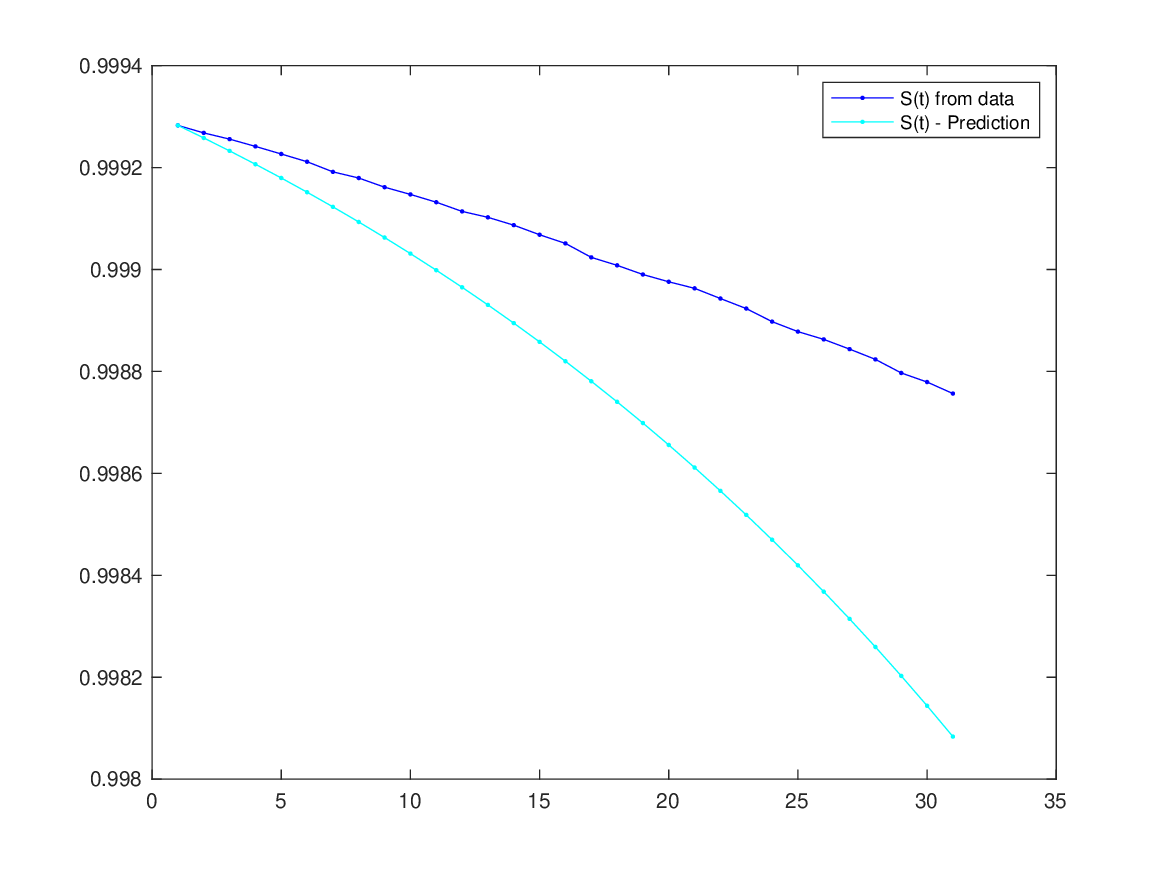}\vspace{-2mm}
\caption{t1}
\end{subfigure}%
\begin{subfigure}{.5\linewidth}
\centering
\includegraphics[scale=.32]{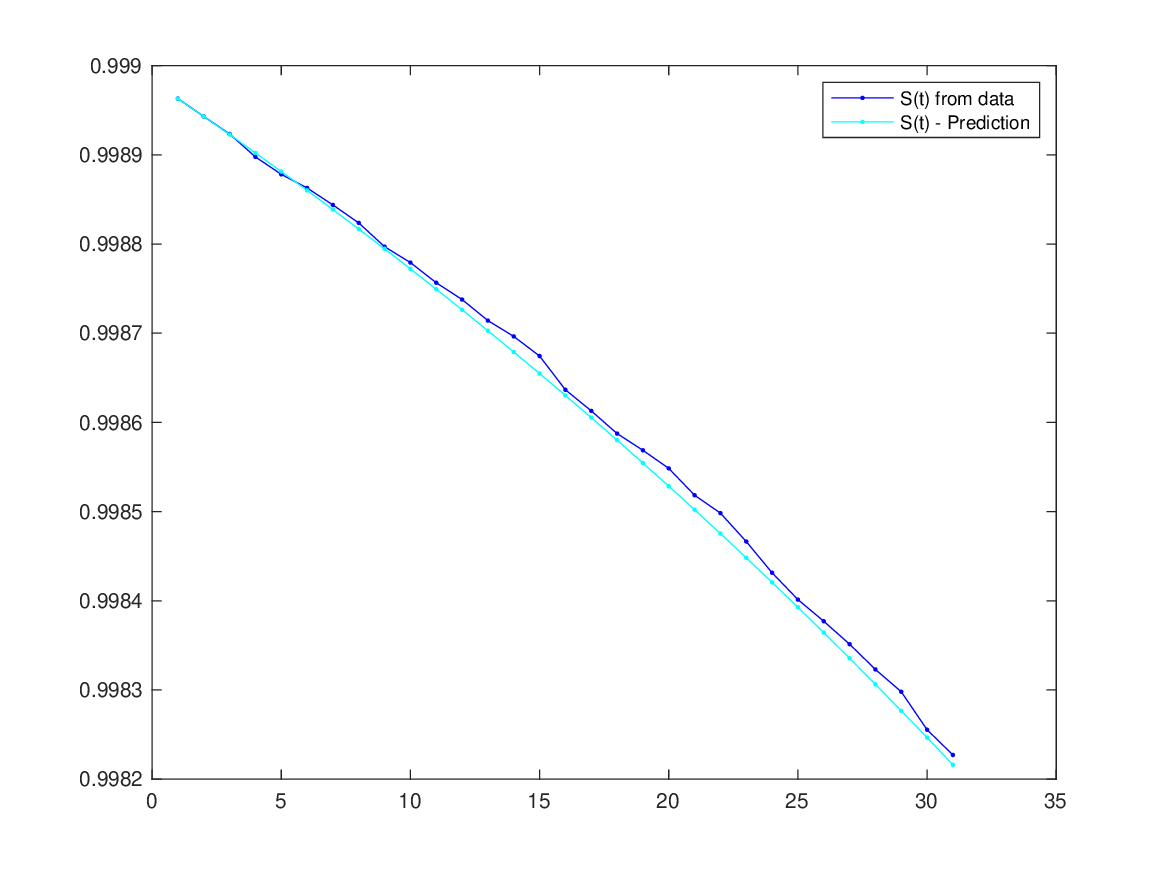}\vspace{-2mm}
\caption{t2}
\end{subfigure}
\begin{subfigure}{0.5\linewidth}
\centering
\includegraphics[scale=.32]{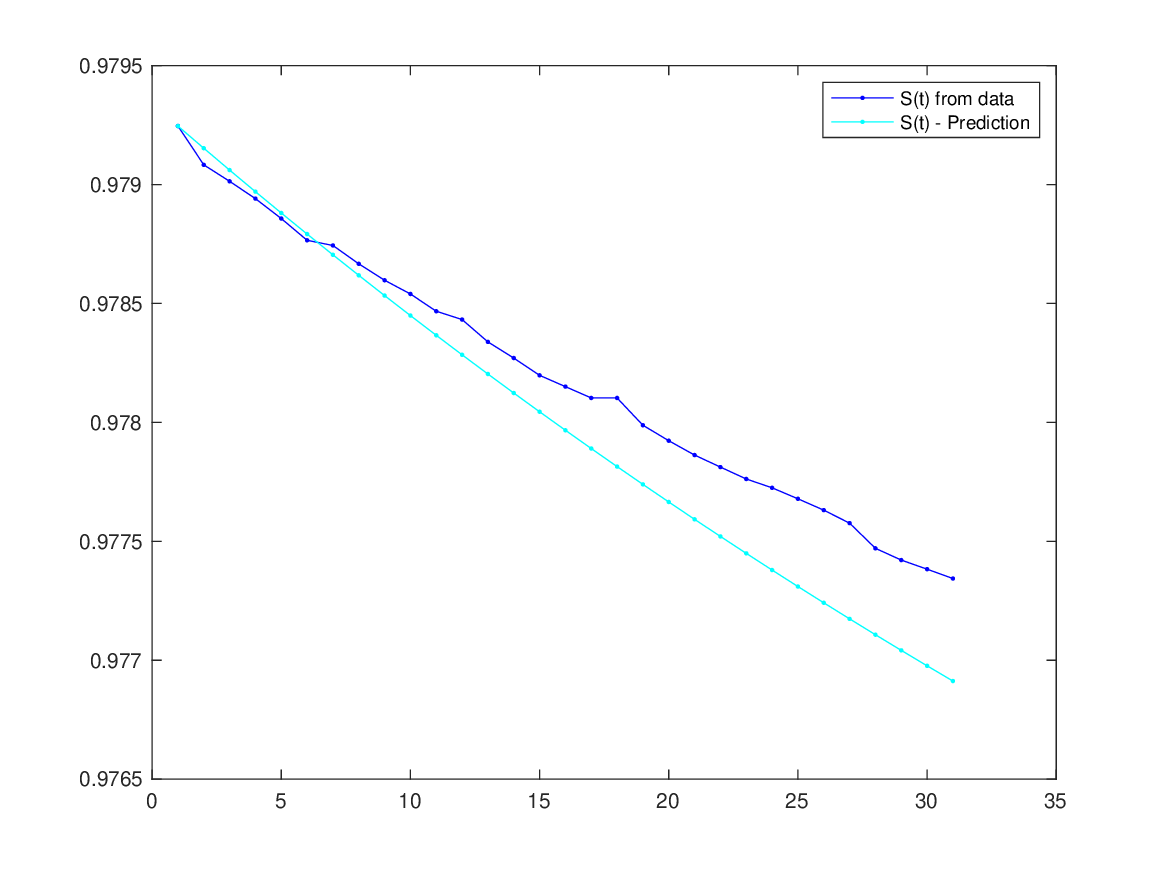}\vspace{-2mm}
\caption{t3}
\end{subfigure}
\begin{subfigure}{0.5\linewidth}
\centering
\includegraphics[scale=.32]{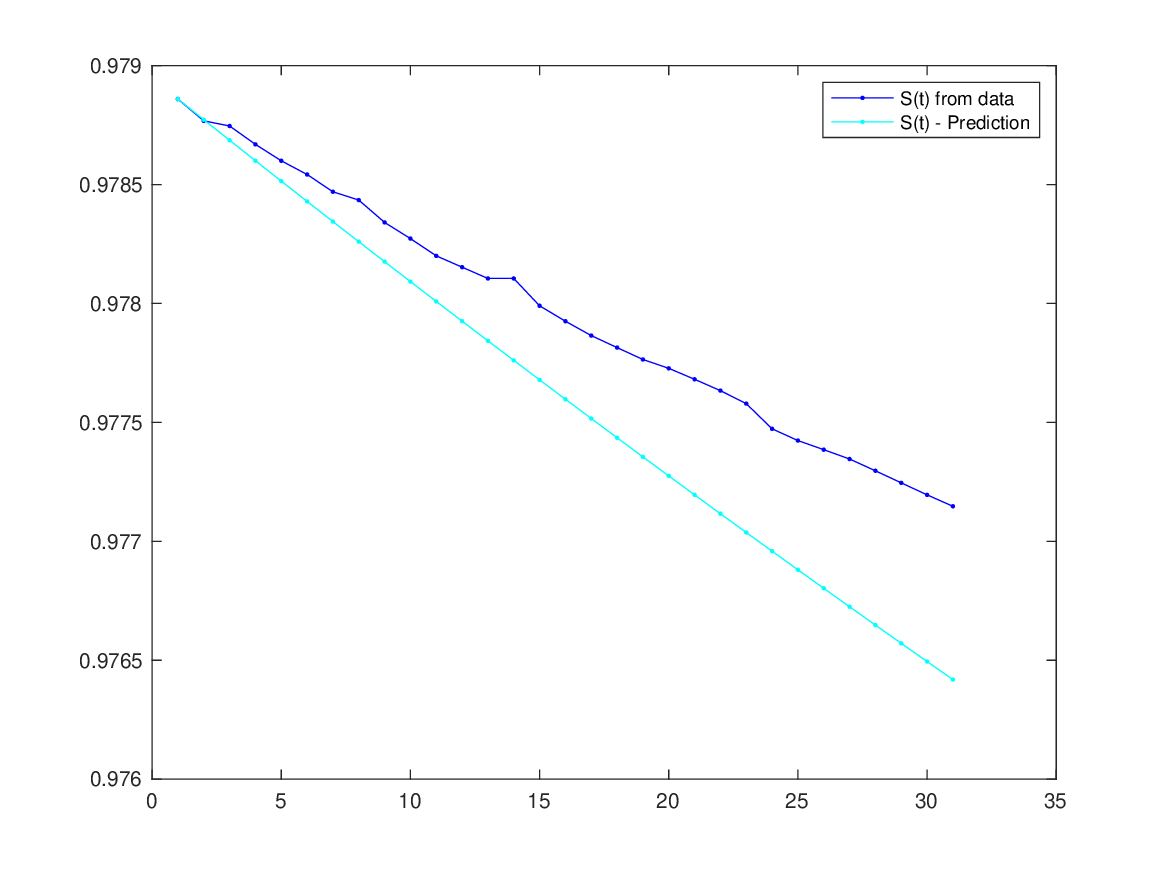}\vspace{-2mm}
\caption{t4}
\end{subfigure}
\caption{Prediction of $s(t)$ for the 4 starting dates and duration of 30 days
}\vspace{-.5cm}
\label{fig:predictionS30}
\end{figure}

\begin{figure}[H]
\begin{subfigure} {0.5\textwidth}
\centering
\includegraphics[scale=0.32]{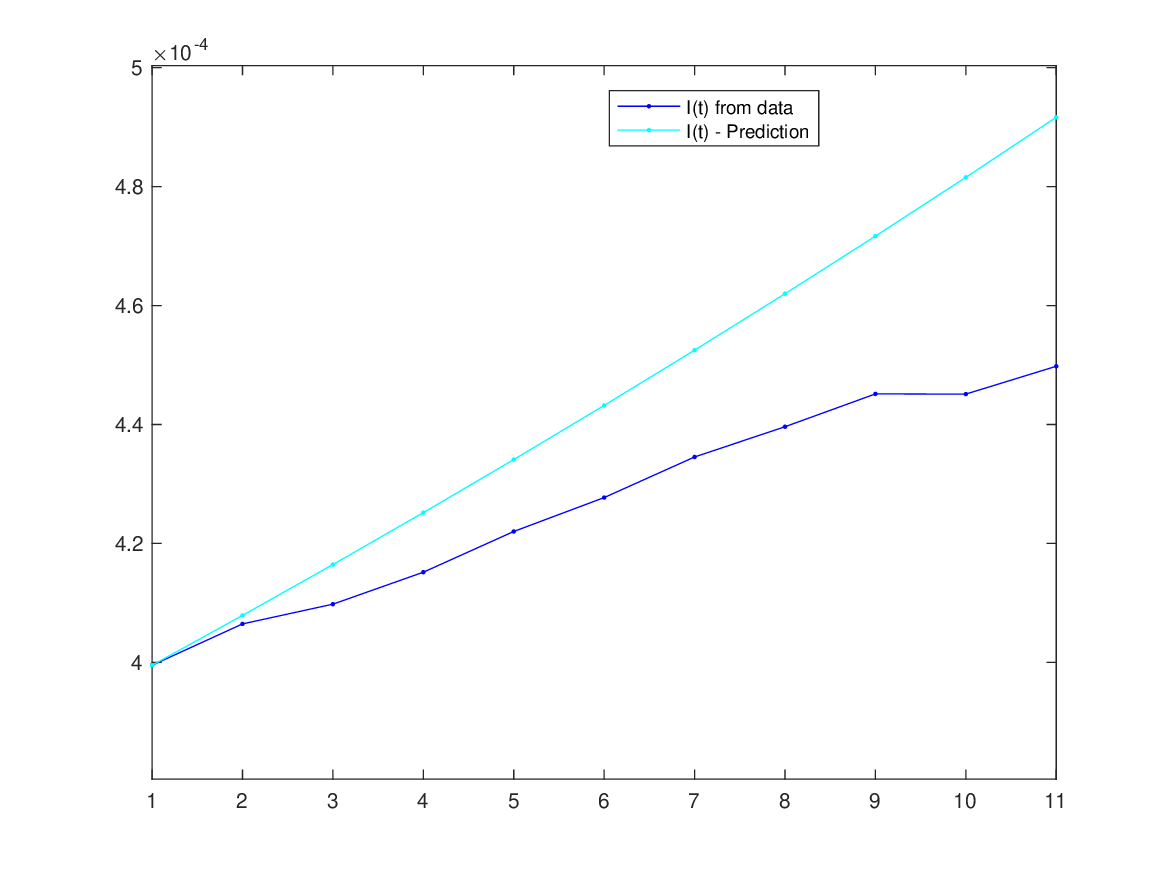}\vspace{-2mm}
\caption{t1}
\end{subfigure}%
\begin{subfigure}{.5\linewidth}
\centering
\includegraphics[scale=.32]{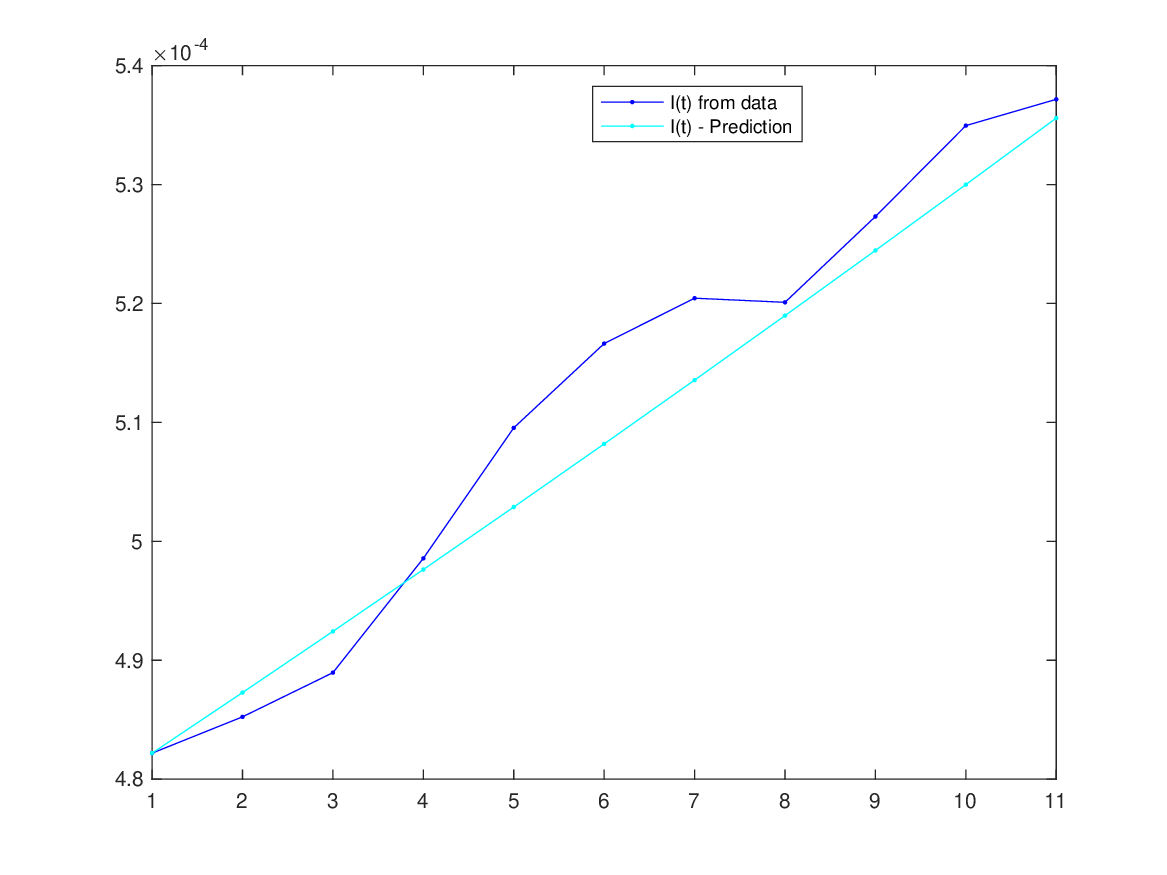}\vspace{-2mm}
\caption{t2}
\end{subfigure}
\begin{subfigure}{0.5\linewidth}
\centering
\includegraphics[scale=.32]{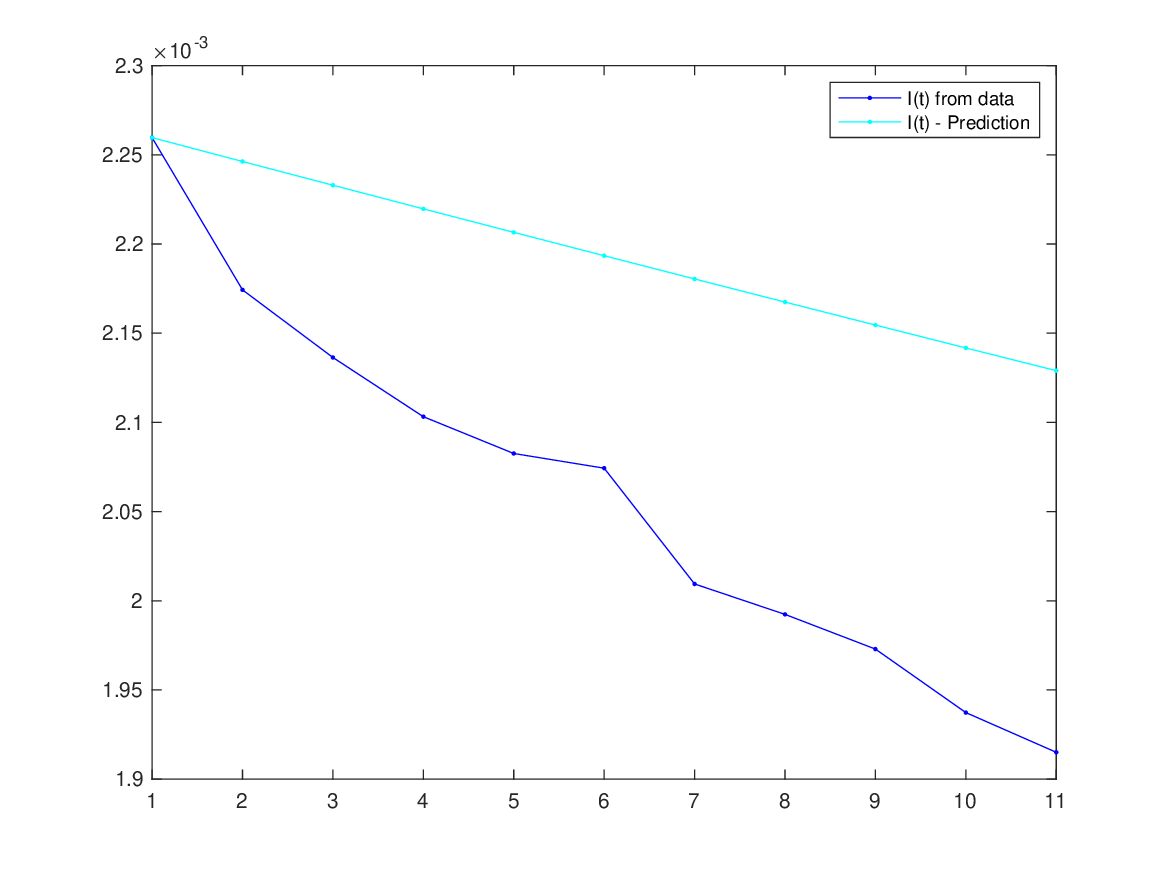}\vspace{-2mm}
\caption{t3}
\end{subfigure}
\begin{subfigure}{0.5\linewidth}
\centering
\includegraphics[scale=.32]{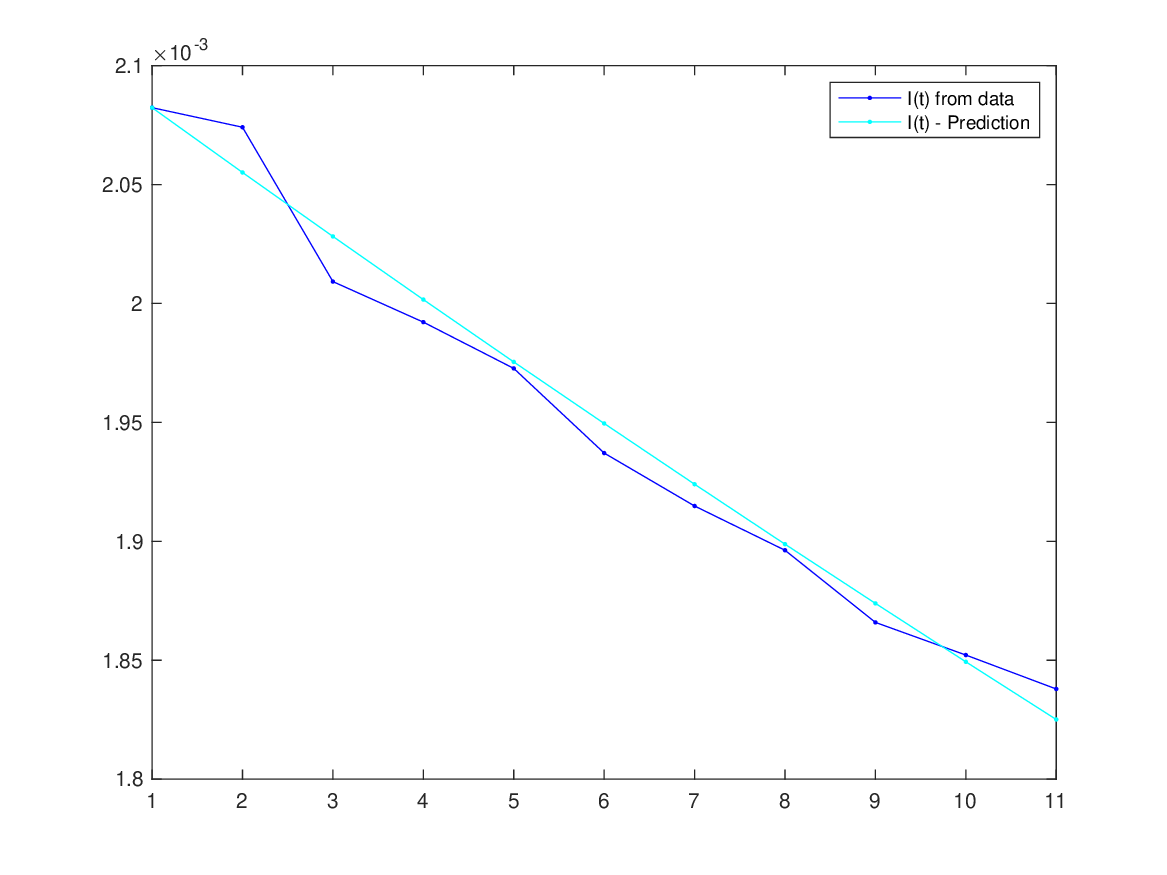}\vspace{-2mm}
\caption{t4}
\end{subfigure}
\caption{Prediction of $i(t)$ for the 4 starting dates and duration of 10 days
}\vspace{-.5cm}
\label{fig:predictionI10}
\end{figure}

\begin{figure}[H]
\begin{subfigure} {0.5\textwidth}
\centering
\includegraphics[scale=0.32]{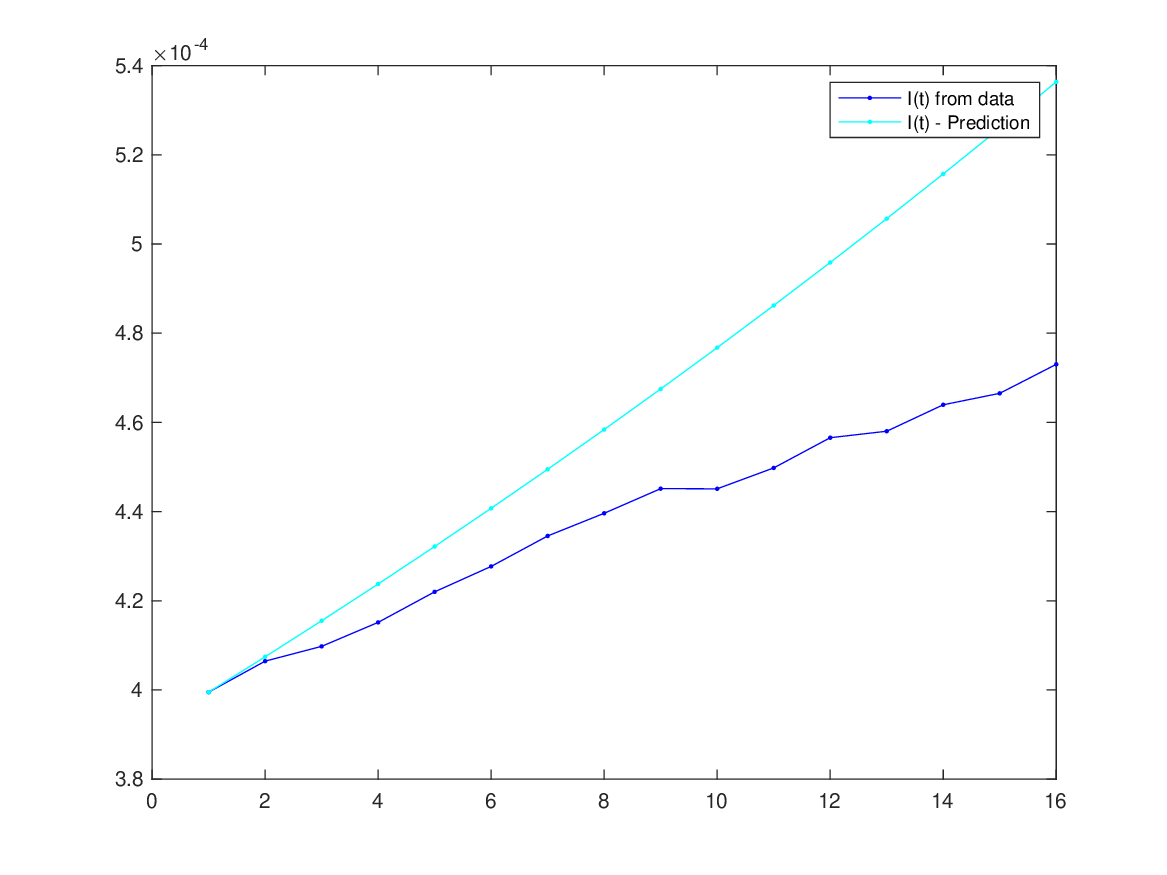}\vspace{-2mm}
\caption{t1}
\end{subfigure}%
\begin{subfigure}{.5\linewidth}
\centering
\includegraphics[scale=.32]{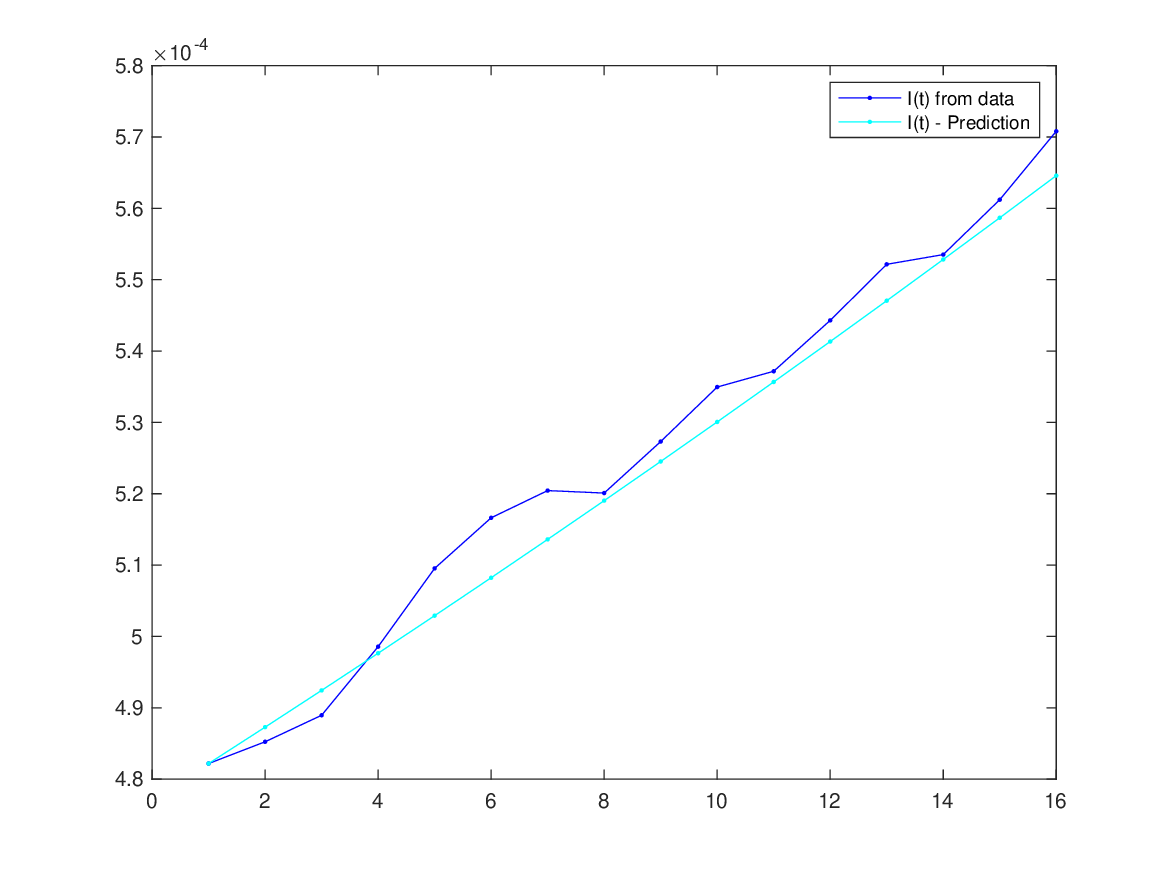}\vspace{-2mm}
\caption{t2}
\end{subfigure}
\begin{subfigure}{0.5\linewidth}
\centering
\includegraphics[scale=.32]{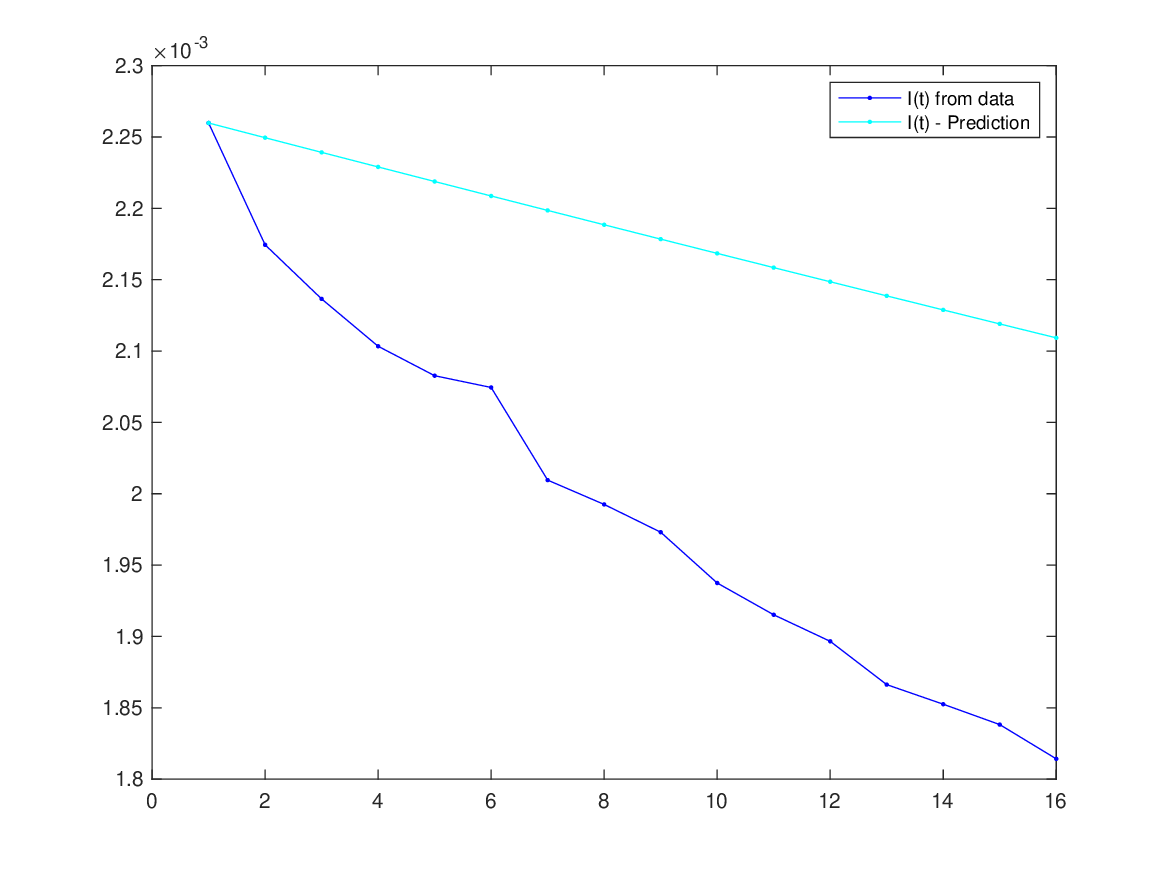}\vspace{-2mm}
\caption{t3}
\end{subfigure}
\begin{subfigure}{0.5\linewidth}
\centering
\includegraphics[scale=.32]{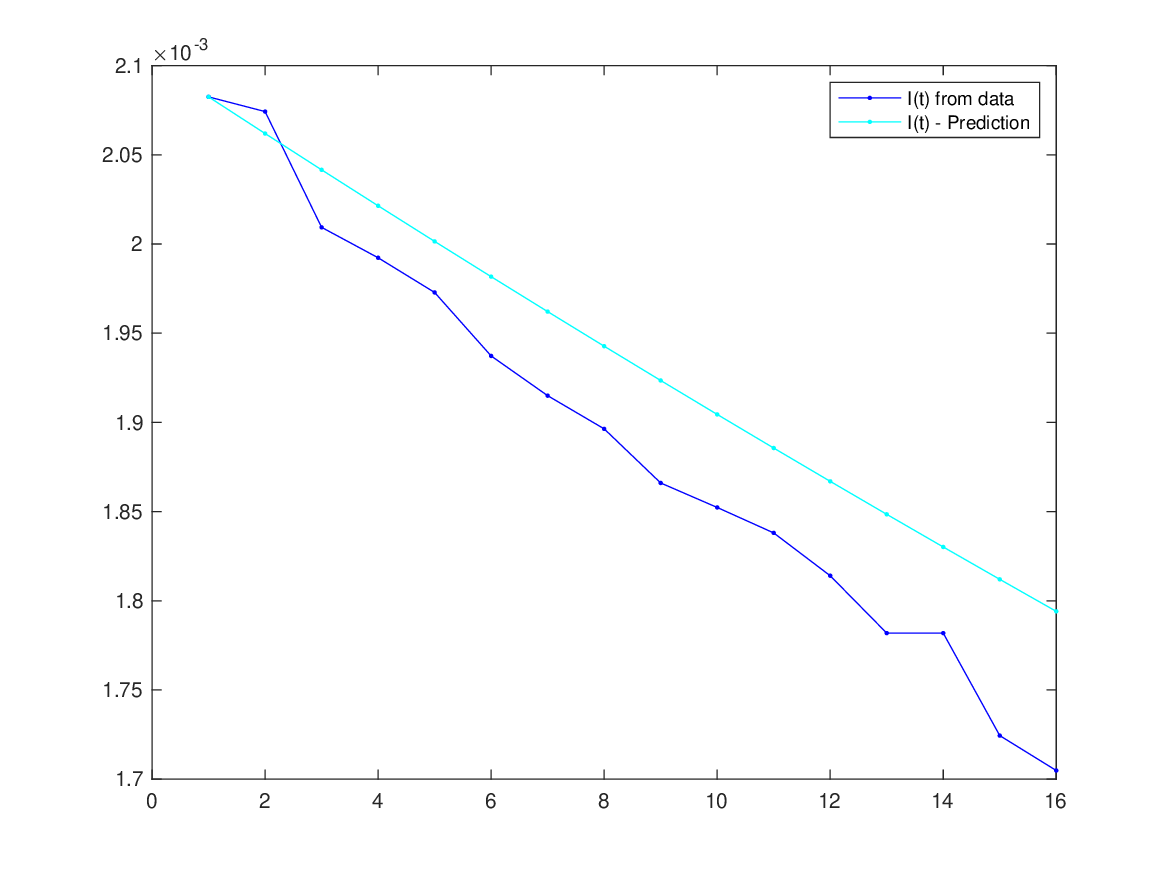}\vspace{-2mm}
\caption{t4}
\end{subfigure}
\caption{Prediction of $i(t)$ for the 4 starting dates and duration of 15 days
}\vspace{-.5cm}
\label{fig:predictionI15}
\end{figure}

\begin{figure}[H]
\begin{subfigure} {0.5\textwidth}
\centering
\includegraphics[scale=0.32]{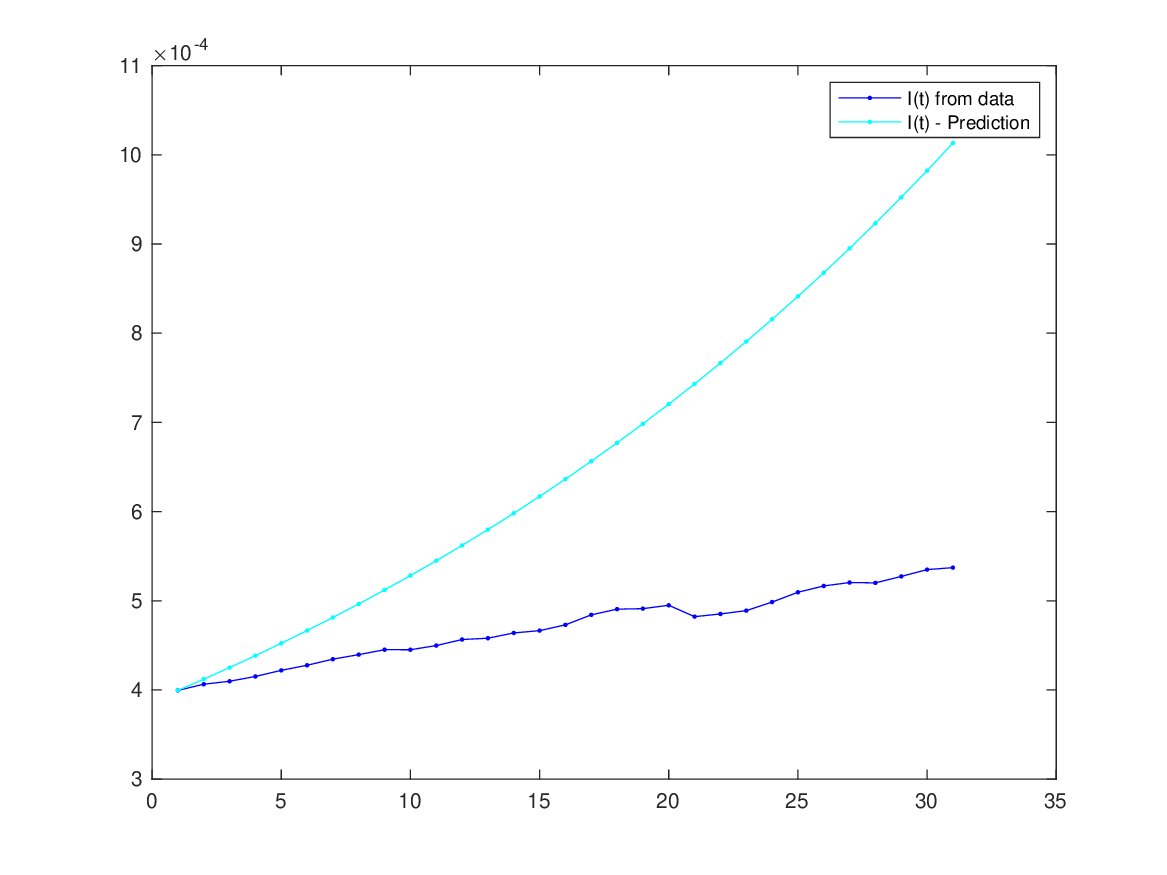}\vspace{-2mm}
\caption{t1}
\end{subfigure}%
\begin{subfigure}{.5\linewidth}
\centering
\includegraphics[scale=.32]{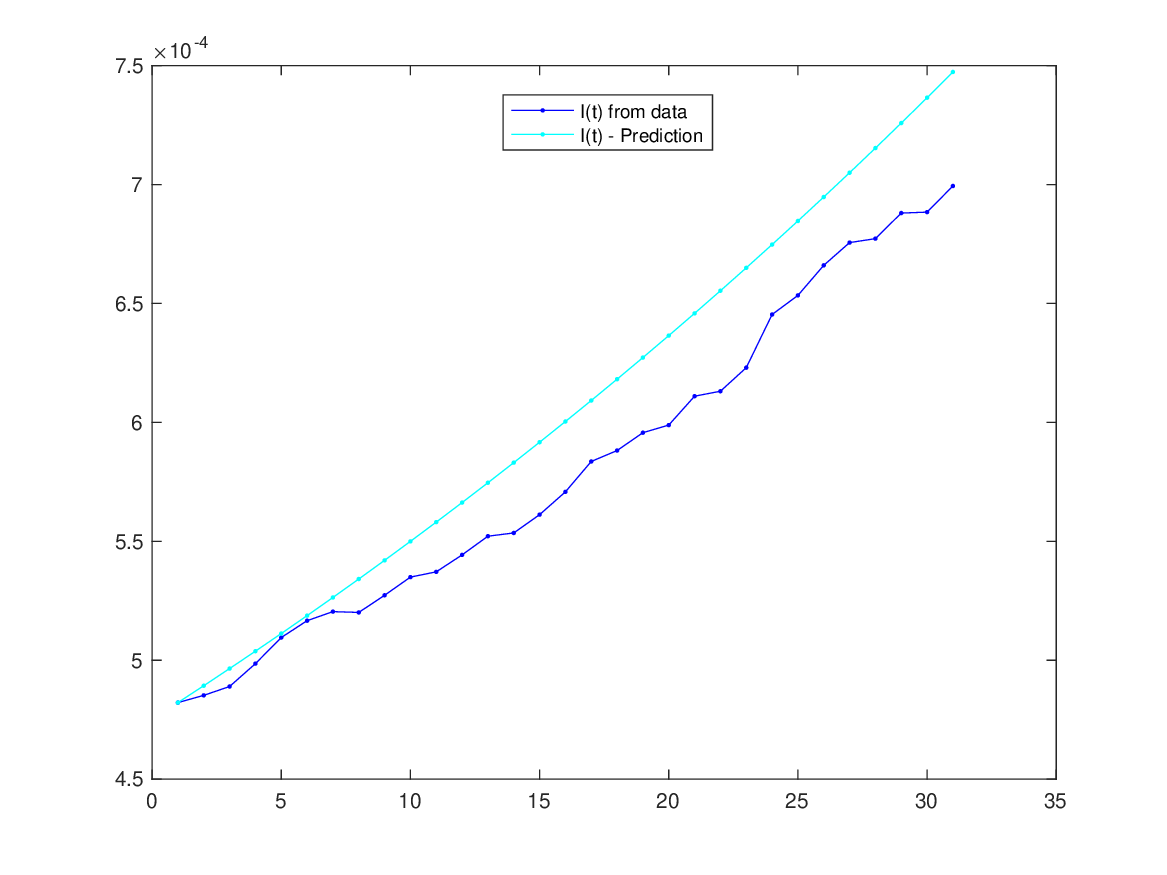}\vspace{-2mm}
\caption{t2}
\end{subfigure}
\begin{subfigure}{0.5\linewidth}
\centering
\includegraphics[scale=.32]{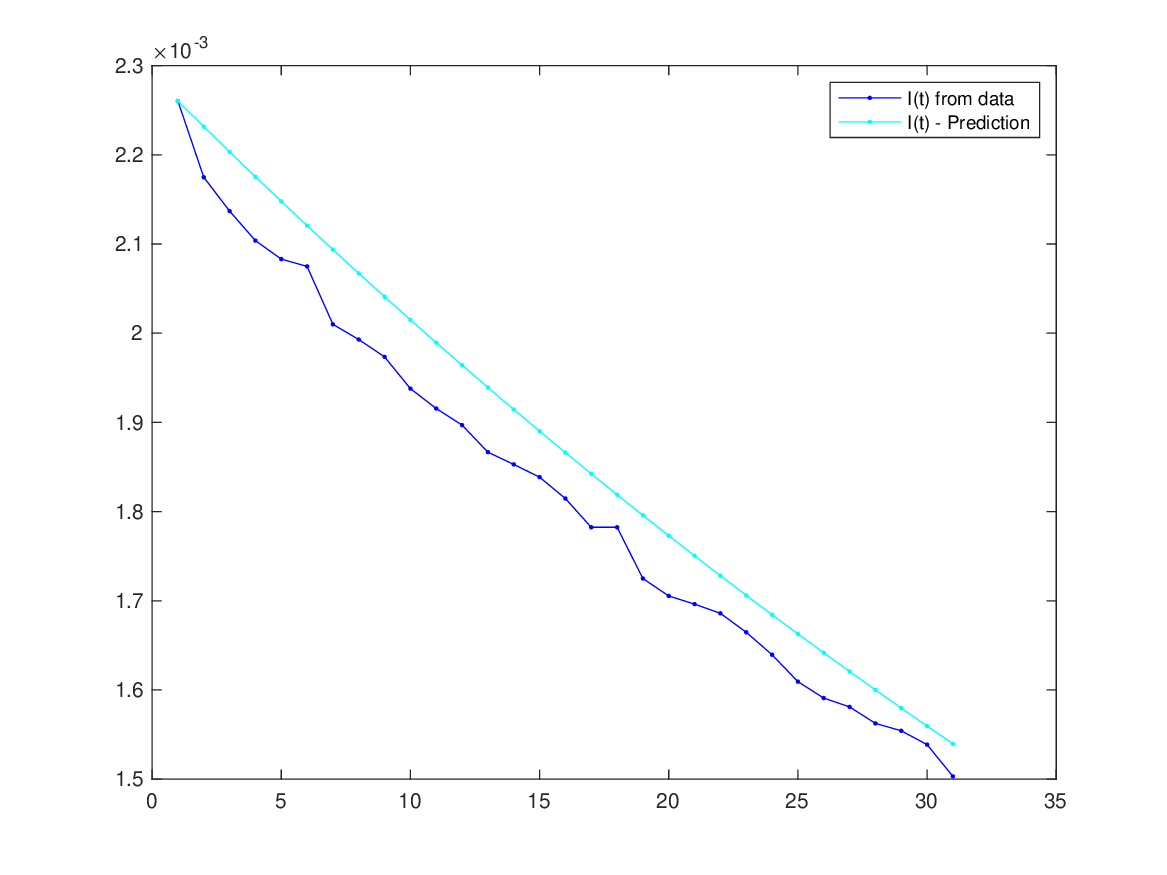}\vspace{-2mm}
\caption{t3}
\end{subfigure}
\begin{subfigure}{0.5\linewidth}
\centering
\includegraphics[scale=.32]{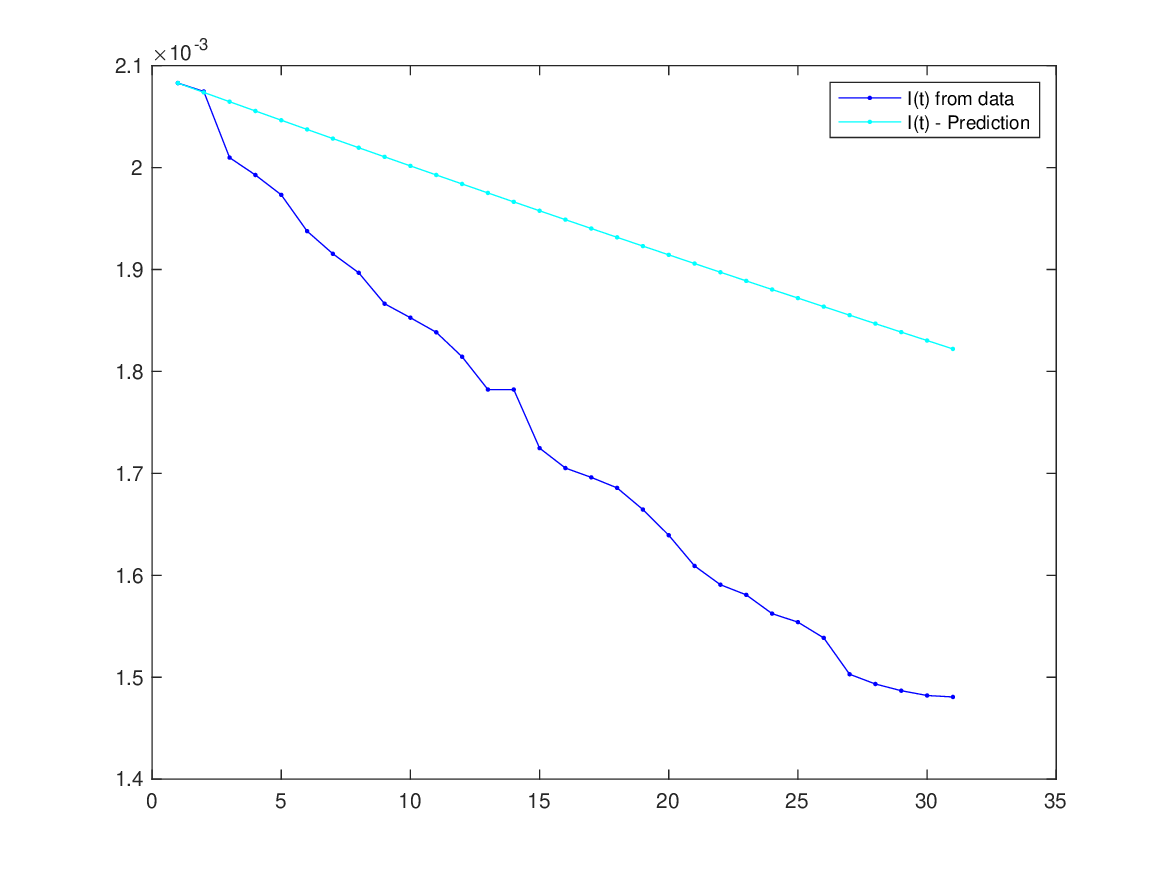}\vspace{-2mm}
\caption{t4}
\end{subfigure}
\caption{Prediction of $i(t)$ for the 4 starting dates and duration of 30 days
}\vspace{-.5cm}
\label{fig:predictionI30}
\end{figure}

\begin{figure}[H]
\begin{subfigure} {0.5\textwidth}
\centering
\includegraphics[scale=0.32]{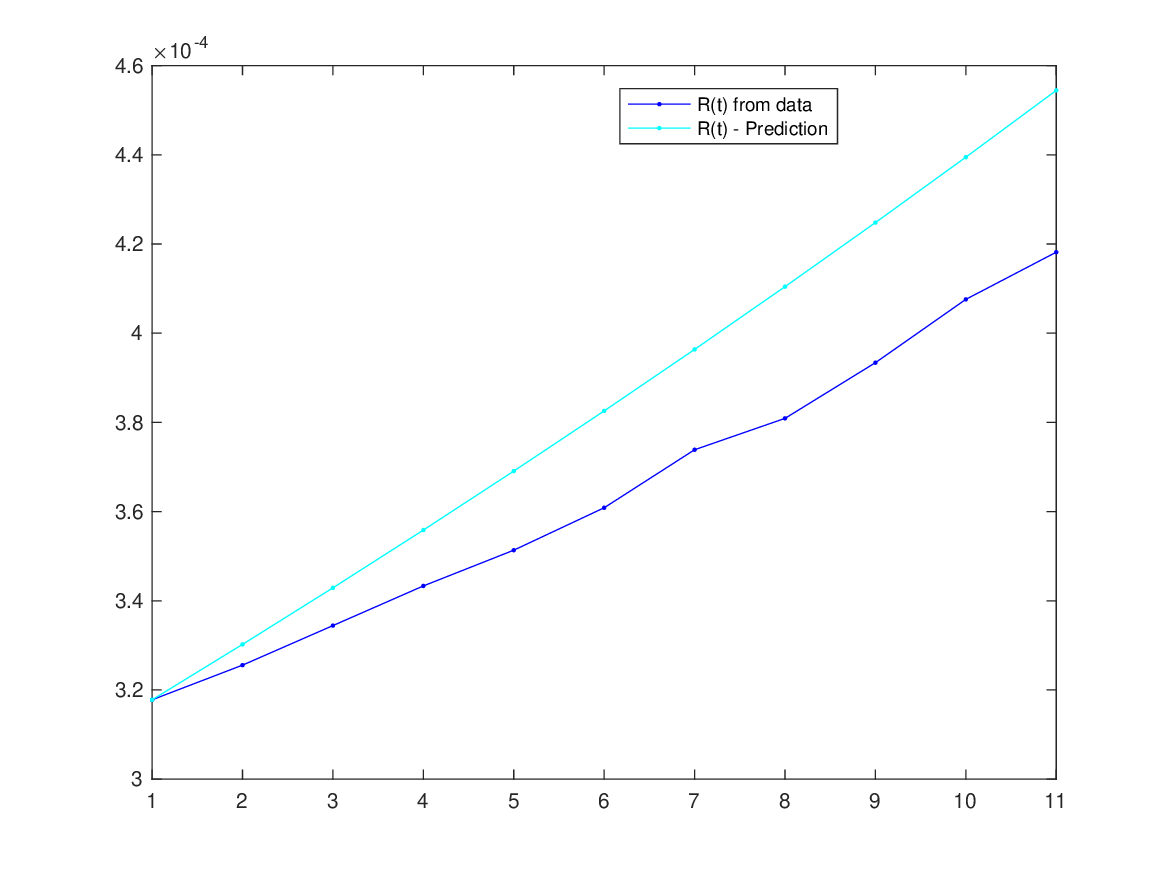}\vspace{-2mm}
\caption{t1}
\end{subfigure}%
\begin{subfigure}{.5\linewidth}
\centering
\includegraphics[scale=.32]{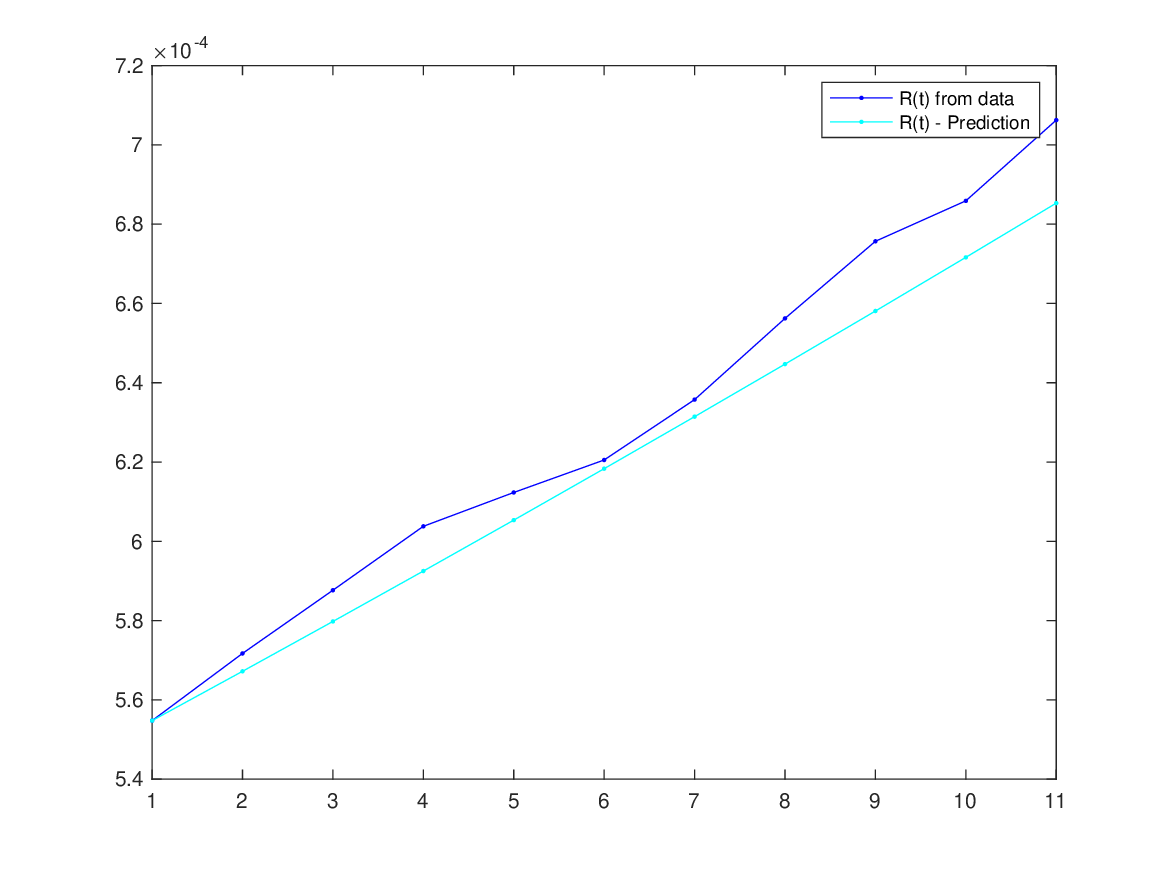}\vspace{-2mm}
\caption{t2}
\end{subfigure}
\begin{subfigure}{0.5\linewidth}
\centering
\includegraphics[scale=.32]{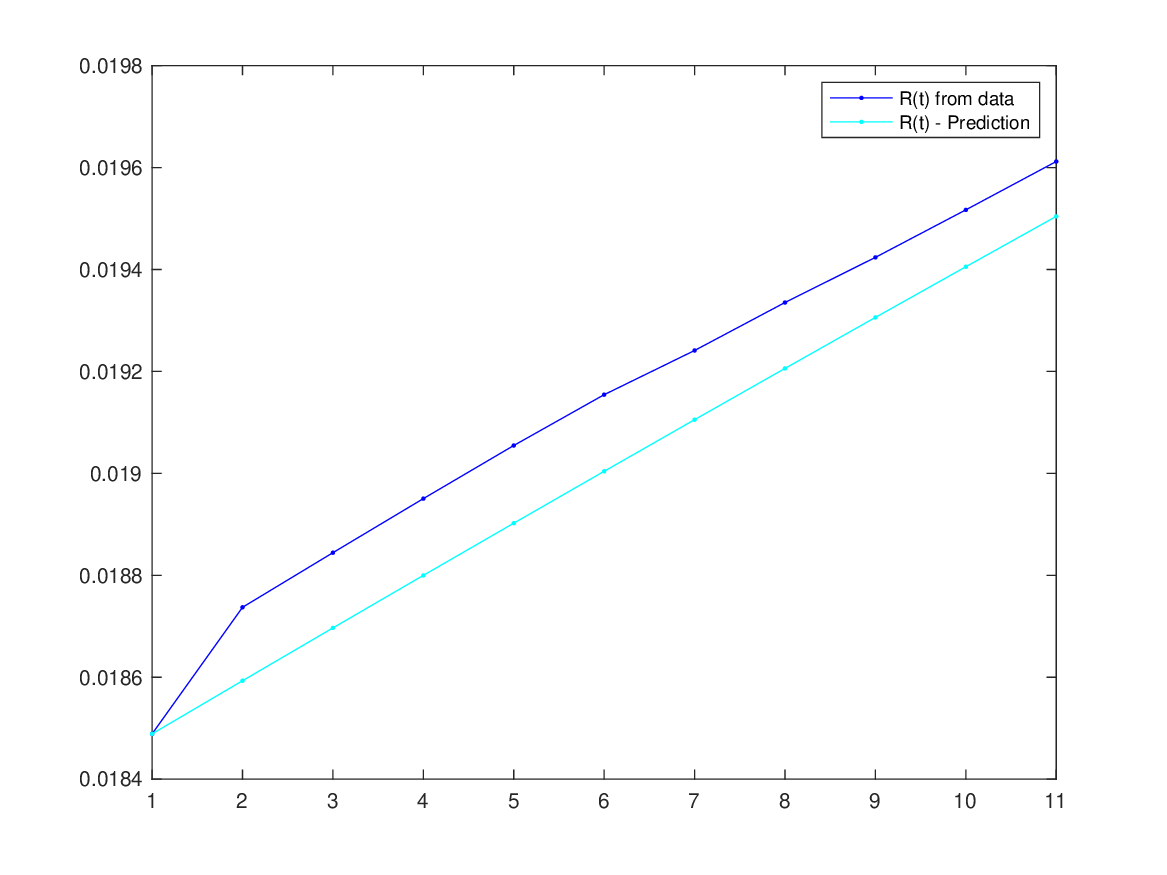}\vspace{-2mm}
\caption{t3}
\end{subfigure}
\begin{subfigure}{0.5\linewidth}
\centering
\includegraphics[scale=.32]{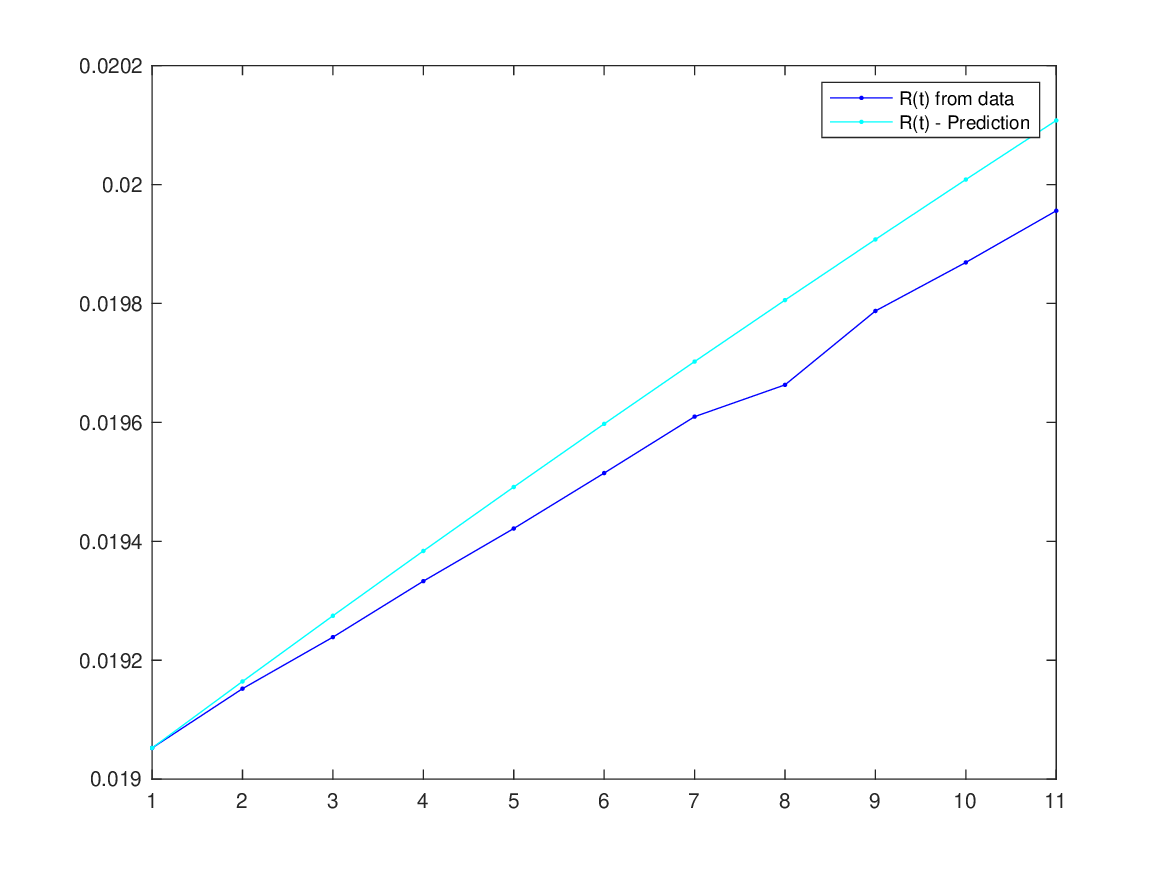}\vspace{-2mm}
\caption{t4}
\end{subfigure}
\caption{Prediction of $r(t)$ for the 4 starting dates and duration of 10 days
}\vspace{-.5cm}
\label{fig:predictionR10}
\end{figure}

\begin{figure}[H]
\begin{subfigure} {0.5\textwidth}
\centering
\includegraphics[scale=0.32]{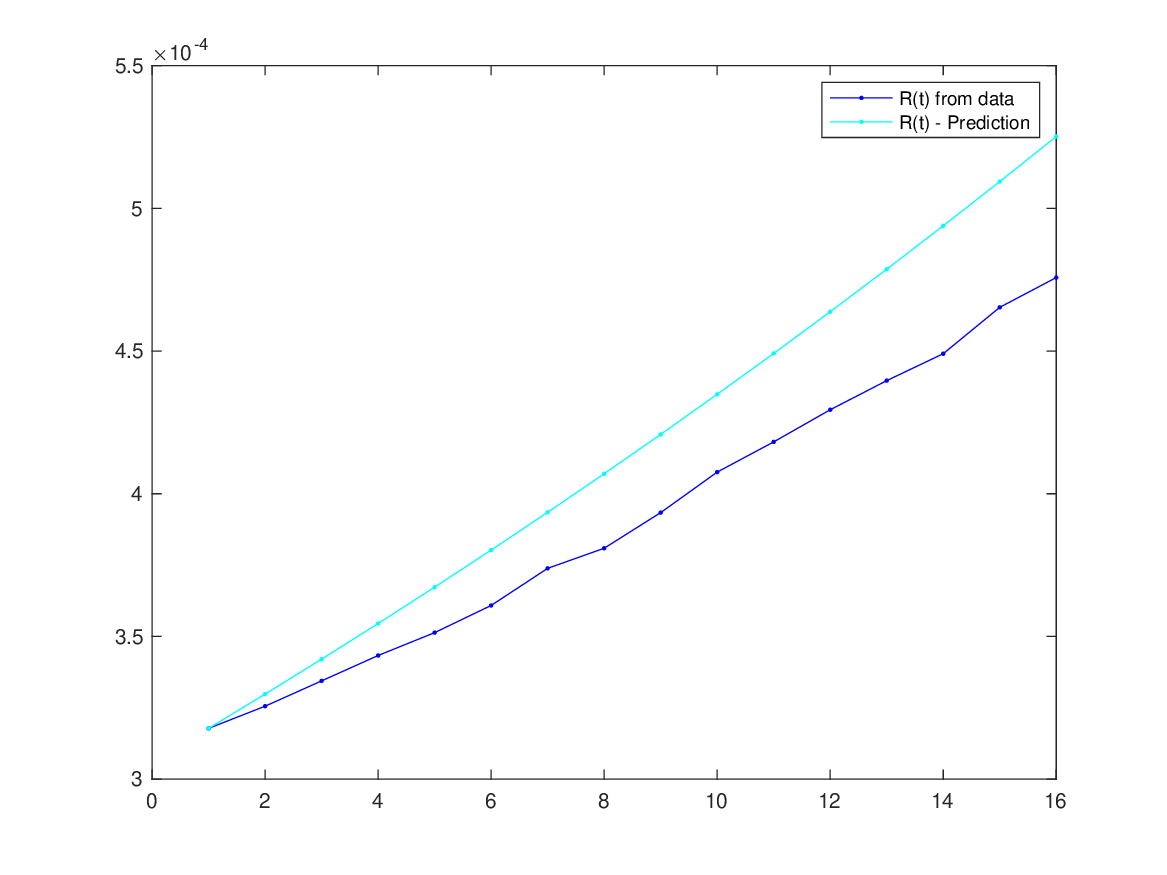}\vspace{-2mm}
\caption{t1}
\end{subfigure}%
\begin{subfigure}{.5\linewidth}
\centering
\includegraphics[scale=.32]{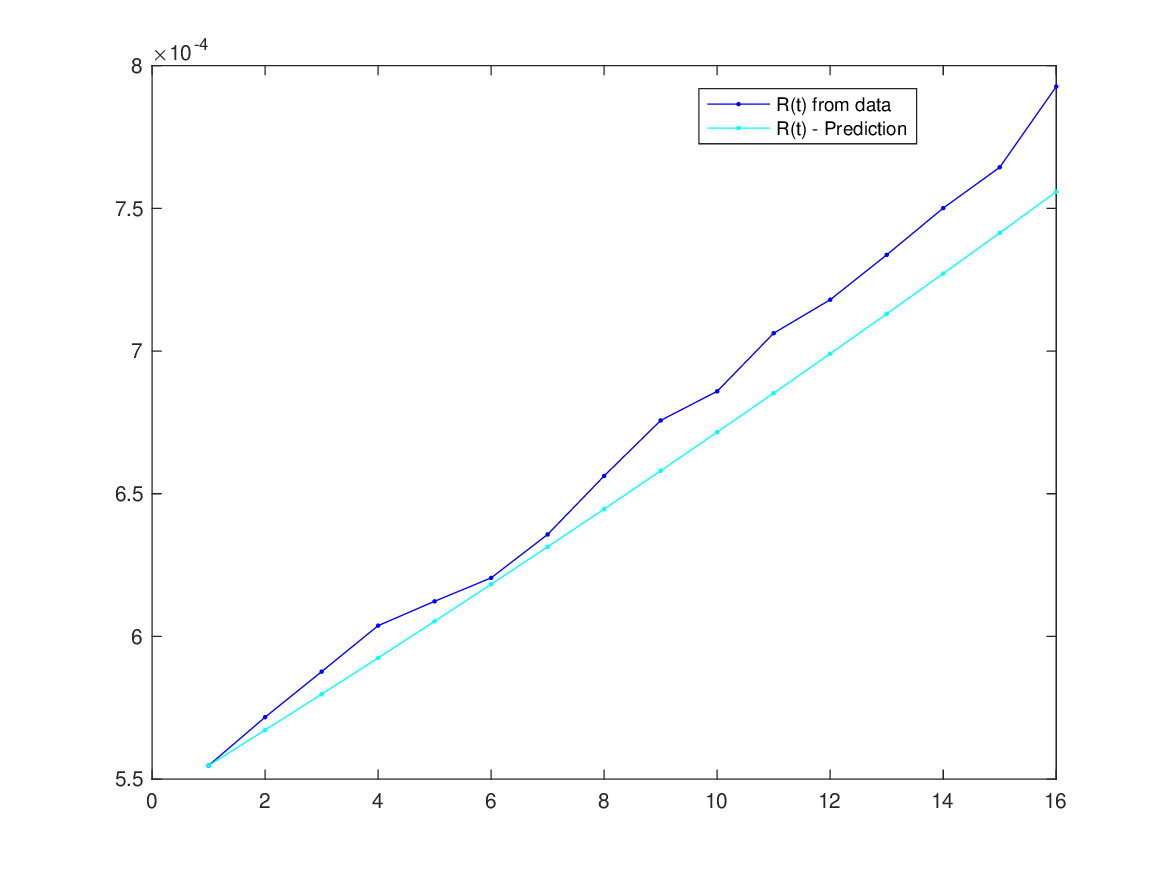}\vspace{-2mm}
\caption{t2}
\end{subfigure}
\begin{subfigure}{0.5\linewidth}
\centering
\includegraphics[scale=.32]{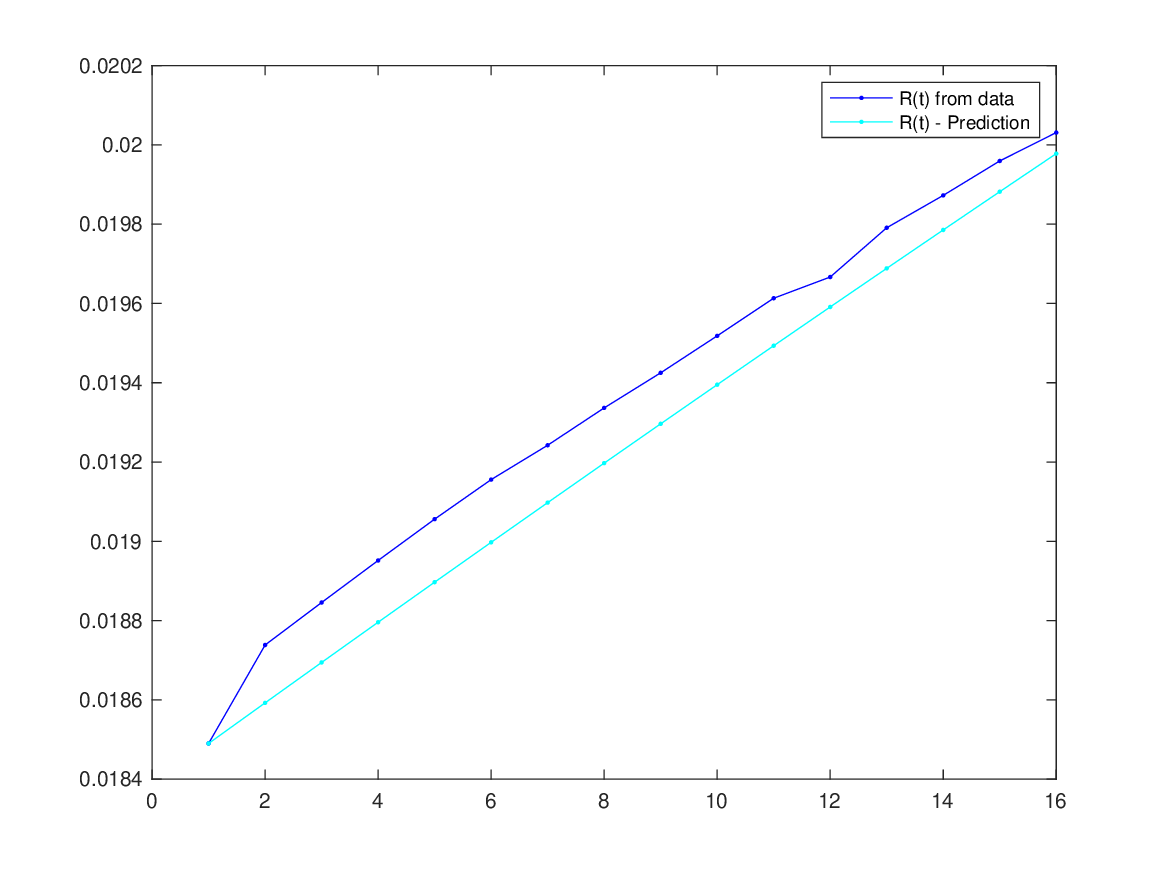}\vspace{-2mm}
\caption{t3}
\end{subfigure}
\begin{subfigure}{0.5\linewidth}
\centering
\includegraphics[scale=.32]{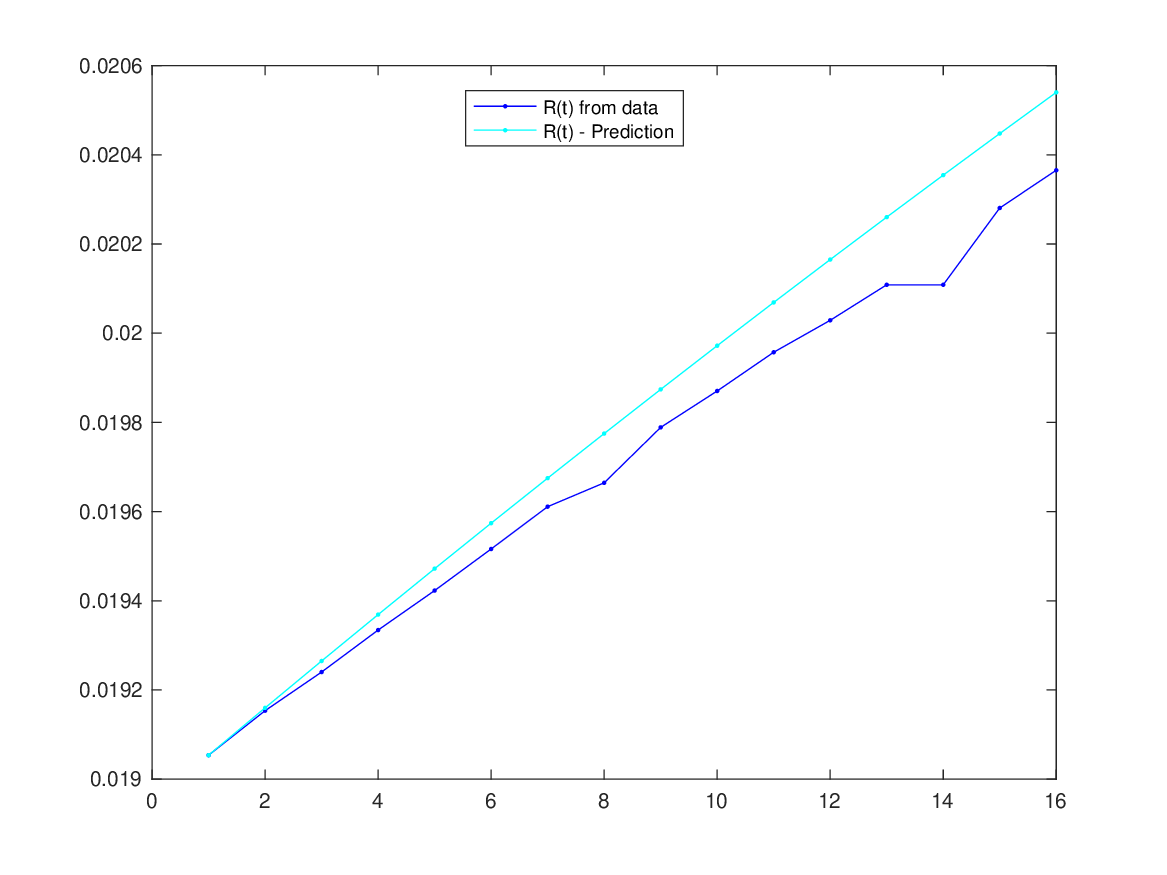}\vspace{-2mm}
\caption{t4}
\end{subfigure}
\caption{Prediction of $r(t)$ for the 4 starting dates and duration of 15 days
}\vspace{-.5cm}
\label{fig:predictionR15}
\end{figure}

\begin{figure}[H]
\begin{subfigure} {0.5\textwidth}
\centering
\includegraphics[scale=0.32]{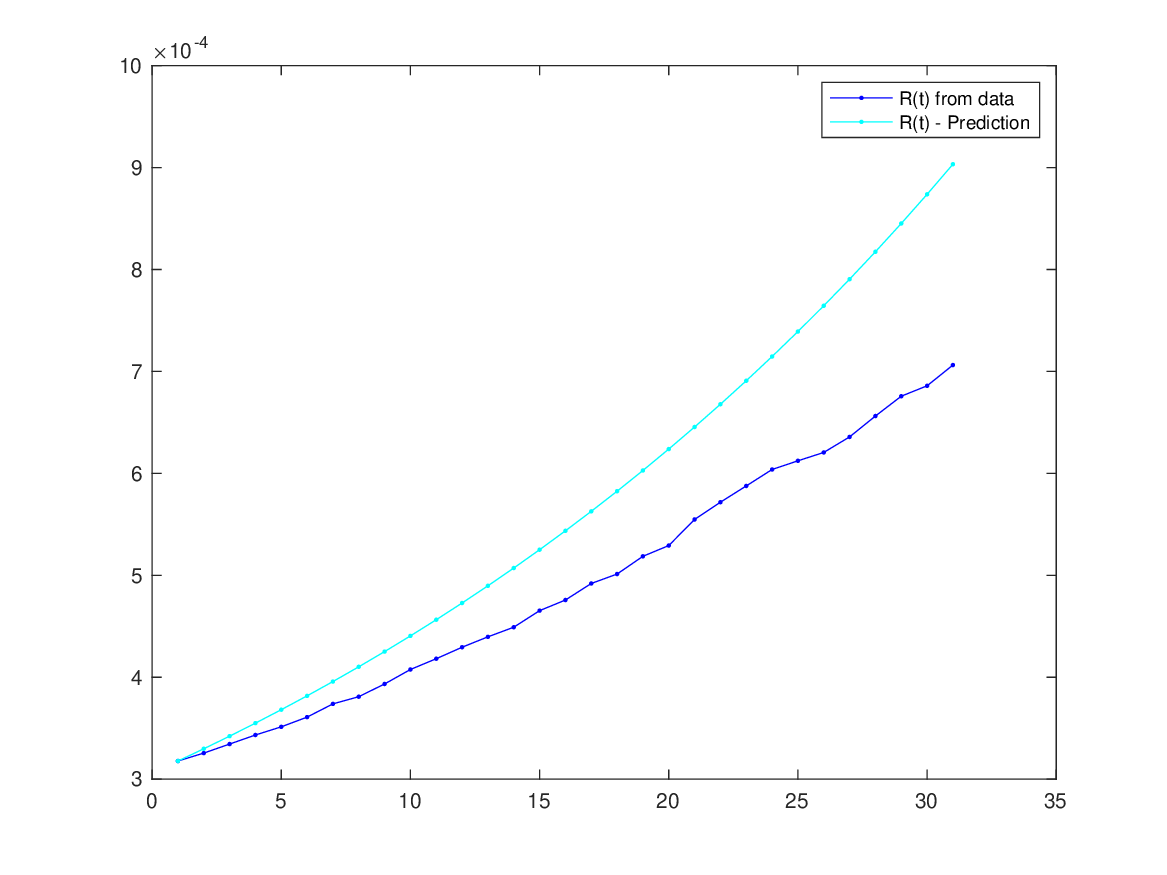}\vspace{-2mm}
\caption{t1}
\end{subfigure}%
\begin{subfigure}{.5\linewidth}
\centering
\includegraphics[scale=.32]{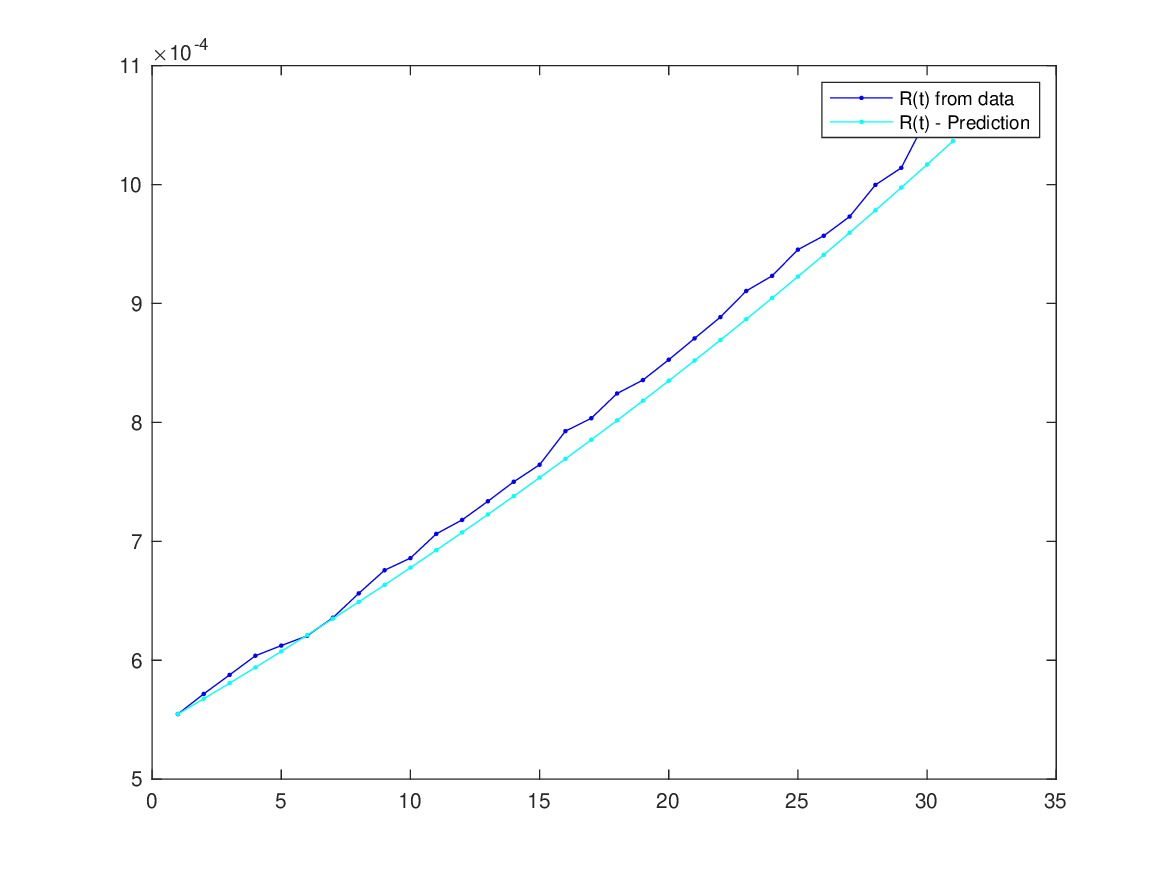}\vspace{-2mm}
\caption{t2}
\end{subfigure}
\begin{subfigure}{0.5\linewidth}
\centering
\includegraphics[scale=.32]{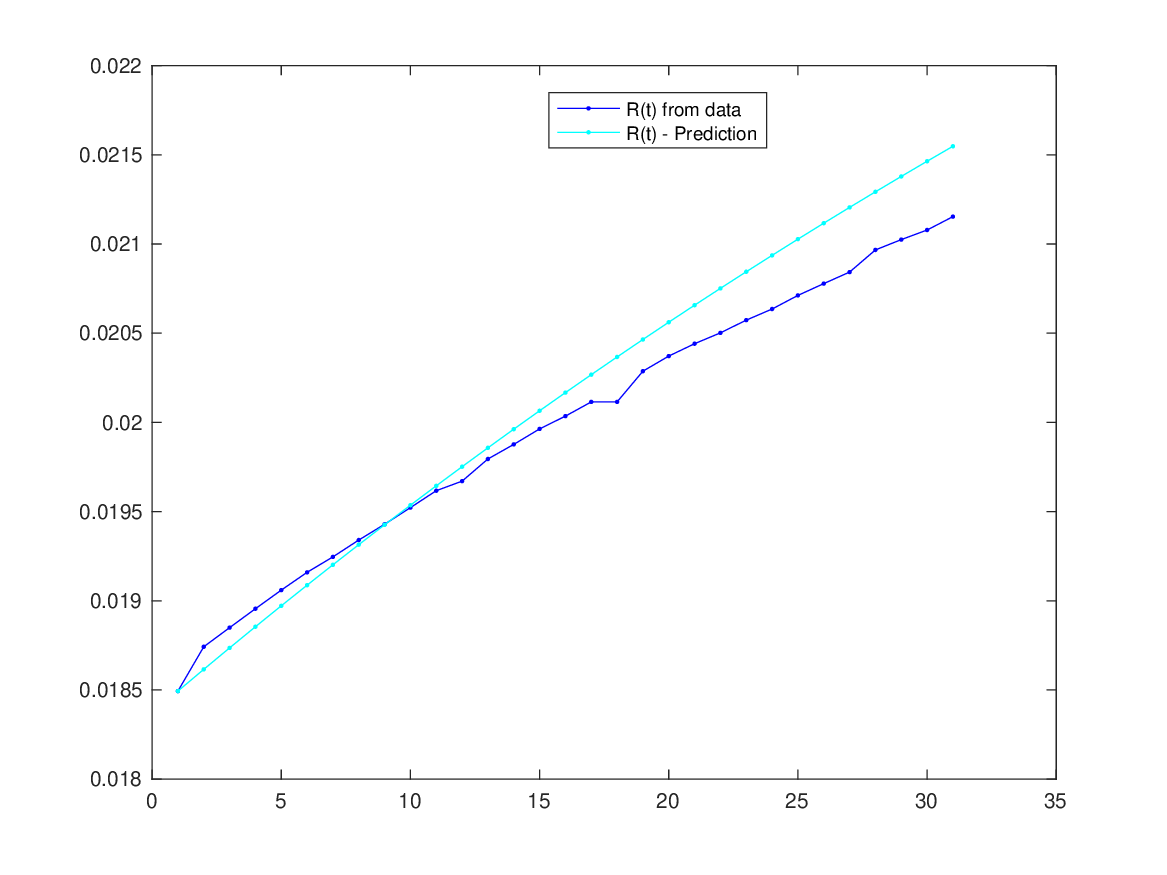}\vspace{-2mm}
\caption{t3}
\end{subfigure}
\begin{subfigure}{0.5\linewidth}
\centering
\includegraphics[scale=.32]{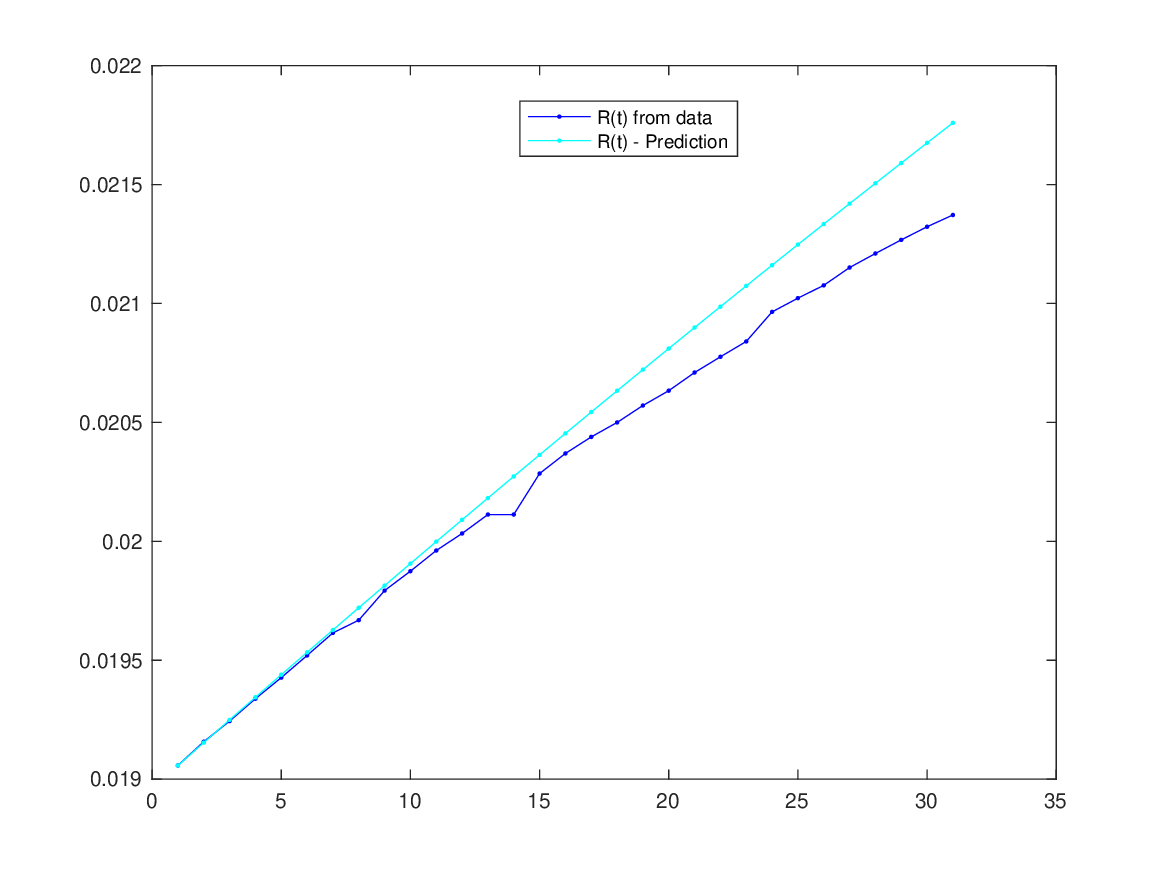}\vspace{-2mm}
\caption{t4}
\end{subfigure}
\caption{Prediction of $r(t)$ for the 4 starting dates and duration of 30 days
}\vspace{-.5cm}
\label{fig:predictionR30}
\end{figure}

We consider 4 starting dates $t_0$ indicated in Table \ref{tab:datesofintervals} and test the efficiency of the constant coefficient model for different durations $k=5,10,15,20,25,30$. Figures \ref{fig:predictionS10}, \ref{fig:predictionS15}, \ref{fig:predictionS30}, show the predicted $s(t)$ for the considered starting dates with a duration of 10, 15 and 30 days respectively.
Figures \ref{fig:predictionI10}, \ref{fig:predictionI15}, \ref{fig:predictionI30}, show the predicted $i(t)$ for the considered starting dates with a duration of 10, 15 and 30 days respectively. Figures \ref{fig:predictionR10}, \ref{fig:predictionR15}, \ref{fig:predictionR30}, show the predicted $r(t)$ for the considered starting dates with a duration of 10, 15 and 30 days respectively. Moreover, Table \ref{tab:RelErrPredM1} summarizes the corresponding relative errors between the predicted compartments and the exact data.

\begin{table}[H]
    \centering
    \setlength{\tabcolsep}{10pt}
{\renewcommand{\arraystretch}{1.4}
    \begin{tabular}{
    |l||c|c|c|c|}\hline
         Label& $t1$ & $t2$& $t3$ &$t4$\\ \hline
 Starting Date&16/05/2020 & 5/6/2020& 16/05/2021 & 20/05/2021 \\
  \hline
    \end{tabular}\vspace{-3mm}
    \caption{The considered starting dates $t_0=t_i$ for testing}\vspace{-3mm}
    \label{tab:datesofintervals}}
\end{table}

Similarly to previous tests, the relative error for $s(t)$ was the smallest and of order $10^{-4}$ or $10^{-5}$ (depending of the starting date) and slightly increases with the increase of the duration. The relative error of $s(t)$ also slightly increases with the increase of the duration, starting with an error of order $10^{-2}$ or $10^{-3}$. However, the relative error is more affected by the increase of the prediction duration $k$, as it increases from $10^{-3}$ to $10^{-2}$ to $10^{-1}$ (t3). 

This difference in sensitivity to the prediction duration can be seen in the figures where for $s(t)$ and $r(t)$, the prediction curves are a good approximation of the exact data for the majority of cases and durations (except for t1 and $k=30$, figure \ref{fig:predictionS30}). On the other hand, given that $i(t)$ varies more than $s(t)$ and $i(t)$, the predicted $i(t)$ may or may not be a good approximation depending on the similarity of the situation where the parameters are approximated to that of the prediction time. For example, the starting date of $t2$  is 20 days after $t1$, and $t4$ is 4 days after $t3$. Yet, the predicted $i(t)$ for duration $k=10$ or $15$ (figures \ref{fig:predictionI10}, \ref{fig:predictionI15}) is a very good approximation of the exact data for $t2$ and $t4$, unlike $t1$ and $t3$. For duration $k=30$ (figures \ref{fig:predictionI30}), the predicted $i(t)$ for $t2$ and $t3$ better approximated the exact data than for $t1$ and $t4$.

\begin{figure}[H]
\begin{subfigure} {0.5\textwidth}
\centering
\includegraphics[scale=0.32]{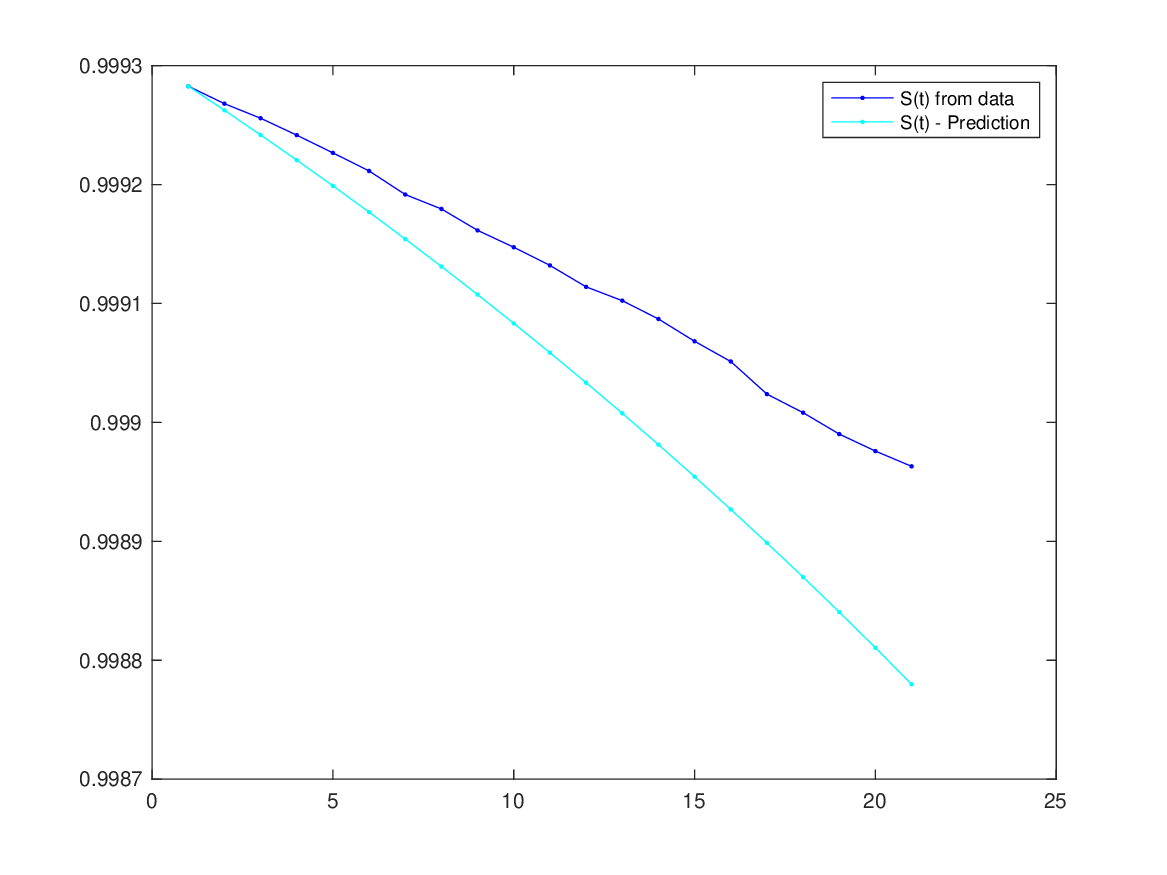}\vspace{-2mm}
\caption{t1}
\end{subfigure}%
\begin{subfigure}{.5\linewidth}
\centering
\includegraphics[scale=.32]{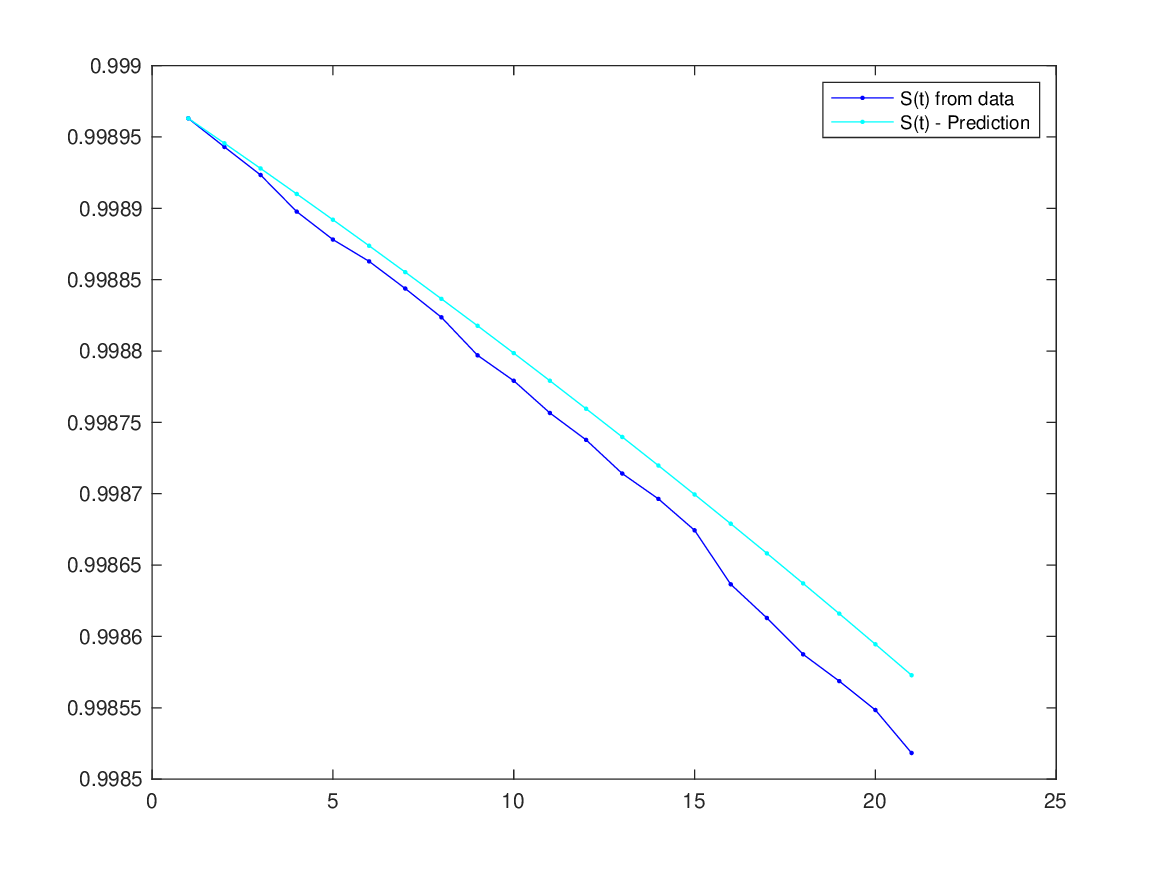}\vspace{-2mm}
\caption{t2}
\end{subfigure}
\begin{subfigure}{0.5\linewidth}
\centering
\includegraphics[scale=.32]{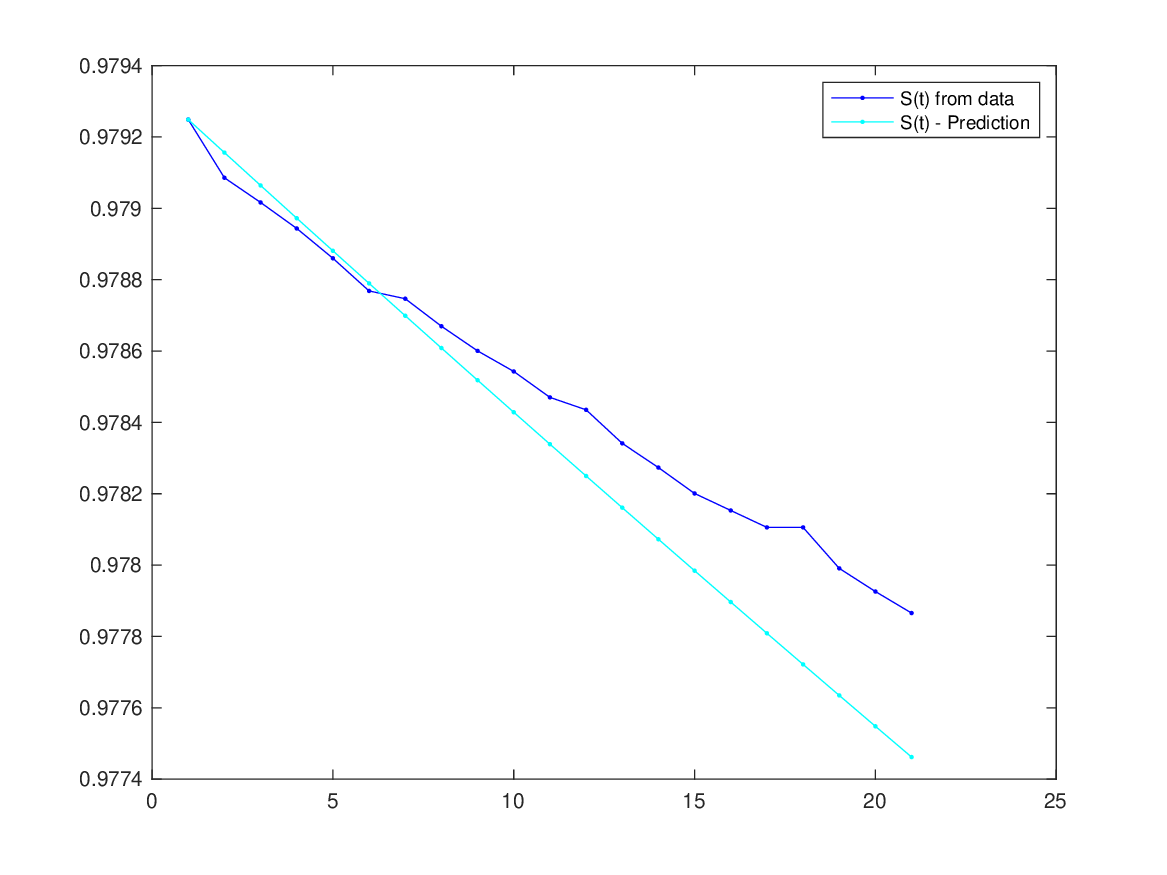}\vspace{-2mm}
\caption{t3}
\end{subfigure}
\begin{subfigure}{0.5\linewidth}
\centering
\includegraphics[scale=.32]{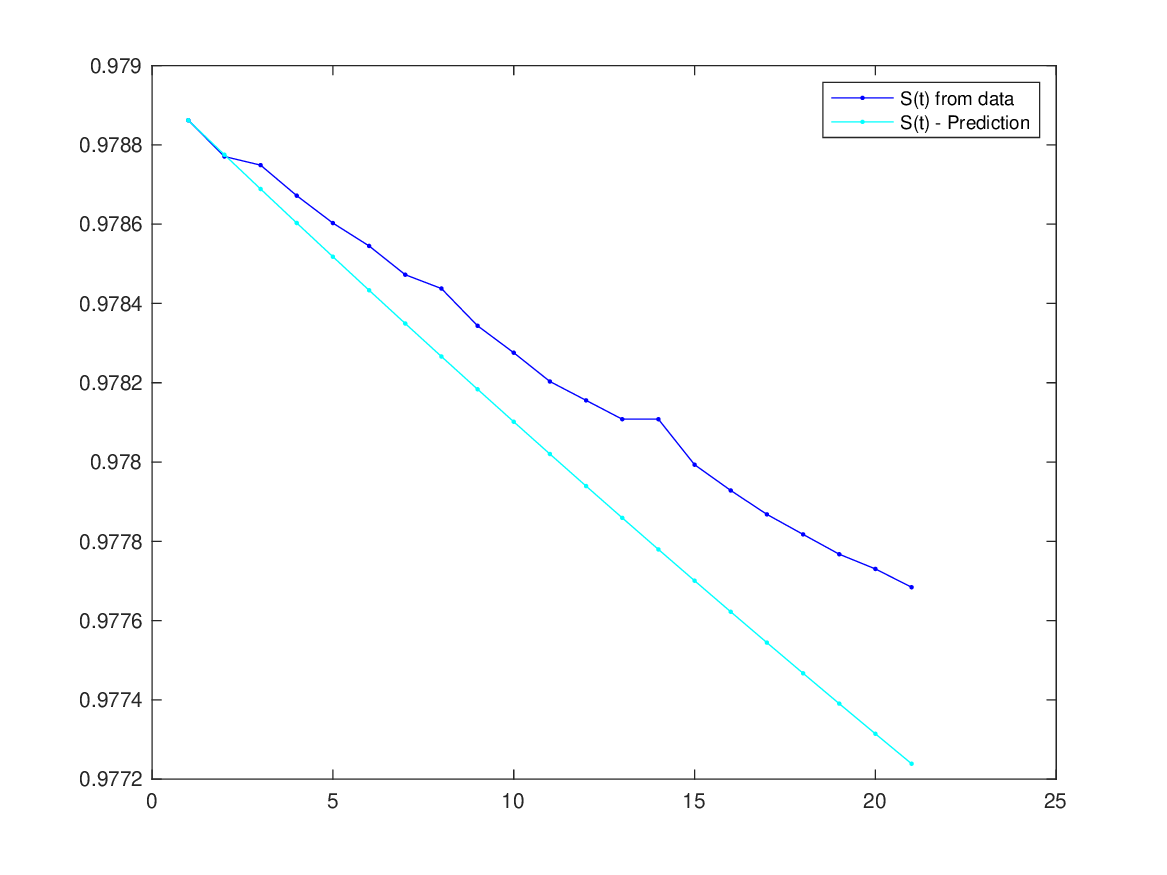}\vspace{-2mm}
\caption{t4}
\end{subfigure}
\caption{Prediction of $s(t)$ for the 4 starting dates and duration of 20 days
}\vspace{-.5cm}
\label{fig:predictionS20}
\end{figure}

\begin{figure}[H]
\begin{subfigure} {0.5\textwidth}
\centering
\includegraphics[scale=0.32]{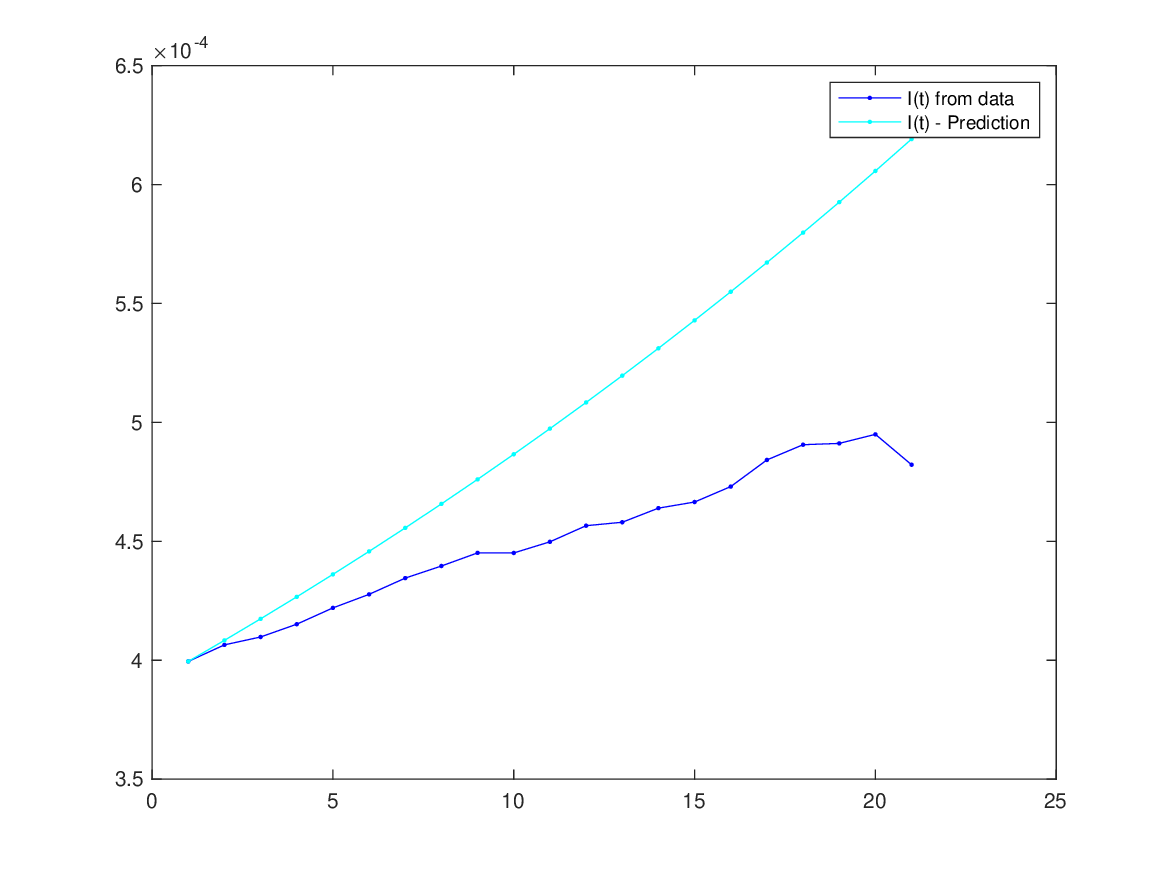}\vspace{-2mm}
\caption{t1}
\end{subfigure}%
\begin{subfigure}{.5\linewidth}
\centering
\includegraphics[scale=.32]{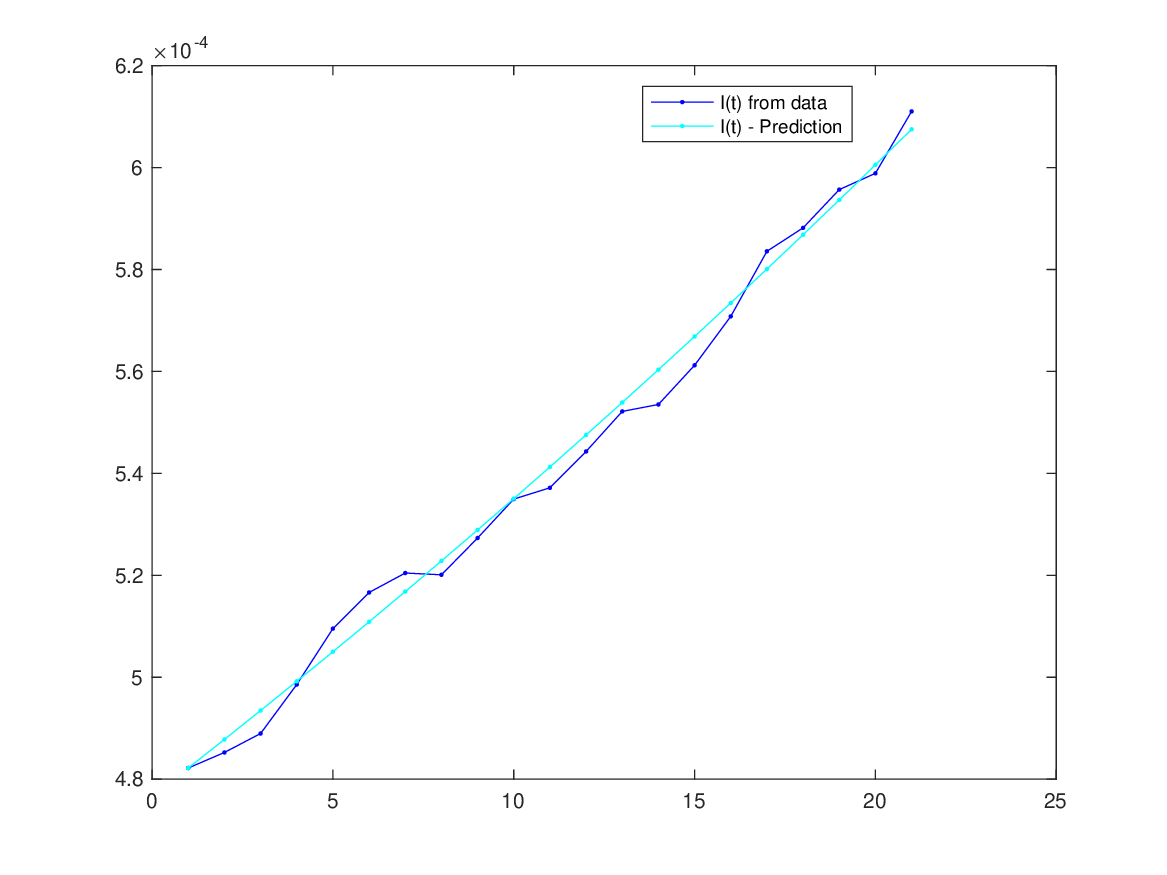}\vspace{-2mm}
\caption{t2}
\end{subfigure}
\begin{subfigure}{0.5\linewidth}
\centering
\includegraphics[scale=.32]{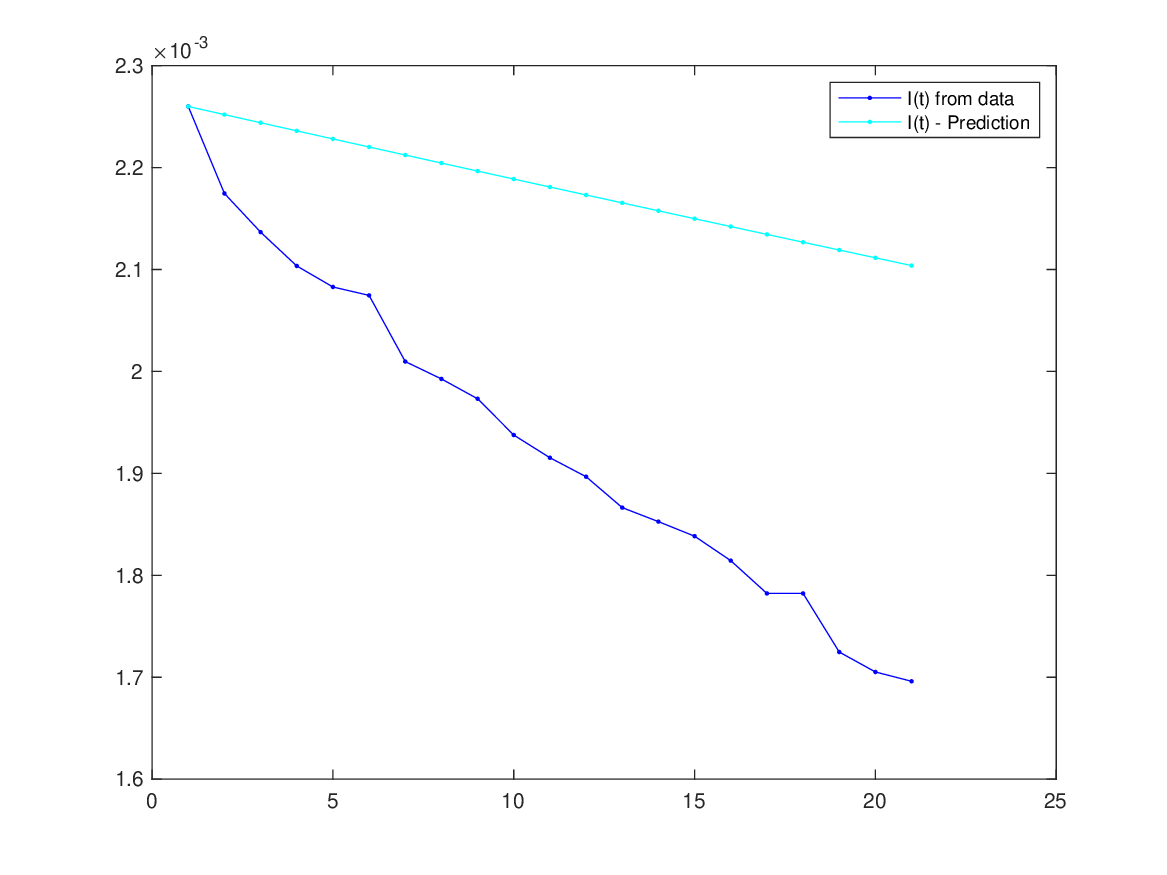}\vspace{-2mm}
\caption{t3}
\end{subfigure}
\begin{subfigure}{0.5\linewidth}
\centering
\includegraphics[scale=.32]{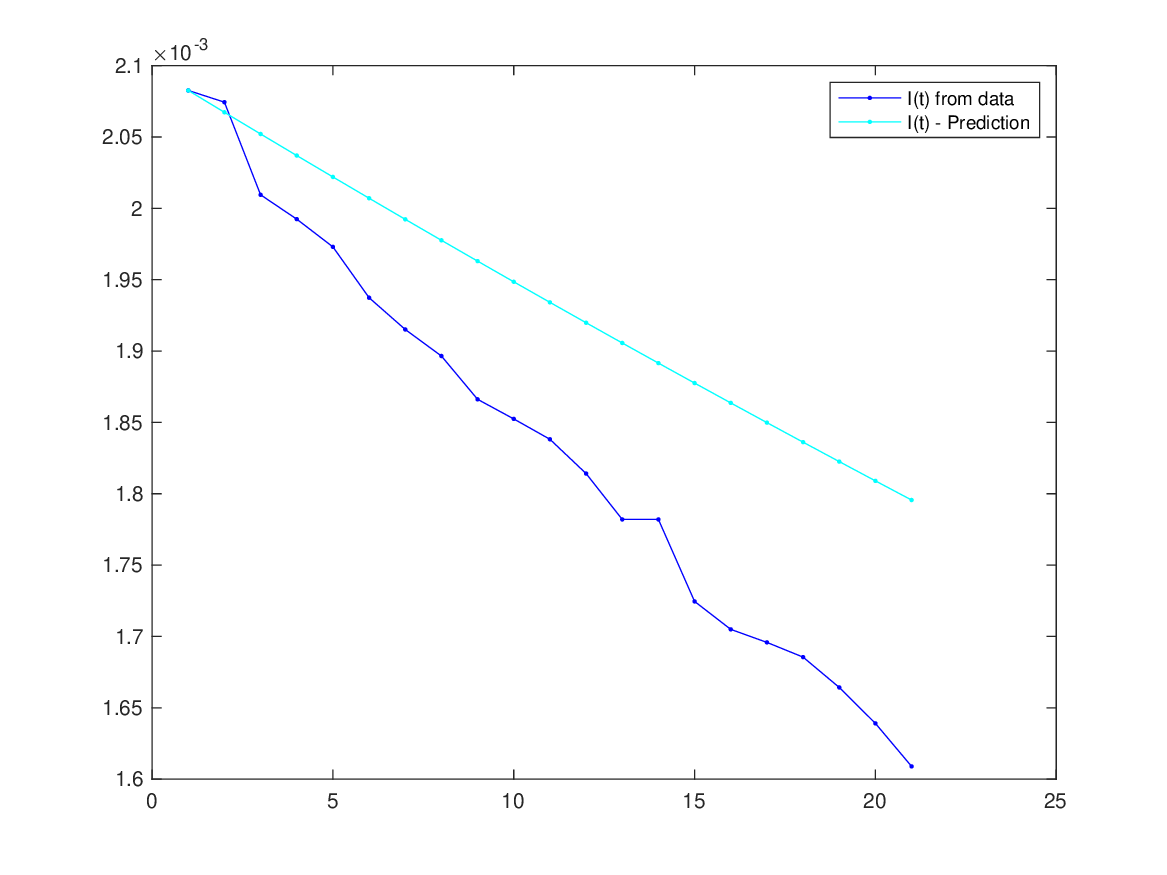}\vspace{-2mm}
\caption{t4}
\end{subfigure}
\caption{Prediction of $i(t)$ for the 4 starting dates and duration of 20 days
}\vspace{-.5cm}
\label{fig:predictionI20}
\end{figure}

\begin{figure}[H]
\begin{subfigure} {0.5\textwidth}
\centering
\includegraphics[scale=0.32]{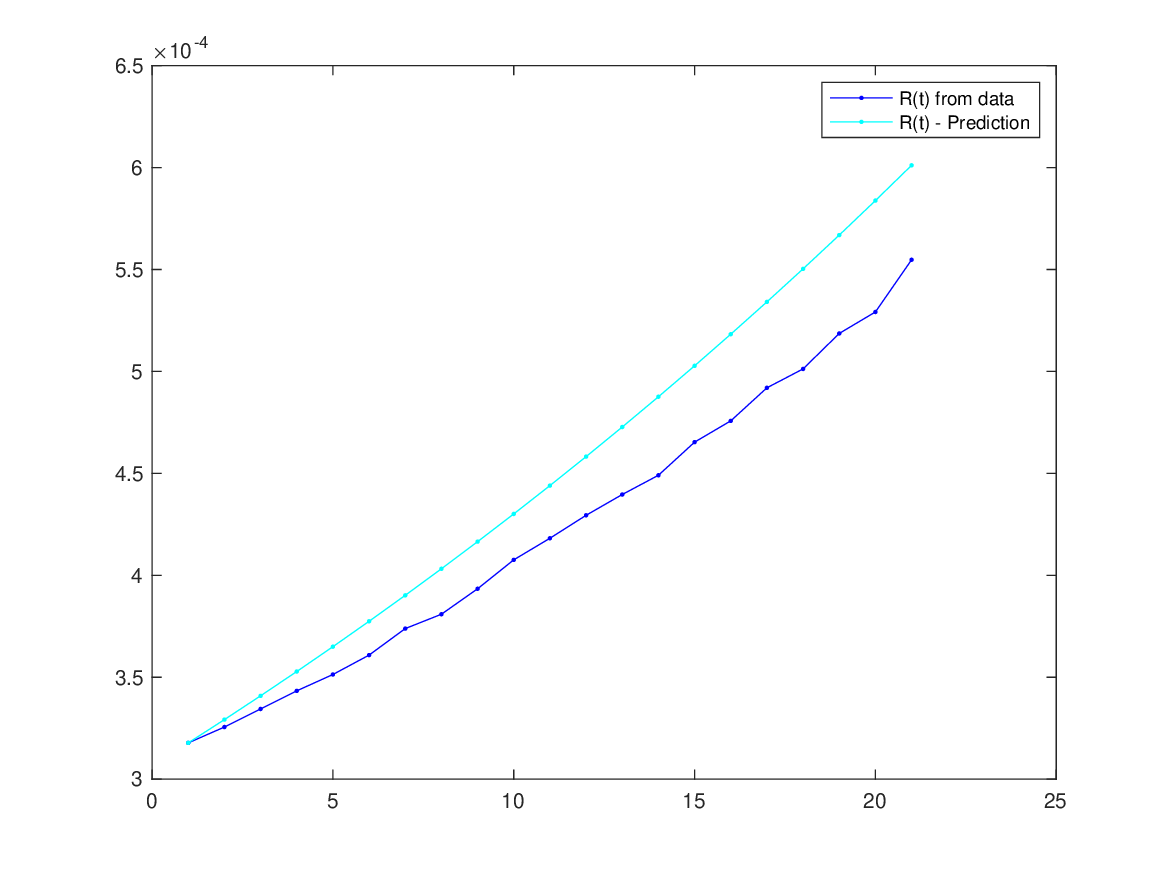}\vspace{-2mm}
\caption{t1}
\end{subfigure}%
\begin{subfigure}{.5\linewidth}
\centering
\includegraphics[scale=.32]{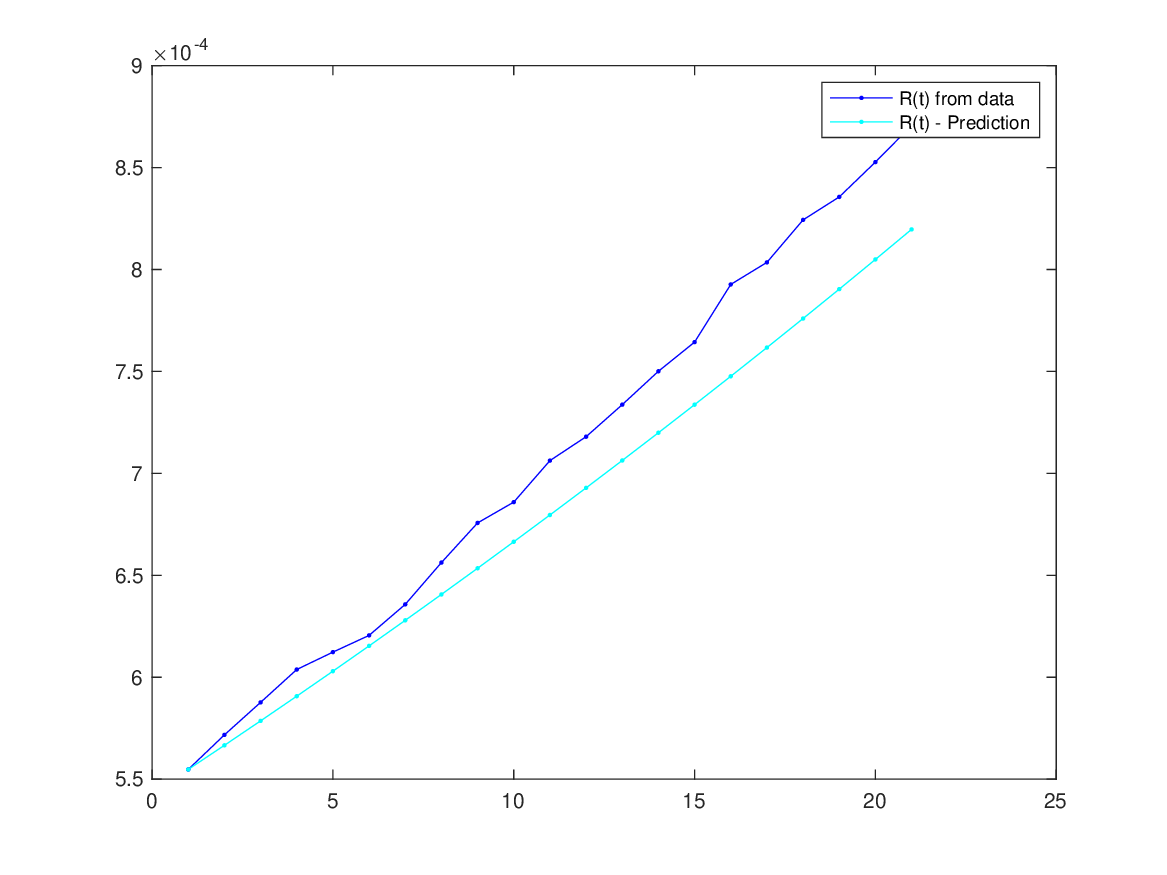}\vspace{-2mm}
\caption{t2}
\end{subfigure}
\begin{subfigure}{0.5\linewidth}
\centering
\includegraphics[scale=.32]{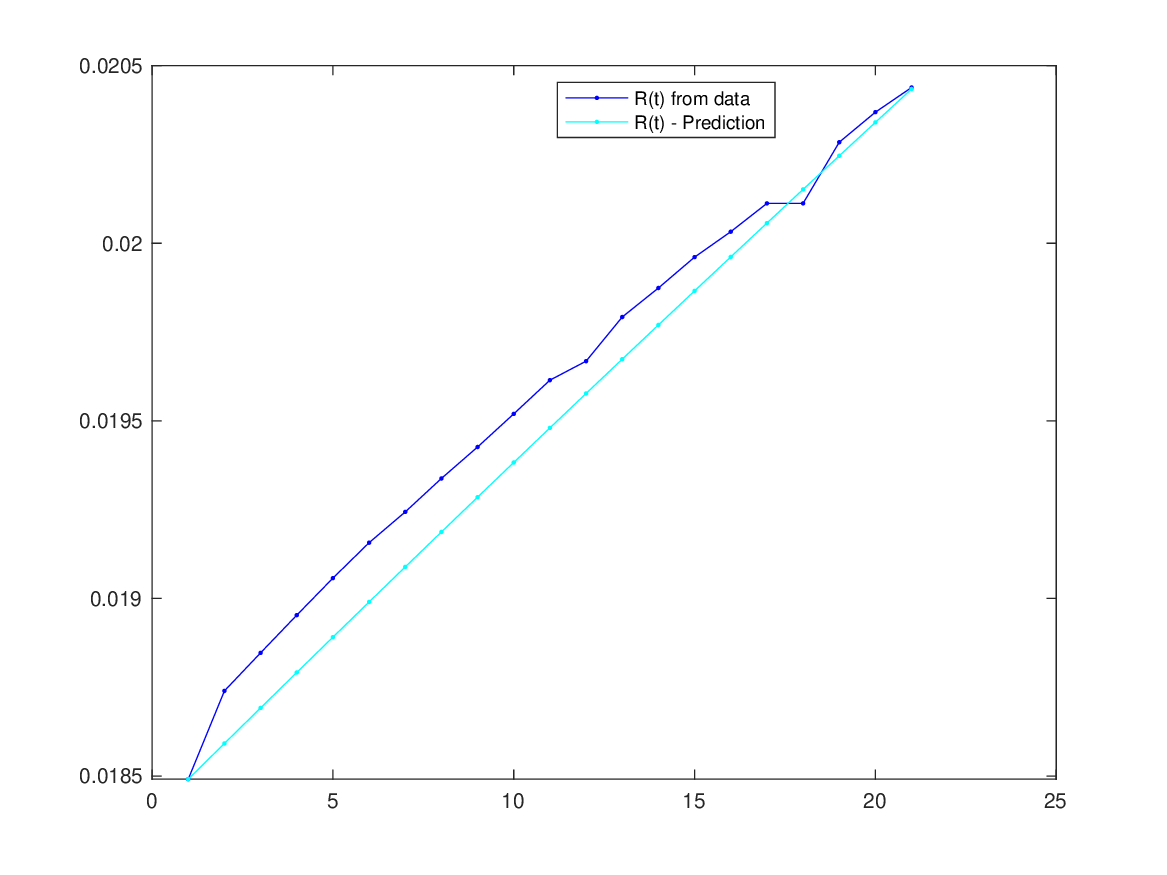}\vspace{-2mm}
\caption{t3}
\end{subfigure}
\begin{subfigure}{0.5\linewidth}
\centering
\includegraphics[scale=.32]{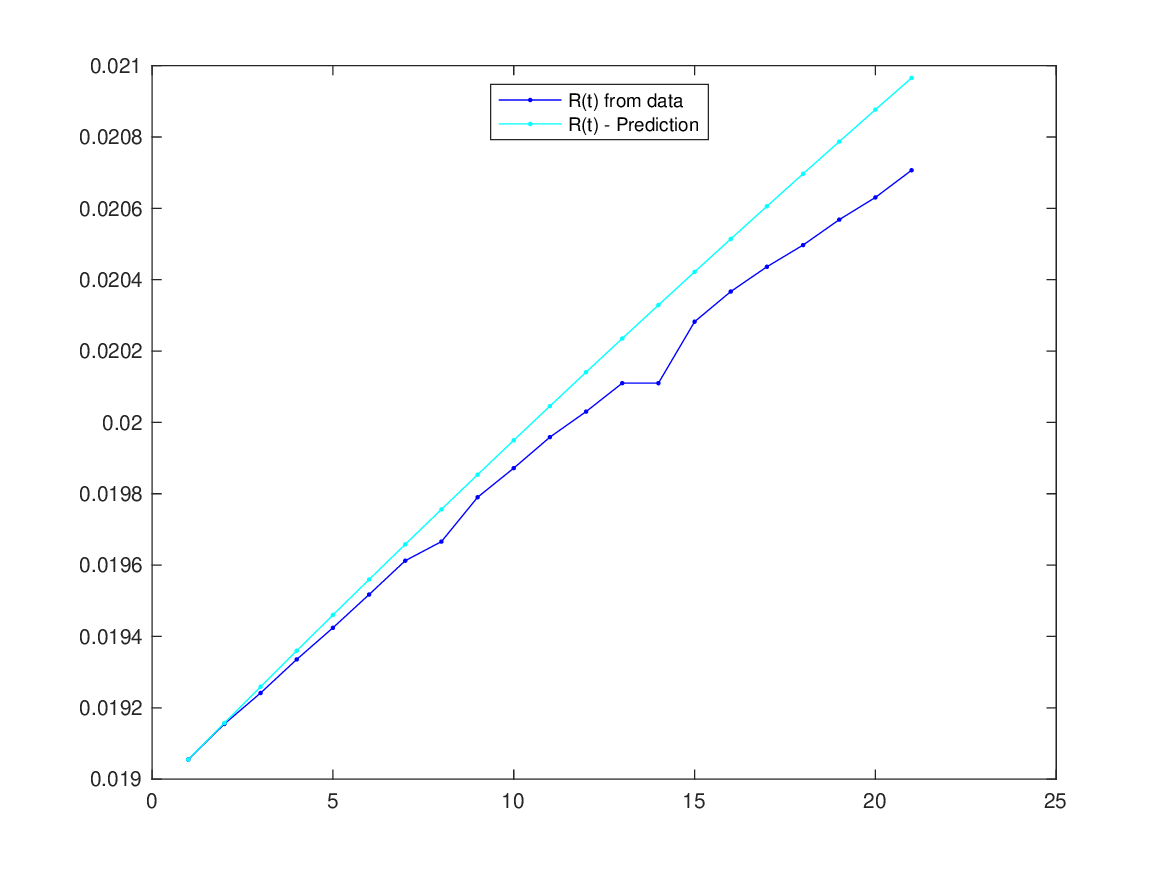}\vspace{-2mm}
\caption{t4}
\end{subfigure}
\caption{Prediction of $r(t)$ for the 4 starting dates and duration of 20 days
}\vspace{-.5cm}
\label{fig:predictionR20}
\end{figure}

\begin{table}[H]
\hspace{-25mm}
\setlength{\tabcolsep}{5pt}
{\renewcommand{\arraystretch}{1.1}
\begin{tabular}{||c|c||c|c|c||c|c|c||c|c|c||c|c|c||}
 \cline{3-14}
  \cline{3-14}
\multicolumn{2}{c||}{}  &\multicolumn{3}{c||}{k = 10 Days}&\multicolumn{3}{c||}{k = 15 Days} & \multicolumn{3}{c||}{k = 20 Days} & \multicolumn{3}{c||}{k = 30 Days} \\
 \cline{3-14}
\multicolumn{2}{c||}{}  & S & I & R& S & I & R& S & I & R& S & I & R \\ 
 \hline\hline
\multirow{2}{*}{$t1$} & $L_2$ &  4.43E-5 & 5.06E-2 & 6.24E-2 & 6.22E-5 & 7.61E-2 & 7.34E-2 &
9.57E-5 & 1.42E-1 & 7.45E-2 & 3.34E-4 & 5.05E-1 & 1.88E-1 \\
\cline{2-14}
 &$L_\infty$ &      7.82E-5 & 9.31E-2 & 8.68E-2 &1.13E-4 & 1.34E-1 & 1.04E-1 &
 1.83E-4 & 2.77E-1 & 9.85E-2 & 6.74E-4 & 8.86E-1 & 2.79E-1 \\ 
 \hline  \hline
\multirow{2}{*}{$t2$} &$L_2$ &  6.01E-5 & 7.13E-2 & 6.76E-3 &
 1.26E-4 & 1.03E-1 & 6.30E-3&
2.16E-4 & 1.40E-1 & 5.95E-3 & 2.51E-4 & 3.10E-2 & 1.07E-2 \\ 
\cline{2-14}
 &$L_\infty$ &  1.09E-4 & 9.47E-2 & 7.77E-3 &  2.47E-4 & 1.31E-1 & 7.94E-3&
 4.12E-4 & 1.80E-1 & 8.16E-3 & 4.40E-4 & 3.70E-2 & 1.87E-2 \\ 
\hline  \hline
 \multirow{2}{*}{$t3$} &$L_2$ & 1.01E-4 & 5.58E-3 & 4.93E-3 & 
 1.68E-4 & 2.73E-2 & 5.87E-3&
 2.54E-4 & 6.20E-2 & 6.86E-3 & 4.20E-4 & 1.38E-1 & 8.68E-3 \\
 \cline{2-14}
 &$L_\infty$ &  1.48E-4 & 9.13E-3 & 7.61E-3 & 3.00E-4 & 4.28E-2 & 1.21E-2&
 4.55E-4 & 8.96E-2 & 1.25E-2 & 7.45E-4 & 1.70E-1 & 1.81E-2 \\
 \hline  \hline
  \multirow{2}{*}{$t4$} &$L_2$ &1.37E-5 & 8.68E-3 & 1.77E-2 &
  1.98E-5 & 8.15E-3 & 2.52E-2 &
  2.94E-5 & 6.39E-3 & 4.16E-2 & 1.22E-5 & 4.82E-2 & 2.15E-2 \\
 \cline{2-14}
 &$L_\infty$ &    2.25E-5 & 1.57E-2 & 2.96E-2 &  4.31E-5 & 1.47E-2 & 4.65E-2 &\
 5.45E-5 & 1.12E-2 & 5.85E-2 & 2.30E-5 & 6.88E-2 & 3.66E-2 \\
 \hline  \hline
\end{tabular}}\caption{Relative error between between exact solution and predicted one.}\label{tab:RelErrPredM1}
\end{table}

Thus clearly, it is possible to predict the evolution of the pandemics using the constant-coefficient SIR model for short periods of time up to 20 days. However, it is preferable to use shorter prediction intervals such as 10 or 15 days, since the increasing or decreasing mode of $i(t)$ is approximated with a relative error of order $10^{-2}$. 
    
  Note that the same tests were performed with $\beta(t)$ and $\rho(t)$ computed using Method 2 (section \ref{sec:ParamM2}) instead of Method 1, and the similar results are shown in Appendix \ref{sec:PredM2}.
  \newpage
  \section{Conclusion}\label{sec:conclusion}
 Since the spread of COVID-19, thousands of papers dealing with different aspects of the pandemics and different types of models, have been published.
In this work, our starting point is the {\bf Worldometer Corona Virus Platform} data,  daily collected  between April 2020 and June 2022. Based on this data, and the basic balance equation that governs the dynamics of a pandemics, we derived a time-dependent IR model \eqref{sys:basicIR} followed by a time-dependent SIR model \eqref{eq:SIR}. For both models,
we have proven local exponential growth (or decay) of the number of infections $I(t)$. We also introduce a time-dependent {\bf reproduction parameter $\sigma_s(t)$}, that can monitor the outbreak of the pandemics ($\sigma_s(t)>1$) or its stabilisation and decay ($\sigma_s(t)\leq 1$). The collected data allows to extract the time-dependent rates  $\beta_s(t)$, $\beta(t)$ and $\rho(t)$ in 2 different ways, in view of computing $\sigma_s(t)$ and $\sigma(t)$, the key indicators of the pandemics' evolution. The accuracy of our approach is validated by the resulting matching between the simulated $i(t)$ and $r(t)$ values with the collected ones. This shows that  the time-dependent SIR model is efficient in describing the pandemics' behavior.

The time-dependent SIR model was then applied over different periods of time for Germany, Italy and the whole world by computing $\beta_s(t)$, $\beta(t)$, $\rho(t)$, $\sigma(t)$ and $\sigma_s(t)$. These parameters fluctuated with respect to time and their statistics varied across countries.  This variation may be probably due to important factors  such as vaccination, confinement, social interaction, immunity,... that depend on the implemented policies in every geographical region. 
Thus, there do not seem to be a way to predict beforehand the evolution of  $\beta_s(t)$, $\beta(t)$ and $\rho(t)$. 

On the other hand, the use of a constant-coefficient SIR model to simulate the evolution of COVID-19, seems also to be inefficient, given the fluctuations in $\beta(t)$ and $\rho(t)$. Yet, it is possible to use it for short-term predictions by setting the constant $\beta$ and $\rho$ as the average of  $\beta(t)$ and $\rho(t)$ which are computed based on available data. 

As for future work, provided additional data is made available to detail the pre-infection stages, then one could consider different variants of the SIR model, for example, the SVIR model which adds a new compartment for vaccinated people, with the aim of analyzing the effect of vaccination on the population in maintaining and controlling the pandemic. Other variants could also be considered, where the susceptible compartment is divided further into smaller sub-compartments that distinguish between isolated, moderately susceptible, and highly susceptible individuals.

\bibliographystyle{abbrv}
\bibliography{sample}

\appendix 
\section{Simulations for Time-dependent SIR Model}

\subsection{Second Outbreak}\label{sec:secondoutb}
\begin{figure}[H]
\begin{subfigure}{0.5\textwidth}
\centering
\includegraphics[scale=0.28]{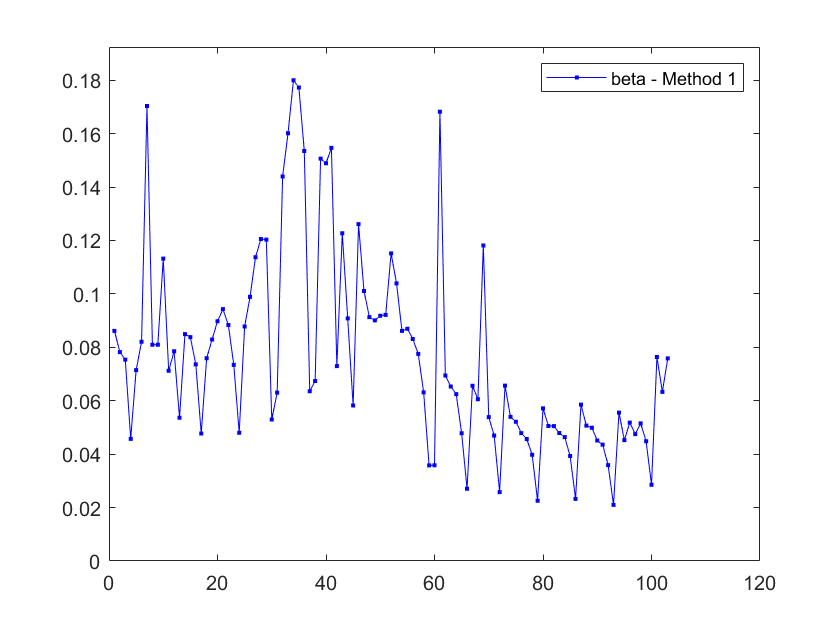} \vspace{-2mm}
\caption{Infection Rate $\beta(t)$ - Method 1}
\label{fig:beta2ndoutbreakitalyM1}
\end{subfigure}
\hfill
\begin{subfigure}{0.5\textwidth}
\centering
\includegraphics[scale=0.28]{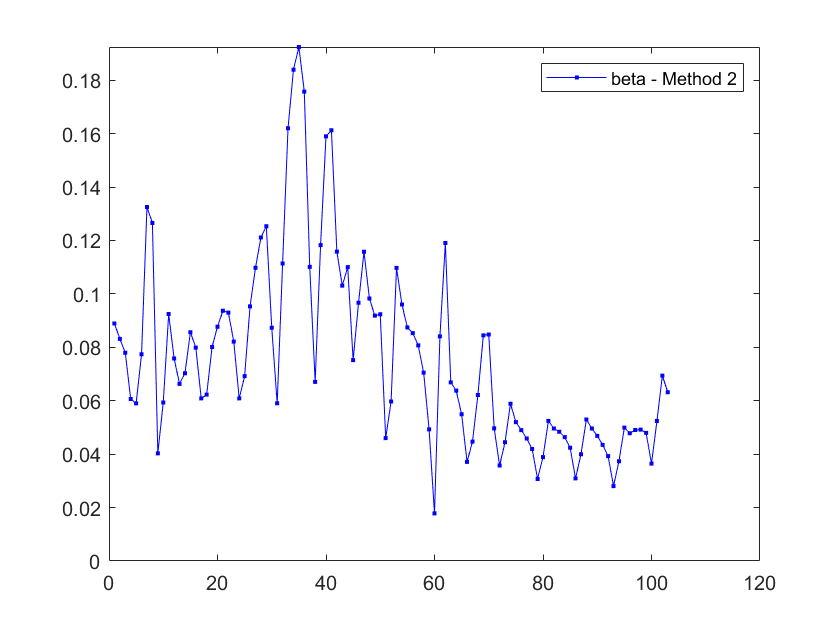} \vspace{-2mm}
\caption{Infection Rate $\beta(t)$ - Method 2}
\label{fig:beta2ndoutbreakitalyM2}
\end{subfigure}
\newline
\begin{subfigure}{0.5\textwidth}
\centering
\includegraphics[scale=0.28]{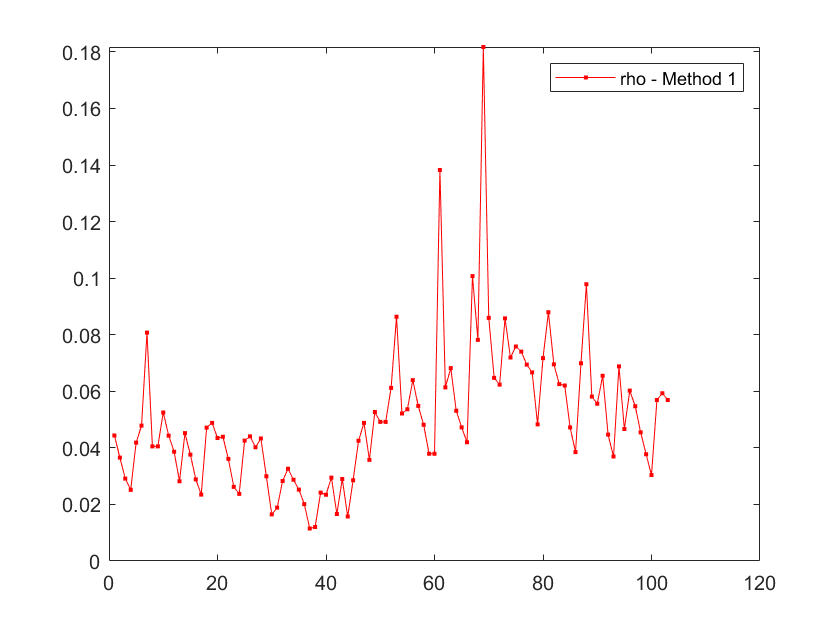} \vspace{-2mm}
\caption{Removal Rate $\rho(t)$ - Method 1 }
\label{fig:rho2ndoutbreakitalyM1}
\end{subfigure}
\hfill
\begin{subfigure}{0.5\textwidth}
\centering
\includegraphics[scale=0.28]{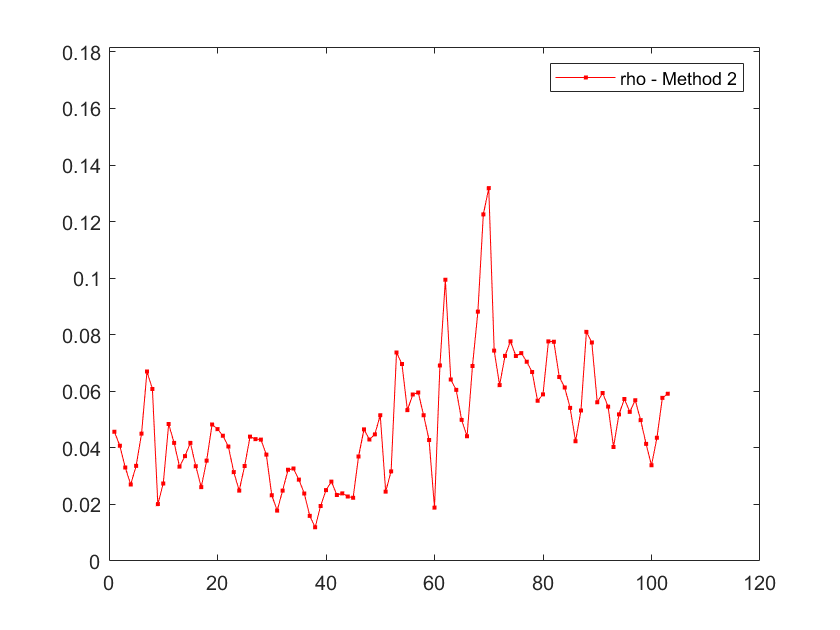} \vspace{-2mm}
\caption{Removal Rate $\rho(t)$ - Method 2 }
\label{fig:rho2ndoutbreakitalyM2}
\end{subfigure}
\newline
\begin{subfigure}{0.5\textwidth}
\centering
\includegraphics[scale=0.28]{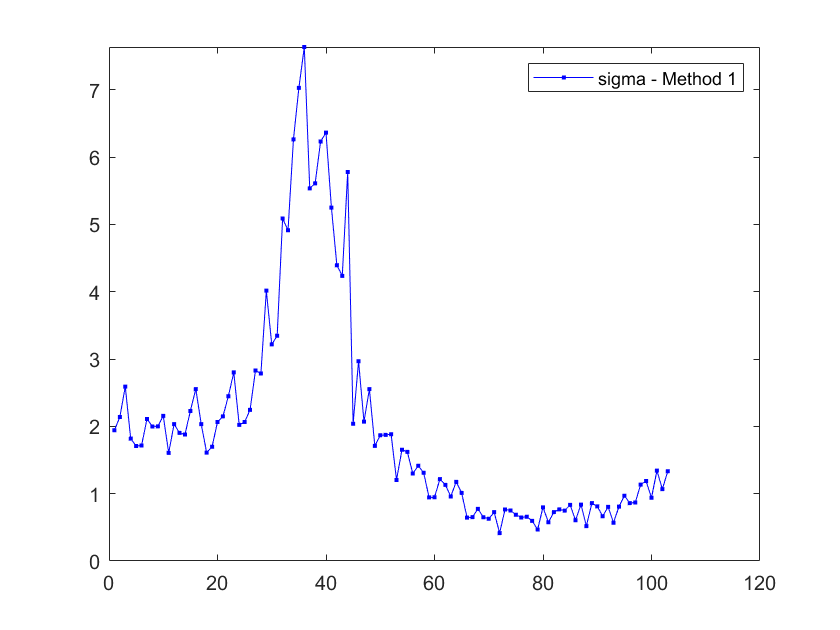} \vspace{-2mm}
\caption{Reproduction Factor $\sigma(t)$ - Method 1}
\label{fig:sigma2ndoutbreakitalyM1}
\end{subfigure}
\hfill
\begin{subfigure}{0.5\textwidth}
\centering
\includegraphics[scale=0.28]{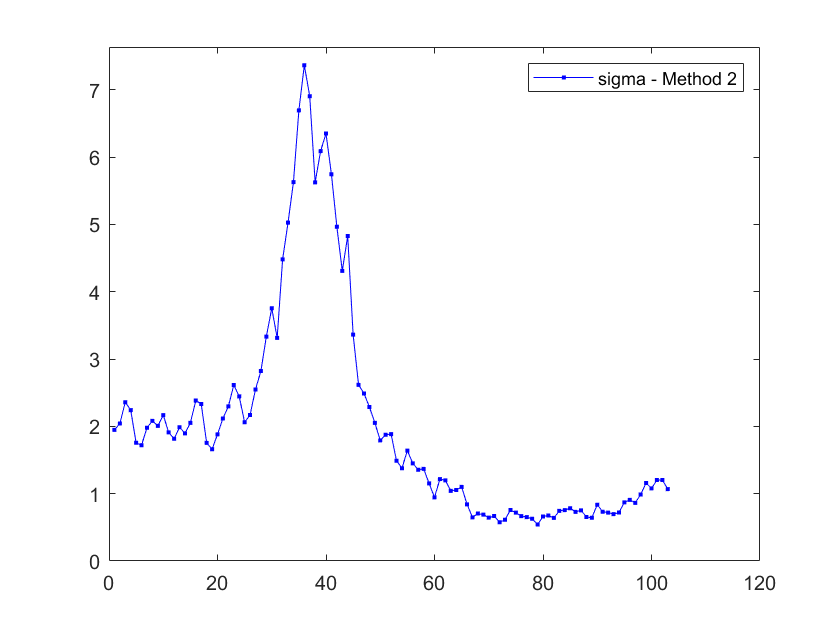} \vspace{-2mm}
\caption{Reproduction Factor $\sigma(t)$ - Method 2}
\label{fig:sigma2ndoutbreakitalyM2}
\end{subfigure}
\newline
\begin{subfigure}{0.5\textwidth}
\centering
\includegraphics[scale=0.28]{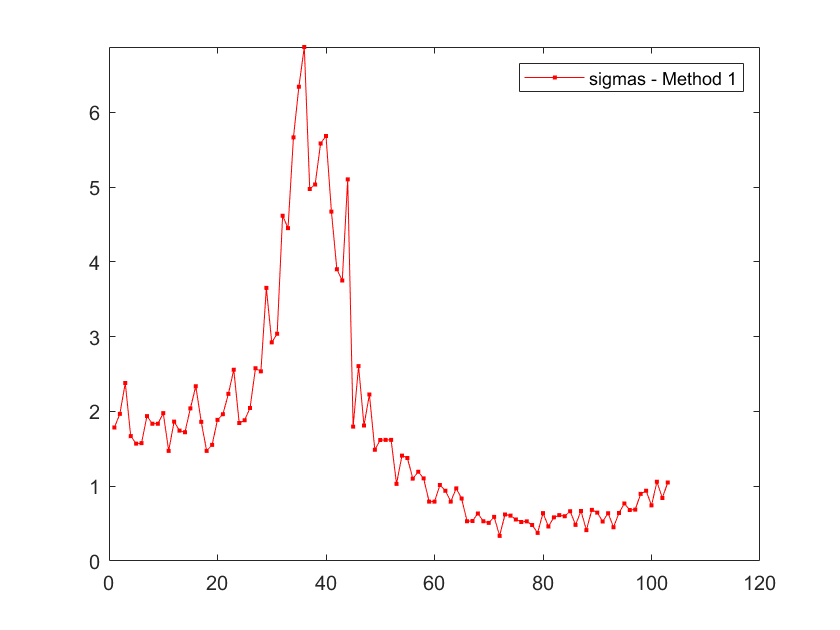}\vspace{-2mm} 
\caption{Replacement Number $\sigma_s(t)$ - Method 1}
\label{fig:sigmas2ndoutbreakitalyM1}
\end{subfigure}\hfill
\begin{subfigure}{0.5\textwidth}
\centering
\includegraphics[scale=0.28]{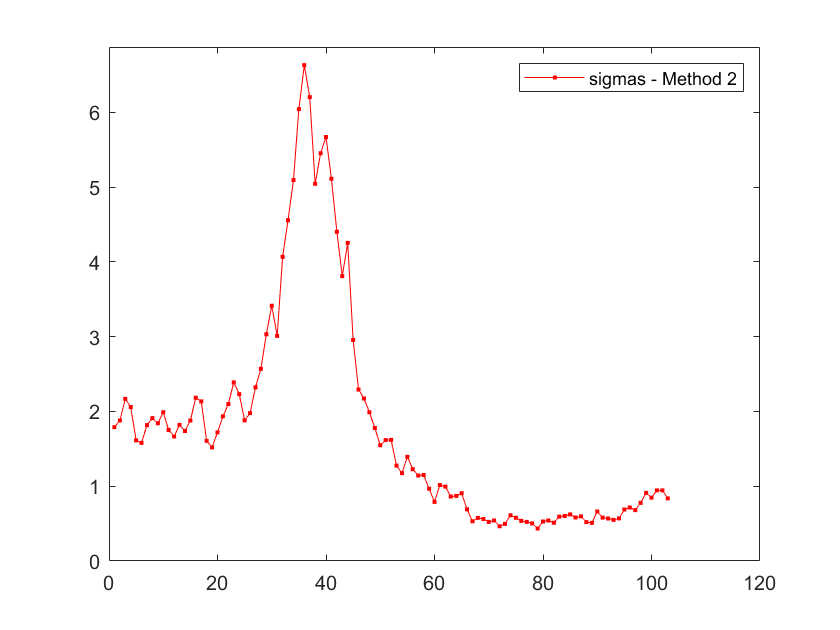}\vspace{-2mm} 
\caption{Replacement Number $\sigma_s(t)$ - Method 2}
\label{fig:sigmas2ndoutbreakitalyM2}
\end{subfigure}
\caption{Parameters of SIR Model during Second Outbreak in Italy }
\label{fig:parameters2ndoutbreakitaly}\vspace{0mm}
\end{figure}

\begin{figure}[H]
\begin{subfigure}{0.5\textwidth}
\centering
\includegraphics[scale=0.28]{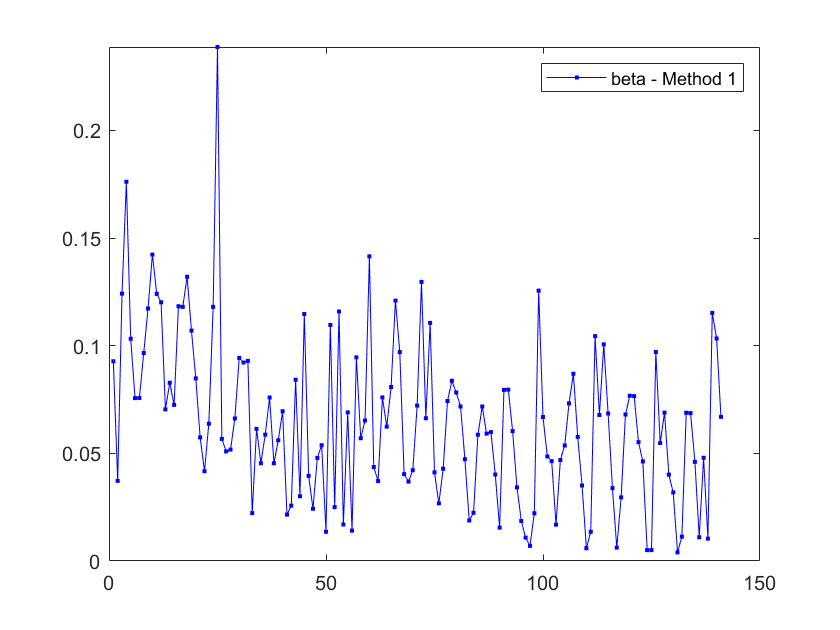} \vspace{-2mm}
\caption{Infection Rate $\beta(t)$ - Method 1}
\label{fig:beta2ndoutbreakgermanyM1}
\end{subfigure}
\hfill
\begin{subfigure}{0.5\textwidth}
\centering
\includegraphics[scale=0.28]{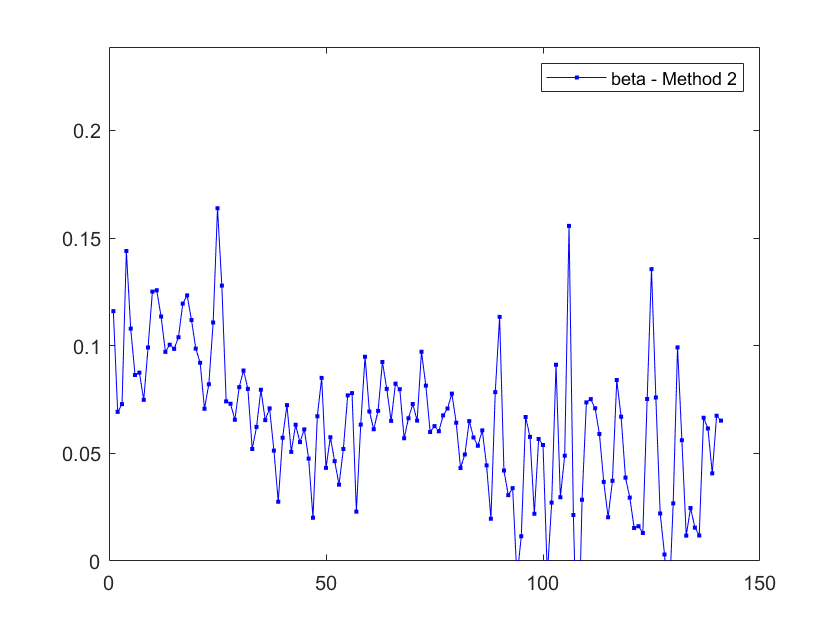} \vspace{-2mm}
\caption{Infection Rate $\beta(t)$ - Method 2}
\label{fig:beta2ndoutbreakgermanyM2}
\end{subfigure}
\newline
\begin{subfigure}{0.5\textwidth}
\centering
\includegraphics[scale=0.28]{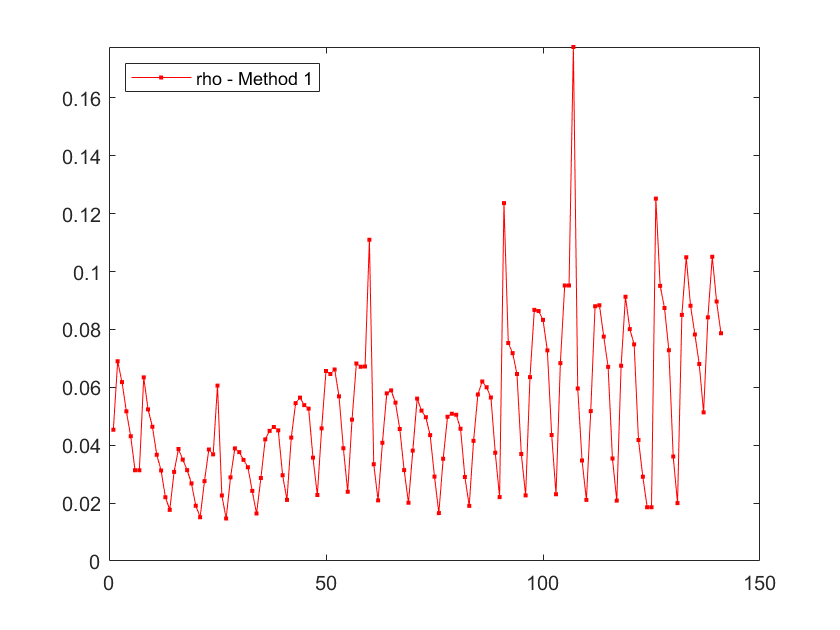} \vspace{-2mm}
\caption{Removal Rate $\rho(t)$ - Method 1 }
\label{fig:rho2ndoutbreakgermanyM1}
\end{subfigure}
\hfill
\begin{subfigure}{0.5\textwidth}
\centering
\includegraphics[scale=0.28]{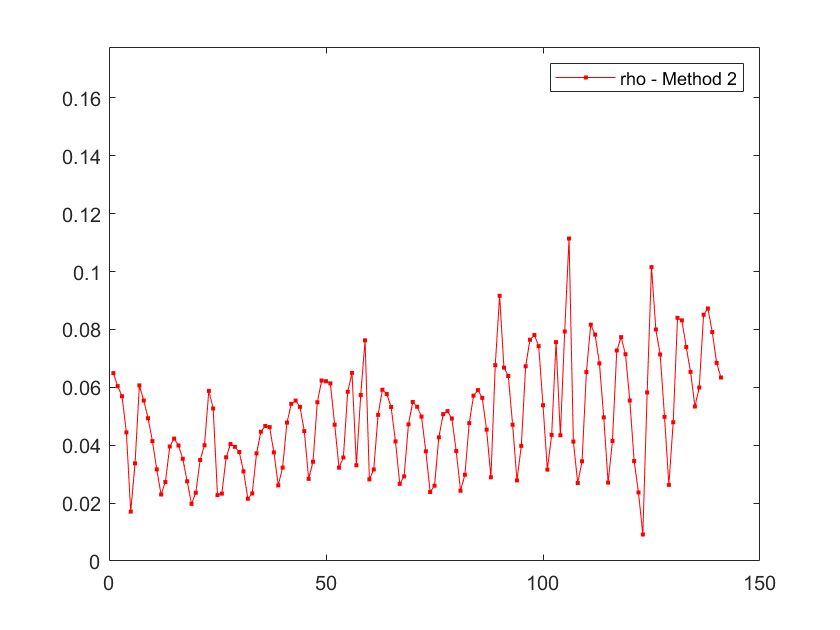} \vspace{-2mm}
\caption{Removal Rate $\rho(t)$ - Method 2 }
\label{fig:rho2ndoutbreakgermanyM2}
\end{subfigure}
\newline
\begin{subfigure}{0.5\textwidth}
\centering
\includegraphics[scale=0.28]{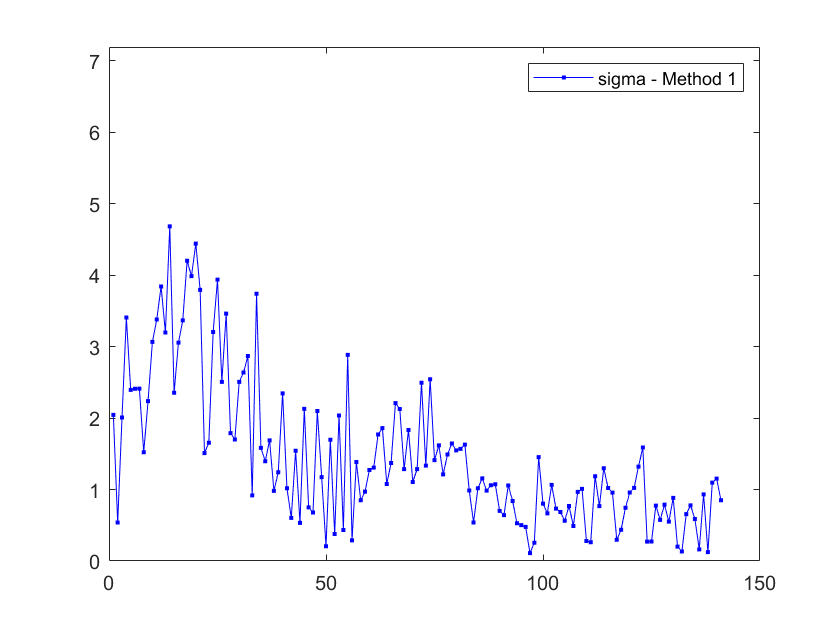} \vspace{-2mm}
\caption{Reproduction Factor $\sigma(t)$ - Method 1}
\label{fig:sigma2ndoutbreakgermanyM1}
\end{subfigure}
\hfill
\begin{subfigure}{0.5\textwidth}
\centering
\includegraphics[scale=0.28]{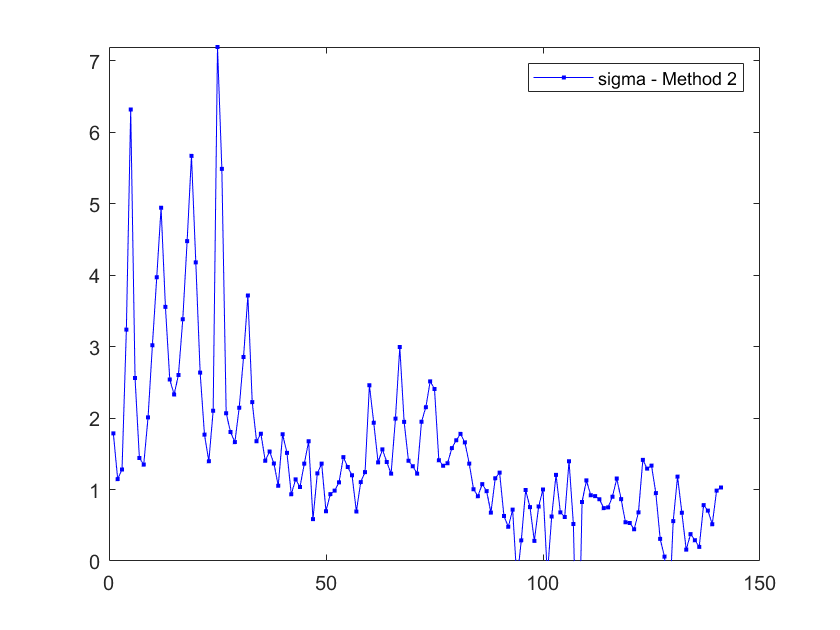} \vspace{-2mm}
\caption{Reproduction Factor $\sigma(t)$ - Method 2}
\label{fig:sigma2ndoutbreakgermanyM2}
\end{subfigure}
\newline
\begin{subfigure}{0.5\textwidth}
\centering
\includegraphics[scale=0.28]{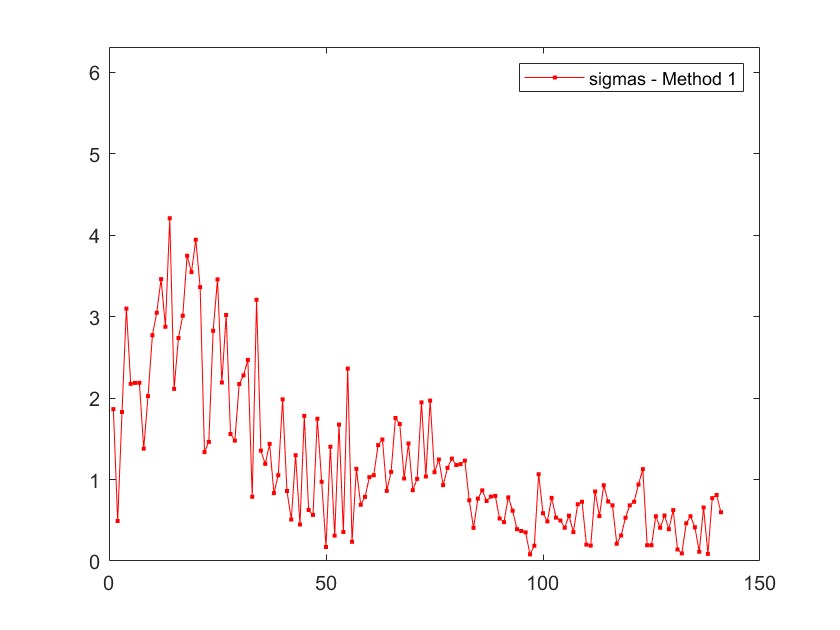}\vspace{-2mm} 
\caption{Replacement Number $\sigma_s(t)$ - Method 1}
\label{fig:sigmas2ndoutbreakgermanyM1}
\end{subfigure}\hfill
\begin{subfigure}{0.5\textwidth}
\centering
\includegraphics[scale=0.28]{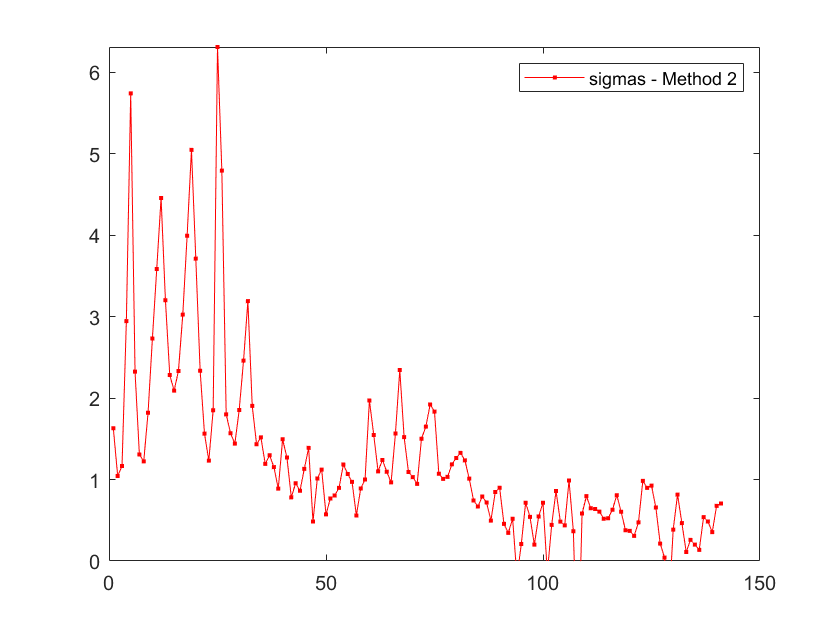}\vspace{-2mm} 
\caption{Replacement Number $\sigma_s(t)$ - Method 2}
\label{fig:sigmas2ndoutbreakgermanyM2}
\end{subfigure}
\caption{Parameters of SIR Model during Second Outbreak in Germany }
\label{fig:parameters2ndoutbreakgermany}\vspace{0mm}
\end{figure}

\begin{figure}[H]
\begin{subfigure}{0.5\textwidth}
\centering
\includegraphics[scale=0.28]{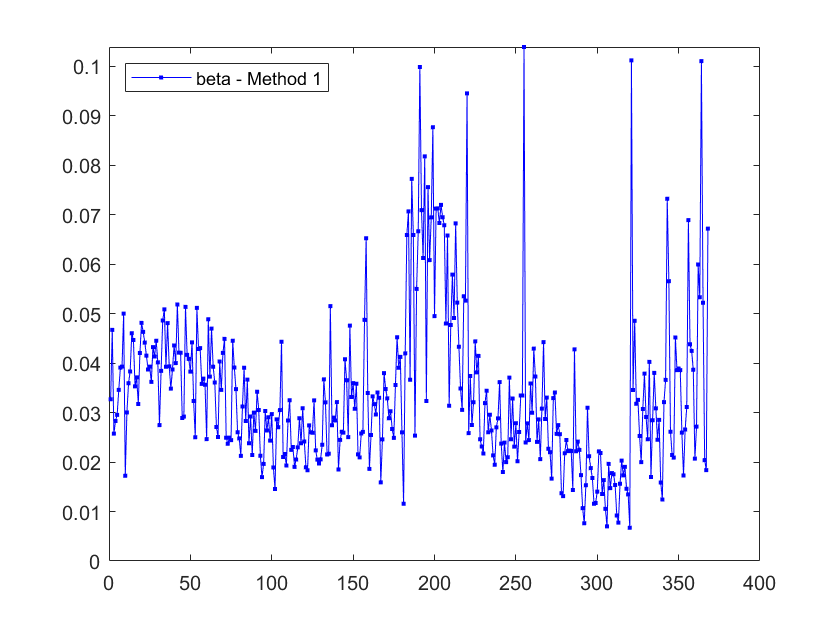}\vspace{-2mm}
\caption{Infection Rate $\beta(t)$ - Method 1}
\label{fig:beta2ndoutbreakworldM1}
\end{subfigure}
\hfill
\begin{subfigure}{0.5\textwidth}
\centering
\includegraphics[scale=0.28]{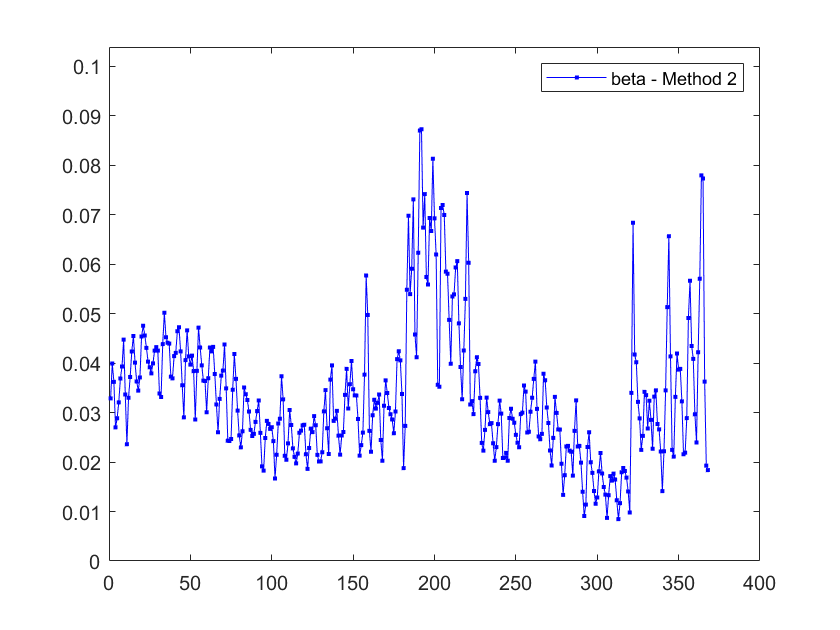} \vspace{-2mm}
\caption{Infection Rate $\beta(t)$ - Method 2}
\label{fig:beta2ndoutbreakworldM2}
\end{subfigure}
\newline
\begin{subfigure}{0.5\textwidth}
\centering
\includegraphics[scale=0.28]{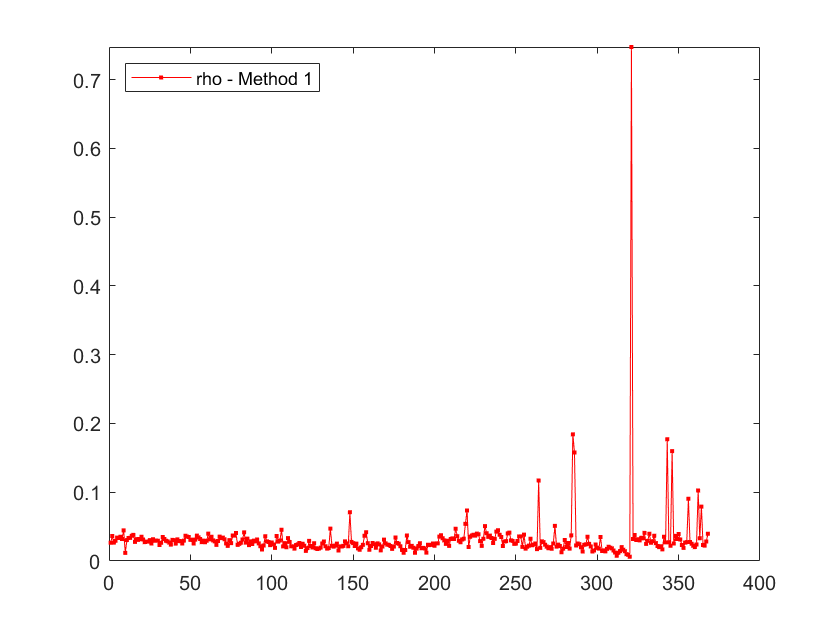} \vspace{-2mm}
\caption{Removal Rate $\rho(t)$ - Method 1 }
\label{fig:rho2ndoutbreakworldM1}
\end{subfigure}
\hfill
\begin{subfigure}{0.5\textwidth}
\centering
\includegraphics[scale=0.28]{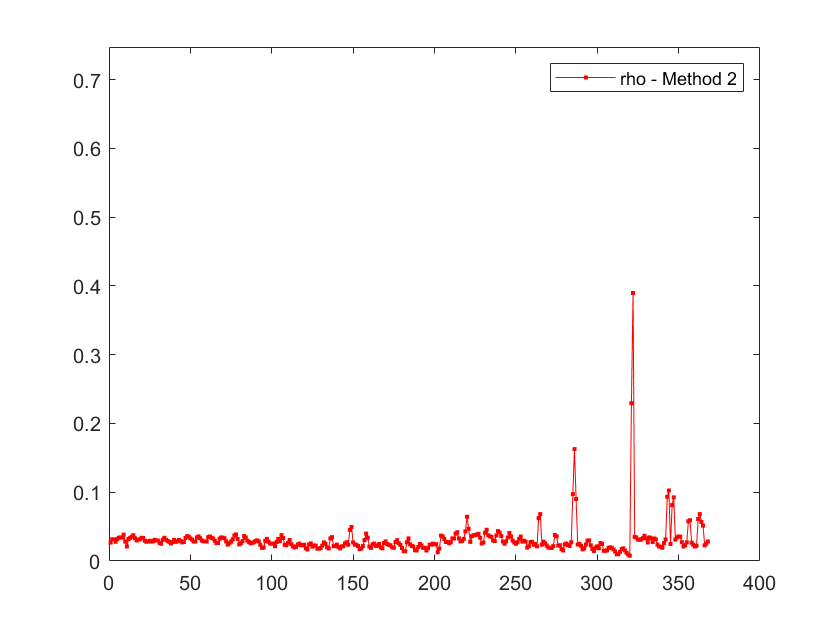} \vspace{-2mm}
\caption{Removal Rate $\rho(t)$ - Method 2 }
\label{fig:rho2ndoutbreakworldM2}
\end{subfigure}
\newline
\begin{subfigure}{0.5\textwidth}
\centering
\includegraphics[scale=0.28]{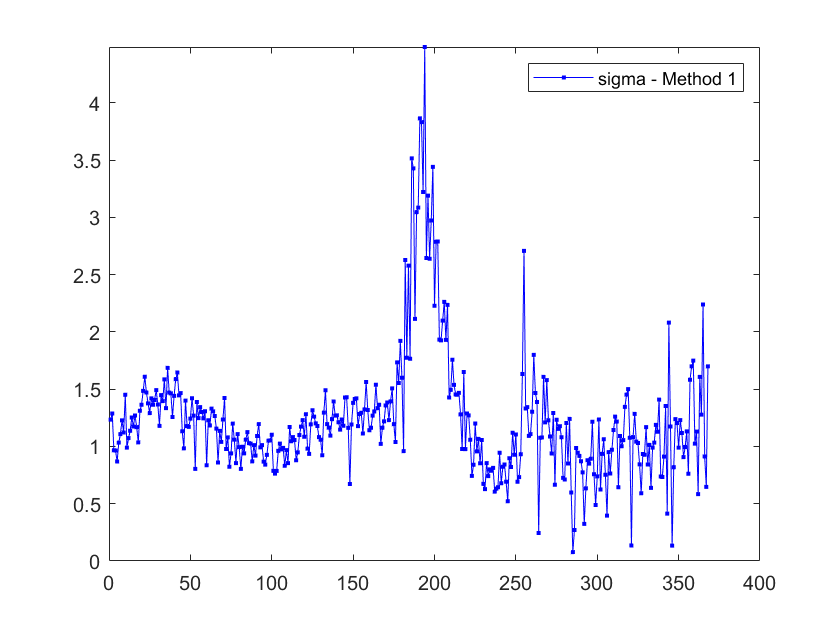} \vspace{-2mm}
\caption{Reproduction Factor $\sigma(t)$ - Method 1}
\label{fig:sigma2ndoutbreakworldM1}
\end{subfigure}
\hfill
\begin{subfigure}{0.5\textwidth}
\centering
\includegraphics[scale=0.28]{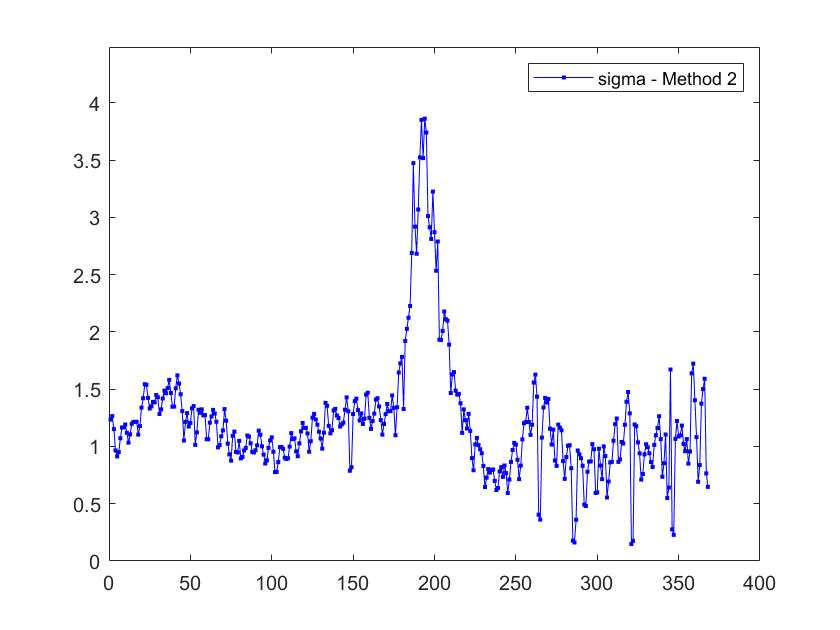} \vspace{-2mm}
\caption{Reproduction Factor $\sigma(t)$ - Method 2}
\label{fig:sigma2ndoutbreakworldM2}
\end{subfigure}
\newline
\begin{subfigure}{0.5\textwidth}
\centering
\includegraphics[scale=0.28]{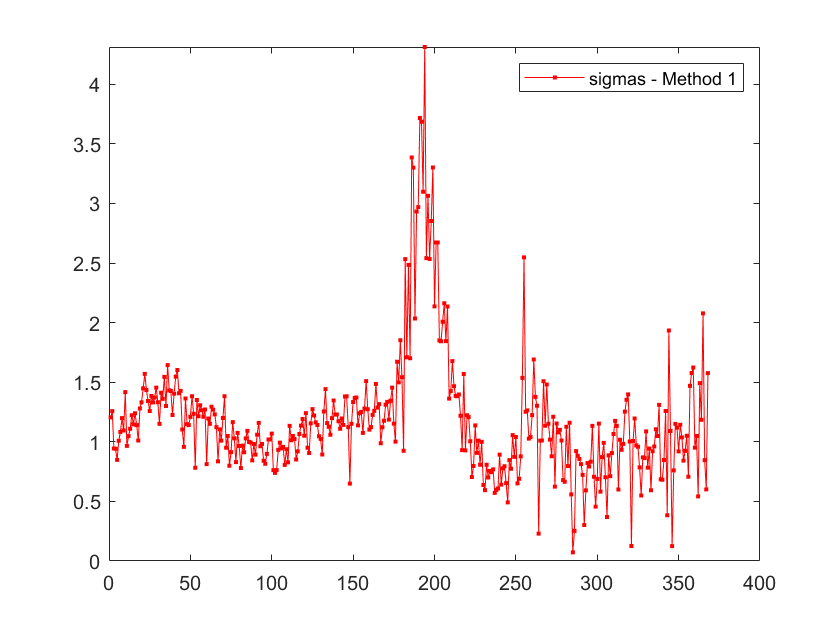}\vspace{-2mm} 
\caption{Replacement Number $\sigma_s(t)$ - Method 1}
\label{fig:sigmas2ndoutbreakworldM1}
\end{subfigure}\hfill
\begin{subfigure}{0.5\textwidth}
\centering
\includegraphics[scale=0.28]{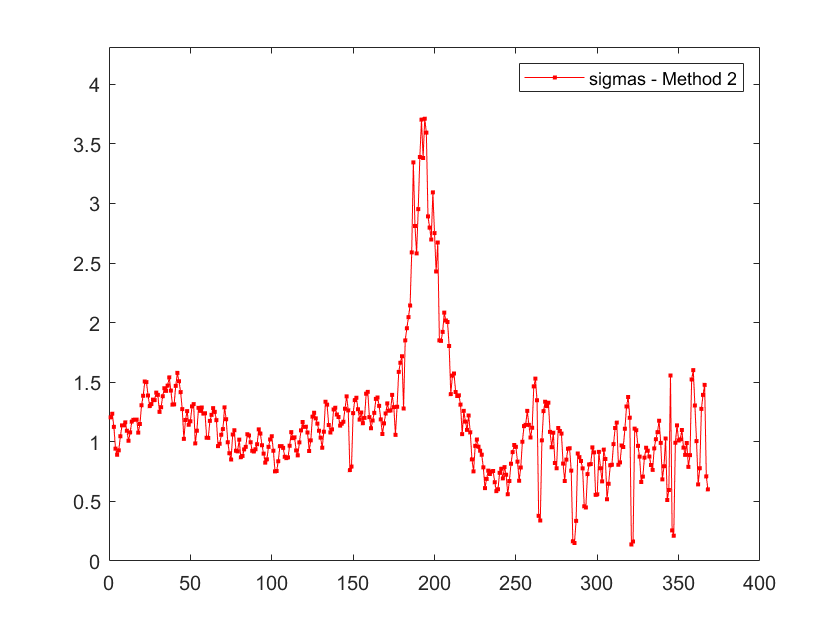}\vspace{-2mm} 
\caption{Replacement Number $\sigma_s(t)$ - Method 2}
\label{fig:sigmas2ndoutbreakworldM2}
\end{subfigure}
\caption{Parameters of SIR Model during Second Outbreak in the World }
\label{fig:parameters2ndoutbreakworld}\vspace{0mm}
\end{figure}

For the second outbreak, 
Figures \ref{fig:parameters2ndoutbreakitaly},  \ref{fig:parameters2ndoutbreakgermany}, and \ref{fig:parameters2ndoutbreakworld} plot the approximated parameters $\beta(t), \rho(t), \sigma(t)$, and $\sigma_s(t)$ using Method 1 (section \ref{sec:ParamM1}) and Method 2 (section \ref{sec:ParamM2}) in Italy, Germany, and the world respectively. 
Similarly to the first outbreak, the results of both methods are comparable in global behavior, with some minor variations. 
 Tables \ref{table:statbetaSecond}, \ref{table:statrhoSecond} \ref{table:statsigmaSecond}, and \ref{table:statsigmasSecond}  summarize the statistical properties of the approximated parameters $\beta(t)$, $\rho(t)$, $\sigma(t)$, and  $\sigma_s(t)$ respectively. 

 Moreover, Table \ref{table:table2ndrelparam} computes the relative L2 and Linfinity errors between the parameters computed using both methods. These errors are again of order $10^{-1}$ for all parameters and all three countries. As for the absolute errors of the parameters at some time $t_i$, then by comparing the means, it is clear that it is of order $10^{-3}$ for $\beta$ and $\rho$, $10^{-3}$ for $\sigma$ (except for Italy, $10^{-2}$), and $10^{-2}$ for $\sigma_s$ (except for Italy, $10^{-1}$).
\begin{table}[H]
\centering
\setlength{\tabcolsep}{10pt}
{\renewcommand{\arraystretch}{1.2}
\begin{tabular}{||c||c| c| c |c| c|c| c |c| c| c||} 
\cline{2-11}
  \multicolumn{1}{c||}{}& \multicolumn{10}{c||}{$\beta(t)$}\\
 \cline{2-11}
 \multicolumn{1}{c||}{}& \multicolumn{5}{c|}{Method 1} &  \multicolumn{5}{|c||}{Method 2}\\
 \cline{2-11}
  \multicolumn{1}{c||}{}& Me&Md&SD&Min&Max& Me&Md&SD &Min&Max\\
  \hline\hline
 Italy&0.08&0.07&0.04 &0.02&0.18&0.08&0.07&0.04  &0.02&0.19 \\ 
\hline
  Germany&0.06&0.06&0.04&0.004&0.24& 0.06&0.07&0.03 &-0.06& 0.16 \\ \hline
  World &0.03&0.03&0.02&$0.006$&0.1&0.03 &0.03 & 0.01&0.008&0.09 \\ 
\hline \hline
\end{tabular}\vspace{-3mm}
\caption{The Mean (Me), Median (Md) Standard Deviation (SD), Minimum (Min) and Maximum (Max) of the computed $\beta(t)$ using the two Methods during the second outbreak.}\label{table:statbetaSecond}}\vspace{-3mm}
\end{table}

\begin{table}[H]
\centering
\setlength{\tabcolsep}{10pt}
{\renewcommand{\arraystretch}{1.2}
\begin{tabular}{||c||c| c|c| c|c|c| c |c| c| c||} 
 \cline{2-11}
  \multicolumn{1}{c||}{}& \multicolumn{10}{c||}{$\rho(t)$}\\
 \cline{2-11}
 \multicolumn{1}{c||}{}& \multicolumn{5}{c|}{Method 1} &  \multicolumn{5}{|c||}{Method 2}\\
 \cline{2-11}
  \multicolumn{1}{c||}{}& Me&Md&SD&Min&Max& Me&Md&SD &Min&Max\\
  \hline\hline
 Italy&0.05&0.05&0.02&0.01&0.18 & 0.05&0.04&0.02  &0.01& 0.13\\ 
\hline
  Germany&0.05&0.05&0.03&0.01&0.18& 0.05&0.05&0.02
  &0.001& 0.11 \\ \hline
  World &0.03&0.03&0.04&$0.01$&0.75& 0.03& 0.03& 0.03&0.01&  0.03\\ 
\hline \hline
\end{tabular}\vspace{-3mm}
\caption{The Mean (Me), Median (Md), Standard Deviation (SD), Minimum (Min) and Maximum (Max) of the computed $\rho(t)$ using the two Methods during the second outbreak.}\label{table:statrhoSecond}}\vspace{-3mm}
\end{table}

\begin{table}[H]
\centering
\setlength{\tabcolsep}{10pt}
{\renewcommand{\arraystretch}{1.2}
\begin{tabular}{||c||c| c|c| c|c|c| c |c| c| c||} 
 \cline{2-11}
  \multicolumn{1}{c||}{}& \multicolumn{10}{c||}{$\sigma(t)$}\\
  \cline{2-11}
 \multicolumn{1}{c||}{}& \multicolumn{5}{c|}{Method 1} &  \multicolumn{5}{|c||}{Method 2}\\
 \cline{2-11}
  \multicolumn{1}{c||}{}& Me&Md&SD&Min&Max& Me&Md&SD &Min&Max\\
  \hline\hline
 Italy&2.01&1.62&1.63&0.41&7.63& 2.05&1.82&1.48 &0.54& 7.36 \\ 
\hline
  Germany&1.47&1.17&1.03&0.11&4.68& 1.46&1.22&0.92& -2.12&7.19\\ \hline
  World &1.24&1.16&0.56&0.078&4.49& 1.22 &1.13&  0.54&0.15&3.86 \\ 
\hline \hline
\end{tabular}\vspace{-3mm}
\caption{The Mean (Me), Median (Md), Standard Deviation (SD), Minimum (Min) and Maximum (Max) of the computed $\sigma(t)$ using the two Methods during the second outbreak.}\label{table:statsigmaSecond}}\vspace{-3mm}
\end{table}

\begin{table}[H]
\centering
\setlength{\tabcolsep}{10pt}
{\renewcommand{\arraystretch}{1.2}
\begin{tabular}{||c||c| c|c| c|c|c| c |c| c| c||} 
 \cline{2-11}
  \multicolumn{1}{c||}{}& \multicolumn{10}{c||}{$\sigma_s(t)$}\\
 \cline{2-11}
 \multicolumn{1}{c||}{}& \multicolumn{5}{c|}{Method 1} &  \multicolumn{5}{|c||}{Method 2}\\
 \cline{2-11}
  \multicolumn{1}{c||}{}& Me&Md&SD&Min&Max& Me&Md&SD &Min&Max\\
  \hline\hline Italy&1.77&1.47&1.49&3.36&6.87 & 1.83&1.67&1.35&0.43&6.63 \\ 
\hline
Germany&1.21&0.87&0.94&0.08&4.21& 1.20&0.97&0.85
  & -1.50&6.31 \\ \hline
  World &1.18&1.11&0.54&0.07&4.31&1.16 & 1.09&0.52  &0.14& 3.71\\ 
\hline \hline
\end{tabular}\vspace{-3mm}
\caption{The Mean (Me), Median (Md), Standard Deviation (SD), Minimum (Min) and Maximum (Max) of the computed $\sigma_s(t)$ using the two Methods during the second outbreak.}\label{table:statsigmasSecond}}\vspace{-3mm}
\end{table}

\begin{figure}[H]
\begin{subfigure} {0.5\textwidth}
\centering
\includegraphics[scale=0.3]{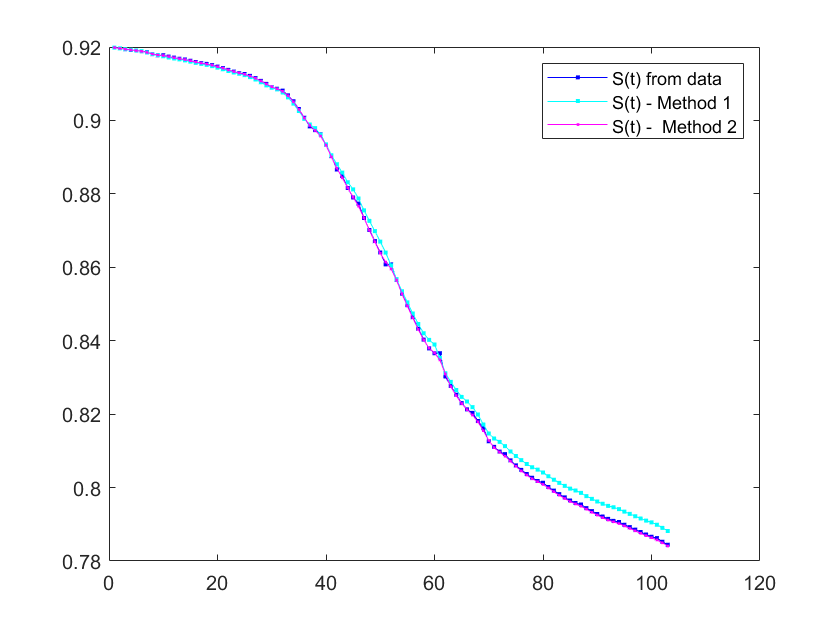}\vspace{-2mm}
\caption{Ratio of Susceptible s(t)}
\label{fig:SusceptiblesecondoutbreakItaly}
\end{subfigure}%
\begin{subfigure}{.5\linewidth}
\centering
\includegraphics[scale=.3]{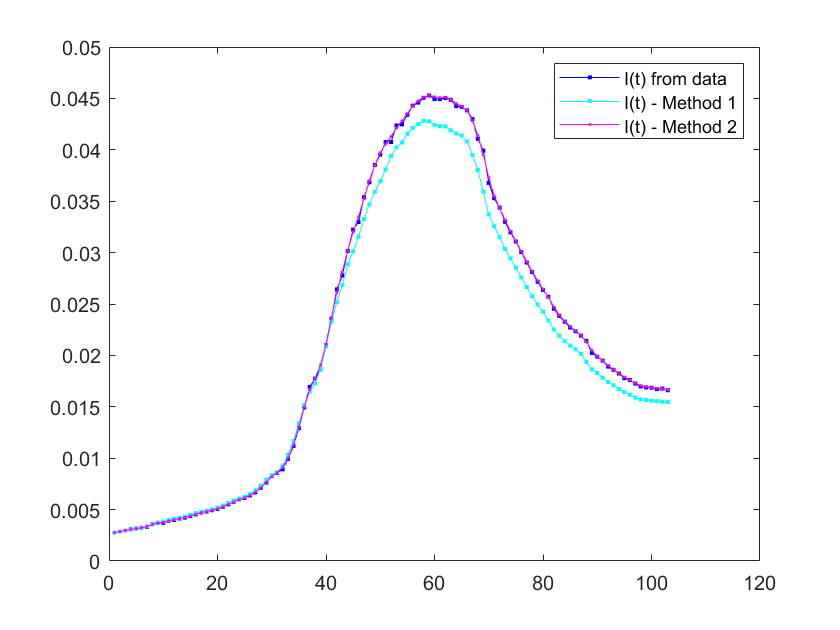}\vspace{-2mm}
\caption{Ratio of Infected i(t)}\label{fig:InfectedsecondoutbreakItaly}
\end{subfigure}
\begin{subfigure}{1.0\linewidth}
\centering
\includegraphics[scale=.3]{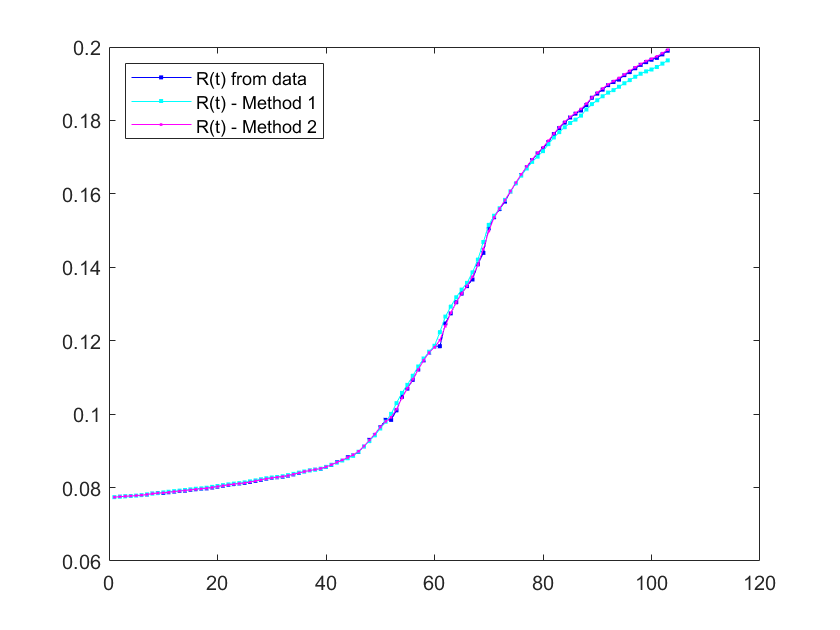}\vspace{-2mm}
\caption{Ratio of Removed r(t)}\label{fig:RemovedsecondoutbreakItaly}
\end{subfigure}

\caption{Comparison of Compartments' ratios from real data to those obtained using SIR model with approximated parameters, during the Second Outbreak in Italy }
\label{fig:SIRsecondoutbreakitaly}\vspace{-0.5cm}
\end{figure}

\begin{figure}[H]
\begin{subfigure} {0.5\textwidth}
\centering
\includegraphics[scale=0.3]{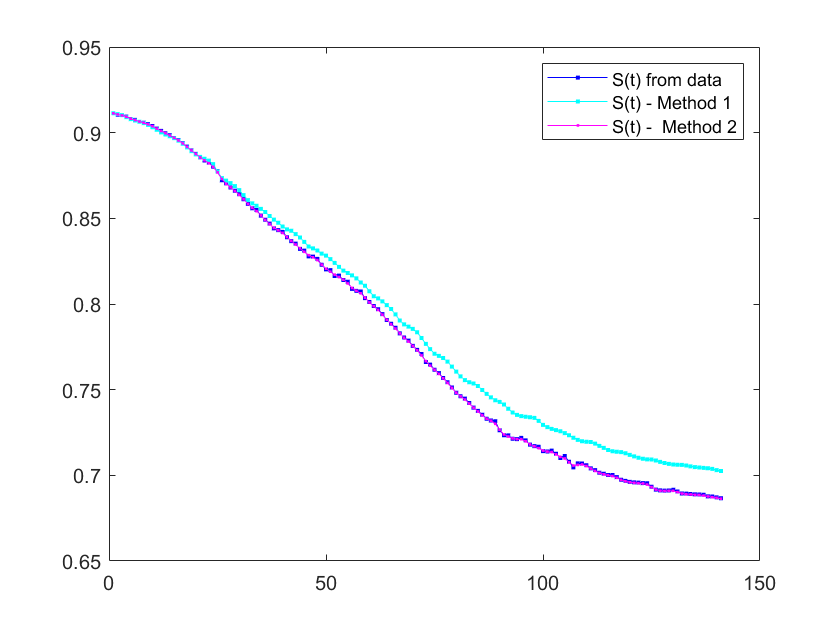}\vspace{-2mm}
\caption{Ratio of Susceptible s(t)}
\label{fig:Susceptiblesecondoutbreakgermany}
\end{subfigure}%
\begin{subfigure}{.5\linewidth}
\centering
\includegraphics[scale=.3]{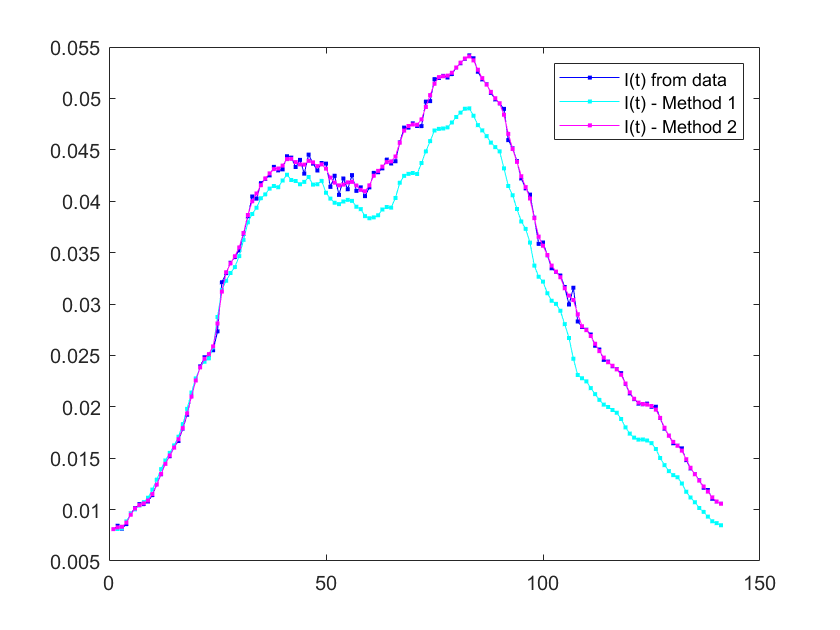}\vspace{-2mm}
\caption{Ratio of Infected i(t)}\label{fig:Infectedsecondoutbreakgermany}
\end{subfigure}
\begin{subfigure}{1.0\linewidth}
\centering
\includegraphics[scale=.3]{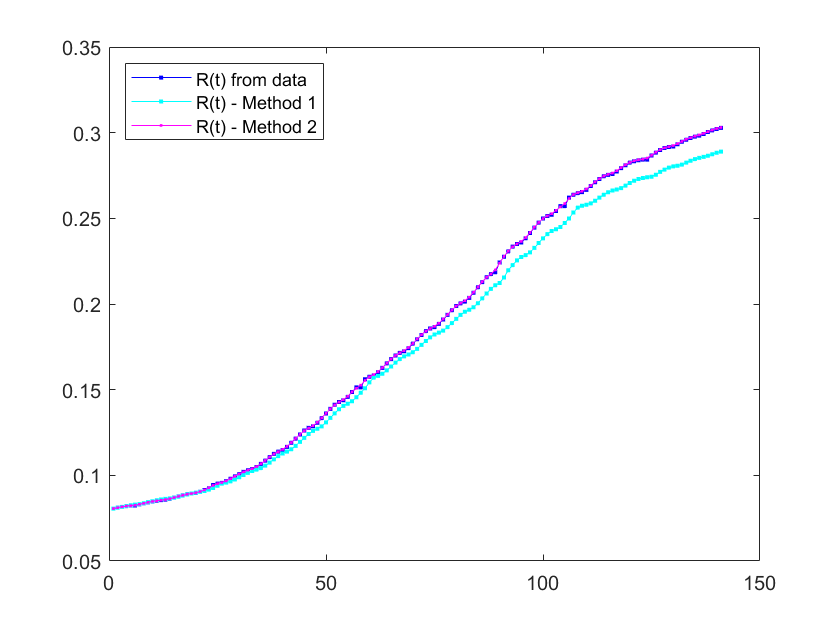}\vspace{-2mm}
\caption{Ratio of Removed r(t)}\label{fig:Removedsecondoutbreakgermany}
\end{subfigure}

\caption{Comparison of Compartments' ratios from real data to those obtained using SIR model with approximated parameters, during the Second Outbreak in Germany}
\label{fig:SIRsecondoutbreakgermany}\vspace{-0.5cm}
\end{figure}

\begin{figure}[H]
\begin{subfigure} {0.5\textwidth}
\centering
\includegraphics[scale=0.3]{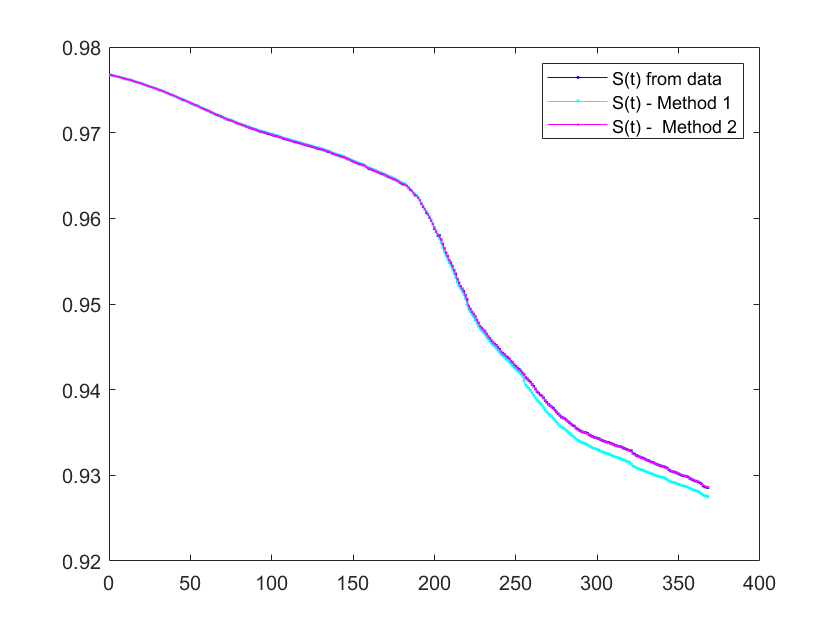}\vspace{-2mm}
\caption{Ratio of Susceptible s(t)}
\label{fig:Susceptible2ndoutbreakworld}
\end{subfigure}%
\begin{subfigure}{.5\linewidth}
\centering
\includegraphics[scale=.3]{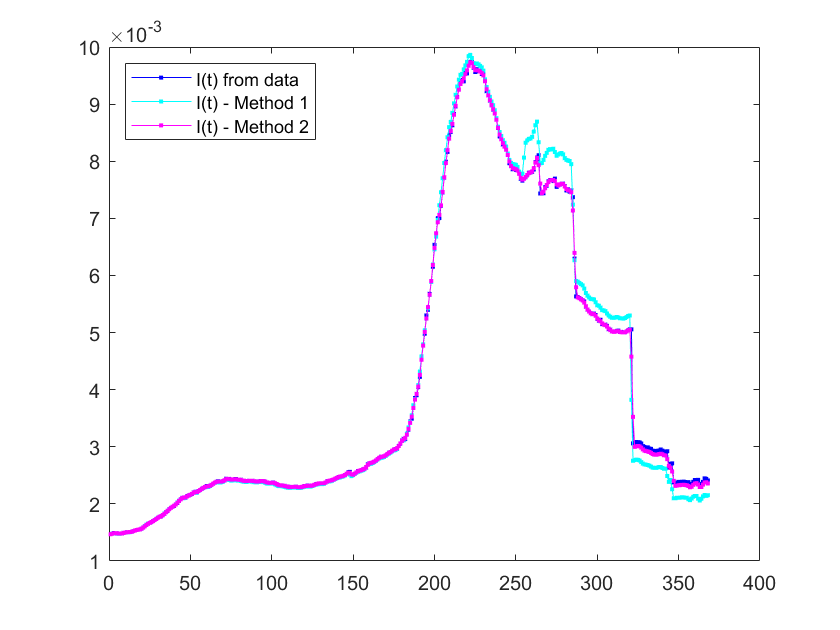}\vspace{-2mm}
\caption{Ratio of Infected i(t)}\label{fig:Infected2ndoutbreakworld}
\end{subfigure}
\begin{subfigure}{1.0\linewidth}
\centering
\includegraphics[scale=.3]{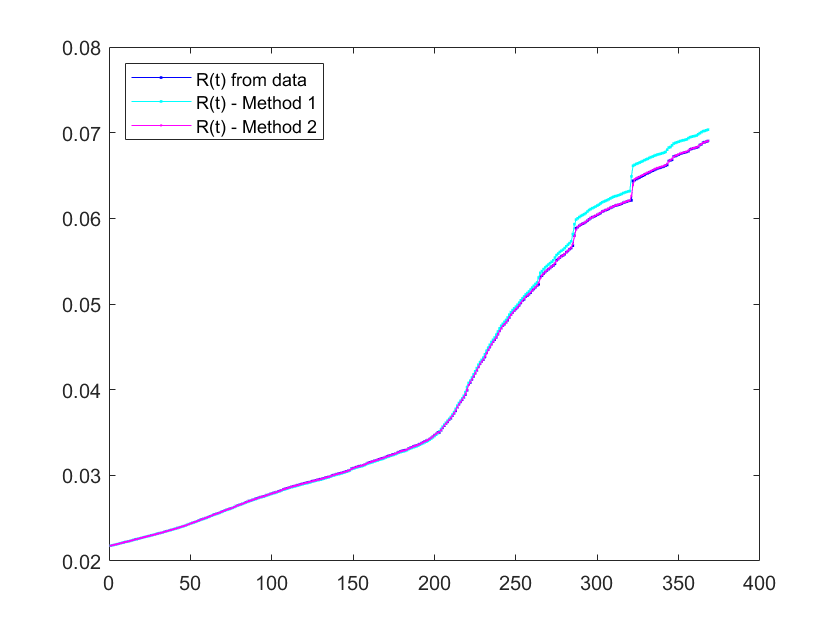}\vspace{-2mm}
\caption{Ratio of Removed r(t)}\label{fig:Removed2ndoutbreakworld}
\end{subfigure}

\caption{Comparison of Compartments' ratios from real data to those obtained using SIR model with approximated parameters, during the Second Outbreak in the World }
\label{fig:SIRsecondoutbreakworld}
\end{figure}

 \begin{table}[H]
\centering
\setlength{\tabcolsep}{10pt}
{\renewcommand{\arraystretch}{1.4}
\begin{tabular}{||c| c| c| c |c||} 
 \hline
  &Norm & Italy  & Germany  & World  \\ 
 \hline\hline
\multirow{2}{*}{$\beta$} &$L_2$ & $ 2.380* 10^{-1}$ &$  5.518* 10^{-1}$ &  $ 2.506 * 10^{-1}$  \\ 
 \cline{2-5}
  & $L_\infty$ & $4.515 * 10^{-1}$  & $6.485 * 10^{-1}$ & $7.159 * 10^{-1}$ \\ 
\hline
\multirow{2}{*}{$\rho$} &$L_2$ & $ 2.604* 10^{-1}$ &$  4.858* 10^{-1}$ &  $ 7.602 * 10^{-1}$ \\ 
\cline{2-5}
& $L_\infty$ & $ 4.407* 10^{-1}$ & $   9.434* 10^{-1}$ & $   9.112 * 10^{-1}$ \\ 
\hline
 \multirow{2}{*}{$\sigma$} & $L_2$ & $ 1.146* 10^{-1}$ &$4.485* 10^{-1}$ &  $   1.737 * 10^{-1}$ \\
\cline{2-5}
 & $L_\infty$ & $ 1.825* 10^{-1}$ & $6.608* 10^{-1}$ & $  3.606 * 10^{-1}$ \\  
 \hline
\multirow{2}{*}{$\sigma_s$} &$L_2$ & $1.137* 10^{-1}$ &$  4.368* 10^{-1}$ &  $ 1.711 * 10^{-1}$ \\ 
 \cline{2-5}
  & $L_\infty$ & $1.820* 10^{-1}$  & $6.782 * 10^{-1}$ & $ 3.529 * 10^{-1}$ \\ 
 \hline
\end{tabular}
\caption{The $L_2$ and $L_\infty$ relative errors between the computed parameters based on Methods 1 and 2, for the second outbreak.}
\label{table:table2ndrelparam}}
\end{table}
Again, we validate these results, by running the SIR model with the obtained parameters $\beta(t), \rho(t)$ using method 1 or method 2, and the corresponding initial values as discussed in section \ref{sec:valid}. Figures \ref{fig:SIRsecondoutbreakitaly}, \ref{fig:SIRsecondoutbreakgermany},  and \ref{fig:SIRsecondoutbreakworld}  compare  the behavior of the ratio of compartments $s(t), i(t), r(t)$ of the real data to the simulated values. Moreover, Table \ref{table:tablesecondSIRrel} shows the relative errors between the real data and simulated values. The trends of the three compartments in all three simulations are as expected and similar to the first outbreak: the number of infected people increases then decreases, the number of susceptible people decreases, and the number of removed people increases. Moreover, we also see a slight difference in the simulated results using methods 1 and 2, where the second is more accurate with an L2 relative error 10 times less than the first method (Table \ref{table:tablesecondSIRrel}). Note that for this period the simulated data using method 1 for Germany is comparable to that of Italy, since there is no data inconsistencies, unlike in the first outbreak.

By comparing the table of relative errors for the second outbreak (table \ref{table:tablesecondSIRrel}) to that for the first outbreak (table \ref{table:table1stSIRrel}), one can note that the errors using method 2 in the second outbreak are of the same order with some values slightly higher (example s and r errors for Italy ). 

\begin{table}[H]
\centering
\setlength{\tabcolsep}{10pt}
{\renewcommand{\arraystretch}{1.4}
\begin{tabular}{||c|c| c| c| c |c||} 
 \hline
 Method& &Norm & Italy  & Germany  & World  \\ 
 \hline\hline
\multirow{6}{*}{1}&\multirow{2}{*}{S} &$L_2$ & $ 2.411* 10^{-3}$ &$ 1.374* 10^{-2}$ &  $  7.396* 10^{-4}$  \\ 
 \cline{3-6}
 & & $L_\infty$ & $4.248 * 10^{-3}$  & $ 1.963* 10^{-2}$ & $ 1.600  * 10^{-3}$ \\ 
 \cline{2-6}
&\multirow{2}{*}{I} &$L_2$ & $6.544* 10^{-2}$ &$9.349 * 10^{-2}$ &  $ 4.534  * 10^{-2}$ \\ 
 \cline{3-6}
&& $L_\infty$ & $ 8.801* 10^{-2}$ & $ 1.275* 10^{-1}$ & $  1.267 * 10^{-1}$ \\ 
  \cline{2-6}
 &\multirow{2}{*}{R} & $L_2$ & $ 9.119* 10^{-3}$ &$3.875* 10^{-2}$ &  $ 1.615 * 10^{-2}$ \\\cline{3-6}
 && $L_\infty$ & $1.930 * 10^{-2}$ & $4.596* 10^{-2}$ & $  4.053 * 10^{-2}$ \\  
 \hline 
 \multirow{6}{*}{2}&\multirow{2}{*}{S} &$L_2$ & $3.936* 10^{-4}$ &$5.576* 10^{-4}$&  $2.368* 10^{-5}$\\ 
 \cline{3-6}
 & & $L_\infty$ & $1.971 *10^{-3}$   &  $1.616 * 10^{-3}$ & $1.530 * 10^{-4}$ \\ 
 \cline{2-6}
&\multirow{2}{*}{I} &$L_2$ &  $  6.022*10^{-3}$& $ 9.316* 10^{-3}$& $ 9.996 * 10^{-3}$ \\ 
 \cline{3-6}
&& $L_\infty$ & $1.171*10^{-2}$  & $2.199 * 10^{-2}$ & $4.902 * 10^{-2}$ \\ 
  \cline{2-6}
 &\multirow{2}{*}{R} & $L_2$ &$2.301* 10^{-3}$&$ 1.607* 10^{-3}$  &  $1.185 * 10^{-3}$ \\\cline{3-6}
 && $L_\infty$ & $ 8.309 * 10^{-3}$ & $4.946 * 10^{-3}$  & $8.032 * 10^{-3}$ \\  
 \hline
\end{tabular}
\caption{ $L_2$ and $L_\infty$ relative errors of the computed S,I,R from the time-dependent model with the S,I,R collected from data during the second outbreak, where $\beta,\rho$ are computed using Method 1 or 2. }\label{table:tablesecondSIRrel}}
\end{table}

We observe the following for the parameters ($\beta$, $\rho$, $\sigma$, $\sigma_s$).  The infection rate in Italy and Germany has a similar behavior to the first outbreak (figures \ref{fig:beta2ndoutbreakitalyM1},   \ref{fig:beta2ndoutbreakitalyM2}, \ref{fig:beta2ndoutbreakgermanyM1},  and \ref{fig:beta2ndoutbreakgermanyM2} ). On the other hand, the mean and the median of the infection rate for Italy are higher than those for Germany, which could be an indicator that the second outbreak in Italy was stronger and more spread out,
as more people are getting infected and less people were getting removed (i.e. the infection period lasted  longer).
This is validated by the fact that $\sigma$ and $\sigma_s$ are larger for Italy than Germany. 

Moreover, by comparing the parameters obtained in the second outbreak to that of the first outbreak, it is clear that the second outbreak in Italy was not as strong as the first, not because it was less spread (i.e. smaller $\beta$). On the contrary, more people were infected, but also more people were removed (recovered  or died) at a faster rate than the first outbreak. Thus, $\sigma$ and $\sigma_s$ are smaller than the first outbreak.  However, in Germany the situation is reversed. In the second outbreak $\beta$ and $\rho$ decreased, but at different rates, leading to a larger $\sigma$ and $\sigma_s$.

As for the world, some fluctuations are observed (figures \ref{fig:beta2ndoutbreakworldM1} and \ref{fig:beta2ndoutbreakworldM2}), where the maximum infection rate attained at some time between the 192$^{nd}$ day  and the 250$^{th}$ has a value of 0.1 (method 1) or 0.0872 (method 2), that is smaller than that attained during the first outbreak. Similarly, the mean, median and standard deviation are relatively smaller than those of the first outbreak  since the range of $\beta$ values is tighter ([0,0.1]) than that of the first outbreak ([0,0.2]). Whereas the removal rate $\rho$ is comparable to that of the first outbreak, thus the mean of $\sigma$ and $\sigma_s$ in the second outbreak are slightly less than those of the first outbreak.

Moreover, The world's replacement number is naturally smaller than both the one in Italy and Germany as many countries did not provide accurate or adequate data, or some countries were more closed off during the pandemic, causing the global replacement number to fall down.

\subsection{Other Periods}\label{sec:random}
\begin{figure}[H]
\begin{subfigure}{0.5\textwidth}
\centering
\includegraphics[scale=0.28]{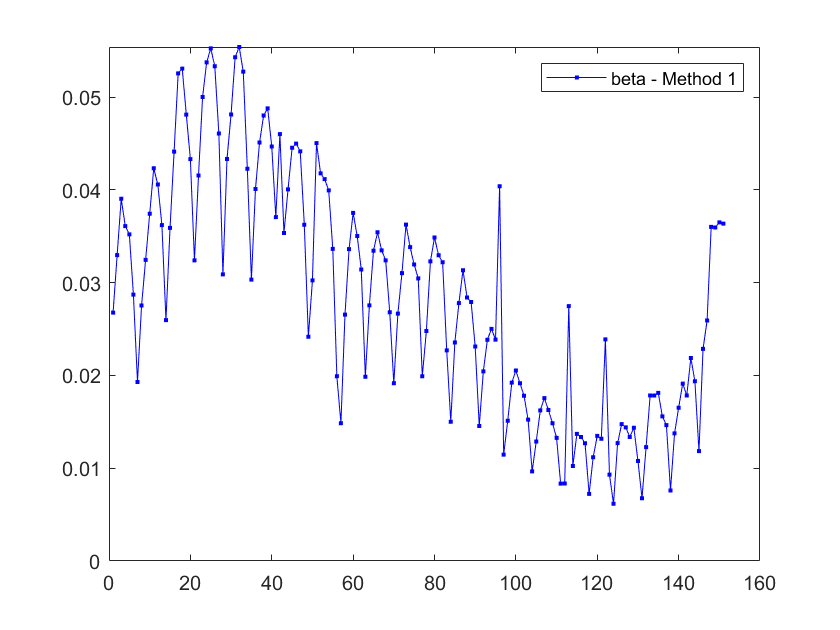} \vspace{-2mm}
\caption{Infection Rate $\beta(t)$ - Method 1}
\label{fig:betaotheritalyM1}
\end{subfigure}
\hfill
\begin{subfigure}{0.5\textwidth}
\centering
\includegraphics[scale=0.28]{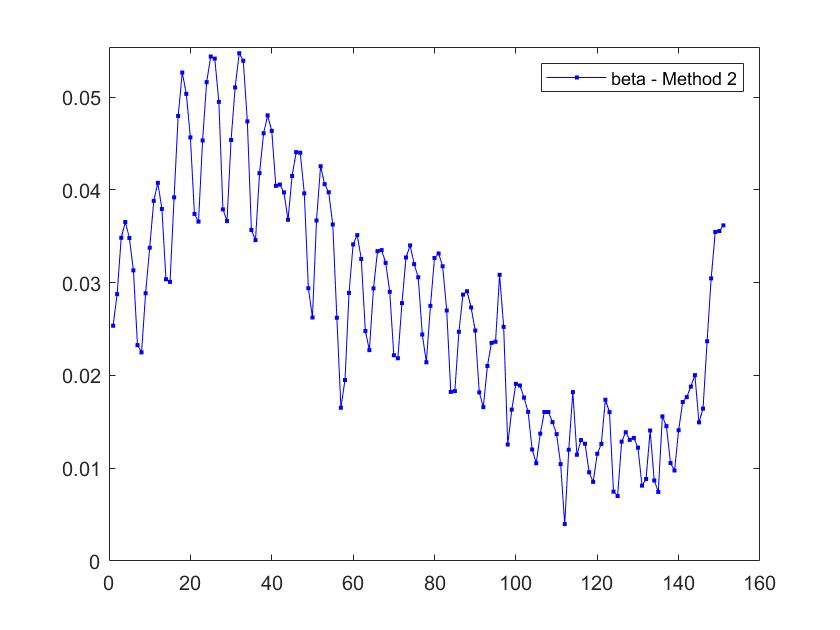} \vspace{-2mm}
\caption{Infection Rate $\beta(t)$ - Method 2}
\label{fig:betaotheritalyM2}
\end{subfigure}
\newline
\begin{subfigure}{0.5\textwidth}
\centering
\includegraphics[scale=0.28]{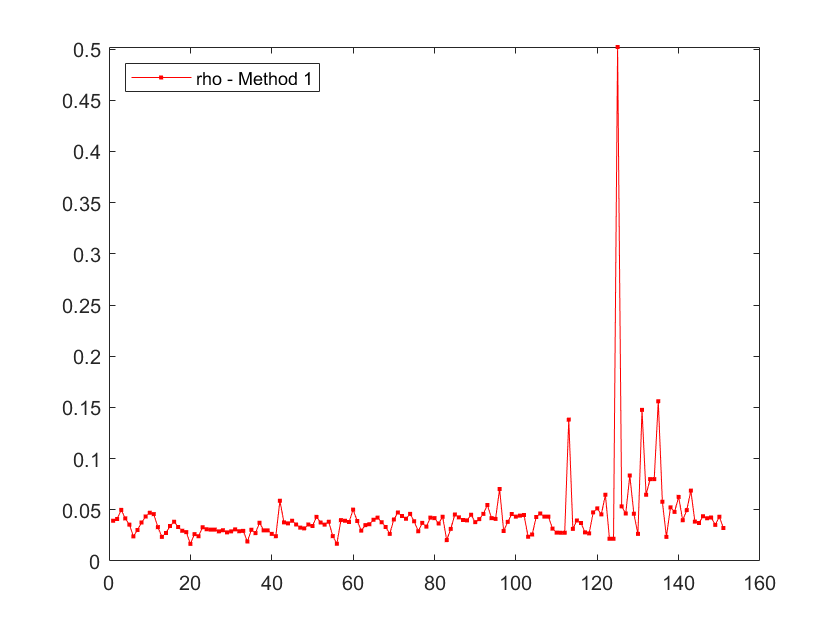} \vspace{-2mm}
\caption{Removal Rate $\rho(t)$ - Method 1 }
\label{fig:rhootheritalyM1}
\end{subfigure}
\hfill
\begin{subfigure}{0.5\textwidth}
\centering
\includegraphics[scale=0.28]{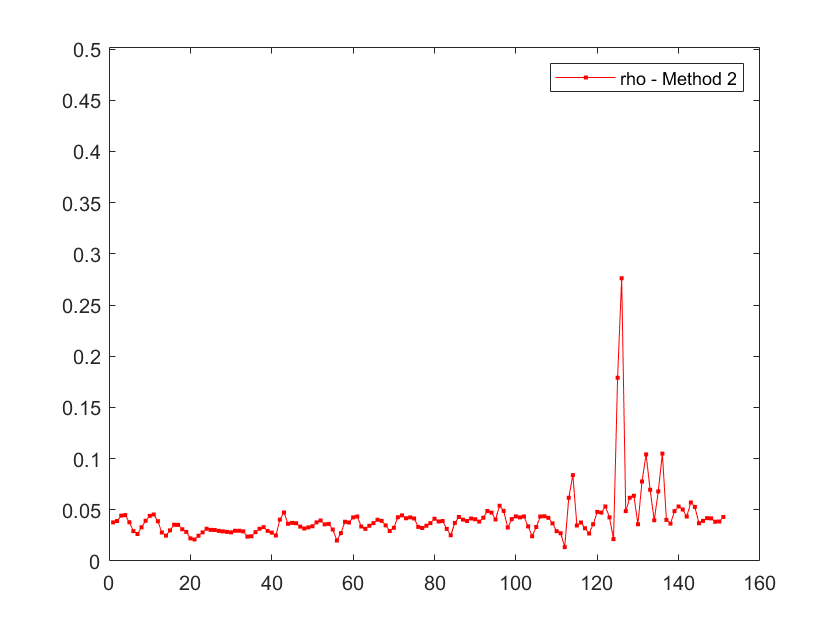} \vspace{-2mm}
\caption{Removal Rate $\rho(t)$ - Method 2 }
\label{fig:rhootheritalyM2}
\end{subfigure}
\newline
\begin{subfigure}{0.5\textwidth}
\centering
\includegraphics[scale=0.28]{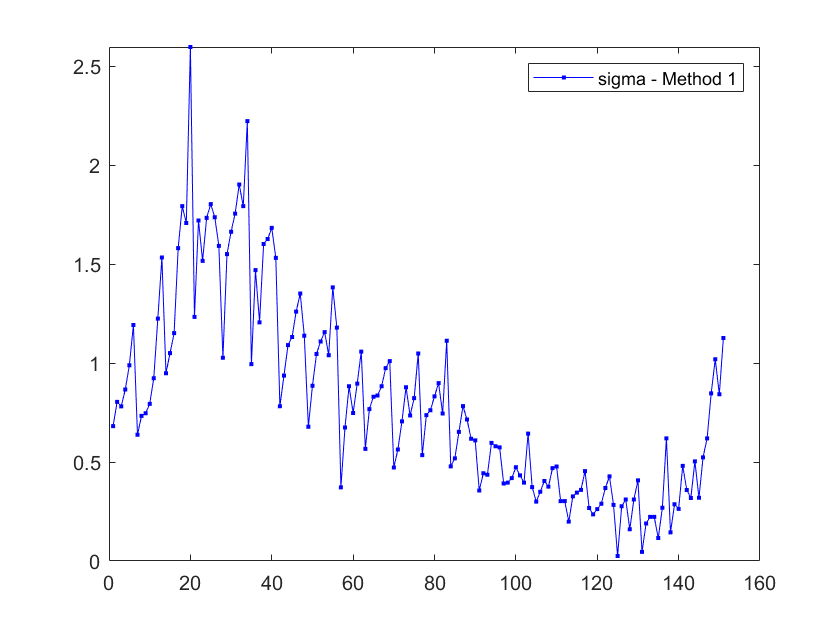} \vspace{-2mm}
\caption{Reproduction Factor $\sigma(t)$ - Method 1}
\label{fig:sigmaotherkitalyM1}
\end{subfigure}
\hfill
\begin{subfigure}{0.5\textwidth}
\centering
\includegraphics[scale=0.28]{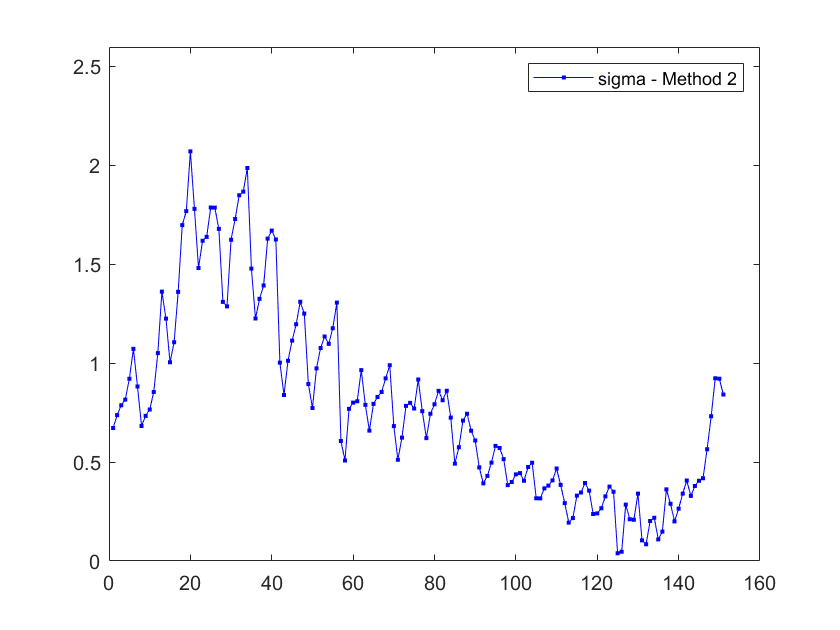} \vspace{-2mm}
\caption{Reproduction Factor $\sigma(t)$ - Method 2}
\label{fig:sigmaotheritalyM2}
\end{subfigure}
\newline
\begin{subfigure}{0.5\textwidth}
\centering
\includegraphics[scale=0.28]{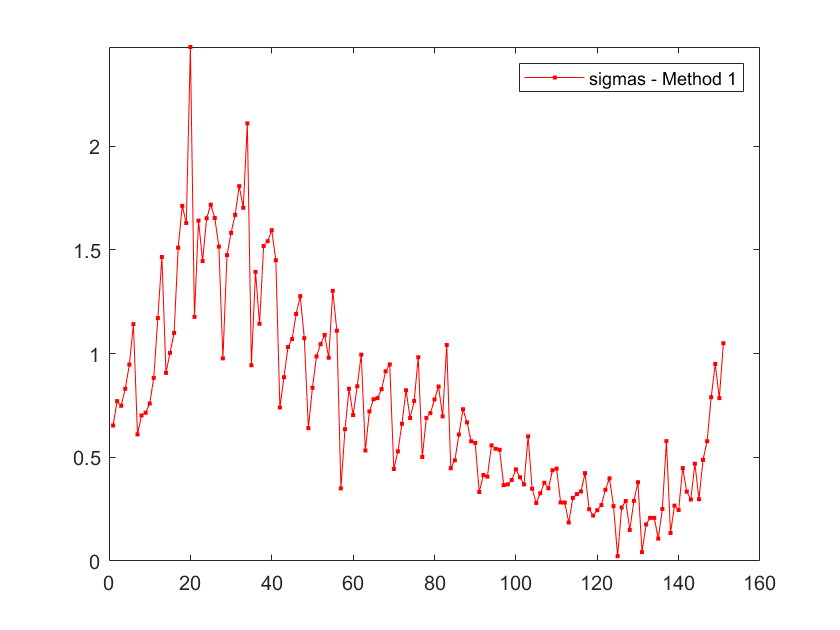}\vspace{-2mm} 
\caption{Replacement Number $\sigma_s(t)$ - Method 1}
\label{fig:sigmasotheritalyM1}
\end{subfigure}\hfill
\begin{subfigure}{0.5\textwidth}
\centering
\includegraphics[scale=0.28]{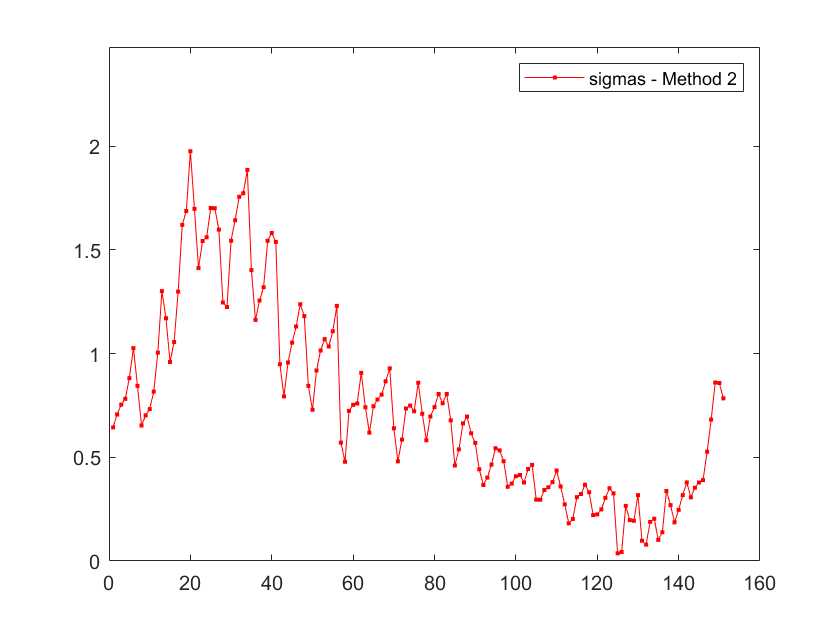}\vspace{-2mm} 
\caption{Replacement Number $\sigma_s(t)$ - Method 2}
\label{fig:sigmasotheritalyM2}
\end{subfigure}
\caption{Parameters of SIR Model during the Other Period in Italy }
\label{fig:parametersotheritaly}\vspace{0mm}
\end{figure}

\begin{figure}[H]
\begin{subfigure}{0.5\textwidth}
\centering
\includegraphics[scale=0.28]{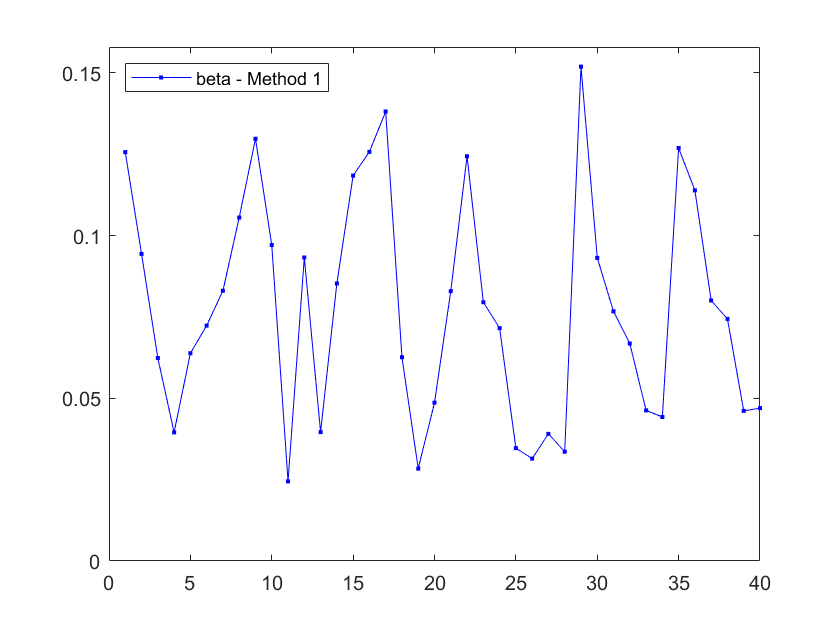} \vspace{-2mm}
\caption{Infection Rate $\beta(t)$ - Method 1}
\label{fig:betaothergermanyM1}
\end{subfigure}
\hfill
\begin{subfigure}{0.5\textwidth}
\centering
\includegraphics[scale=0.28]{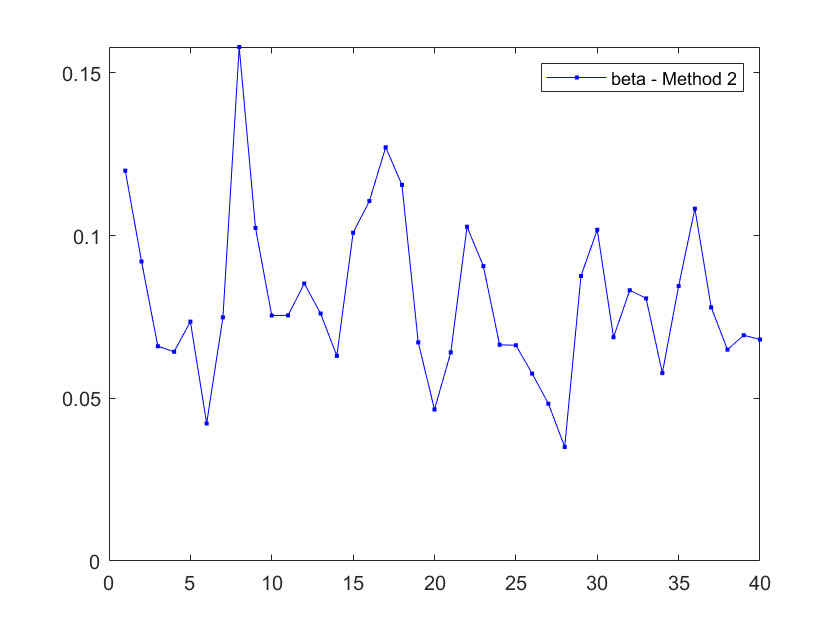} \vspace{-2mm}
\caption{Infection Rate $\beta(t)$ - Method 2}
\label{fig:betaothergermanyM2}
\end{subfigure}
\newline
\begin{subfigure}{0.5\textwidth}
\centering
\includegraphics[scale=0.28]{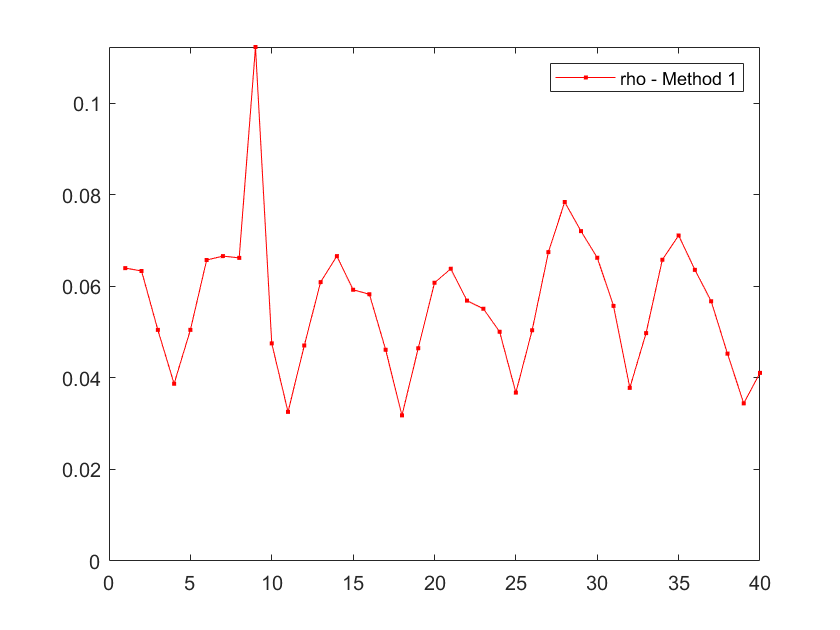} \vspace{-2mm}
\caption{Removal Rate $\rho(t)$ - Method 1 }
\label{fig:rhoothergermanyM1}
\end{subfigure}
\hfill
\begin{subfigure}{0.5\textwidth}
\centering
\includegraphics[scale=0.28]{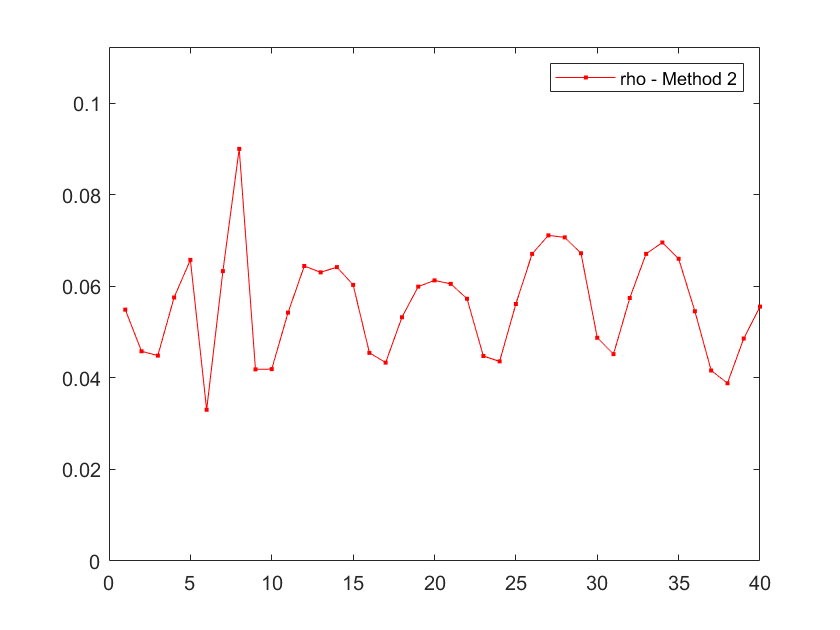} \vspace{-2mm}
\caption{Removal Rate $\rho(t)$ - Method 2 }
\label{fig:rhoothergermanyM2}
\end{subfigure}
\newline
\begin{subfigure}{0.5\textwidth}
\centering
\includegraphics[scale=0.28]{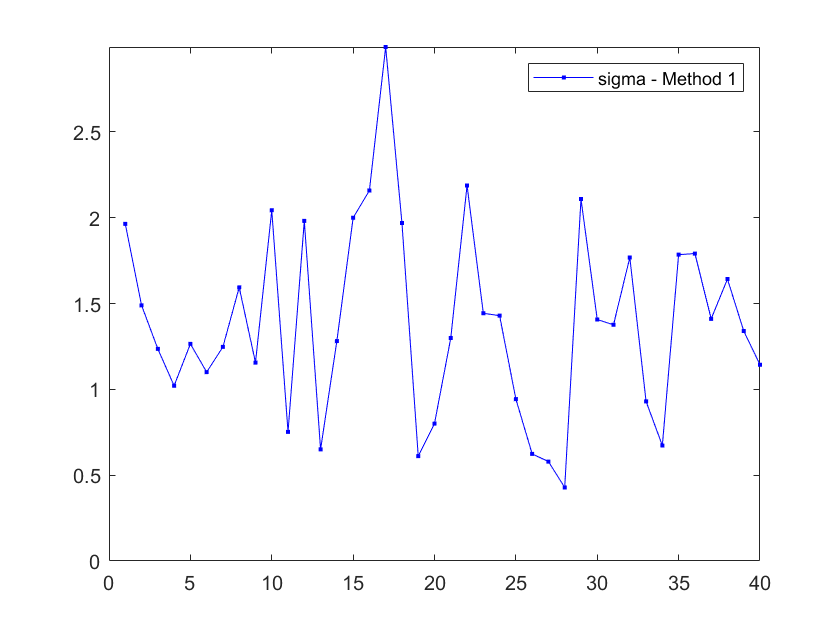} \vspace{-2mm}
\caption{Reproduction Factor $\sigma(t)$ - Method 1}
\label{fig:sigmaothergermanyM1}
\end{subfigure}
\hfill
\begin{subfigure}{0.5\textwidth}
\centering
\includegraphics[scale=0.28]{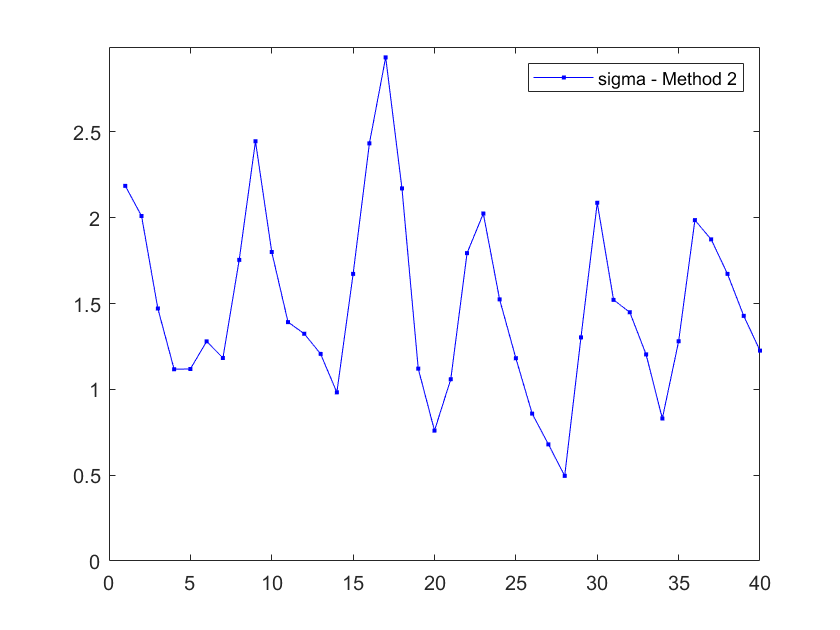} \vspace{-2mm}
\caption{Reproduction Factor $\sigma(t)$ - Method 2}
\label{fig:sigmaothergermanyM2}
\end{subfigure}
\newline
\begin{subfigure}{0.5\textwidth}
\centering
\includegraphics[scale=0.28]{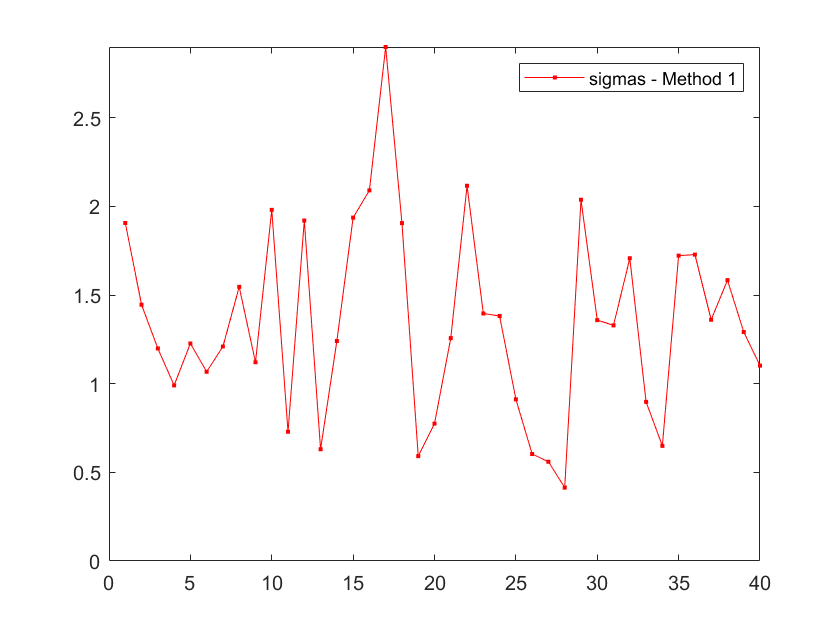}\vspace{-2mm} 
\caption{Replacement Number $\sigma_s(t)$ - Method 1}
\label{fig:sigmasothergermanyM1}
\end{subfigure}\hfill
\begin{subfigure}{0.5\textwidth}
\centering
\includegraphics[scale=0.28]{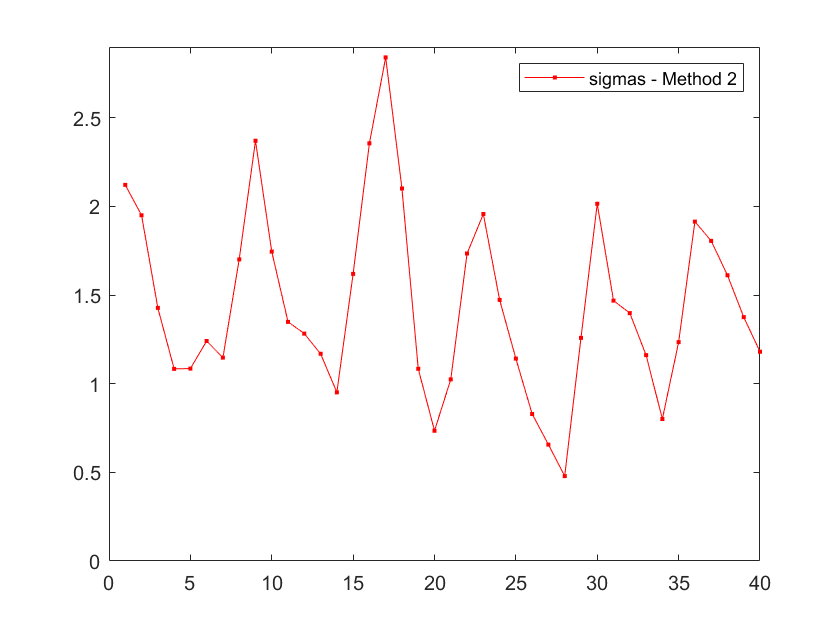}\vspace{-2mm} 
\caption{Replacement Number $\sigma_s(t)$ - Method 2}
\label{fig:sigmasothergermanyM2}
\end{subfigure}
\caption{Parameters of SIR Model during Other Period in Germany }
\label{fig:parametersothergermany}\vspace{0mm}
\end{figure}

\begin{figure}[H]
\begin{subfigure}{0.5\textwidth}
\centering
\includegraphics[scale=0.28]{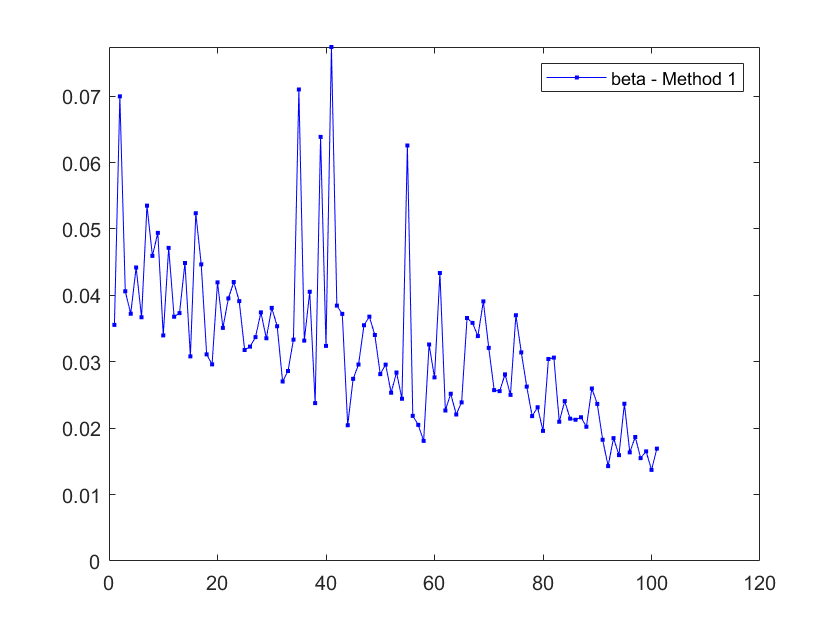}\vspace{-2mm}
\caption{Infection Rate $\beta(t)$ - Method 1}
\label{fig:betaotherworldM1}
\end{subfigure}
\hfill
\begin{subfigure}{0.5\textwidth}
\centering
\includegraphics[scale=0.28]{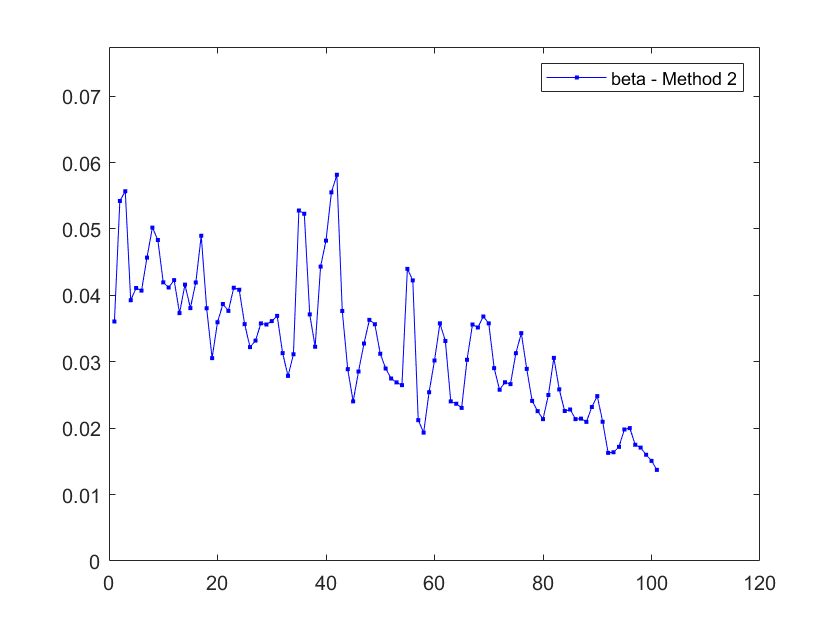} \vspace{-2mm}
\caption{Infection Rate $\beta(t)$ - Method 2}
\label{fig:betaotherworldM2}
\end{subfigure}
\newline
\begin{subfigure}{0.5\textwidth}
\centering
\includegraphics[scale=0.28]{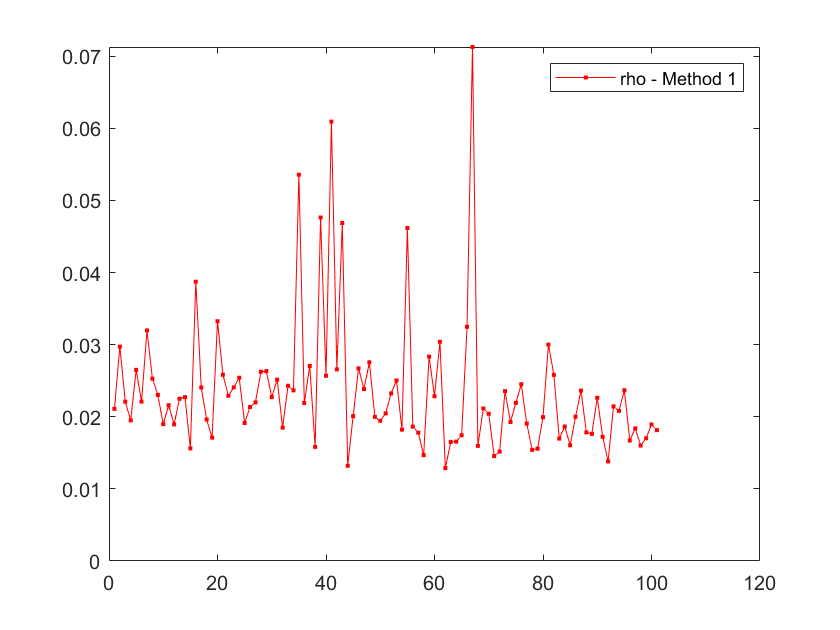} \vspace{-2mm}
\caption{Removal Rate $\rho(t)$ - Method 1 }
\label{fig:rhootherworldM1}
\end{subfigure}
\hfill
\begin{subfigure}{0.5\textwidth}
\centering
\includegraphics[scale=0.28]{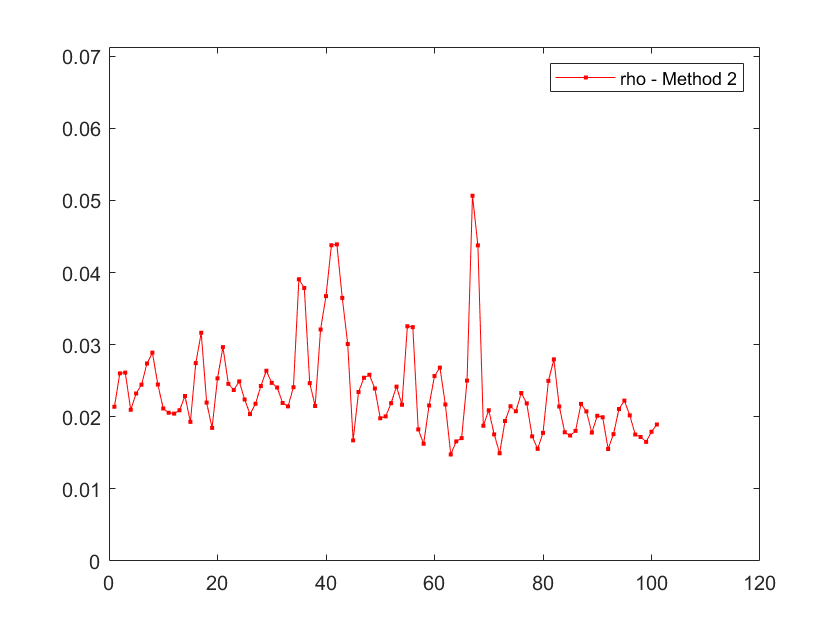} \vspace{-2mm}
\caption{Removal Rate $\rho(t)$ - Method 2 }
\label{fig:rhootherworldM2}
\end{subfigure}
\newline
\begin{subfigure}{0.5\textwidth}
\centering
\includegraphics[scale=0.28]{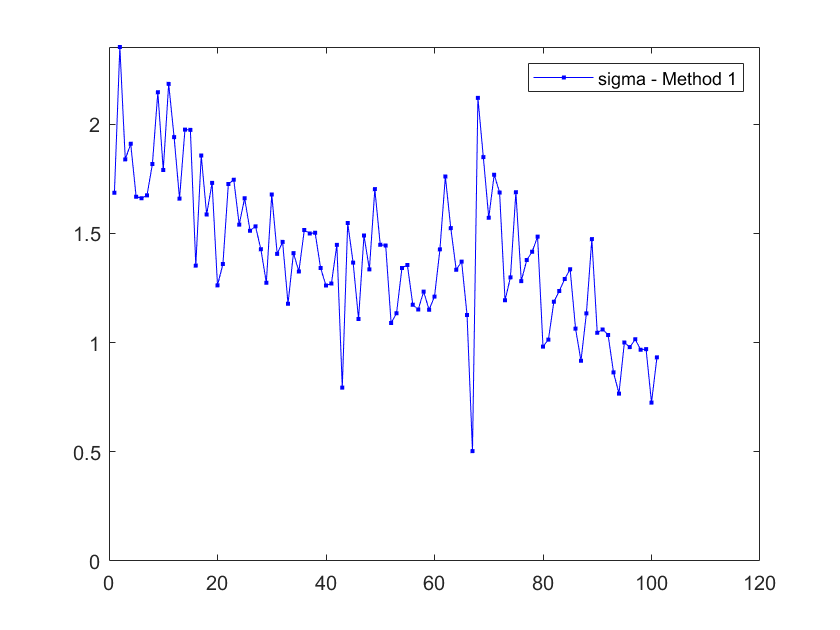} \vspace{-2mm}
\caption{Reproduction Factor $\sigma(t)$ - Method 1}
\label{fig:sigmaotherworldM1}
\end{subfigure}
\hfill
\begin{subfigure}{0.5\textwidth}
\centering
\includegraphics[scale=0.28]{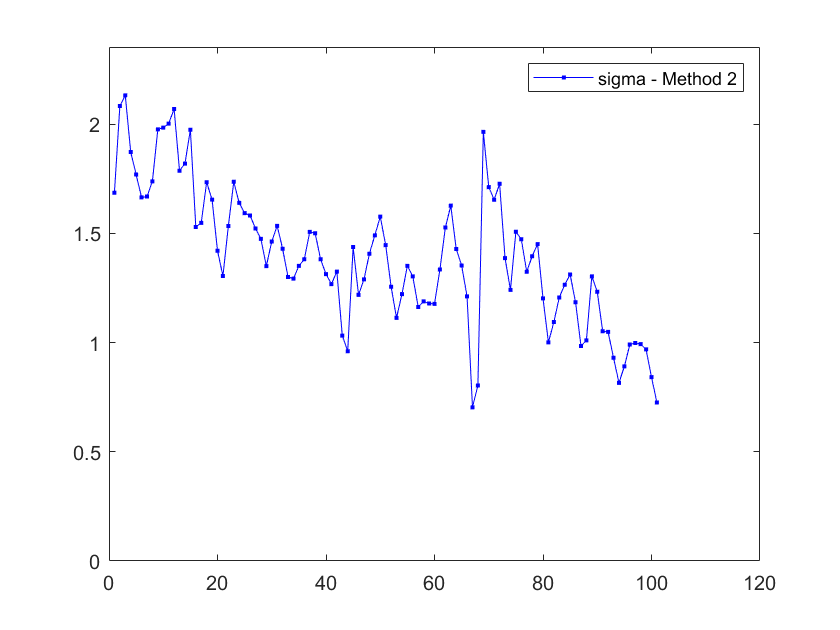} \vspace{-2mm}
\caption{Reproduction Factor $\sigma(t)$ - Method 2}
\label{fig:sigmaotherworldM2}
\end{subfigure}
\newline
\begin{subfigure}{0.5\textwidth}
\centering
\includegraphics[scale=0.28]{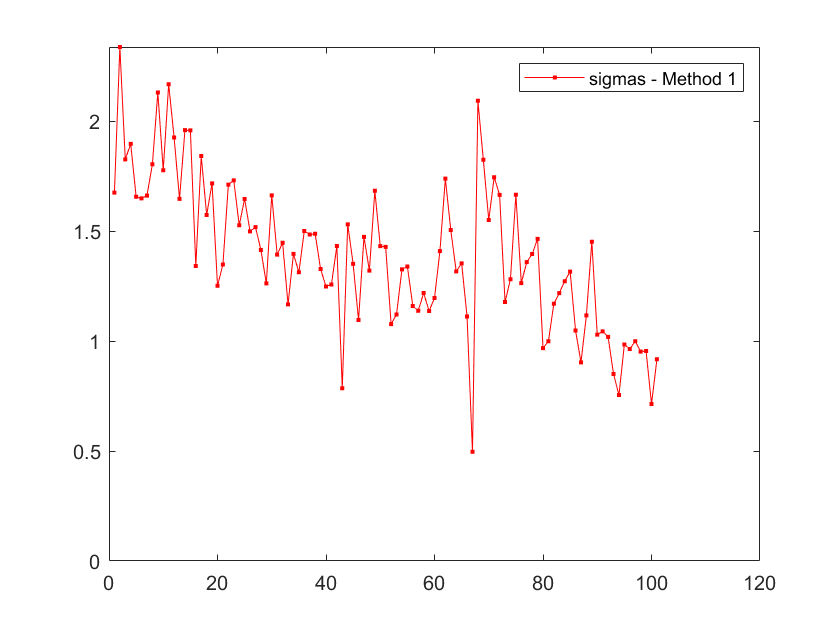}\vspace{-2mm} 
\caption{Replacement Number $\sigma_s(t)$ - Method 1}
\label{fig:sigmasotherworldM1}
\end{subfigure}\hfill
\begin{subfigure}{0.5\textwidth}
\centering
\includegraphics[scale=0.28]{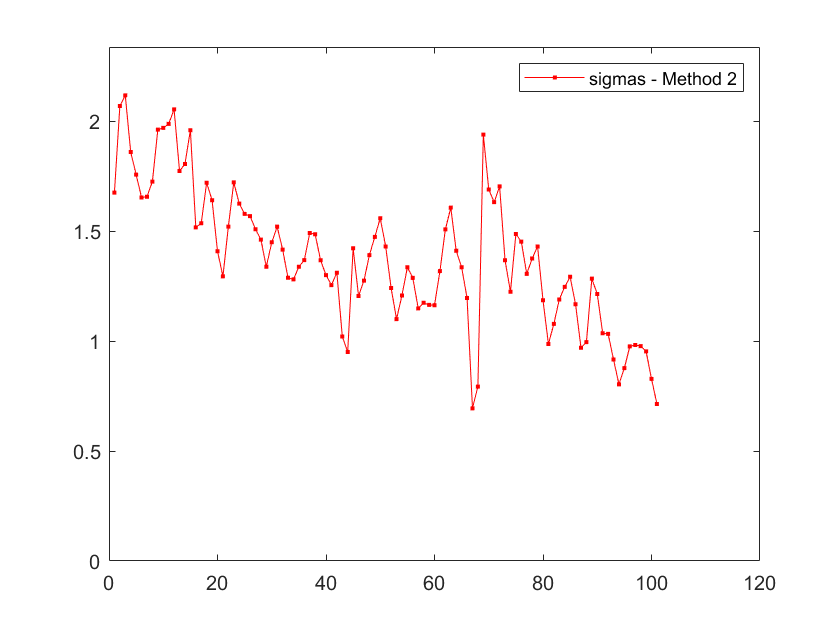}\vspace{-2mm} 
\caption{Replacement Number $\sigma_s(t)$ - Method 2}
\label{fig:sigmasotherworldM2}
\end{subfigure}
\caption{Parameters of SIR Model during Other Period in the World }
\label{fig:parametersotherworld}\vspace{0mm}
\end{figure}

Figures \ref{fig:parametersotheritaly},  \ref{fig:parametersothergermany}, and \ref{fig:parametersotherworld} plot the approximated parameters $\beta(t), \rho(t), \sigma(t)$, and $\sigma_s(t)$ using Method 1 (section \ref{sec:ParamM1}) and Method 2 (section \ref{sec:ParamM2}) during the first outbreak in Italy, Germany, and the world respectively. Clearly the results of both methods are similar in global behavior, with some different variations. These differences are detailed in  
 Tables \ref{table:statbetaother}, \ref{table:statrhoother} \ref{table:statsigmaother}, and \ref{table:statsigmasother} that summarize the statistical properties (mean, median, standard deviation) of the approximated parameters $\beta(t)$, $\rho(t)$, $\sigma(t)$, and  $\sigma_s(t)$ respectively. 
 
Moreover, Table \ref{table:tableotherrelparam} computes the relative L2 and Linfinity errors between the parameters computed using both methods. These errors are of order $10^{-1}$ for all parameters and all three countries. 
As for the absolute errors of the parameters at some time $t_i$, then by comparing the means, it is clear that it is of orders $10^{-4}$, $10^{-2}$, and $10^{-3}$ for $\beta$ in Italy, Germany, and the world respectively. Whereas, it is of the order $10^{-3}$ for $\rho$, and $10^{-1}$ for $\sigma$ and $\sigma_s$ (except for the world, $10^{-2}$).

\begin{table}[H]
\centering
\setlength{\tabcolsep}{10pt}
{\renewcommand{\arraystretch}{1.2}
\begin{tabular}{||c||c| c| c |c| c|c| c |c| c| c||} 
\cline{2-11}
  \multicolumn{1}{c||}{}& \multicolumn{10}{c||}{$\beta(t)$}\\
 \cline{2-11}
 \multicolumn{1}{c||}{}& \multicolumn{5}{c|}{Method 1} &  \multicolumn{5}{|c||}{Method 2}\\
 \cline{2-11}
  \multicolumn{1}{c||}{}& Me&Md&SD&Min&Max& Me&Md&SD &Min&Max\\
  \hline\hline
 Italy&0.03&0.03&0.01&0.01&0.05&0.03&0.03&0.01&0.004&0.05 \\ 
\hline
  Germany&0.08&0.08&0.04&0.002&0.15& 0.08&0.08&0.03  &0.04& 0.16\\ \hline
  World &0.03&0.03&0.01&0.01&0.08&  0.03&0.03& 0.01&0.01&  0.06\\ 
\hline \hline
\end{tabular}\vspace{-3mm}
\caption{The Mean (Me), Median (Md) Standard Deviation (SD), Minimum (Min) and Maximum (Max) of the computed $\beta(t)$ using the two Methods,  for a random period. }\label{table:statbetaother}}\vspace{-3mm}
\end{table}

\begin{table}[H]
\centering
\setlength{\tabcolsep}{10pt}
{\renewcommand{\arraystretch}{1.2}
\begin{tabular}{||c||c| c|c| c|c|c| c |c| c| c||} 
 \cline{2-11}
  \multicolumn{1}{c||}{}& \multicolumn{10}{c||}{$\rho(t)$}\\
 \cline{2-11}
 \multicolumn{1}{c||}{}& \multicolumn{5}{c|}{Method 1} &  \multicolumn{5}{|c||}{Method 2}\\
 \cline{2-11}
  \multicolumn{1}{c||}{}& Me&Md&SD&Min&Max& Me&Md&SD &Min&Max\\
  \hline\hline
 Italy&0.04&0.04&0.04&0.04&0.50& 0.04&0.04& 0.03 &0.01&0.28 \\ 
\hline
  Germany&0.05&0.05&0.03&0.02&0.18& 0.06&0.06&0.01
  &0.03&0.09\\ \hline
  World &0.02&0.02&0.009&0.01&0.07& 0.02&  0.02& 0.006&0.01& 0.05\\ 
\hline \hline
\end{tabular}\vspace{-3mm}
\caption{The Mean (Me), Median (Md), Standard Deviation (SD), Minimum (Min) and Maximum (Max) of the computed $\rho(t)$ using the two Methods,  for a random period. }\label{table:statrhoother}}\vspace{-3mm}
\end{table}

\begin{table}[H]
\centering
\setlength{\tabcolsep}{10pt}
{\renewcommand{\arraystretch}{1.2}
\begin{tabular}{||c||c| c|c| c|c|c| c |c| c| c||} 
 \cline{2-11}
  \multicolumn{1}{c||}{}& \multicolumn{10}{c||}{$\sigma(t)$}\\
 \cline{2-11}
 \multicolumn{1}{c||}{}& \multicolumn{5}{c|}{Method 1} &  \multicolumn{5}{|c||}{Method 2}\\
 \cline{2-11}
  \multicolumn{1}{c||}{}& Me&Md&SD&Min&Max& Me&Md&SD &Min&Max\\
  \hline\hline
 Italy&0.81&0.74&0.49&0.03&2.59 & 0.79 &0.75&0.48 &0.04& 2.07\\ 
\hline
  Germany&1.47&1.17&1.03&0.11&4.68& 1.50&1.46&0.52
  &0.49&2.93\\ \hline
  World &1.41&1.38&0.34&0.50&2.35&1.39 &1.38&  0.32&0.70& 2.13\\ 
\hline \hline
\end{tabular}\vspace{-3mm}
\caption{The Mean (Me), Median (Md), Standard Deviation (SD), Minimum (Min) and Maximum (Max) of the computed $\sigma(t)$ using the two Methods,  for a random period. }\label{table:statsigmaother}}\vspace{-3mm}
\end{table}

\begin{table}[H]
\centering
\setlength{\tabcolsep}{10pt}
{\renewcommand{\arraystretch}{1.2}
\begin{tabular}{||c||c| c|c| c|c|c| c |c| c| c||} 
 \cline{2-11}
  \multicolumn{1}{c||}{}& \multicolumn{10}{c||}{$\sigma_s(t)$}\\
 \cline{2-11}
 \multicolumn{1}{c||}{}& \multicolumn{5}{c|}{Method 1} &  \multicolumn{5}{|c||}{Method 2}\\
 \cline{2-11}
  \multicolumn{1}{c||}{}& Me&Md&SD&Min&Max& Me&Md&SD &Min&Max\\
  \hline\hline
 Italy&0.76&0.69&0.47&0.02& 2.48&0.58&0.70&0.46&0.036&1.97\\ 
\hline
  Germany&1.21&0.87&0.94&0.08&4.21& 1.46&1.41&0.50 
  &0.48&2.84\\ \hline
  World &1.39&1.36&0.34&0.49&2.34&1.38&  1.37&0.32 &0.69& 2.12\\ 
\hline \hline
\end{tabular}\vspace{-3mm}
\caption{The Mean (Me), Median (Md), Standard Deviation (SD), Minimum (Min) and Maximum (Max) of the computed $\sigma_s(t)$ using the two Methods,  for a random period. }\label{table:statsigmasother}}\vspace{-3mm}
\end{table}

 \begin{table}[H]
\centering
\setlength{\tabcolsep}{10pt}
{\renewcommand{\arraystretch}{1.4}
\begin{tabular}{||c| c| c| c |c||} 
 \hline
  &Norm & Italy  & Germany  & World  \\ 
 \hline\hline
\multirow{2}{*}{$\beta$} &$L_2$ & $ 1.283 * 10^{-1}$ &$3.058 * 10^{-1}$ &  $  1.992 * 10^{-1}$  \\ 
 \cline{2-5}
  & $L_\infty$ & $ 2.811 * 10^{-1}$  & $4.151 * 10^{-1}$ & $3.229 * 10^{-1}$ \\ 
\hline
\multirow{2}{*}{$\rho$} &$L_2$ & $ 6.429* 10^{-1}$ &$3.009 * 10^{-1}$ &  $2.618 * 10^{-1} $ \\ 
\cline{2-5}
& $L_\infty$ & $ 
   8.302* 10^{-1}$ & $ 6.960* 10^{-1}$ & $ 4.559 * 10^{-1}$ \\ 
\hline
 \multirow{2}{*}{$\sigma$} & $L_2$ & $ 1.436* 10^{-1}$ &$2.609* 10^{-1}$ &  $ 1.307 * 10^{-1}$ \\
\cline{2-5}
 & $L_\infty$ & $2.339 * 10^{-1}$ & $ 4.349* 10^{-1}$ & $5.871 * 10^{-1}$ \\  
 \hline
\multirow{2}{*}{$\sigma_s$} &$L_2$ & $ 1.433* 10^{-1}$ &$2.610* 10^{-1}$ &  $1.305 * 10^{-1}$ \\ 
 \cline{2-5}
  & $L_\infty$ & $2.339* 10^{-1}$  & $  4.354* 10^{-1}$ & $ 5.836 * 10^{-1}$ \\ 
 \hline
\end{tabular}
\caption{The $L_2$ and $L_\infty$ relative errors between the computed parameters based on Methods 1 and 2,  for a random period.}
\label{table:tableotherrelparam}}
\end{table}

\begin{figure}[H]
\begin{subfigure} {0.5\textwidth}
\centering
\includegraphics[scale=0.3]{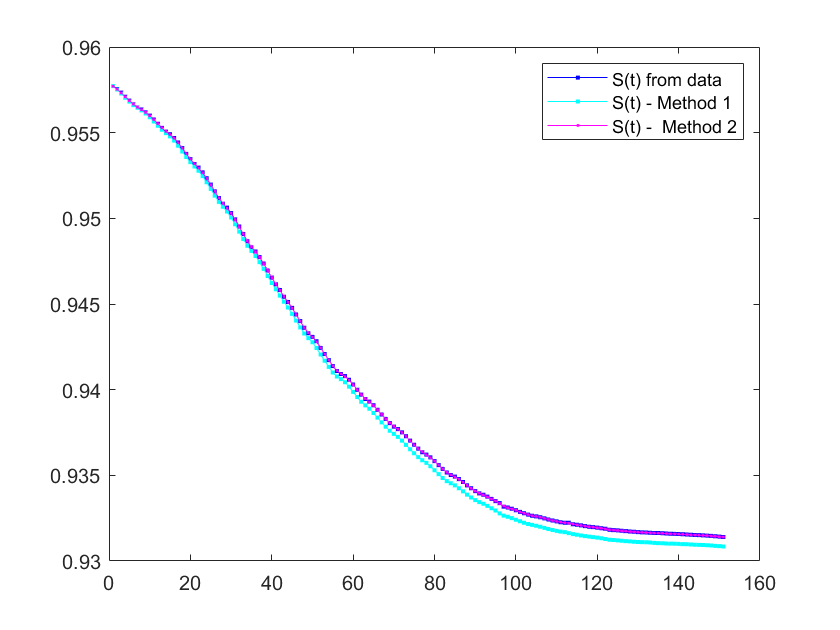}\vspace{-2mm}
\caption{Ratio of Susceptible s(t)}
\label{fig:Susceptibleotheritaly}
\end{subfigure}%
\begin{subfigure}{.5\linewidth}
\centering
\includegraphics[scale=.3]{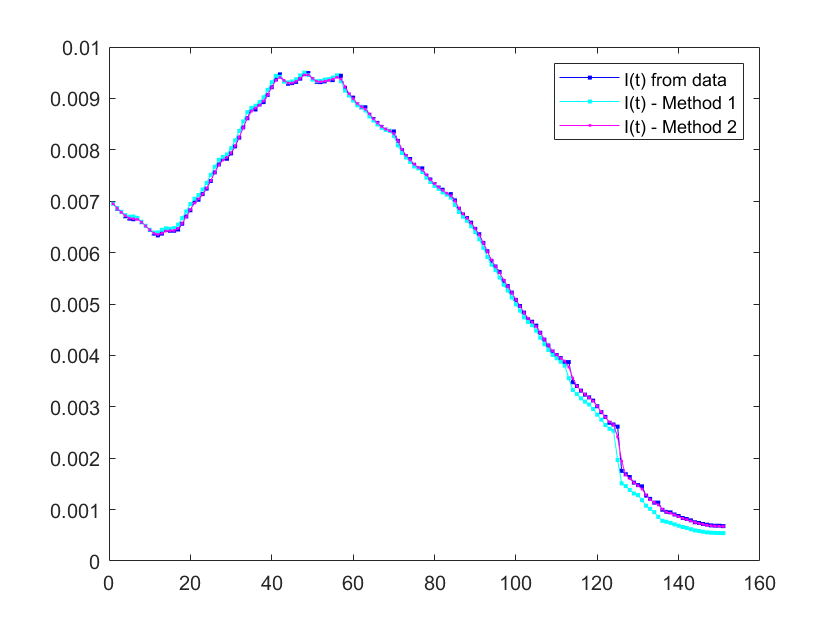}\vspace{-2mm}
\caption{Ratio of Infected i(t)}\label{fig:Infectedotheritaly}
\end{subfigure}
\begin{subfigure}{1.0\linewidth}
\centering
\includegraphics[scale=.3]{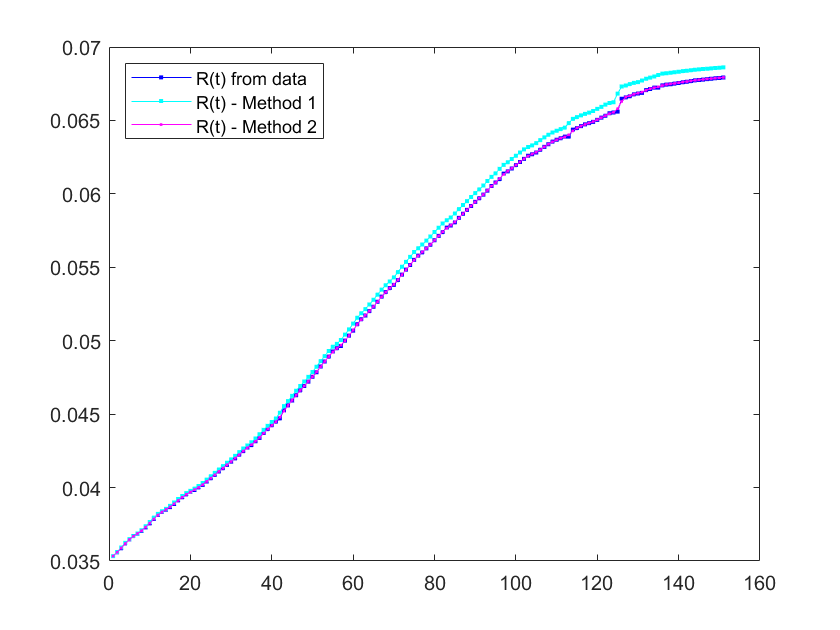}\vspace{-2mm}
\caption{Ratio of Removed r(t)}\label{fig:Removedotheritaly}
\end{subfigure}

\caption{Comparison of Compartments' ratios from real data to those obtained using SIR model with approximated parameters, during the Other Period in Italy }
\label{fig:SIRotheritaly}\vspace{-0.5cm}
\end{figure}

\begin{figure}[H]
\begin{subfigure} {0.5\textwidth}
\centering
\includegraphics[scale=0.3]{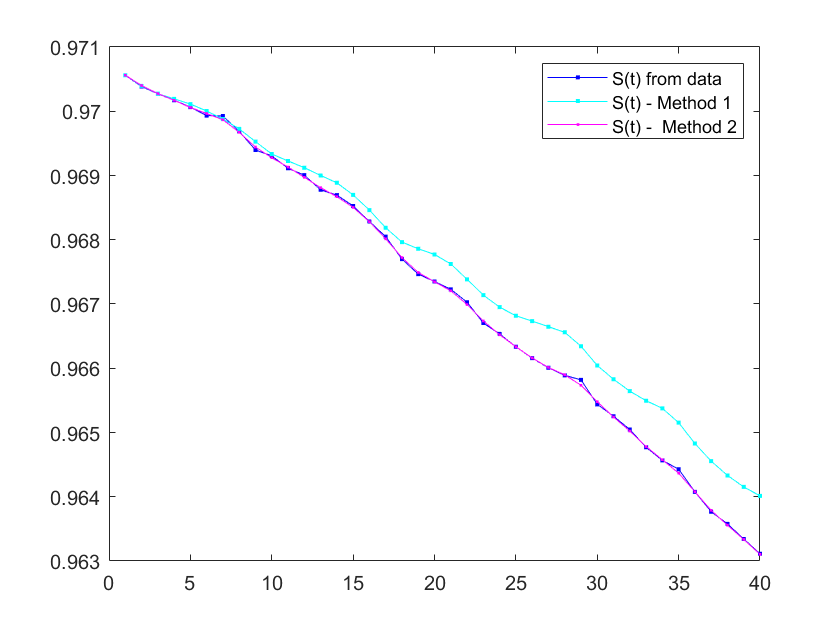}\vspace{-2mm}
\caption{Ratio of Susceptible s(t)}
\label{fig:Susceptiblothergermany}
\end{subfigure}%
\begin{subfigure}{.5\linewidth}
\centering
\includegraphics[scale=.3]{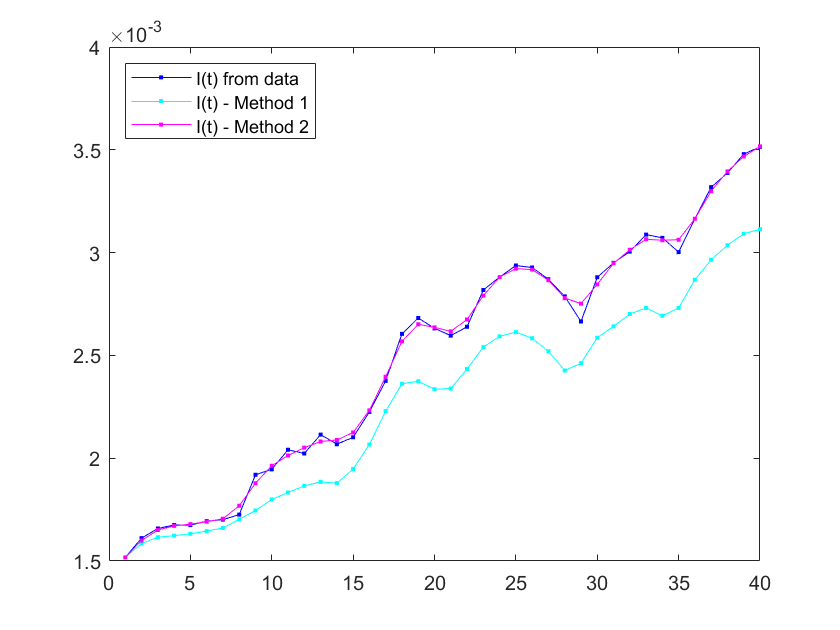}\vspace{-2mm}
\caption{Ratio of Infected i(t)}\label{fig:Infectedothergermany}
\end{subfigure}
\begin{subfigure}{1.0\linewidth}
\centering
\includegraphics[scale=.3]{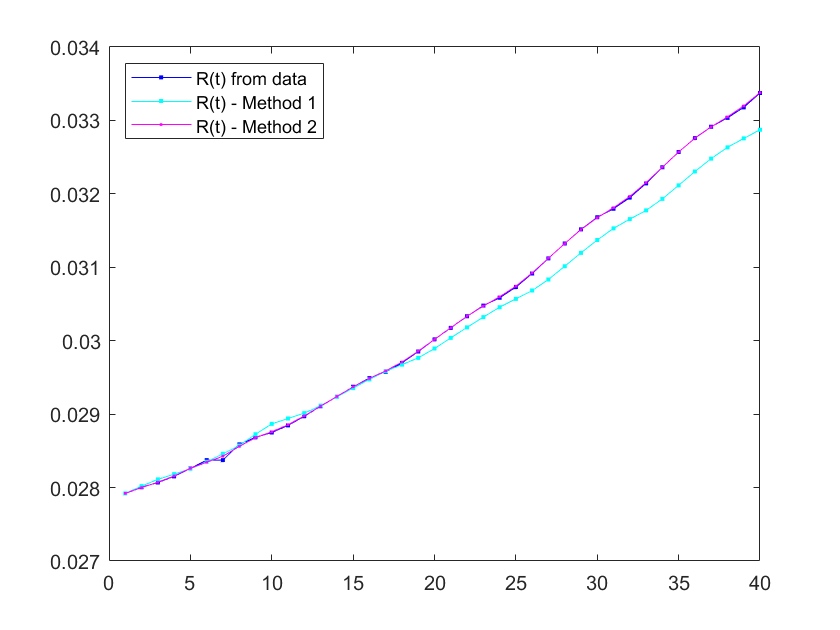}\vspace{-2mm}
\caption{Ratio of Removed r(t)}\label{fig:Removedothergermany}
\end{subfigure}

\caption{Comparison of Compartments' ratios from real data to those obtained using SIR model with approximated parameters, during the Other Period in Germany}
\label{fig:SIRothergermany}\vspace{-.5cm}
\end{figure}

\begin{figure}[H]
\begin{subfigure} {0.5\textwidth}
\centering
\includegraphics[scale=0.3]{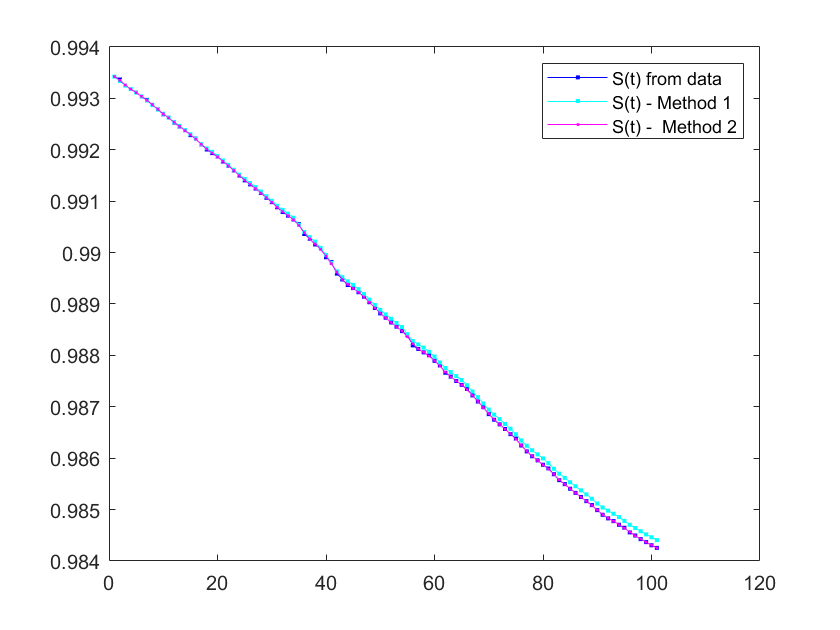}\vspace{-2mm}
\caption{Ratio of Susceptible s(t)}
\label{fig:Susceptibleotherworld}
\end{subfigure}%
\begin{subfigure}{.5\linewidth}
\centering
\includegraphics[scale=.3]{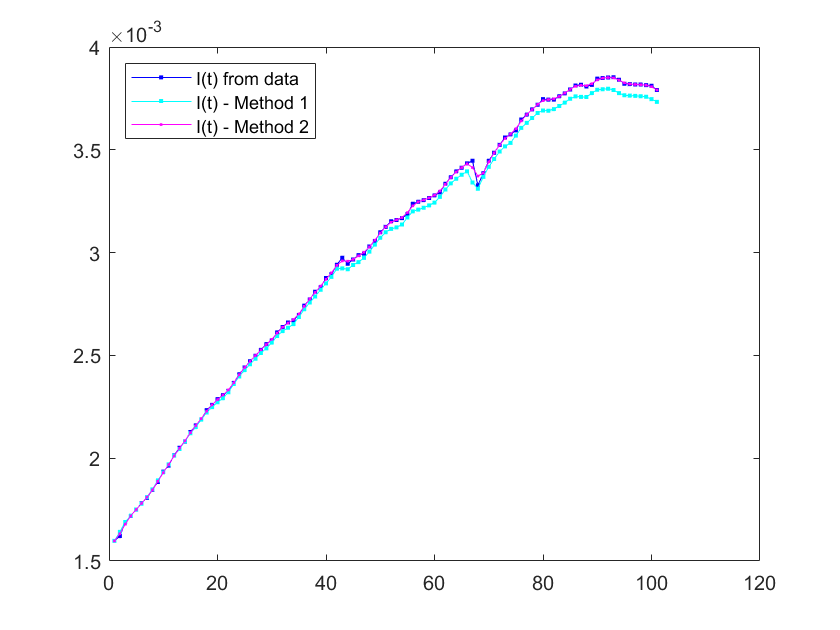}\vspace{-2mm}
\caption{Ratio of Infected i(t)}\label{fig:Infectedotherworld}
\end{subfigure}
\begin{subfigure}{1.0\linewidth}
\centering
\includegraphics[scale=.3]{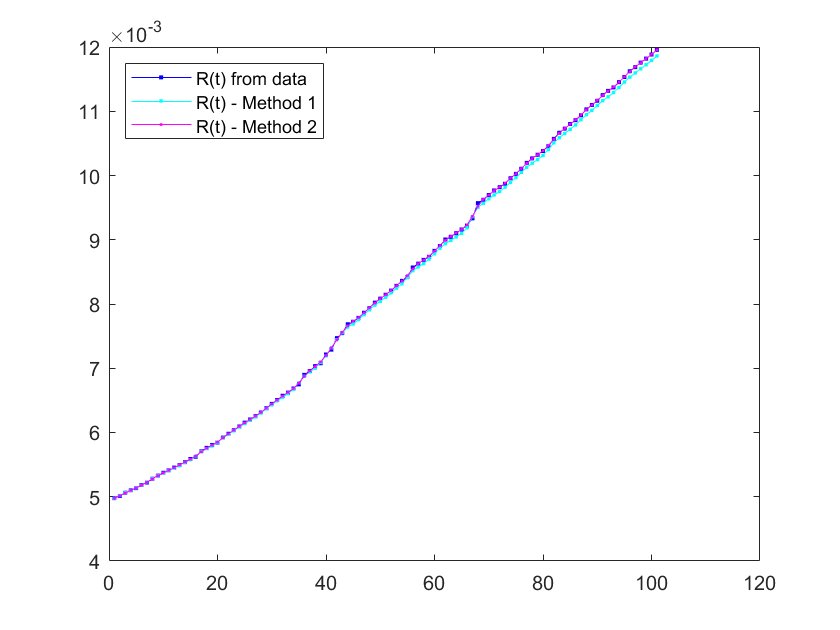}\vspace{-2mm}
\caption{Ratio of Removed r(t)}\label{fig:Removedotherworld}
\end{subfigure}

\caption{Comparison of Compartments' ratios from real data to those obtained using SIR model with approximated parameters, during the Other Period in the World }
\label{fig:SIRotherworld}
\end{figure}

Similarly to the first and second outbreaks, we validate the results by running the SIR model with the obtained parameters $\beta(t), \rho(t)$ using method 1 or method 2, and the corresponding initial values as discussed in section \ref{sec:valid}. Figures \ref{fig:SIRotheritaly}, \ref{fig:SIRothergermany},  and \ref{fig:SIRotherworld}  compare  the behavior of the ratio of compartments $s(t), i(t), r(t)$ of the real data to the simulated values. Moreover, Table \ref{table:tableotherSIRrel} 
shows the relative errors between the real data and simulated values. The number of infected people in both Germany and the world is only increasing in the respective random periods taken, whereas the number of infected people increases then decreases during the random period taken in Italy. Naturally, the number of susceptible people decreases during the random period, showing a somewhat exponential decay in Italy and linear decays in Germany and the world. The number of removed people increases during the random period, similar to its behavior in the first and second outbreaks for the three countries. It is clear from the figures and table of relative errors that there is a slight difference in the simulated results using methods 1 and 2, where the second is more accurate with an L2 relative error 10 times less than the first method (Table \ref{table:tableotherSIRrel}).


\begin{table}[H]
\centering
\setlength{\tabcolsep}{10pt}
{\renewcommand{\arraystretch}{1.4}
\begin{tabular}{||c|c| c| c| c |c||} 
 \hline
 Method& &Norm & Italy  & Germany  & World  \\ 
 \hline\hline
\multirow{6}{*}{1}&\multirow{2}{*}{S} &$L_2$ & $ 4.794* 10^{-4}$ &$ 4.889* 10^{-4}$ &  $8.306 * 10^{-5}$  \\ 
 \cline{3-6}
 & & $L_\infty$ & $ 6.351* 10^{-4}$  & $9.285 * 10^{-4}$ & $  1.587* 10^{-4}$ \\ 
 \cline{2-6}
&\multirow{2}{*}{I} &$L_2$ & $1.868* 10^{-2}$ &$9.902* 10^{-2}$ &  $ 1.134  * 10^{-2}$ \\ 
 \cline{3-6}
&& $L_\infty$ & $6.796 * 10^{-2}$ & $1.141 * 10^{-1}$ & $ 2.743  * 10^{-2}$ \\ 
  \cline{2-6}
 &\multirow{2}{*}{R} & $L_2$ & $ 9.745* 10^{-3}$ &$7.847* 10^{-3}$ &  $ 5.862  * 10^{-3}$ \\\cline{3-6}
 && $L_\infty$ & $1.808 * 10^{-2}$ & $1.499* 10^{-2}$ & $ 8.029 * 10^{-3}$ \\  
 \hline 

 \multirow{6}{*}{2}&\multirow{2}{*}{S} &$L_2$ & $ 1.202* 10^{-5}$	&$2.897 * 10^{-5}$&	$9.783* 10^{-6}$	\\ 
 \cline{3-6}
 & & $L_\infty$ & $ 3.637 * 10^{-5}$	&$8.686 * 10^{-5}$	&$3.363* 10^{-5}$					 \\ 
 \cline{2-6}
&\multirow{2}{*}{I} &$L_2$ &  $ 4.556* 10^{-3}$	&$  1.013 * 10^{-2}$&	$ 2.279* 10^{-3}$				 \\ 
 \cline{3-6}
&& $L_\infty$ & $2.079*10^{-2}$&	$2.479 * 10^{-2}$ & $1.168 * 10^{-2}$ \\ 
  \cline{2-6}
 &\multirow{2}{*}{R} & $L_2$ &$5.582 * 10^{-4}$	&$4.337 * 10^{-4}$&	$ 1.161 * 10^{-3}$	 \\\cline{3-6}
 && $L_\infty$ & $ 2.981 * 10^{-3}$&	$ 1.603 * 10^{-3}$&	$  3.856 * 10^{-3}$ \\  
 \hline
\end{tabular}
\caption{The $L_2$ and $L_\infty$ relative errors of the computed S,I,R from the time-dependent model with the S,I,R collected from data for a random period, where $\beta,\rho$ are computed using Method 1 or 2. }\label{table:tableotherSIRrel}}
\end{table}

\subsection{Two-year Period}\label{sec:twoywar}

\begin{figure}[H]
\begin{subfigure}{0.5\textwidth}
\centering
\includegraphics[scale=0.28]{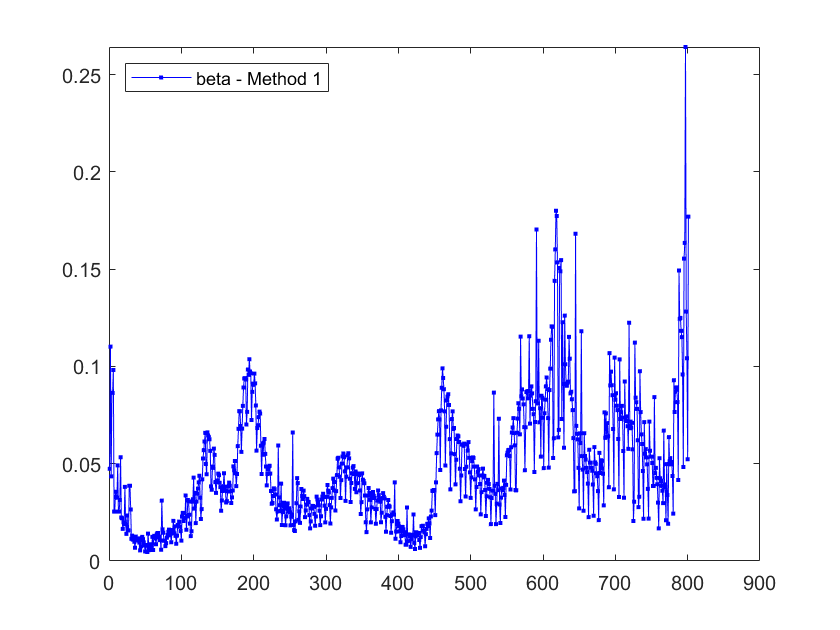} \vspace{-2mm}
\caption{Infection Rate $\beta(t)$ - Method 1}
\label{fig:betatwoyearitalyM1}
\end{subfigure}
\hfill
\begin{subfigure}{0.5\textwidth}
\centering
\includegraphics[scale=0.28]{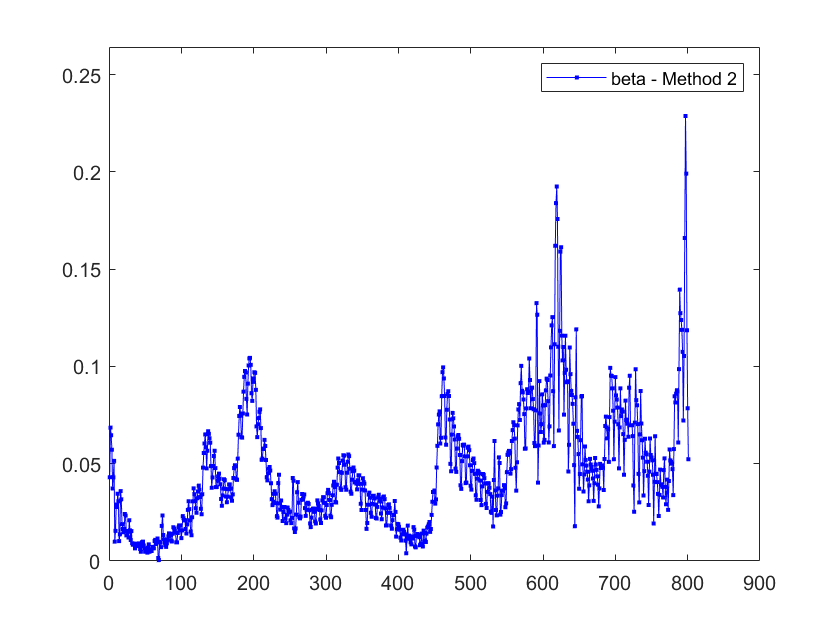} \vspace{-2mm}
\caption{Infection Rate $\beta(t)$ - Method 2}
\label{fig:betatwoyearitalyM2}
\end{subfigure}
\newline
\begin{subfigure}{0.5\textwidth}
\centering
\includegraphics[scale=0.28]{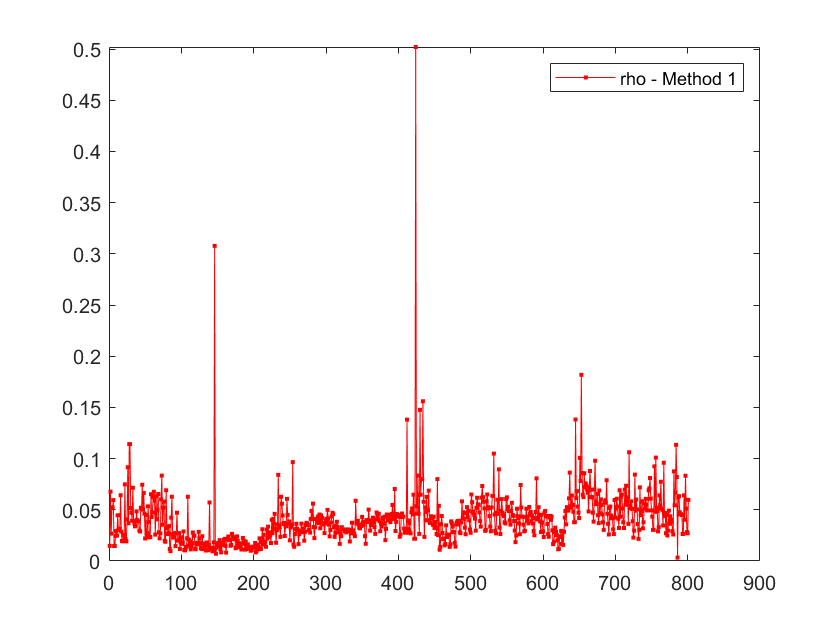} \vspace{-2mm}
\caption{Removal Rate $\rho(t)$ - Method 1 }
\label{fig:rhotwoyearitalyM1}
\end{subfigure}
\hfill
\begin{subfigure}{0.5\textwidth}
\centering
\includegraphics[scale=0.28]{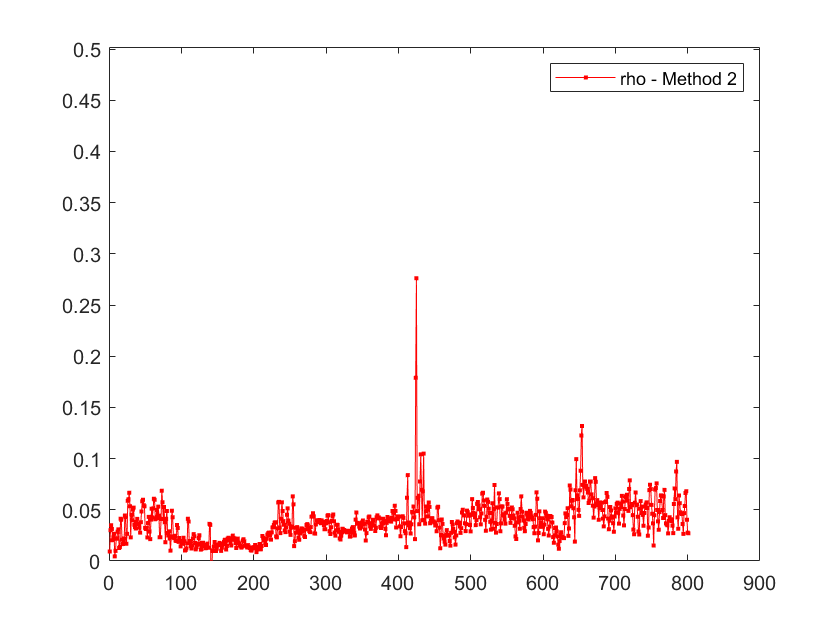} \vspace{-2mm}
\caption{Removal Rate $\rho(t)$ - Method 2 }
\label{fig:rhotwoyearitalyM2}
\end{subfigure}
\newline
\begin{subfigure}{0.5\textwidth}
\centering
\includegraphics[scale=0.28]{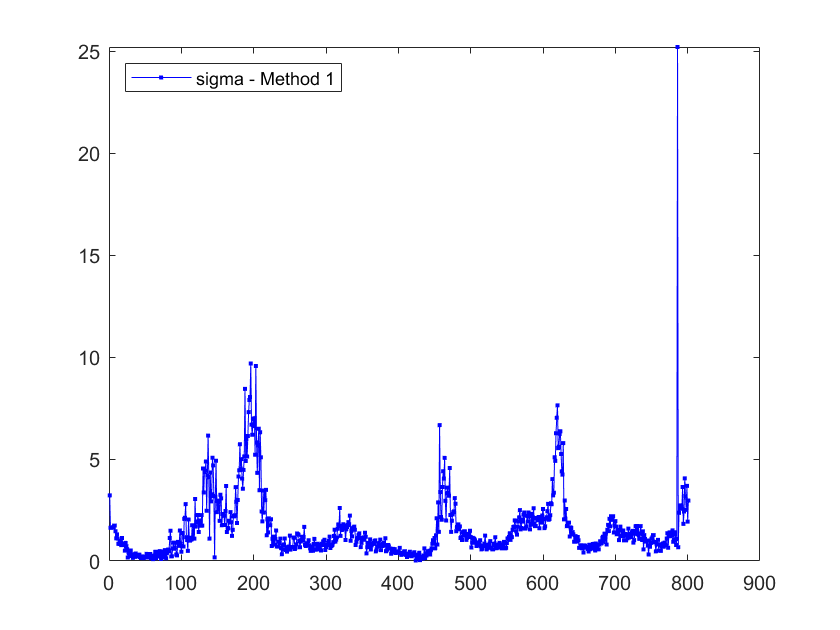} \vspace{-2mm}
\caption{Reproduction Factor $\sigma(t)$ - Method 1}
\label{fig:sigmatwoyearitalyM1}
\end{subfigure}
\hfill
\begin{subfigure}{0.5\textwidth}
\centering
\includegraphics[scale=0.28]{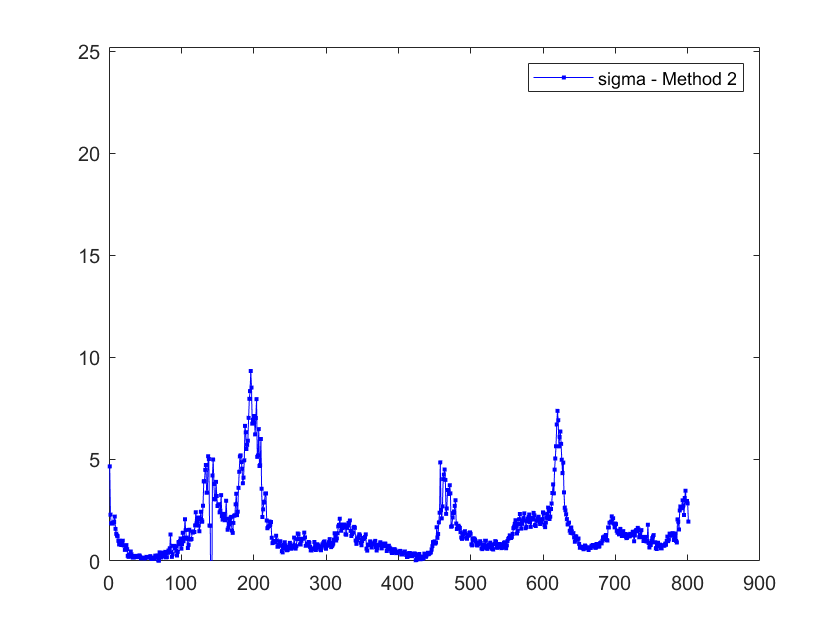} \vspace{-2mm}
\caption{Reproduction Factor $\sigma(t)$ - Method 2}
\label{fig:sigmatwoyearitalyM2}
\end{subfigure}
\newline
\begin{subfigure}{0.5\textwidth}
\centering
\includegraphics[scale=0.28]{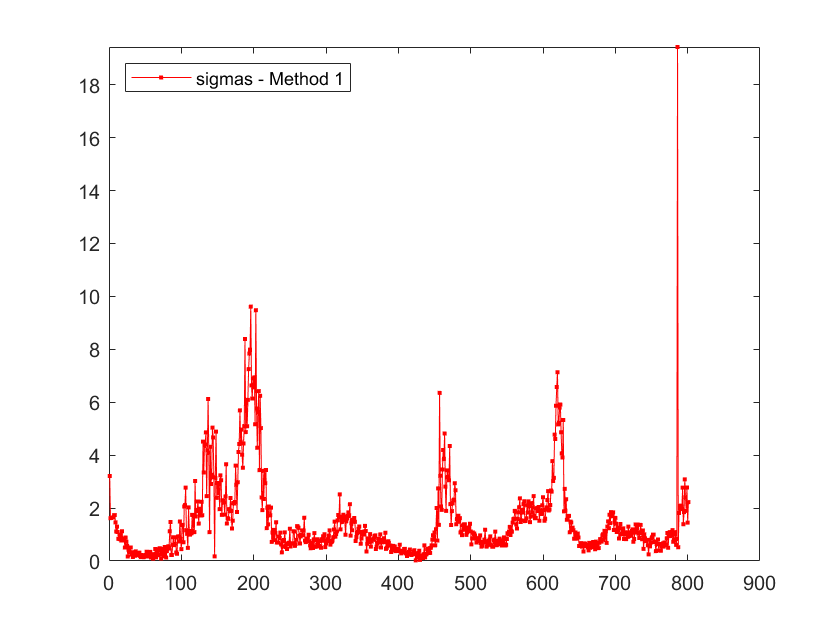}\vspace{-2mm} 
\caption{Replacement Number $\sigma_s(t)$ - Method 1}
\label{fig:sigmastwoyearitalyM1}
\end{subfigure}\hfill
\begin{subfigure}{0.5\textwidth}
\centering
\includegraphics[scale=0.28]{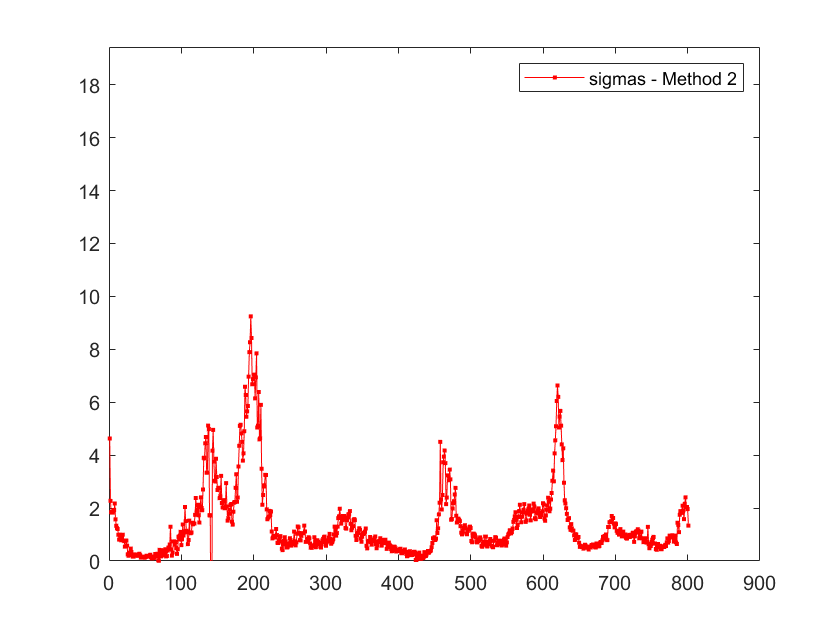}\vspace{-2mm} 
\caption{Replacement Number $\sigma_s(t)$ - Method 2}
\label{fig:sigmastwoyearitalyM2}
\end{subfigure}
\caption{Parameters of SIR Model during the Two-Year Period in Italy }
\label{fig:parameterstwoyearitaly}\vspace{0mm}
\end{figure}

\begin{figure}[H]
\begin{subfigure}{0.5\textwidth}
\centering
\includegraphics[scale=0.28]{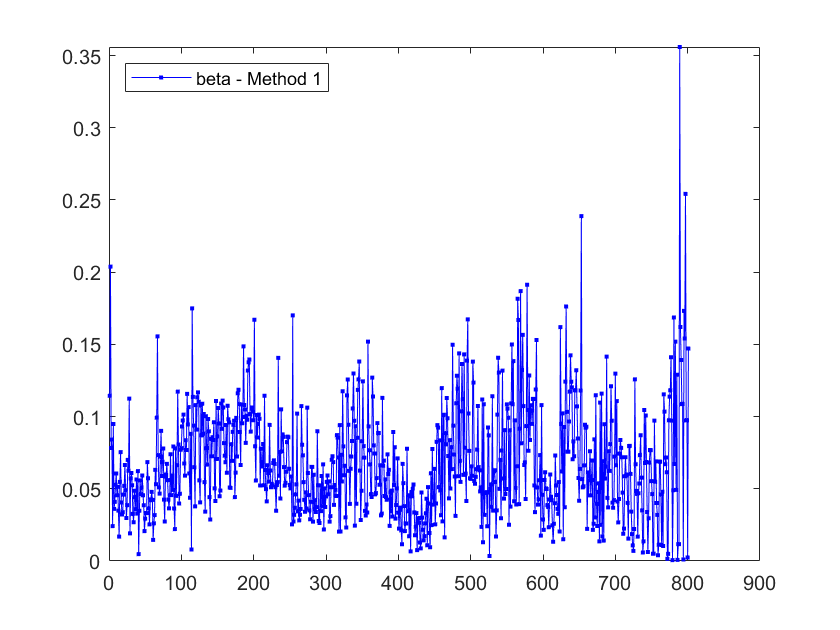} \vspace{-2mm}
\caption{Infection Rate $\beta(t)$ - Method 1}
\label{fig:betatwoyeargermanyM1}
\end{subfigure}
\hfill
\begin{subfigure}{0.5\textwidth}
\centering
\includegraphics[scale=0.28]{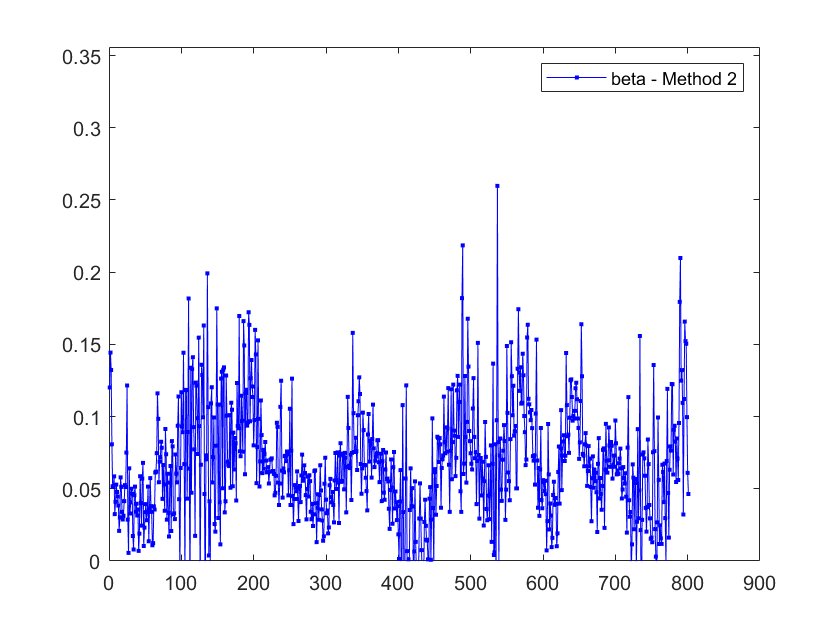} \vspace{-2mm}
\caption{Infection Rate $\beta(t)$ - Method 2}
\label{fig:betatwoyeargermanyM2}
\end{subfigure}
\newline
\begin{subfigure}{0.5\textwidth}
\centering
\includegraphics[scale=0.28]{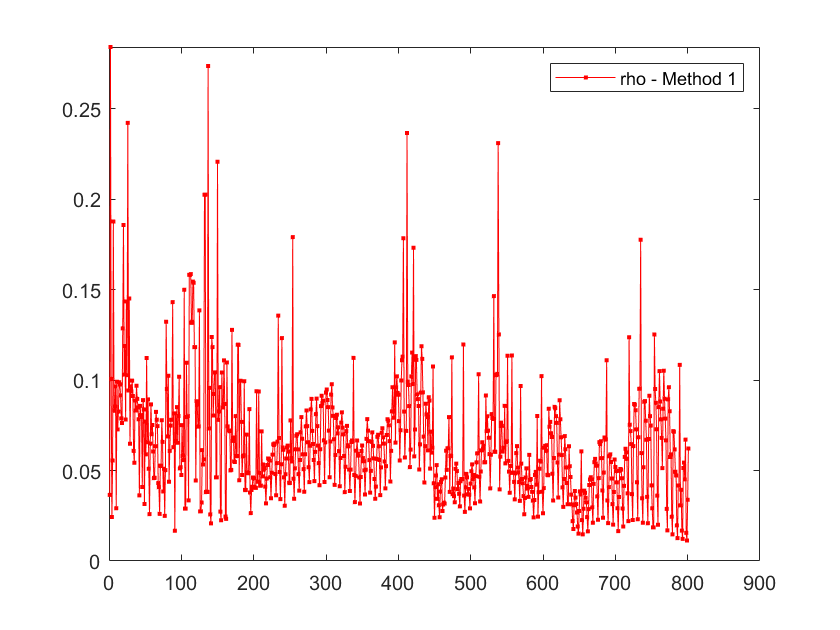} \vspace{-2mm}
\caption{Removal Rate $\rho(t)$ - Method 1 }
\label{fig:rhotwoyeargermanyM1}
\end{subfigure}
\hfill
\begin{subfigure}{0.5\textwidth}
\centering
\includegraphics[scale=0.28]{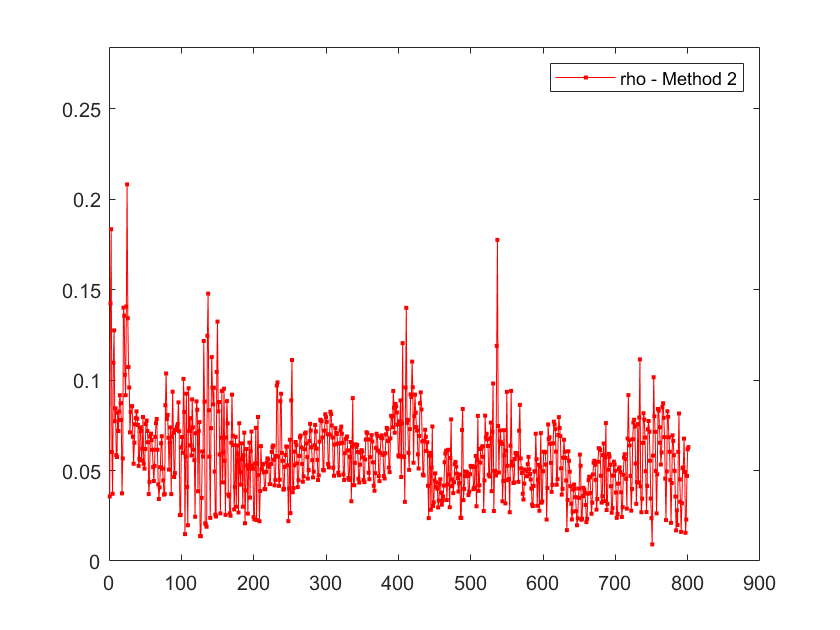} \vspace{-2mm}
\caption{Removal Rate $\rho(t)$ - Method 2 }
\label{fig:rhotwoyeargermanyM2}
\end{subfigure}
\newline
\begin{subfigure}{0.5\textwidth}
\centering
\includegraphics[scale=0.28]{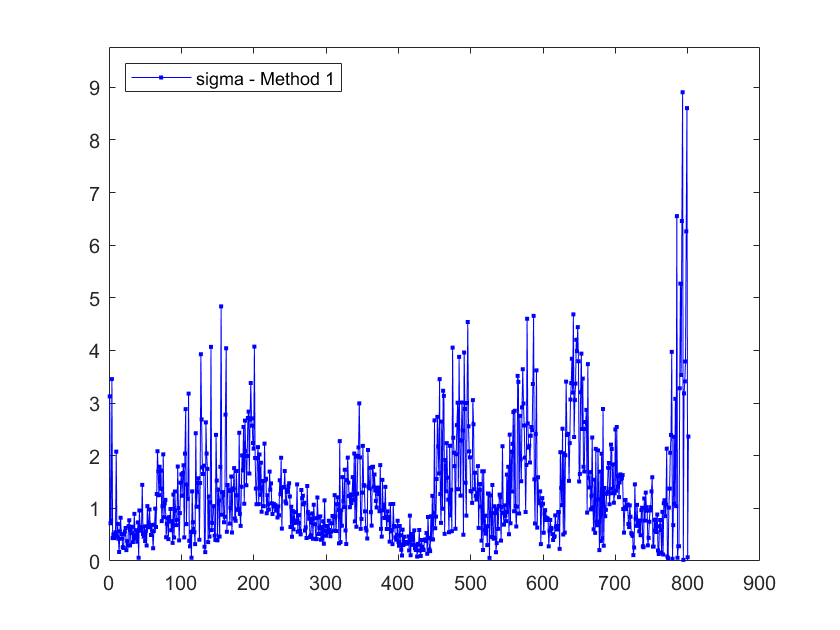} \vspace{-2mm}
\caption{Reproduction Factor $\sigma(t)$ - Method 1}
\label{fig:sigmatwoyeargermanyM1}
\end{subfigure}
\hfill
\begin{subfigure}{0.5\textwidth}
\centering
\includegraphics[scale=0.28]{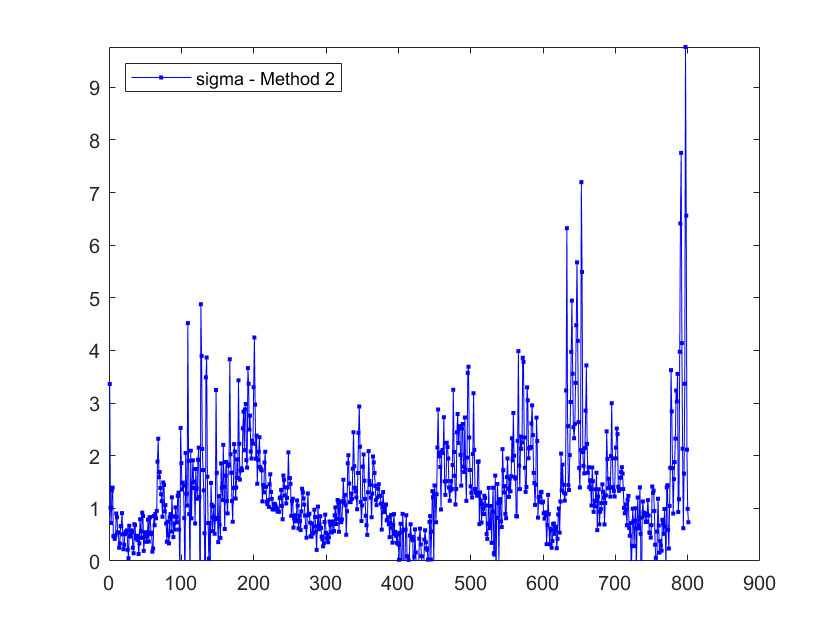} \vspace{-2mm}
\caption{Reproduction Factor $\sigma(t)$ - Method 2}
\label{fig:sigmatwoyeargermanyM2}
\end{subfigure}
\newline
\begin{subfigure}{0.5\textwidth}
\centering
\includegraphics[scale=0.28]{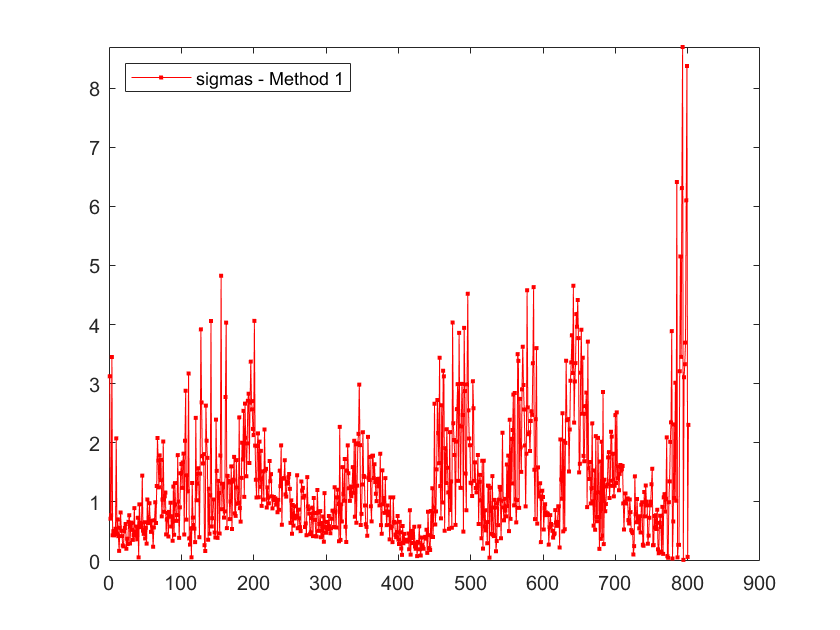}\vspace{-2mm} 
\caption{Replacement Number $\sigma_s(t)$ - Method 1}
\label{fig:sigmastwoyeargermanyM1}
\end{subfigure}\hfill
\begin{subfigure}{0.5\textwidth}
\centering
\includegraphics[scale=0.28]{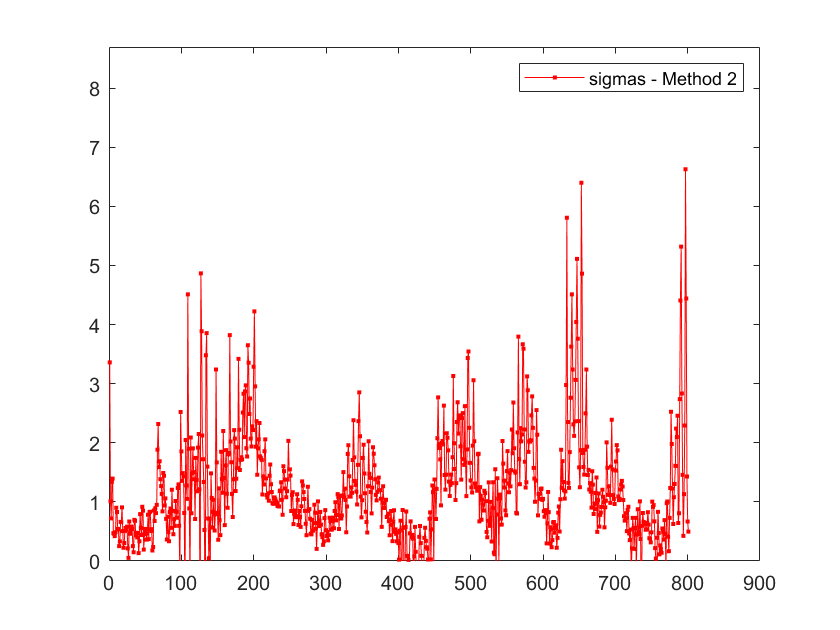}\vspace{-2mm} 
\caption{Replacement Number $\sigma_s(t)$ - Method 2}
\label{fig:sigmastwoyeargermanyM2}
\end{subfigure}
\caption{Parameters of SIR Model during Two-Year Period in Germany }
\label{fig:parameterstwoyeargermany}\vspace{0mm}
\end{figure}

\begin{figure}[H]
\begin{subfigure}{0.5\textwidth}
\centering
\includegraphics[scale=0.28]{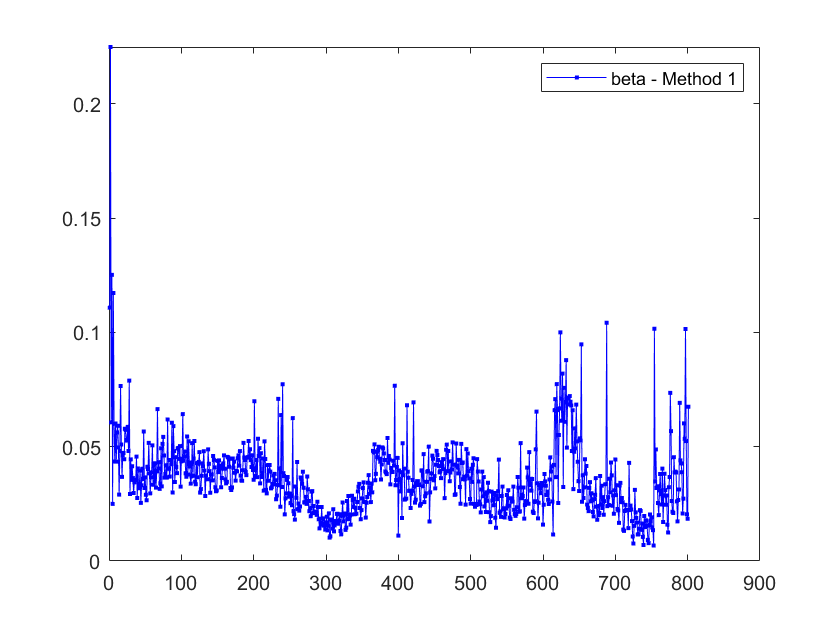}\vspace{-2mm}
\caption{Infection Rate $\beta(t)$ - Method 1}
\label{fig:betatwoyearworldM1}
\end{subfigure}
\hfill
\begin{subfigure}{0.5\textwidth}
\centering
\includegraphics[scale=0.28]{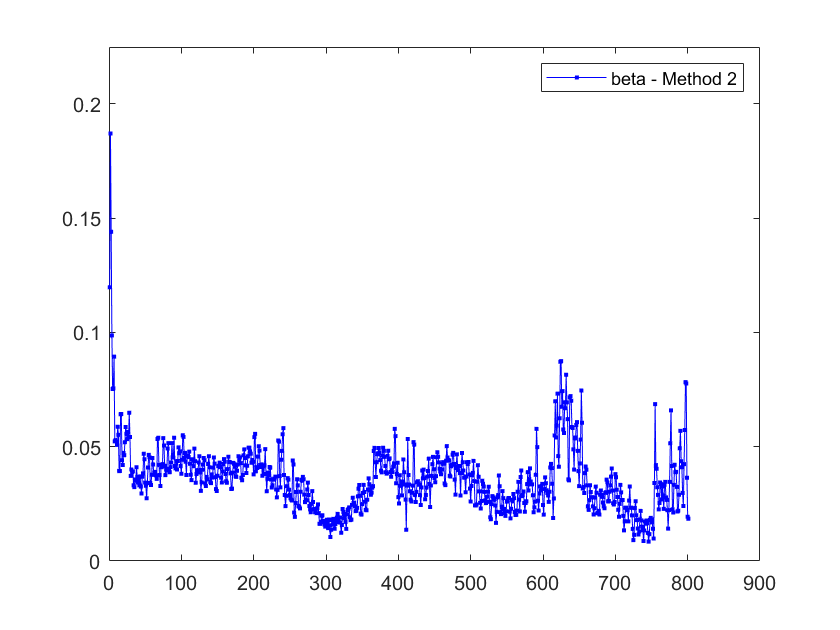} \vspace{-2mm}
\caption{Infection Rate $\beta(t)$ - Method 2}
\label{fig:betatwoyearworldM2}
\end{subfigure}
\newline
\begin{subfigure}{0.5\textwidth}
\centering
\includegraphics[scale=0.28]{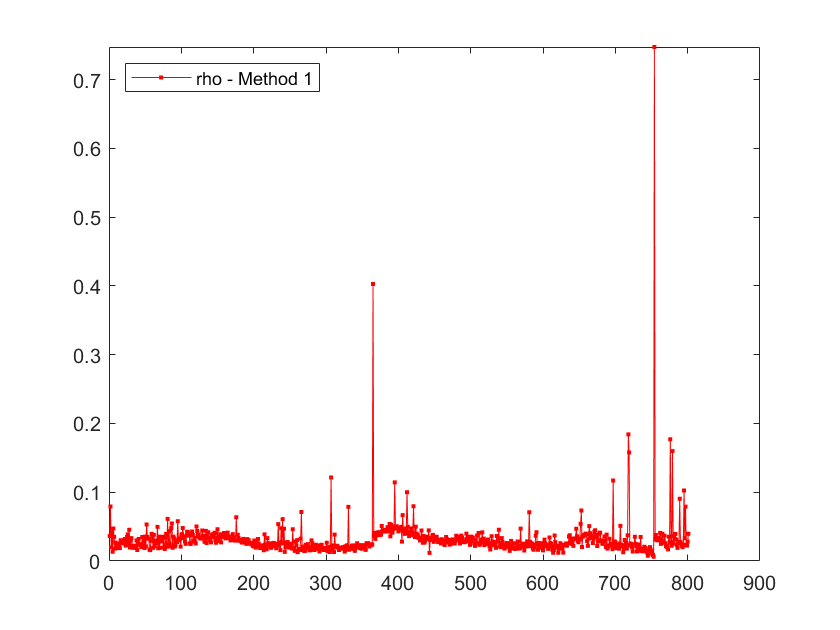} \vspace{-2mm}
\caption{Removal Rate $\rho(t)$ - Method 1 }
\label{fig:rhotwoyearworldM1}
\end{subfigure}
\hfill
\begin{subfigure}{0.5\textwidth}
\centering
\includegraphics[scale=0.28]{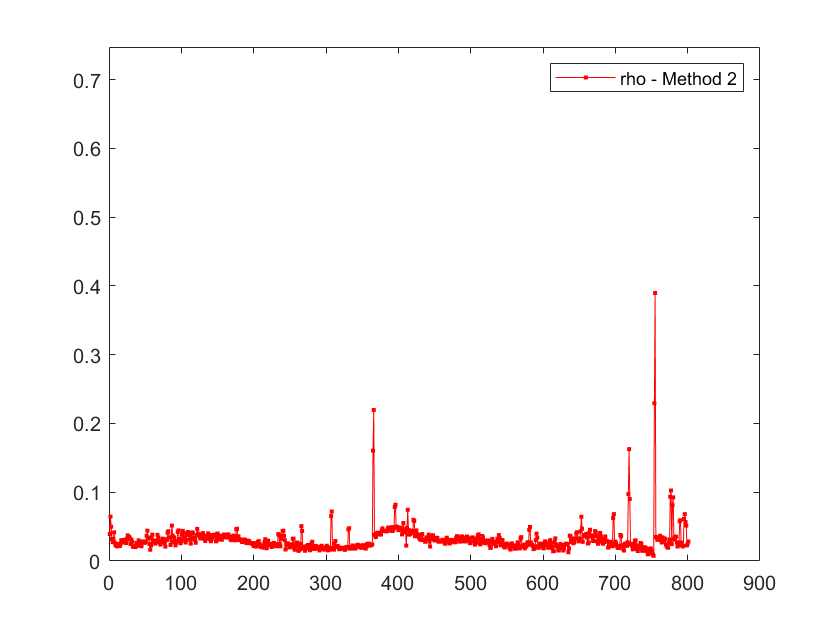} \vspace{-2mm}
\caption{Removal Rate $\rho(t)$ - Method 2 }
\label{fig:rhotwoyearworldM2}
\end{subfigure}
\newline
\begin{subfigure}{0.5\textwidth}
\centering
\includegraphics[scale=0.28]{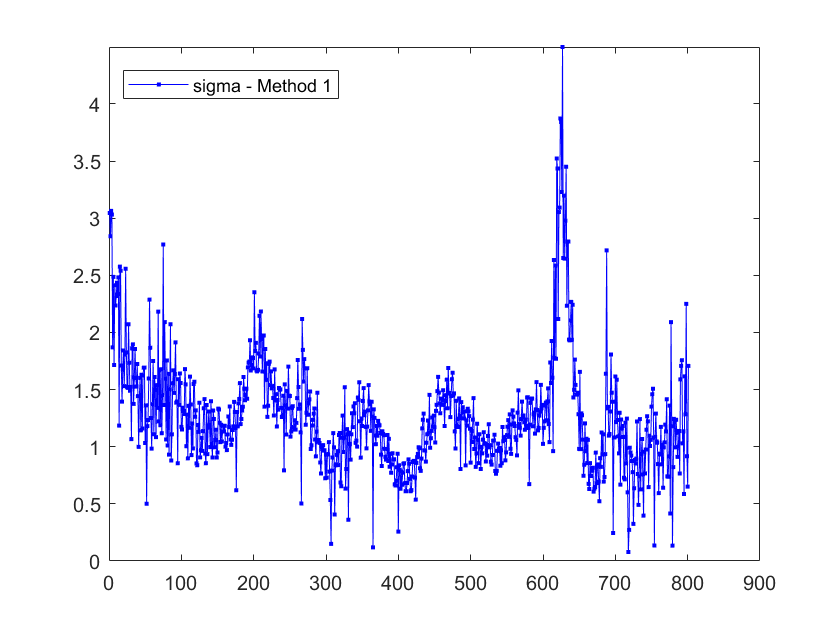} \vspace{-2mm}
\caption{Reproduction Factor $\sigma(t)$ - Method 1}
\label{fig:sigmatwoyearworldM1}
\end{subfigure}
\hfill
\begin{subfigure}{0.5\textwidth}
\centering
\includegraphics[scale=0.28]{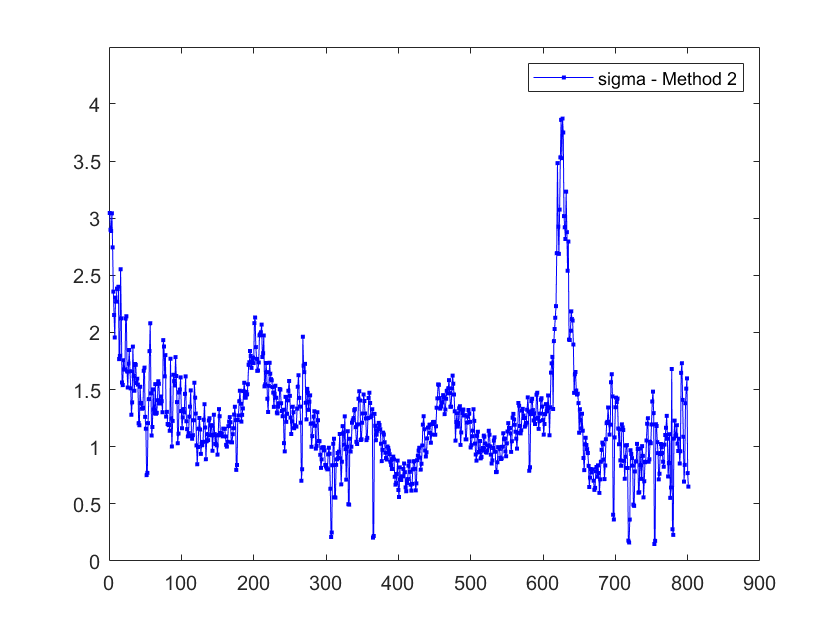} \vspace{-2mm}
\caption{Reproduction Factor $\sigma(t)$ - Method 2}
\label{fig:sigmatwoyearworldM2}
\end{subfigure}
\newline
\begin{subfigure}{0.5\textwidth}
\centering
\includegraphics[scale=0.28]{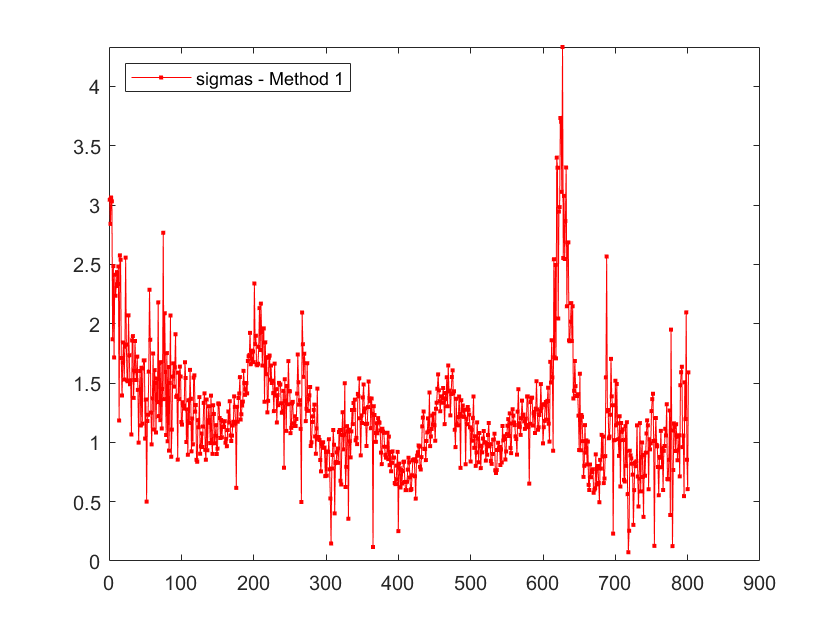}\vspace{-2mm} 
\caption{Replacement Number $\sigma_s(t)$ - Method 1}
\label{fig:sigmastwoyearworldM1}
\end{subfigure}\hfill
\begin{subfigure}{0.5\textwidth}
\centering
\includegraphics[scale=0.28]{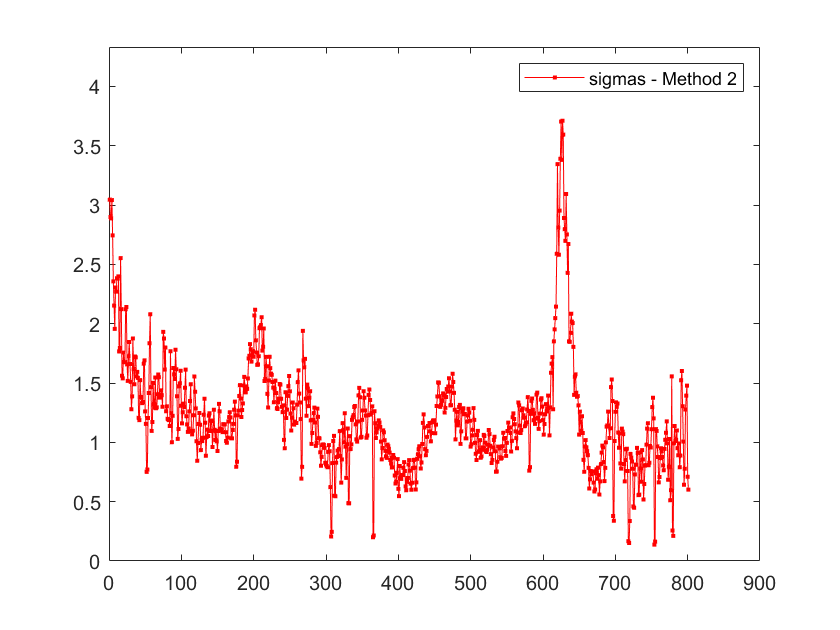}\vspace{-2mm} 
\caption{Replacement Number $\sigma_s(t)$ - Method 2}
\label{fig:sigmastwoyearworldM2}
\end{subfigure}
\caption{Parameters of SIR Model during Two-Year Period in the World }
\label{fig:parameterstwoyearworld}\vspace{0mm}
\end{figure}

Figures \ref{fig:parameterstwoyearitaly},  \ref{fig:parameterstwoyeargermany}, and \ref{fig:parameterstwoyearworld} plot the approximated parameters $\beta(t), \rho(t), \sigma(t)$, and $\sigma_s(t)$ using Method 1 (section \ref{sec:ParamM1}) and Method 2 (section \ref{sec:ParamM2}) during the first outbreak in Italy, Germany, and the world respectively. Clearly the results of both methods are similar in global behavior, with some different variations. These differences are detailed in  
 Tables \ref{table:statbetawhole}, \ref{table:statrhowhole} \ref{table:statsigmawhole}, and \ref{table:statsigmaswhole} that summarize the statistical properties (mean, median, standard deviation) of the approximated parameters $\beta(t)$, $\rho(t)$, $\sigma(t)$, and  $\sigma_s(t)$ respectively. 

Moreover, Table \ref{table:tablewholerelparam} computes the relative L2 and Linfinity errors between the parameters computed using both methods. These errors are of order $10^{-1}$ for all parameters and all three countries, except for Linfinity for $\sigma$ and $\sigma_s$ in Italy where it is 1.341 and 1.254 respectively. As for the absolute errors of the parameters at some time $t_i$, then by comparing the means, it is clear that it is of orders $10^{-3}$ for $\beta$ (except for the world, $10^{-4}$); whereas, it is of the order $10^{-2}$ for $\rho$ (except for the world, $10^{-3}$). Also, it is of the order $10^{-2}$ for $\sigma$ (except for germany, $10^{-1}$) and $10^{-2}$ for $\sigma_s$ (except for italy, $10^{-1}$).

\begin{table}[H]
\centering
\setlength{\tabcolsep}{10pt}
{\renewcommand{\arraystretch}{1.2}
\begin{tabular}{||c||c| c| c |c| c|c| c |c| c| c||} 
\cline{2-11}
  \multicolumn{1}{c||}{}& \multicolumn{10}{c||}{$\beta(t)$}\\
 \cline{2-11}
 \multicolumn{1}{c||}{}& \multicolumn{5}{c|}{Method 1} &  \multicolumn{5}{|c||}{Method 2}\\
 \cline{2-11}
  \multicolumn{1}{c||}{}& Me&Md&SD&Min&Max& Me&Md&SD &Min&Max\\
  \hline\hline
 Italy&0.05&0.04&0.03&0.005&0.26&0.05&0.04&0.031 &$0.0005$&0.23 \\ 
\hline
  Germany&0.07&0.06&0.04&$0.0006$& 0.36&0.07&0.06&0.04&-0.06&0.26  \\ \hline
  World &0.04&0.03&0.02 &0.007&0.22& 0.04 &0.03& 0.01&0.008&0.19 \\ 
\hline \hline
\end{tabular}\vspace{-3mm}
\caption{The Mean (Me), Median (Md) Standard Deviation (SD), Minimum (Min) and Maximum (Max) of the computed $\beta(t)$ using the two Methods, for the two-year period. }\label{table:statbetawhole}}\vspace{-3mm}
\end{table}

\begin{table}[H]
\centering
\setlength{\tabcolsep}{10pt}
{\renewcommand{\arraystretch}{1.2}
\begin{tabular}{||c||c| c|c| c|c|c| c |c| c| c||} 
 \cline{2-11}
  \multicolumn{1}{c||}{}& \multicolumn{10}{c||}{$\rho(t)$}\\
 \cline{2-11}
 \multicolumn{1}{c||}{}& \multicolumn{5}{c|}{Method 1} &  \multicolumn{5}{|c||}{Method 2}\\
 \cline{2-11}
  \multicolumn{1}{c||}{}& Me&Md&SD&Min&Max& Me&Md&SD &Min&Max\\
  \hline\hline
 Italy&0.04&0.04&0.03&0.003&0.50 &  0.04&0.04& 0.02 &$-0.01$& 0.28\\ 
\hline
  Germany&0.06&0.06&0.03&0.01&0.28&0.06&0.06&0.02
  &0.01& 0.21\\ \hline
  World &0.03&0.03&0.03&$0.01$&0.75& 0.03&0.03& 0.02 &0.01&0.39 \\ 
\hline \hline
\end{tabular}\vspace{-3mm}
\caption{The Mean (Me), Median (Md), Standard Deviation (SD), Minimum (Min) and Maximum (Max) of the computed $\rho(t)$ using the two Methods, for the two-year period. }\label{table:statrhowhole}}\vspace{-3mm}
\end{table}

\begin{table}[H]
\centering
\setlength{\tabcolsep}{10pt}
{\renewcommand{\arraystretch}{1.2}
\begin{tabular}{||c||c| c|c| c|c|c| c |c| c| c||} 
 \cline{2-11}
  \multicolumn{1}{c||}{}& \multicolumn{10}{c||}{$\sigma(t)$}\\
 \cline{2-11}
 \multicolumn{1}{c||}{}& \multicolumn{5}{c|}{Method 1} &  \multicolumn{5}{|c||}{Method 2}\\
 \cline{2-11}
  \multicolumn{1}{c||}{}& Me&Md&SD&Min&Max& Me&Md&SD &Min&Max\\
  \hline\hline
 Italy&1.57&1.09&1.66&0.03&25.2 & 1.52 &1.09&1.43&-4.13& 9.32\\ 
\hline
  Germany&1.29&1.02&1.04&0.02&8.91& 1.31&1.14&1.07
  &-2.13&9.76\\ \hline
  World &1.27&1.19&0.49&0.08&4.49& 1.25& 1.19 &  0.48&0.15& 3.87\\ 
\hline \hline
\end{tabular}\vspace{-3mm}
\caption{The Mean (Me), Median (Md), Standard Deviation (SD), Minimum (Min) and Maximum (Max) of the computed $\sigma(t)$ using the two Methods, for the two-year period. }\label{table:statsigmawhole}}\vspace{-3mm}
\end{table}

\begin{table}[H]
\centering
\setlength{\tabcolsep}{10pt}
{\renewcommand{\arraystretch}{1.2}
\begin{tabular}{||c||c| c|c| c|c|c| c |c| c| c||} 
 \cline{2-11}
  \multicolumn{1}{c||}{}& \multicolumn{10}{c||}{$\sigma_s(t)$}\\
 \cline{2-11}
 \multicolumn{1}{c||}{}& \multicolumn{5}{c|}{Method 1} &  \multicolumn{5}{|c||}{Method 2}\\
 \cline{2-11}
  \multicolumn{1}{c||}{}& Me&Md&SD&Min&Max& Me&Md&SD &Min&Max\\
  \hline\hline
 Italy&1.47&1.01&1.54&0.02&19.4 & 1.39& 0.94&1.39&-4.12& 9.25\\ 
\hline
  Germany&1.28&1.02&1.02&0.02&8.69& 1.20&1.05&0.93
  &-1.70&6.63\\ \hline
  World &1.24&1.16&0.49&0.073&4.33& 1.22 & 1.16&0.47 &0.14&3.71 \\ 
\hline \hline
\end{tabular}\vspace{-3mm}
\caption{The Mean (Me), Median (Md), Standard Deviation (SD), Minimum (Min) and Maximum (Max) of the computed $\sigma_s(t)$ using the two Methods, for the two-year period.}\label{table:statsigmaswhole}}\vspace{-3mm}
\end{table}

 \begin{table}[H]
\centering
\setlength{\tabcolsep}{10pt}
{\renewcommand{\arraystretch}{1.4}
\begin{tabular}{||c| c| c| c |c||} 
 \hline
  &Norm & Italy  & Germany  & World  \\ 
 \hline\hline
\multirow{2}{*}{$\beta$} &$L_2$ & $ 2.191* 10^{-1}$ &$4.762 * 10^{-1}$ &  $ 2.253 * 10^{-1}$  \\ 
 \cline{2-5}
  & $L_\infty$ & $5.061 * 10^{-1}$  & $  5.732* 10^{-1}$ & $ 4.049 * 10^{-1}$ \\ 
\hline
\multirow{2}{*}{$\rho$} &$L_2$ & $4.631* 10^{-1}$ &$4.650 * 10^{-1}$ &  $ 6.480 * 10^{-1}$ \\ 
\cline{2-5}
& $L_\infty$ & $ 8.302 * 10^{-1}$ & $ 7.374* 10^{-1}$ & $9.112 * 10^{-1}$ \\ 
\hline
 \multirow{2}{*}{$\sigma$} & $L_2$ & $4.476 * 10^{-1}$ &$4.972* 10^{-1}$ &  $ 1.586 * 10^{-1}$ \\
\cline{2-5}
 & $L_\infty$ & $ 1.341$ & $7.254* 10^{-1} $ & $ 3.611 * 10^{-1}$ \\  
 \hline
\multirow{2}{*}{$\sigma_s$} &$L_2$ & $4.012* 10^{-1}$ &$5.050* 10^{-1}$ &  $1.572 * 10^{-1}$ \\ 
 \cline{2-5}
  & $L_\infty$ & $1.254$  & $ 9.447* 10^{-1}$ & $  3.566 * 10^{-1}$ \\ 
 \hline
\end{tabular}
\caption{The $L_2$ and $L_\infty$ relative errors between the computed parameters based on Methods 1 and 2, for the two-year period .}
\label{table:tablewholerelparam}}
\end{table}

\begin{figure}[H]
\begin{subfigure} {0.5\textwidth}
\centering
\includegraphics[scale=0.3]{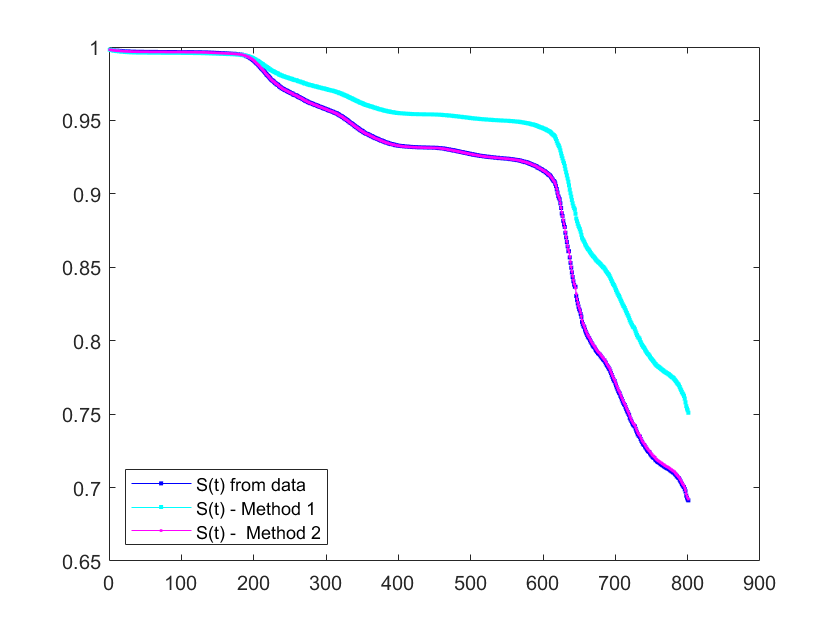}\vspace{-2mm}
\caption{Ratio of Susceptible s(t)}
\label{fig:Susceptibletwoyearitaly}
\end{subfigure}%
\begin{subfigure}{.5\linewidth}
\centering
\includegraphics[scale=.3]{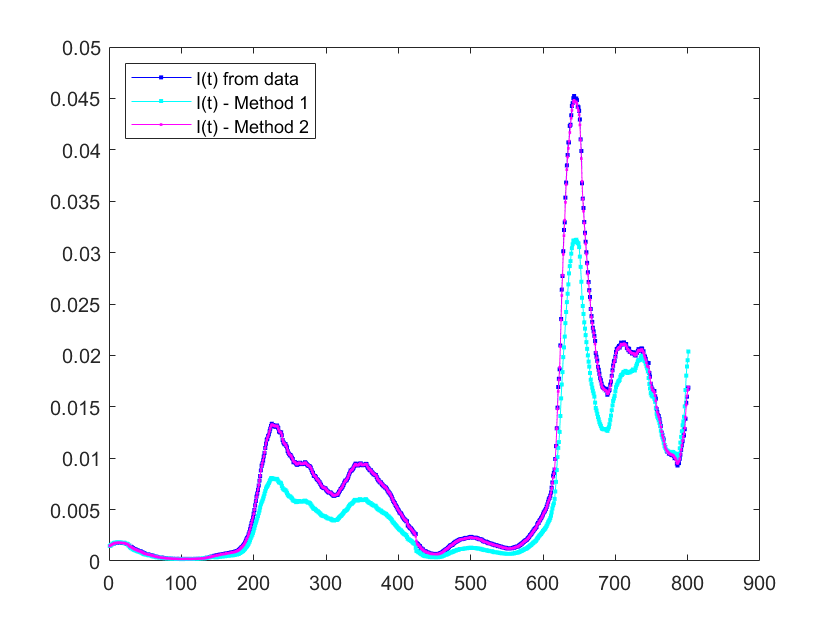}\vspace{-2mm}
\caption{Ratio of Infected i(t)}\label{fig:Infectedtwoyearitaly}
\end{subfigure}
\begin{subfigure}{1.0\linewidth}
\centering
\includegraphics[scale=.3]{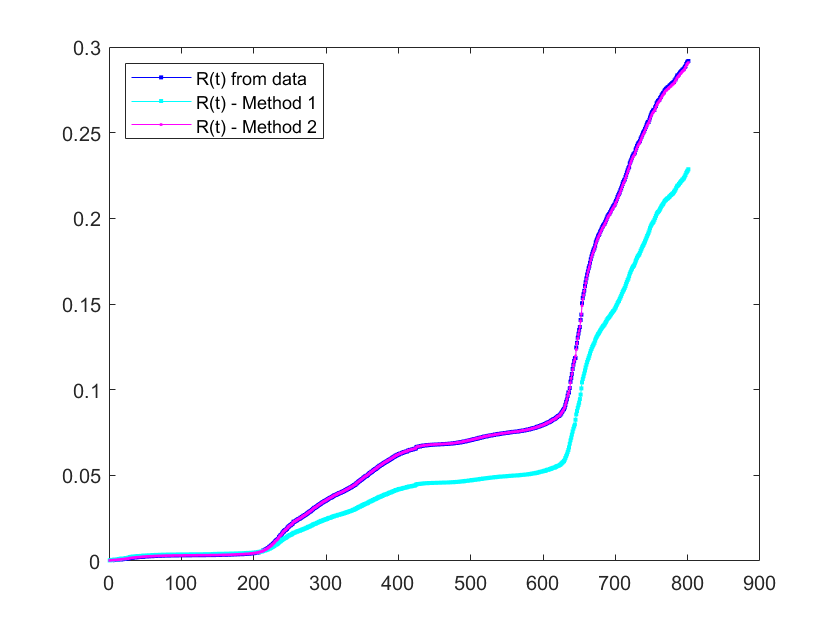}\vspace{-2mm}
\caption{Ratio of Removed r(t)}\label{fig:Removedtwoyearitaly}
\end{subfigure}

\caption{Comparison of Compartments' ratios from real data to those obtained using SIR model with approximated parameters, during the Two-Year Period in Italy }
\label{fig:SIRtwoyearitaly}\vspace{-.5cm}
\end{figure}

\begin{figure}[H]
\begin{subfigure} {0.5\textwidth}
\centering
\includegraphics[scale=0.3]{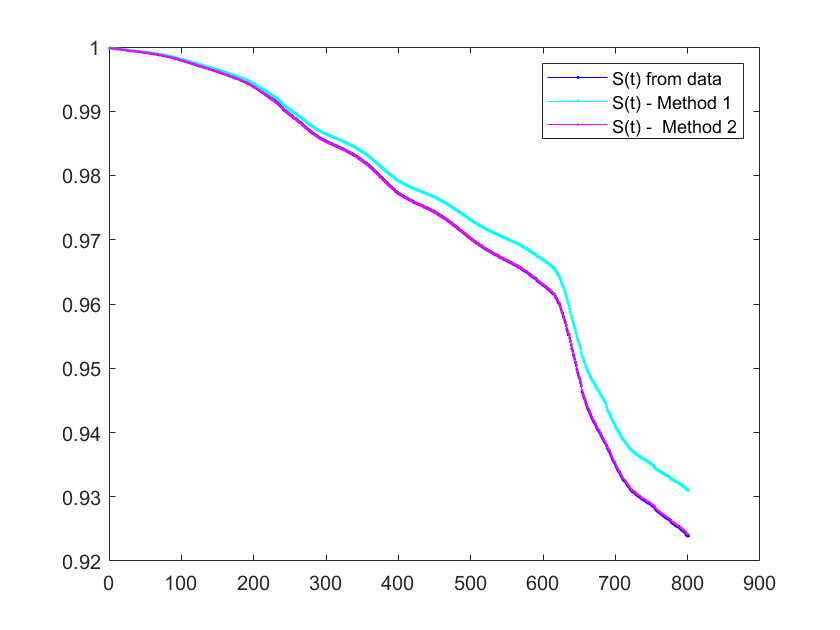}\vspace{-2mm}
\caption{Ratio of Susceptible s(t)}
\label{fig:Susceptibletwoyearworld}
\end{subfigure}%
\begin{subfigure}{.5\linewidth}
\centering
\includegraphics[scale=.3]{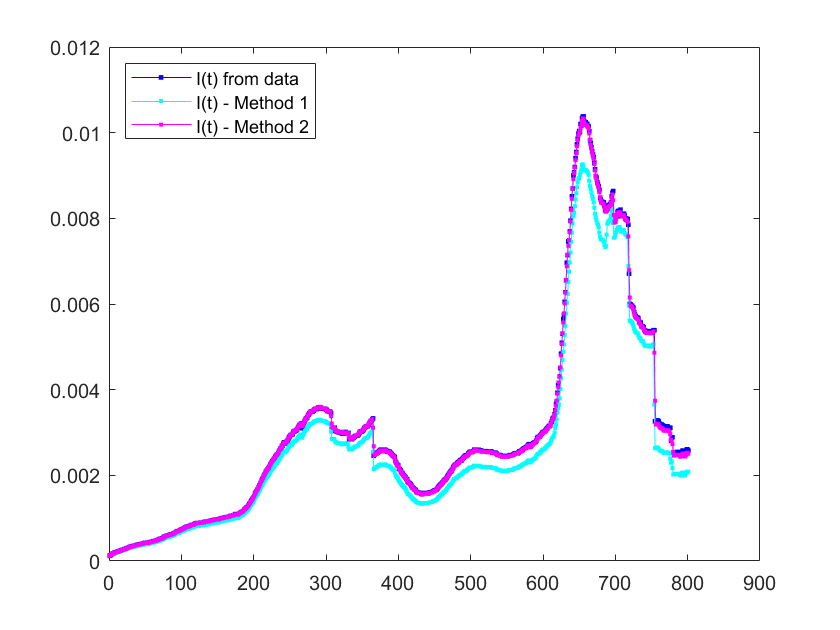}\vspace{-2mm}
\caption{Ratio of Infected i(t)}\label{fig:Infectedtwoyearworld}
\end{subfigure}
\begin{subfigure}{1.0\linewidth}
\centering
\includegraphics[scale=.3]{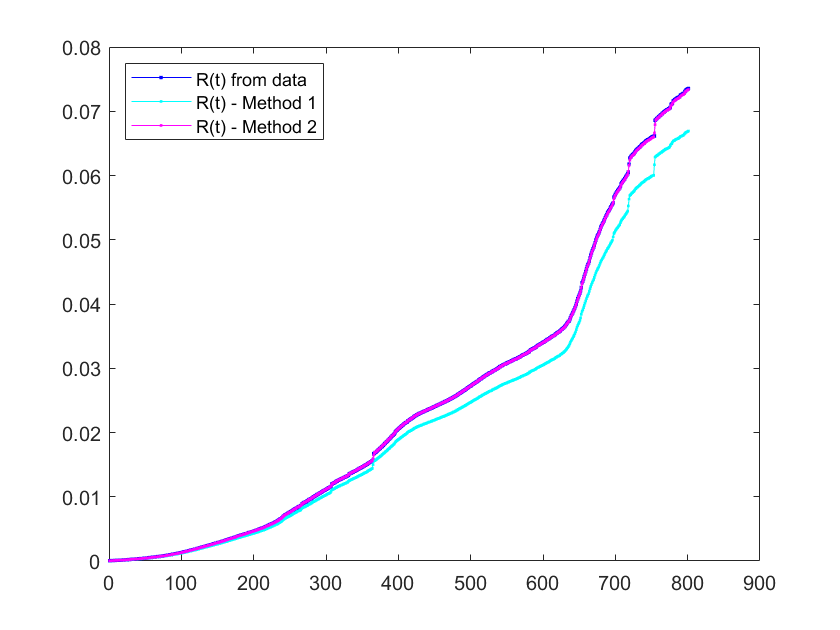}\vspace{-2mm}
\caption{Ratio of Removed r(t)}\label{fig:Removedtwoyearworld}
\end{subfigure}

\caption{Comparison of Compartments' ratios from real data to those obtained using SIR model with approximated parameters, during the Two-Year Period in the World }\vspace{-.5cm}
\label{fig:SIRtwoyearworld}
\end{figure}

Just like what has been done in the previous three cases, we validate the results by running the SIR model with the obtained parameters $\beta(t), \rho(t)$ using method 1 or method 2, and the corresponding initial values as discussed in section \ref{sec:valid}. Figures \ref{fig:SIRtwoyearitaly} and \ref{fig:SIRtwoyearworld}  compare  the behavior of the ratio of compartments $s(t), i(t), r(t)$ of the real data to the simulated values. We do not show the figures of the compartments for Germany as there is some data inconsistency. \\ Table \ref{table:tablewholeSIRrel} 
shows the relative errors between the real data and simulated values. 

A representation of the whole period shows us how both the compartments and the parameters behave throughout these two years in Italy, Germany, and the world. Simulating all the data together helps us visualize more clearly how faster the second outbreak is in comparison to the first in all of the three compartments. Yet it is not stronger than the first, as the $\sigma$ and $\sigma_s$ are not larger. The growths and decays of the infected people in Italy and the world are not perfectly exponential (\ref{fig:Infectedtwoyearitaly} and \ref{fig:Infectedtwoyearworld}); this may be because of the unreported cases, differences in contact rates between people, differences in immunities, and many other assumptions not considered in the model. \\
\noindent Note that there are several time intervals where $I(t)$ fluctuates rather than increases/decreases sharply.\vspace{1mm}\\ It is clear from the figures and table of relative errors that there is a slight difference in the simulated results using methods 1 and 2 where method 2 gives more accurate results as expected.

\begin{table}[H]
\centering
\setlength{\tabcolsep}{10pt}
{\renewcommand{\arraystretch}{1.4}
\begin{tabular}{||c|c| c| c| c |c||} 
 \hline
 Method& &Norm & Italy  & Germany  & World  \\ 
 \hline\hline
\multirow{6}{*}{1}&\multirow{2}{*}{S} &$L_2$ & $ 3.569* 10^{-2}$ &$ 1.216* 10^{-1}$ &  $ 3.453 * 10^{-3}$  \\ 
 \cline{3-6}
 & & $L_\infty$ & $ 6.698* 10^{-2}$  & $ 3.114* 10^{-1}$ & $ 7.156 * 10^{-3}$ \\ 
 \cline{2-6}
&\multirow{2}{*}{I} &$L_2$ & $2.826* 10^{-1}$ &$ 9.454* 10^{-1}$ &  $  1.041 * 10^{-1}$ \\ 
 \cline{3-6}
&& $L_\infty$ & $3.105 * 10^{-1}$ & $9.417 * 10^{-1}$ & $  1.672 * 10^{-1}$ \\ 
  \cline{2-6}
 &\multirow{2}{*}{R} & $L_2$ & $ 2.738* 10^{-1}$ &$9.363* 10^{-1}$ &  $9.349 * 10^{-2}$ \\\cline{3-6}
 && $L_\infty$ & $ 2.252 * 10^{-1}$ & $9.316* 10^{-1}$ & $  9.019  * 10^{-2}$ \\  
 \hline 
 \multirow{6}{*}{2}&\multirow{2}{*}{S} &$L_2$ & $4.444 * 10^{-4}$	&$   8.082  * 10^{-3}$&	$7.158* 10^{-5}$	\\ 
 \cline{3-6}
 & & $L_\infty$ &$  1.544 * 10^{-3}$	&$1.581 * 10^{-2}$&	$2.643 * 10^{-4}$				 \\ 
 \cline{2-6}
&\multirow{2}{*}{I} &$L_2$ &  $7.398 * 10^{-3}$&	$4.142 * 10^{-2}$&	$ 1.043 * 10^{-2}$				 \\ 
 \cline{3-6}
&& $L_\infty$ & $  1.607 * 10^{-2}$	&$5.295 * 10^{-2}$&	$5.057 * 10^{-2}$ \\ 
  \cline{2-6}
 &\multirow{2}{*}{R} & $L_2$ &$3.369 * 10^{-3}$	&$6.599 * 10^{-2}$	&$ 1.908 * 10^{-3}$				 \\\cline{3-6}
 && $L_\infty$ & $4.812 * 10^{-3}$	&$  4.773 * 10^{-2}$&	$9.149 * 10^{-3}$\\  
 \hline
\end{tabular}
\caption{The $L_2$ and $L_\infty$ relative errors of the computed S,I,R from the time-dependent model with the S,I,R collected from data during the two-year period, where $\beta,\rho$ are computed using Method 1 or 2. }\label{table:tablewholeSIRrel}}
\end{table}

\section{Incompatibility of Constant-Coefficient Models using Method2}\label{sec:append}

\begin{figure}[H]
\begin{subfigure} {0.5\textwidth}
\centering
\includegraphics[scale=0.3]{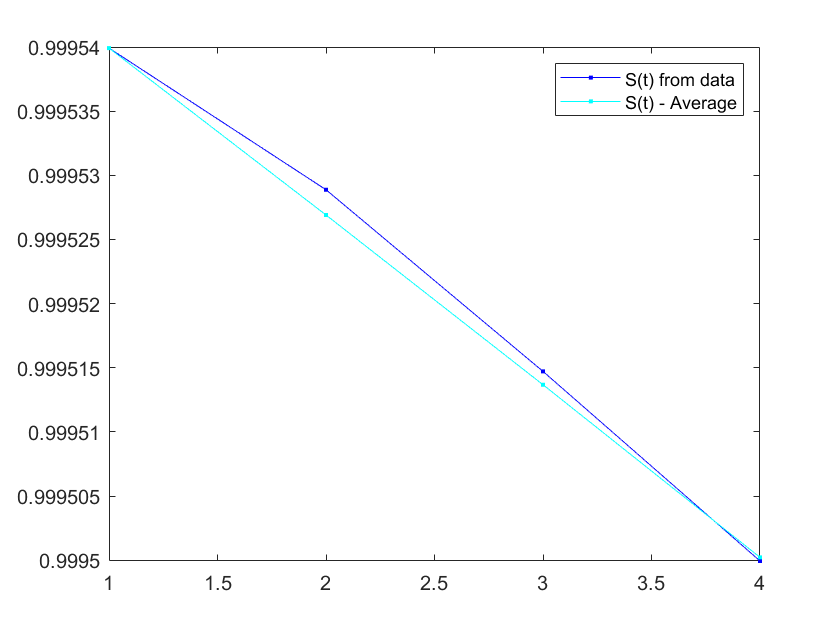}\vspace{-2mm}
\caption{Ratio of Susceptible s(t)}\label{fig:susceptibleavfourmethod2}
\end{subfigure}%
\begin{subfigure}{.5\linewidth}
\centering
\includegraphics[scale=.3]{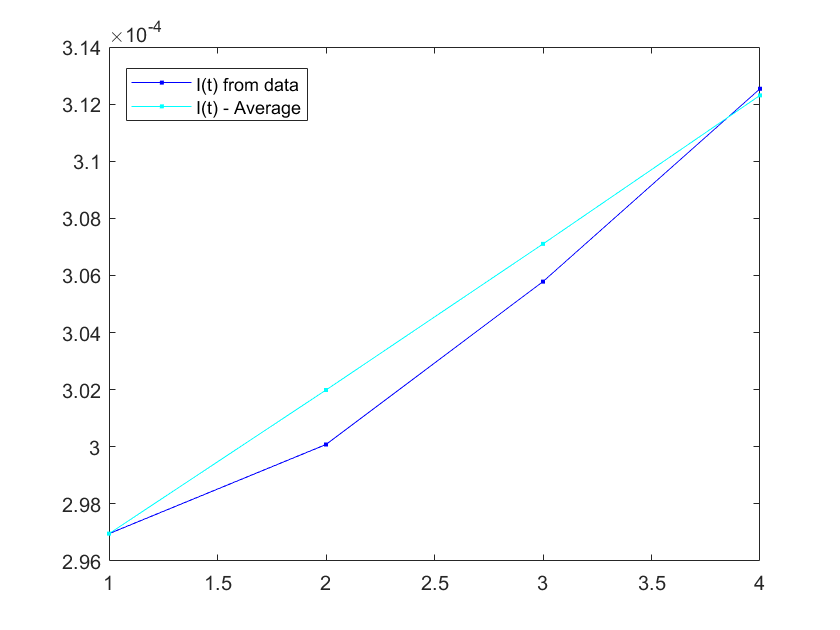}\vspace{-2mm}
\caption{Ratio of Infected i(t)}\label{fig:infectedavfourmethod2}
\end{subfigure}
\begin{subfigure}{1.0\linewidth}
\centering
\includegraphics[scale=.3]{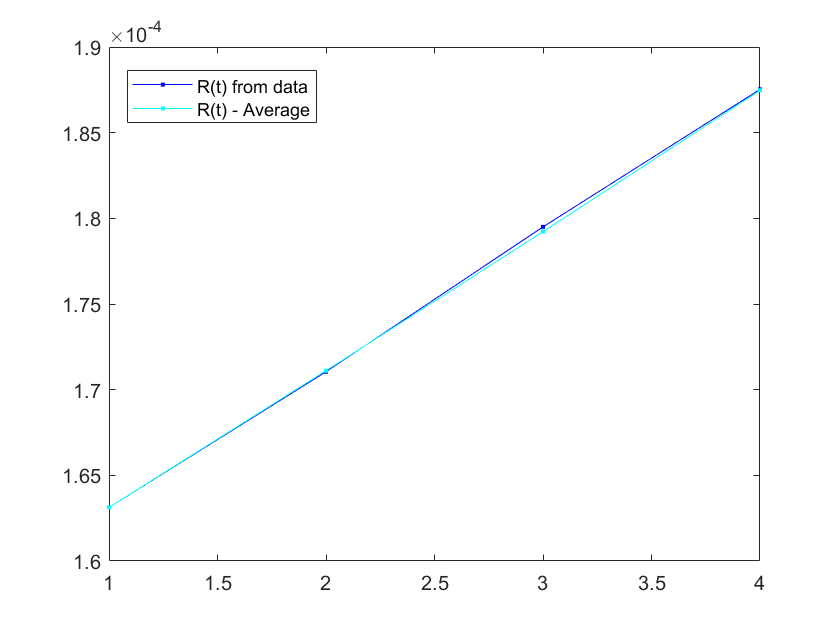}\vspace{-2mm}
\caption{Ratio of Removed r(t)}\label{fig:rav4method2}
\end{subfigure}
\caption{Time-Independent SIR Model Using the Average: First Interval}\vspace{-.5cm}
\label{fig:firstintervalmethod2}
\end{figure}

\begin{figure}[H]
\begin{subfigure}{.5\textwidth}
\centering
\includegraphics[scale=.3]{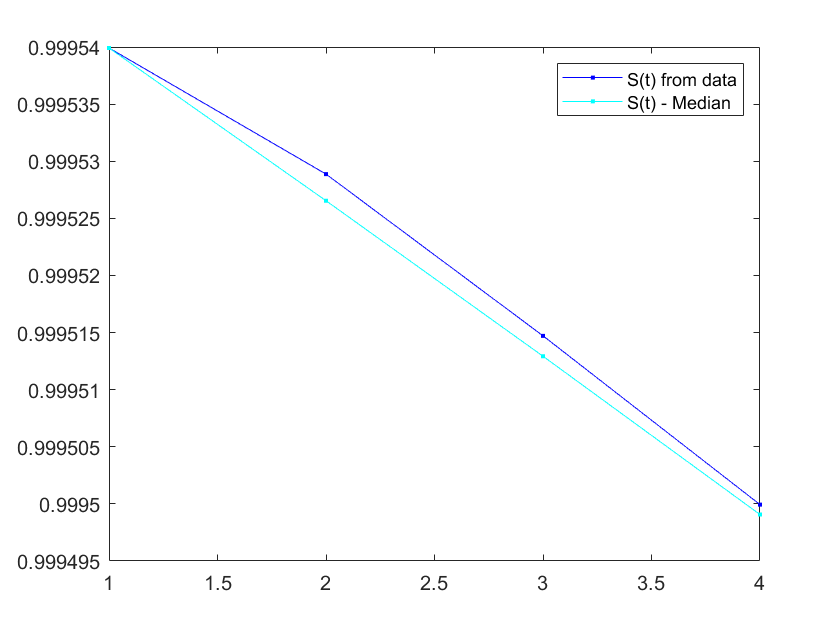}\vspace{-2mm}
\caption{Ratio of Susceptible s(t)}\label{fig:smedian1method2}
\end{subfigure}%
\begin{subfigure}{.5\textwidth}
\centering
\includegraphics[scale=.3]{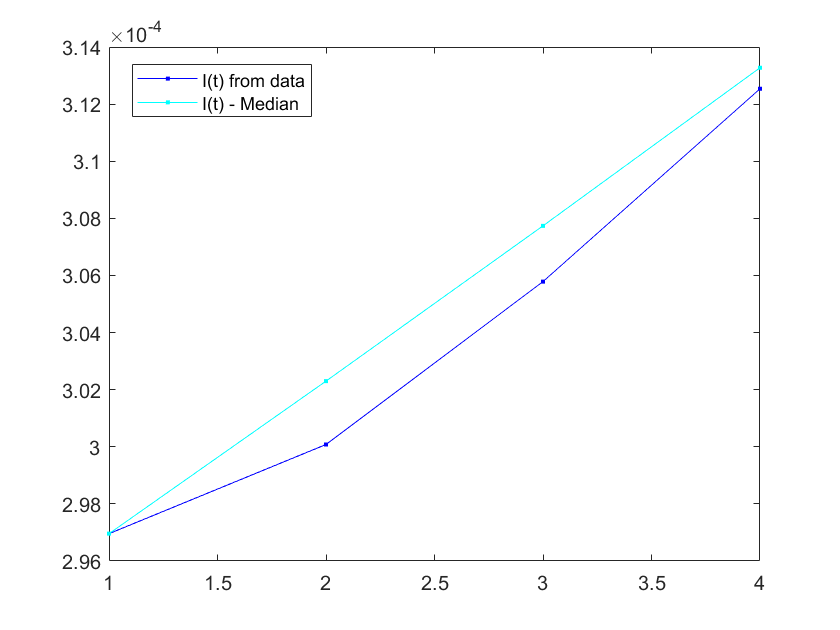}\vspace{-2mm}
\caption{Ratio of Infected i(t)}\label{fig:imedian1method2}
\end{subfigure}
\begin{subfigure}{1.0\linewidth}
\centering
\includegraphics[scale=.3]{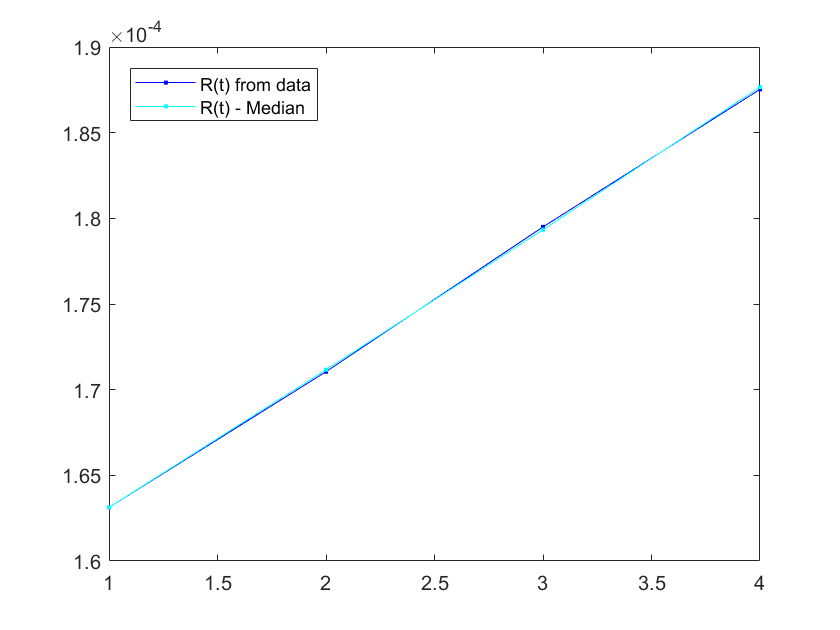}\vspace{-2mm}
\caption{Ratio of Removed r(t)}\label{fig:rmedian1method2}
\end{subfigure}
\caption{Time-Independent SIR Model Using the Median: First Interval}\vspace{-.5cm}
\label{fig:sirmedian1method2
}
\end{figure}

\begin{figure}[H]
\begin{subfigure} {0.5\textwidth}
\centering
\includegraphics[scale=0.3]{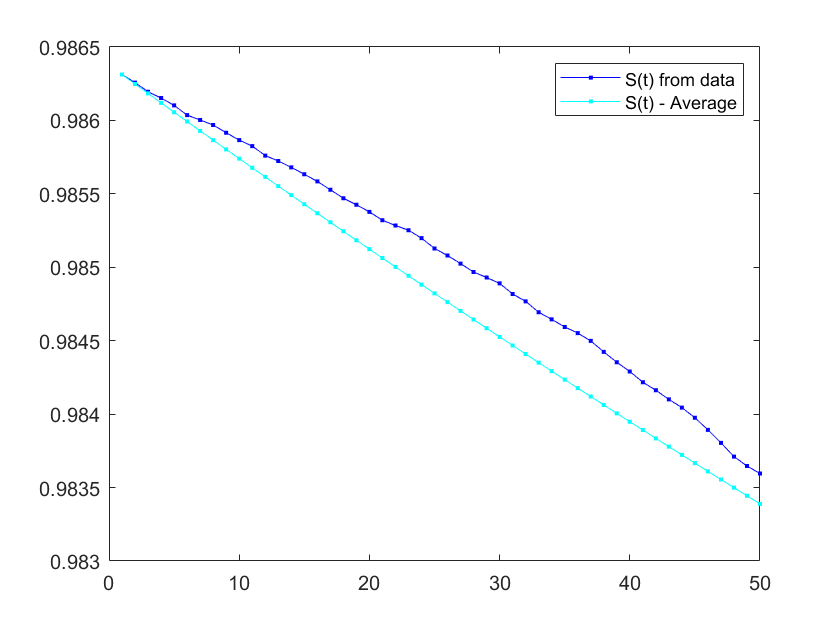}\vspace{-2mm}
\caption{Ratio of Susceptible s(t)}\label{fig:susceptibleav2method2}
\end{subfigure}%
\begin{subfigure}{.5\linewidth}
\centering
\includegraphics[scale=.3]{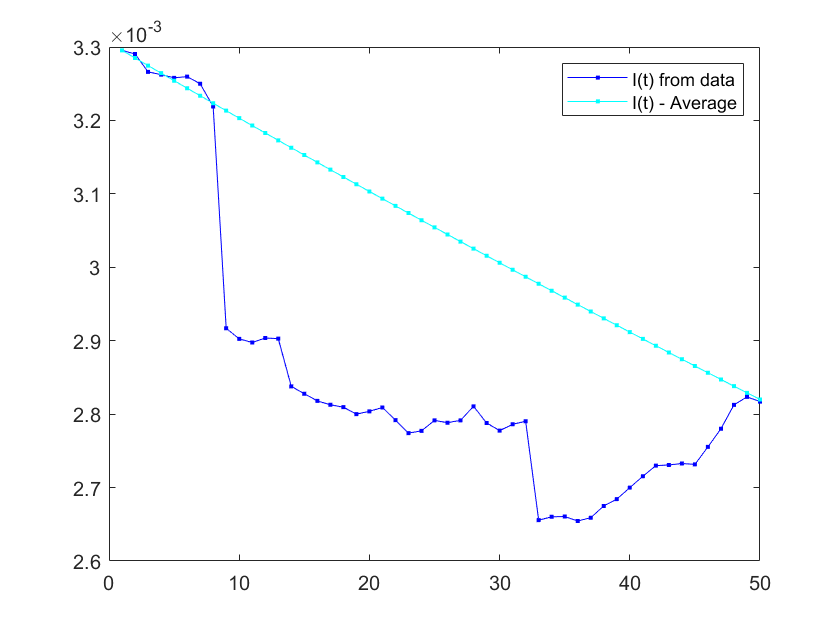}\vspace{-2mm}
\caption{Ratio of Infected i(t)}\label{fig:infectedav2method2}
\end{subfigure}
\begin{subfigure}{1.0\linewidth}
\centering
\includegraphics[scale=.3]{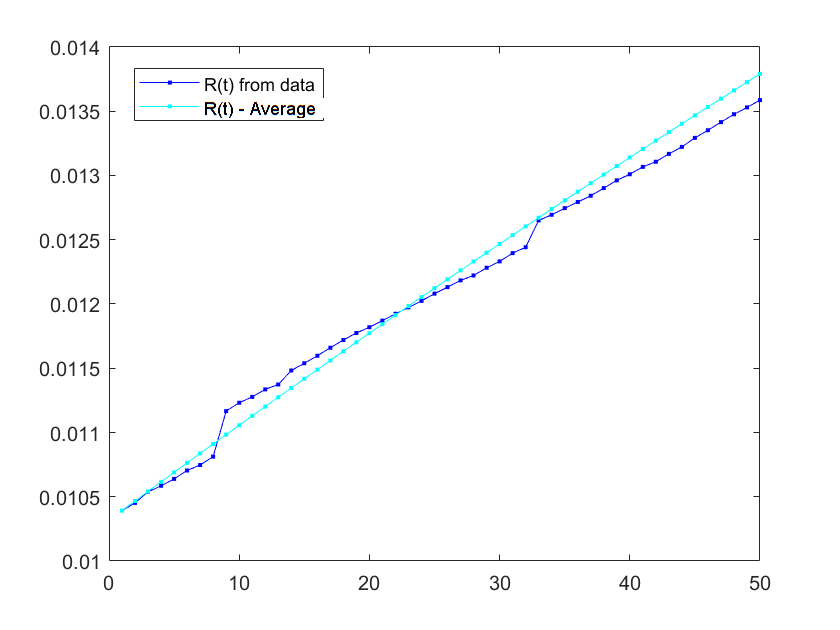}\vspace{-2mm}
\caption{Ratio of Removed r(t)}\label{fig:rav2method2}
\end{subfigure}
\caption{Time-Independent SIR Model Using the Average: Second Interval}\vspace{-.5cm}
\label{fig:siraverage2method2}
\end{figure}

\begin{figure}[H]
\begin{subfigure} {0.5\textwidth}
\centering
\includegraphics[scale=0.3]{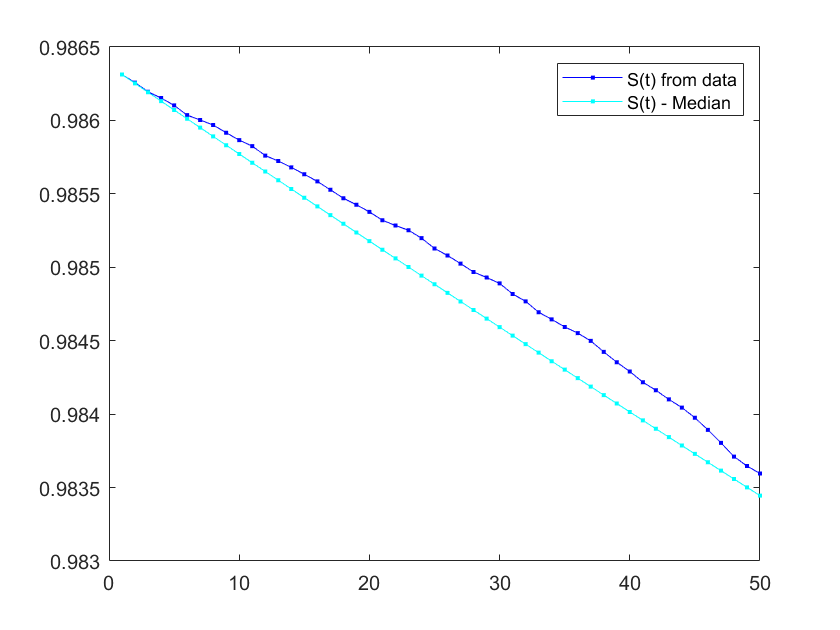}\vspace{-2mm}
\caption{Ratio of Susceptible s(t)}\label{fig:smedian2method2}
\end{subfigure}%
\begin{subfigure}{.5\linewidth}
\centering
\includegraphics[scale=.3]{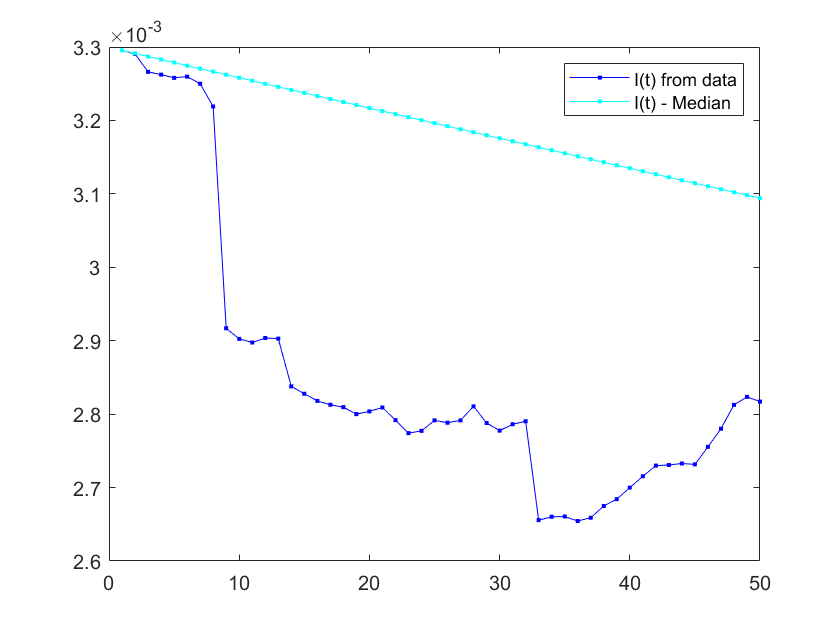}\vspace{-2mm}
\caption{Ratio of Infected i(t)}\label{fig:imedian2method2}
\end{subfigure}
\begin{subfigure}{1.0\linewidth}
\centering
\includegraphics[scale=.3]{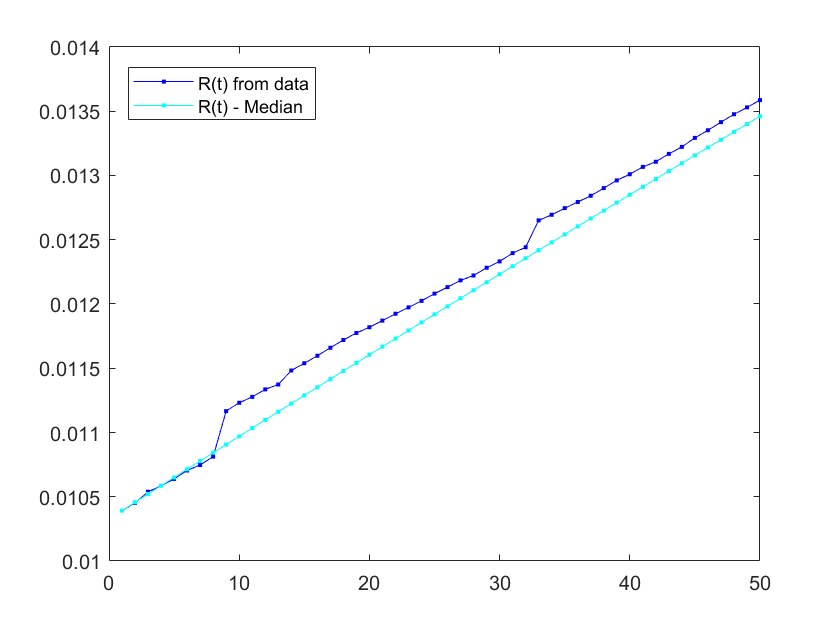}\vspace{-2mm}
\caption{Ratio of Removed r(t)}\label{Rsecondmethodmedianfifty}
\end{subfigure}
\caption{Time-Independent SIR Model Using the Median: Second Interval}\vspace{-.5cm}
\label{fig:sirmedian2method2}
\end{figure}

\begin{figure}[H]
\begin{subfigure} {0.5\textwidth}
\centering
\includegraphics[scale=0.3]{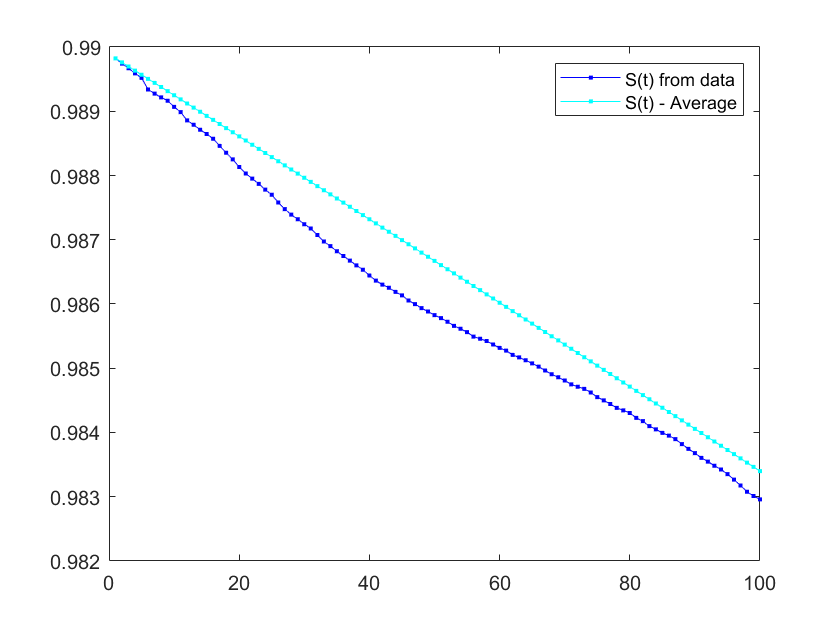}\vspace{-2mm}
\caption{Ratio of Susceptible s(t)}\label{fig:susceptibleav3method2}
\end{subfigure}%
\begin{subfigure}{.5\linewidth}
\centering
\includegraphics[scale=.3]{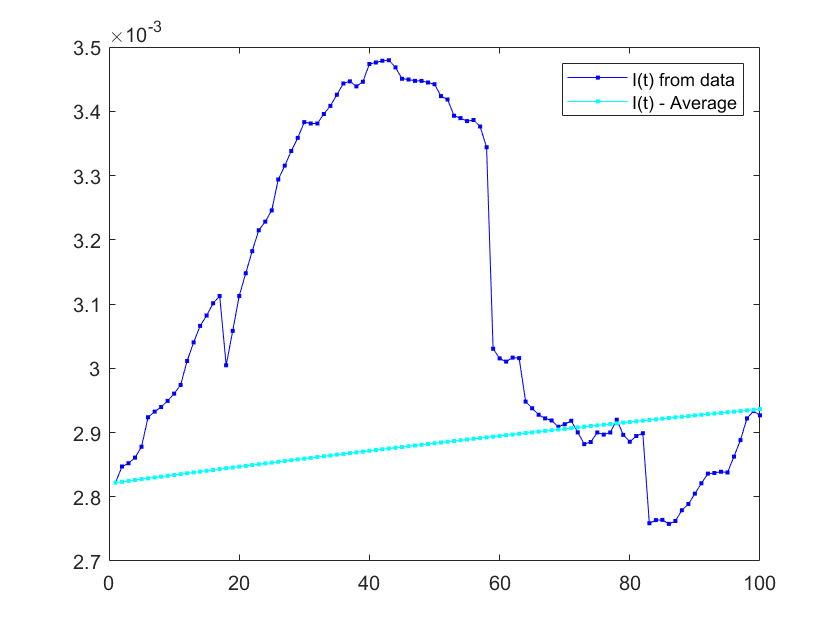}\vspace{-2mm}
\caption{Ratio of Infected i(t)}\label{fig:infectedav3method2}
\end{subfigure}
\begin{subfigure}{1.0\linewidth}
\centering
\includegraphics[scale=.3]{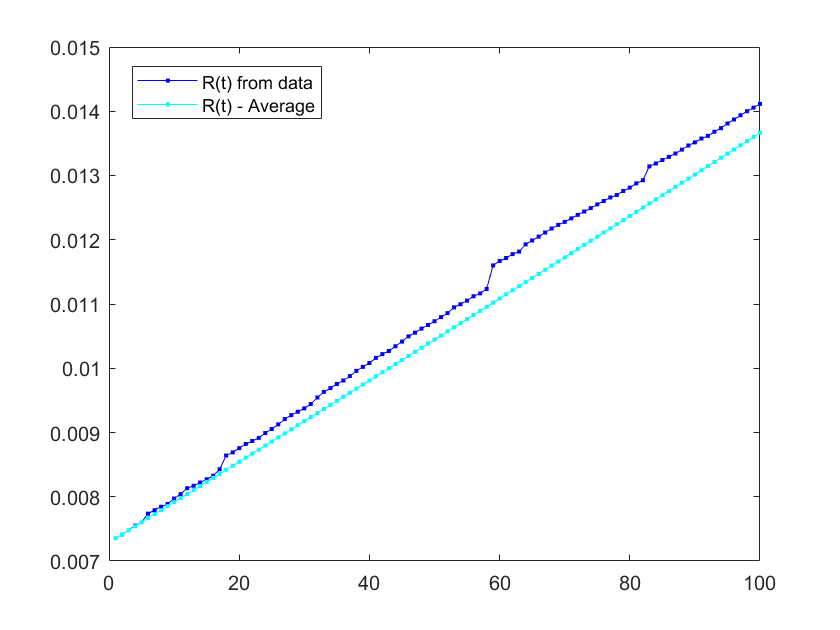}\vspace{-2mm}
\caption{Ratio of Removed r(t)}\label{fig:rav3method2}
\end{subfigure}
\caption{Time-Independent SIR Model Using the Average: Third Interval}\vspace{-.5cm}
\label{fig:sirav3method2}
\end{figure}

\begin{figure}[H]
\begin{subfigure} {0.5\textwidth}
\centering
\includegraphics[scale=0.3]{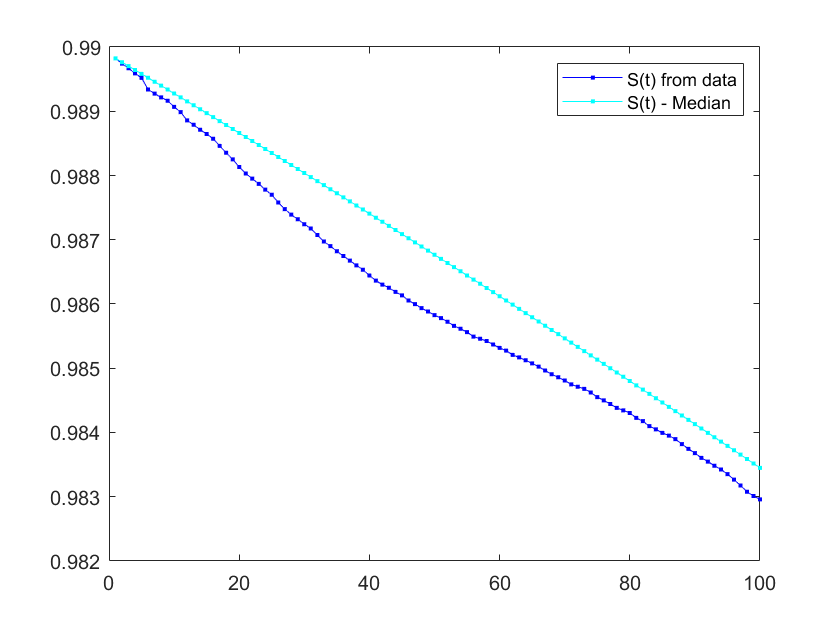}\vspace{-2mm}
\caption{Ratio of Susceptible s(t)}\label{fig:smedian3method2}
\end{subfigure}%
\begin{subfigure}{.5\linewidth}
\centering
\includegraphics[scale=.3]{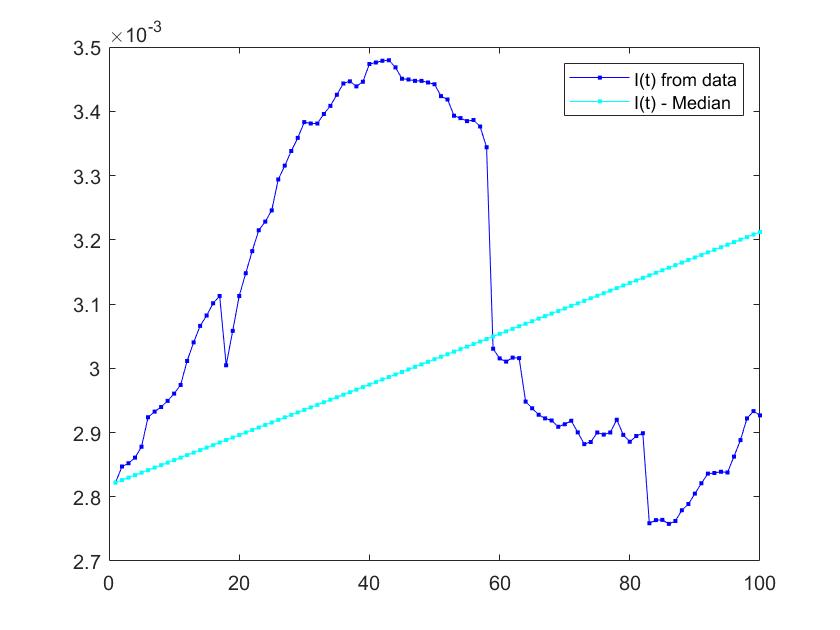}\vspace{-2mm}
\caption{Ratio of Infected i(t)}\label{fig:imedian3method2}
\end{subfigure}
\begin{subfigure}{1.0\linewidth}
\centering
\includegraphics[scale=.3]{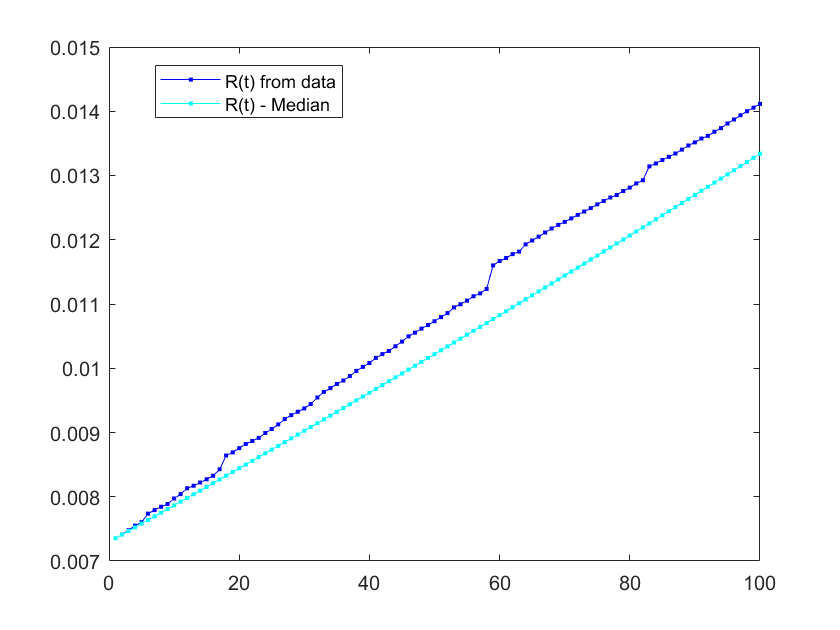}\vspace{-2mm}
\caption{Ratio of Removed r(t)}\label{Rmedianthirdintervalsecondmethod}
\end{subfigure}
\caption{Time-Independent SIR Model Using the Median: Third Interval}\vspace{-.5cm}
\label{fig:sirmedian3method2}
\end{figure}

\begin{table}[H]
\centering
\setlength{\tabcolsep}{5pt}
{\renewcommand{\arraystretch}{1.1}
\begin{tabular}{||c|c|c|c|c|c||} 
 \cline{2-6}
\multicolumn{1}{c|}{} &Period&Norm & S & I & R \\ 
 \hline\hline
\multirow{6}{*}{Average} &\multirow{2}{*}{4\; days} & $L_2$ &  $ 1.122*10^{-6}$ & $3.834* 10^{-3}$ &  $ 8.282*10^{-4}$  \\ 
\cline{3-6}
& &$L_\infty$ &   $1.969* 10^{-6}$ &  $6.104*10^{-3}$ & $ 1.489*10^{-3}$ \\ 
 \cline{2-6}
&\multirow{2}{*}{50\; days} &$L_2$ &  $ 2.695* 10^{-4}$& $   8.037*10^{-2}$ &  $ 9.7102*10^{-3}$ \\ 
\cline{3-6}
& &$L_\infty$ &   $ 3.841 * 10^{-4}$ &  $ 9.865*10^{-2}$ & $  1.503*10^{-2}$\\
 \cline{2-6}
 &\multirow{2}{*}{100\; days} &$L_2$ &  $  5.972*10^{-4}$& $    1.076*10^{-1}$ &  $ 3.395*10^{-2}$ \\ 
 \cline{3-6}
& &$L_\infty$ &   $ 8.996 * 10^{-4}$ &  $  1.737*10^{-1}$ & $4.159*10^{-2}$ \\ 
 \hline

\multirow{6}{*}{Median} &\multirow{2}{*}{4\; days} & $L_2$ &  $ 1.533*10^{-6}$ & $5.012* 10^{-3}$ &  $  6.820*10^{-4}$  \\ 
\cline{3-6}
 &&$L_\infty$ &   $2.333* 10^{-6}$ &  $  7.101*10^{-3}$ & $8.748*10^{-4}$ \\ 
\cline{2-6}
&\multirow{2}{*}{50\; days} &$L_2$ &  $ 2.150* 10^{-4}$& $   1.292*10^{-1}$ &  $ 1.401*10^{-2}$ \\ 
\cline{3-6}
 &&$L_\infty$ &   $ 3.156 * 10^{-4}$ &  $  1.541*10^{-1}$ & $  1.924*10^{-2}$\\
 \cline{2-6}
 &\multirow{2}{*}{100\; days} &$L_2$ &  $  6.735*10^{-4}$& $    1.021*10^{-1}$ &  $  5.451*10^{-2}$ \\ 
 \cline{3-6}
 &&$L_\infty$ &   $  9.895 * 10^{-4}$ &  $ 1.434*10^{-1}$ & $6.285*10^{-2}$ \\ 
 \hline

\multirow{6}{*}{Time-dependent} &\multirow{2}{*}{4\; days} & $L_2$ &  $4.020*10^{-7}$ & $ 1.147* 10^{-3}$ &  $ 4.790*10^{-4}$  \\ 
\cline{3-6}
& &$L_\infty$ &   $7.858* 10^{-7}$ &  $2.067*10^{-3}$ & $ 7.422*10^{-4}$ \\ 
\cline{2-6}
&\multirow{2}{*}{50\; days} &$L_2$ &  $ 3.905* 10^{-6}$& $ 5.653*10^{-3}$ &  $1.414*10^{-3}$ \\ 
\cline{3-6}
& &$L_\infty$ &   $8.646 * 10^{-6}$ &  $  2.194*10^{-2}$ & $  5.350*10^{-3}$\\
 \cline{2-6}
& \multirow{2}{*}{100\; days} &$L_2$ &  $ 6.131*10^{-6}$& $    4.230*10^{-3}$ &  $  1.292*10^{-3}$ \\ 
\cline{3-6}
 &&$L_\infty$ &   $2.965 * 10^{-5}$ &  $  2.158*10^{-2}$ & $ 5.345*10^{-3}$ \\ 
 \hline
\end{tabular}}
\caption{Relative errors between the S,I,R values computed from the constant-coefficient model with average or median $\beta, \rho$ values or time-dependent model; and the S,I,R values collected from the world data for different periods. }\label{tab:averageWorldC}
\end{table}

\subsection{Predicting}\label{sec:PredM2}

\begin{figure}[H]
\begin{subfigure} {0.5\textwidth}
\centering
\includegraphics[scale=0.32]{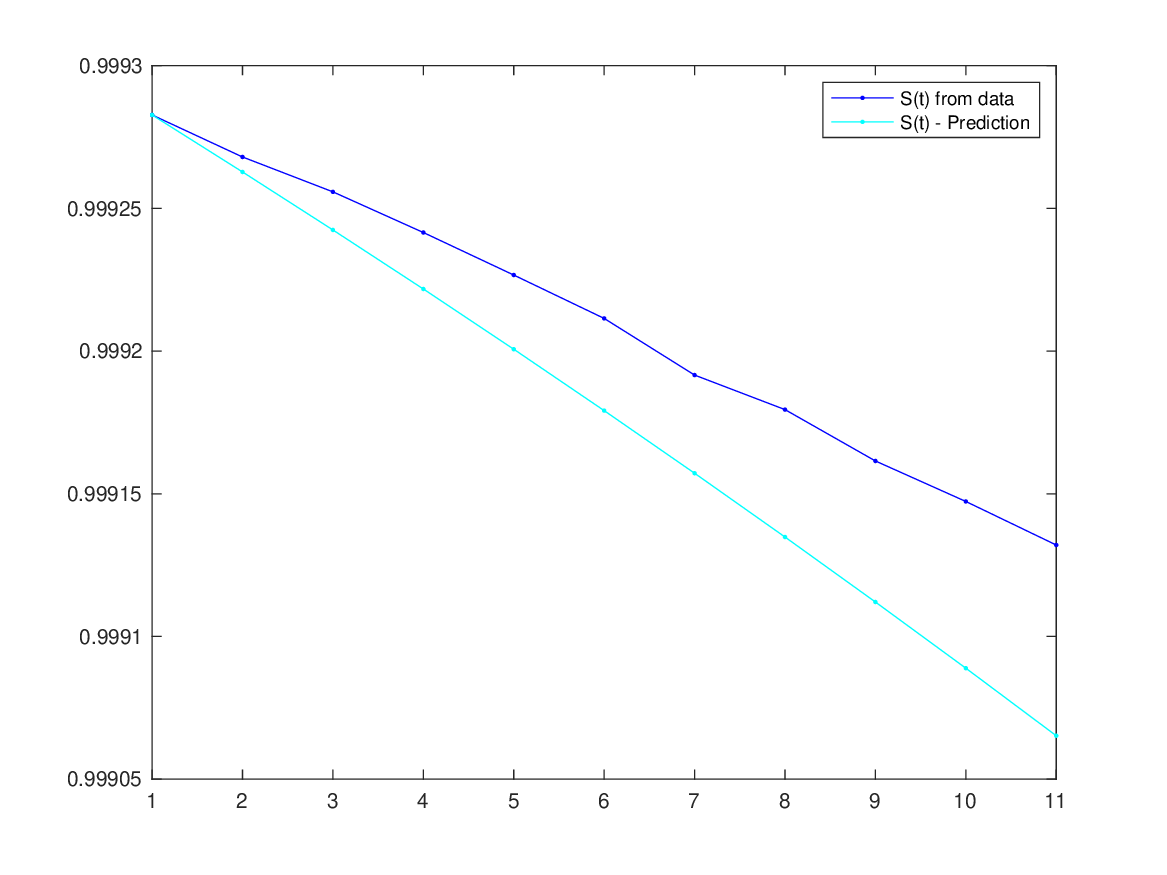}\vspace{-2mm}
\caption{t1}
\end{subfigure}%
\begin{subfigure}{.5\linewidth}
\centering
\includegraphics[scale=.32]{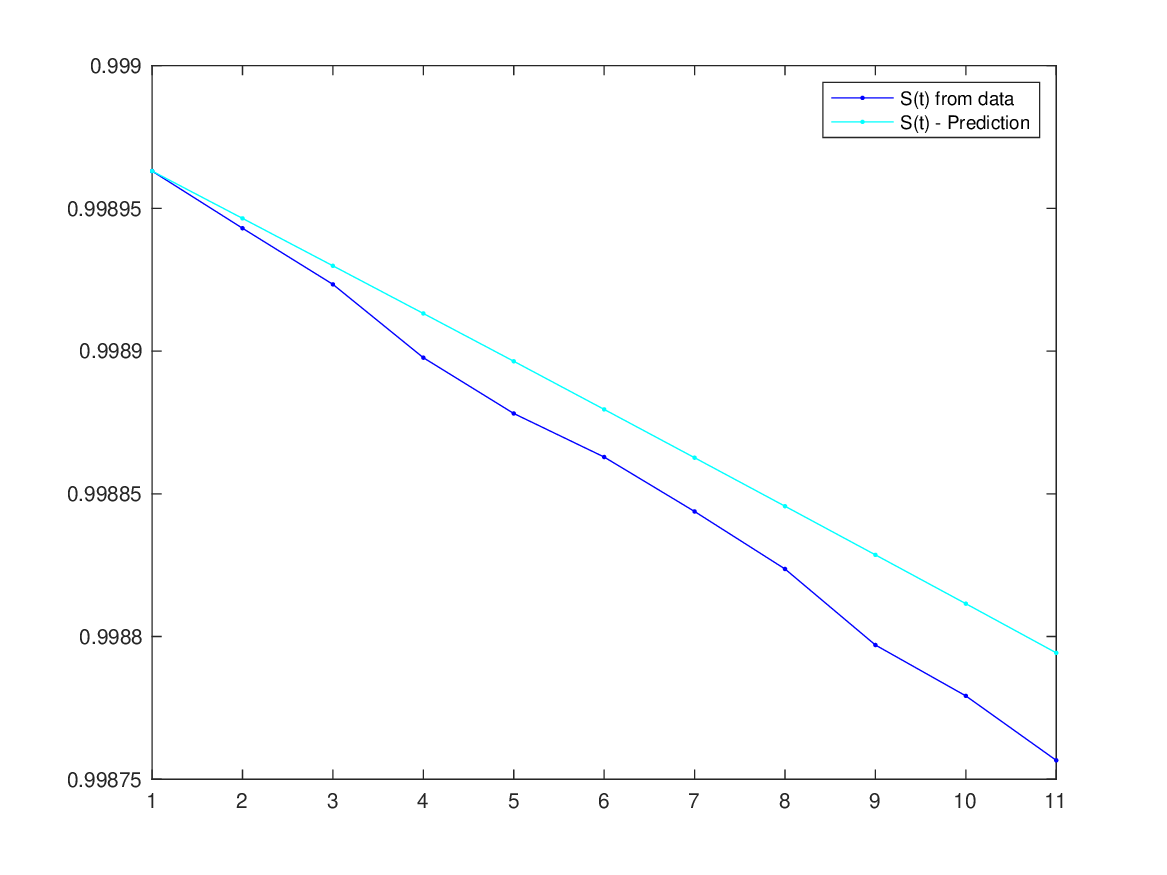}\vspace{-2mm}
\caption{t2}
\end{subfigure}
\begin{subfigure}{0.5\linewidth}
\centering
\includegraphics[scale=.32]{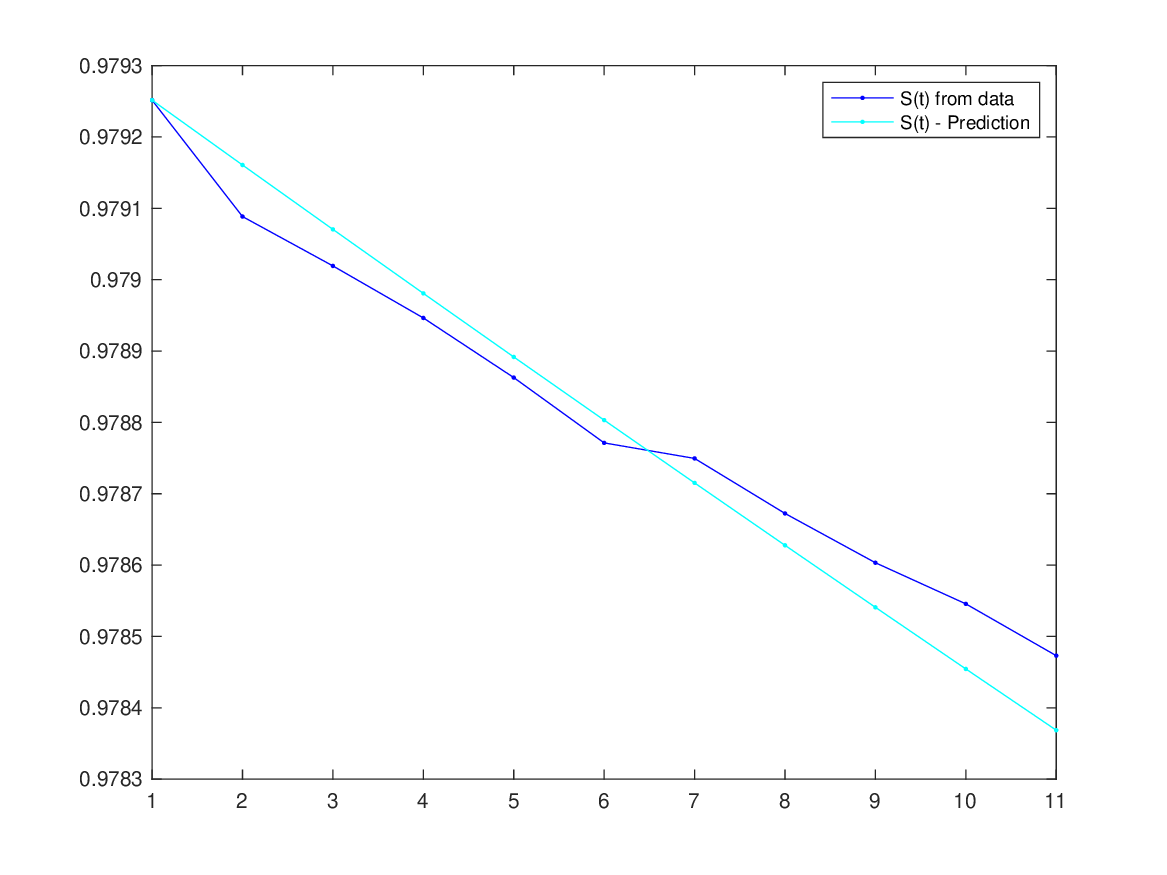}\vspace{-2mm}
\caption{t3}
\end{subfigure}
\begin{subfigure}{0.5\linewidth}
\centering
\includegraphics[scale=.32]{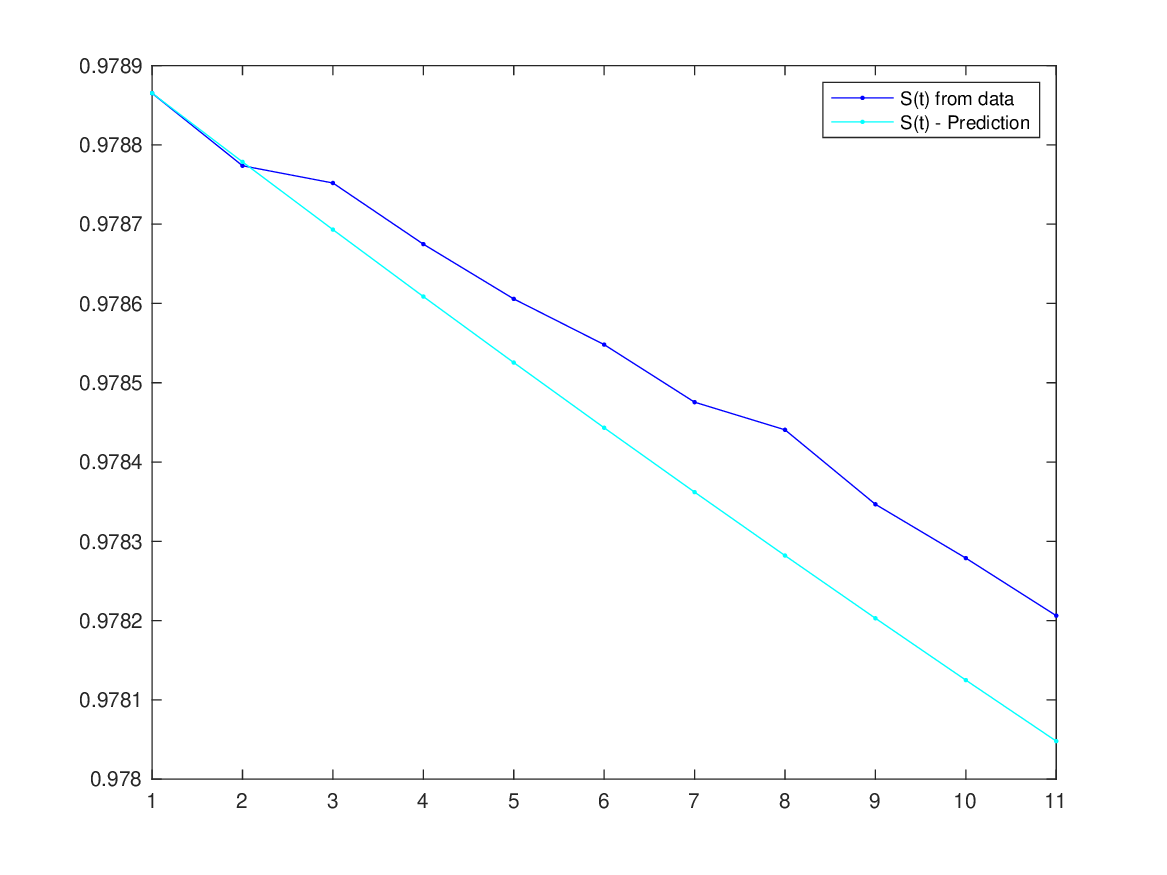}\vspace{-2mm}
\caption{t4}
\end{subfigure}
\caption{Prediction of $s(t)$ for the 4 starting dates and duration of 10 days
}\vspace{-.5cm}
\label{fig:predictionS10M}
\end{figure}

\begin{figure}[H]
\begin{subfigure} {0.5\textwidth}
\centering
\includegraphics[scale=0.32]{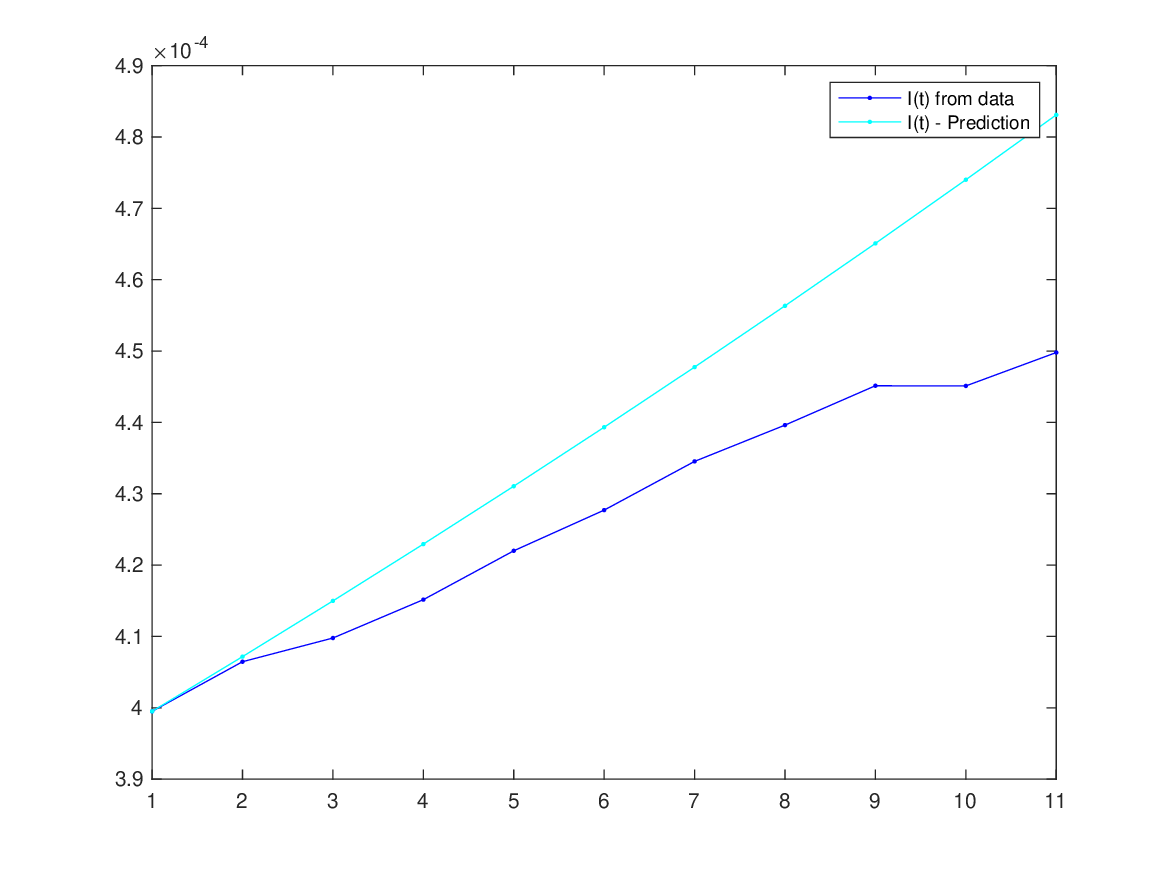}\vspace{-2mm}
\caption{t1}
\end{subfigure}%
\begin{subfigure}{.5\linewidth}
\centering
\includegraphics[scale=.32]{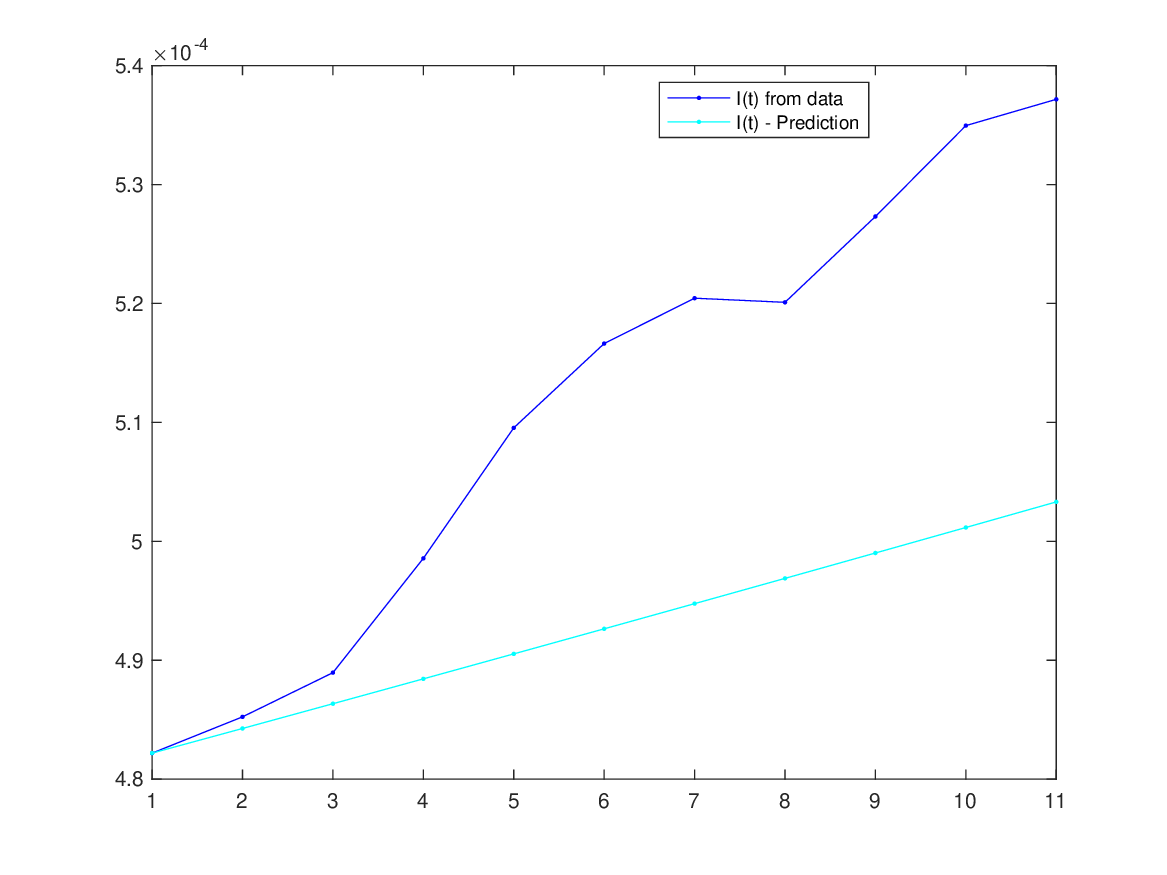}\vspace{-2mm}
\caption{t2}
\end{subfigure}
\begin{subfigure}{0.5\linewidth}
\centering
\includegraphics[scale=.32]{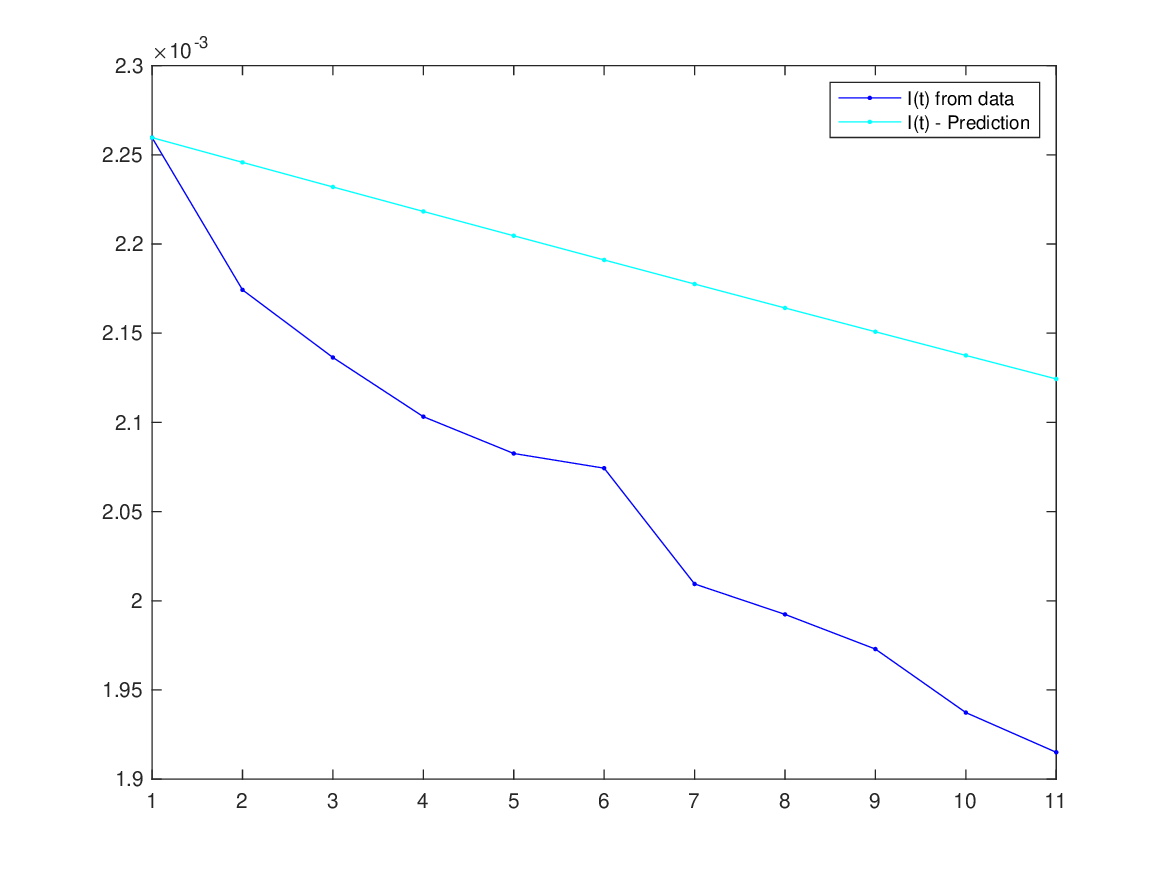}\vspace{-2mm}
\caption{t3}
\end{subfigure}
\begin{subfigure}{0.5\linewidth}
\centering
\includegraphics[scale=.32]{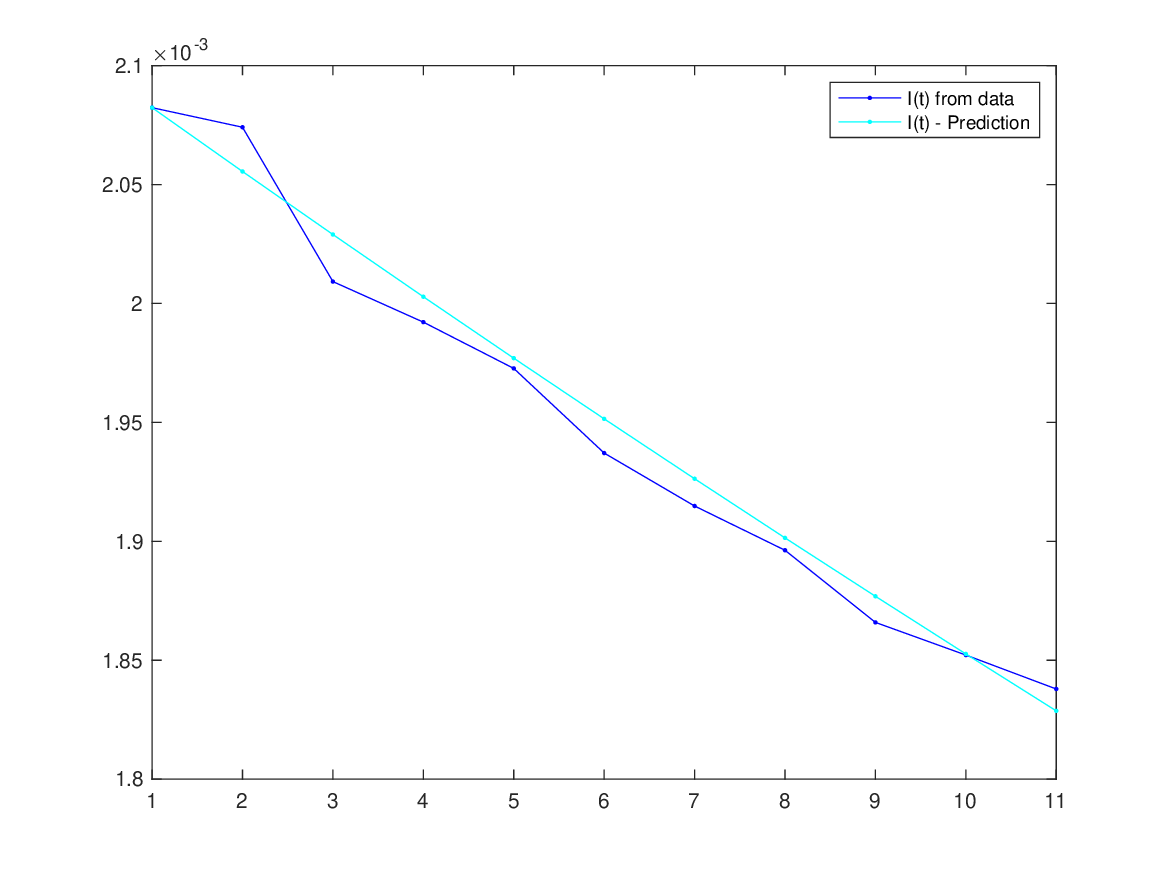}\vspace{-2mm}
\caption{t4}
\end{subfigure}
\caption{Prediction of $i(t)$ for the 4 starting dates and duration of 10 days
}\vspace{-.5cm}
\label{fig:predictionI10M}
\end{figure}

\begin{figure}[H]
\begin{subfigure} {0.5\textwidth}
\centering
\includegraphics[scale=0.32]{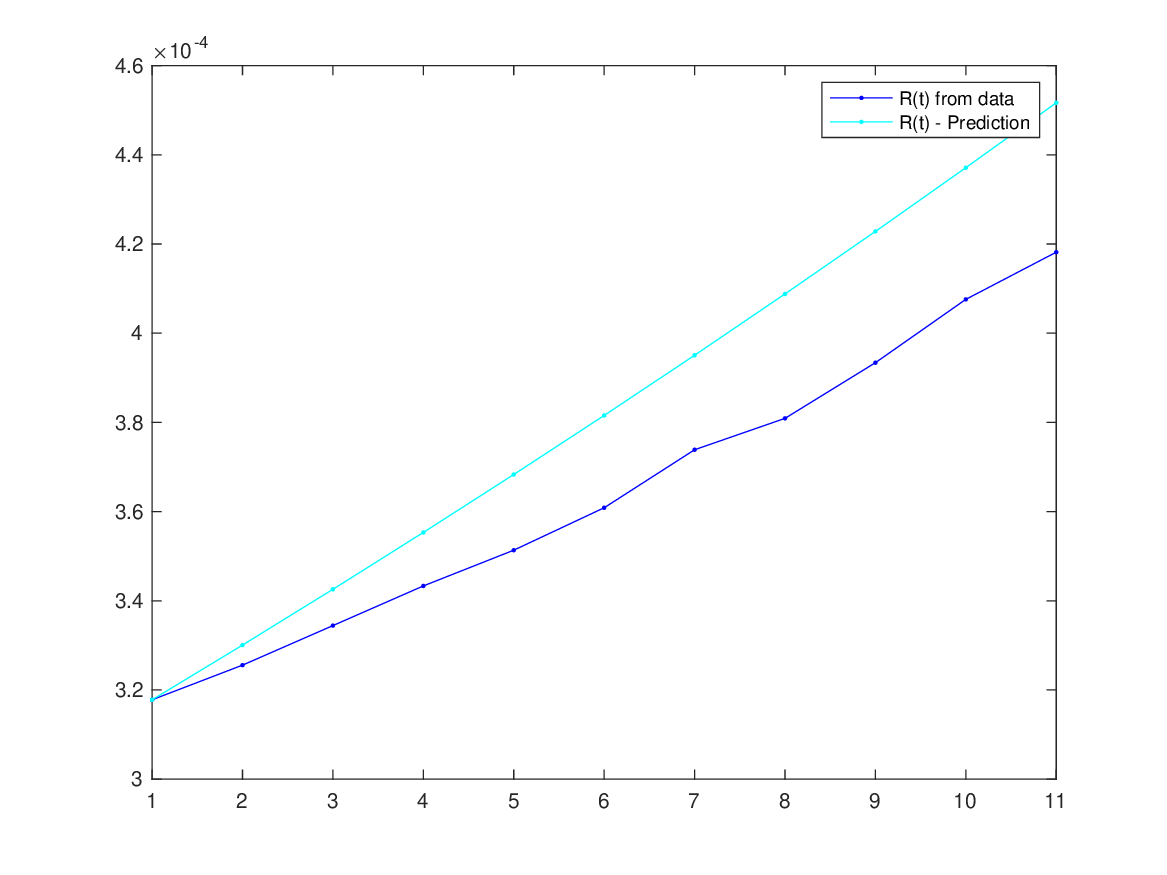}\vspace{-2mm}
\caption{t1}
\end{subfigure}%
\begin{subfigure}{.5\linewidth}
\centering
\includegraphics[scale=.32]{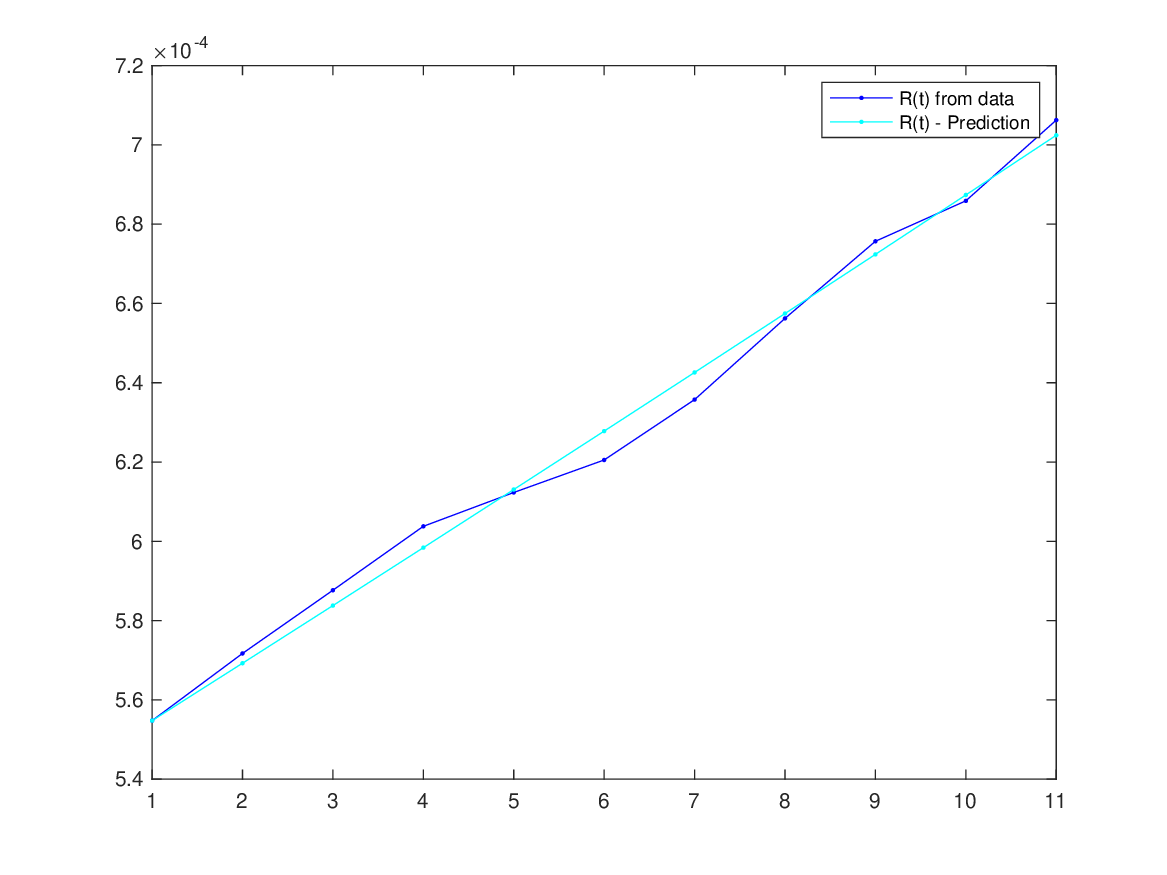}\vspace{-2mm}
\caption{t2}
\end{subfigure}
\begin{subfigure}{0.5\linewidth}
\centering
\includegraphics[scale=.32]{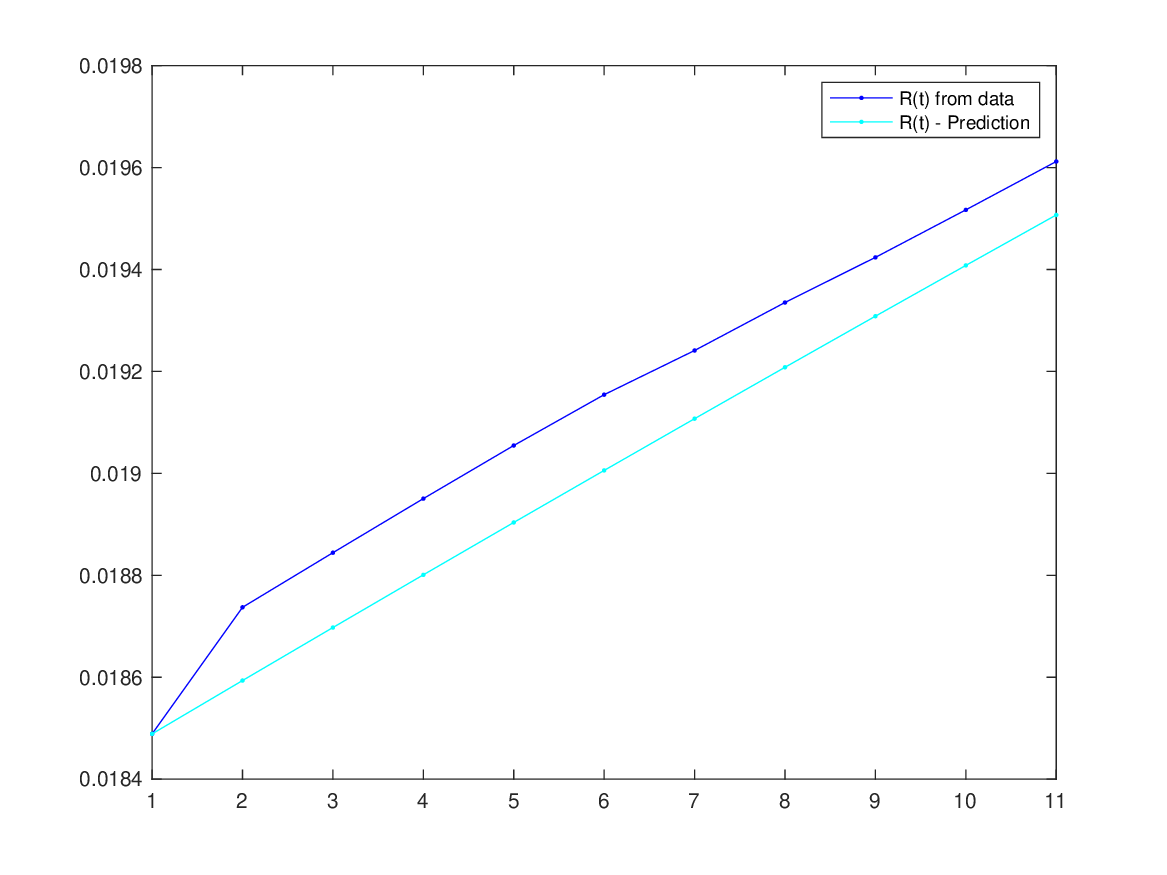}\vspace{-2mm}
\caption{t3}
\end{subfigure}
\begin{subfigure}{0.5\linewidth}
\centering
\includegraphics[scale=.32]{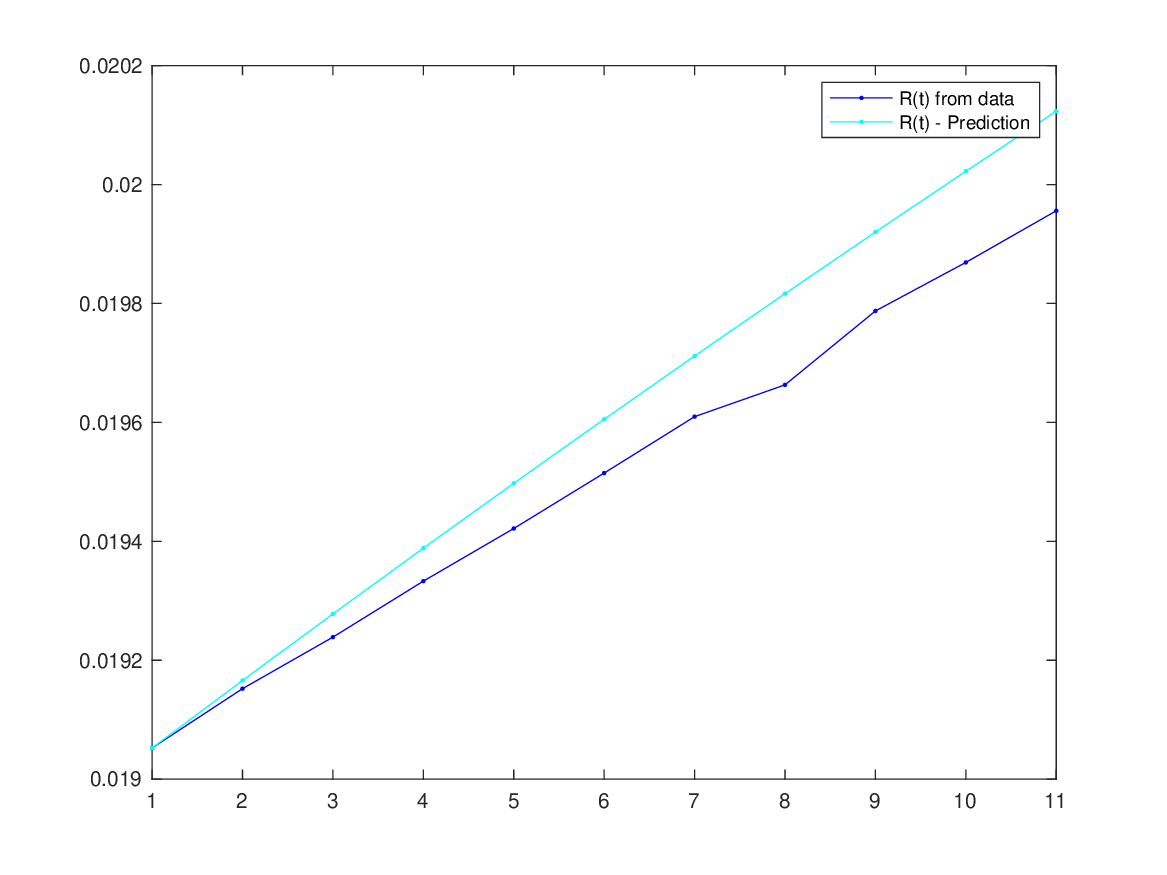}\vspace{-2mm}
\caption{t4}
\end{subfigure}
\caption{Prediction of $r(t)$ for the 4 starting dates and duration of 10 days
}\vspace{-.5cm}
\label{fig:predictionR10M}
\end{figure}

\begin{figure}[H]
\begin{subfigure} {0.5\textwidth}
\centering
\includegraphics[scale=0.32]{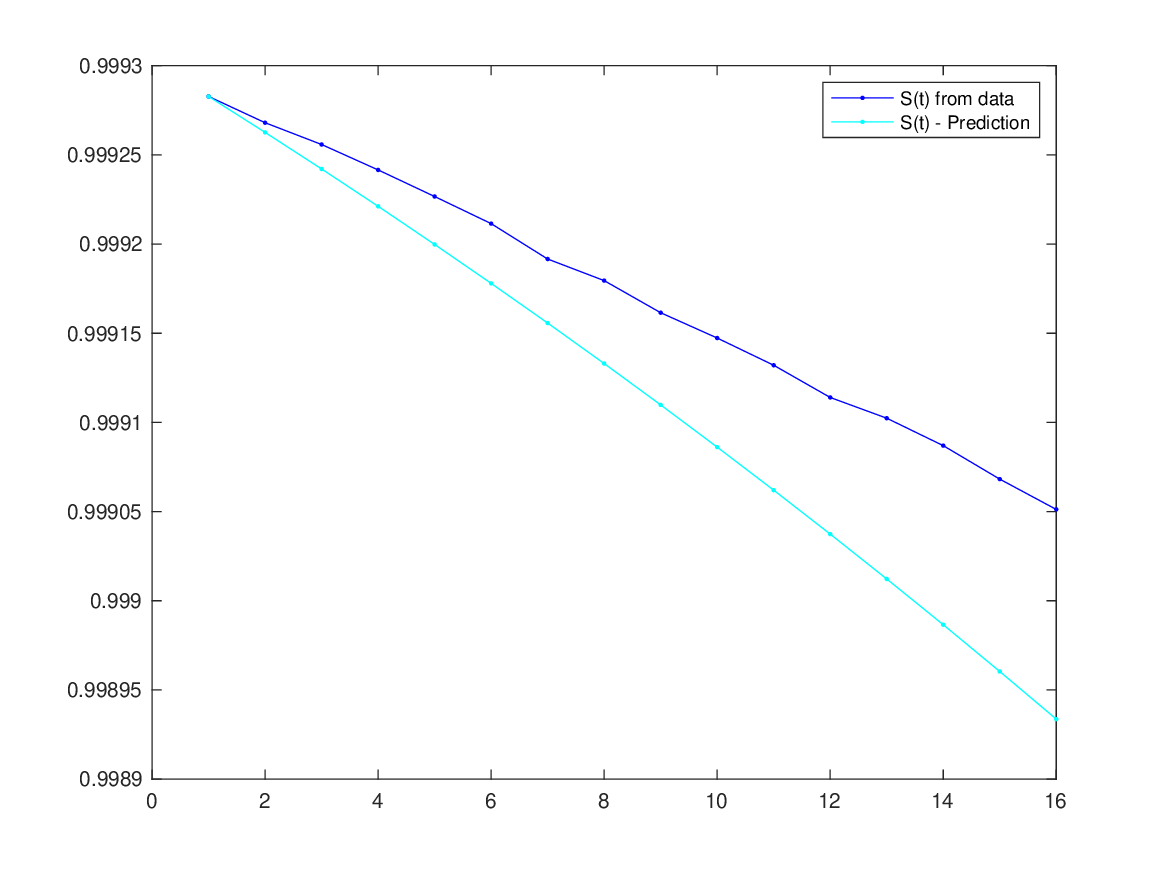}\vspace{-2mm}
\caption{t1}
\end{subfigure}%
\begin{subfigure}{.5\linewidth}
\centering
\includegraphics[scale=.32]{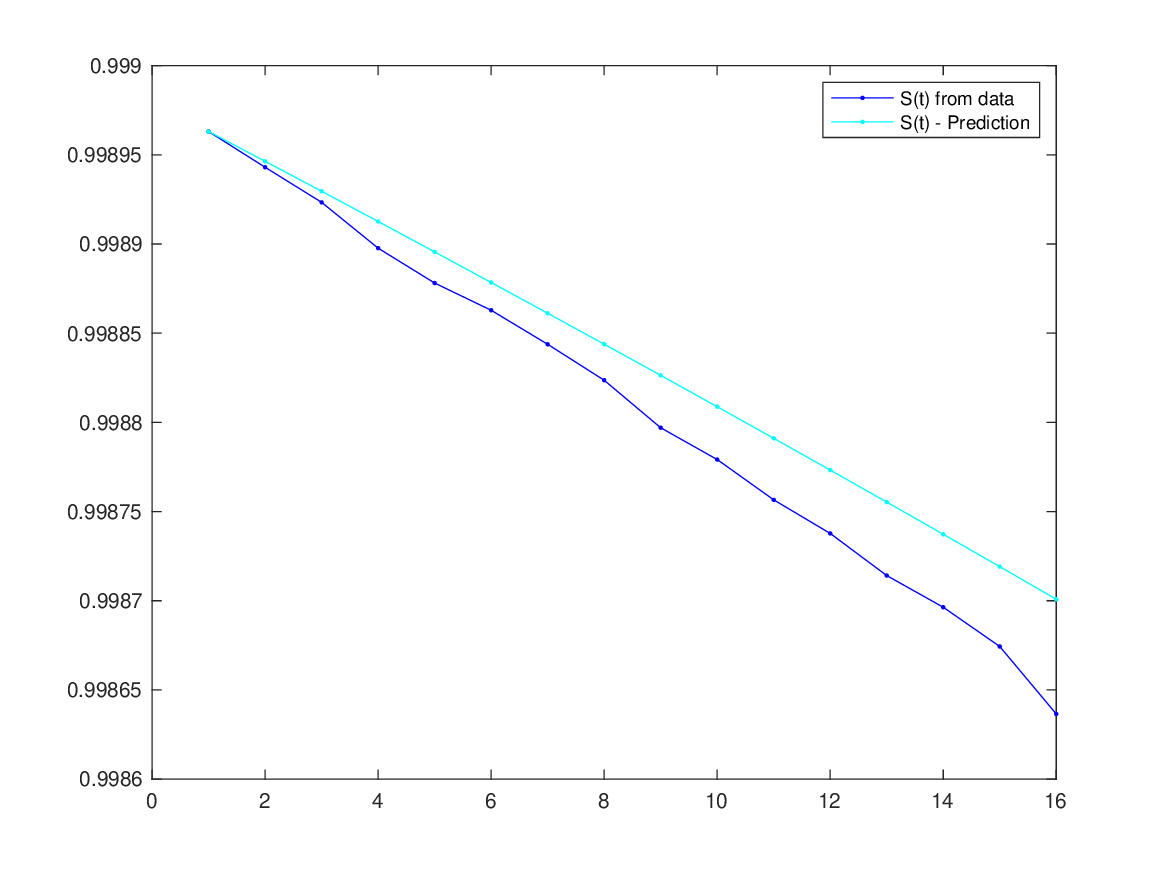}\vspace{-2mm}
\caption{t2}
\end{subfigure}
\begin{subfigure}{0.5\linewidth}
\centering
\includegraphics[scale=.32]{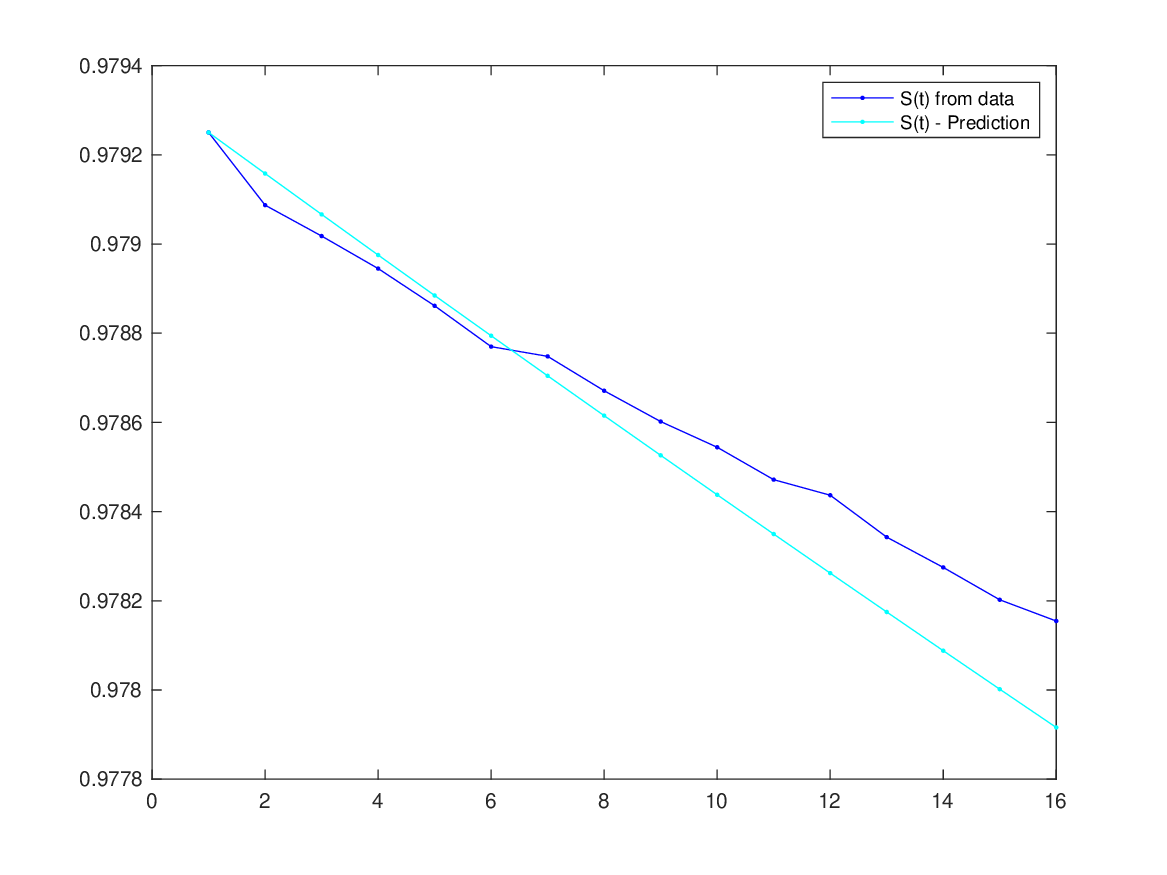}\vspace{-2mm}
\caption{t3}
\end{subfigure}
\begin{subfigure}{0.5\linewidth}
\centering
\includegraphics[scale=.32]{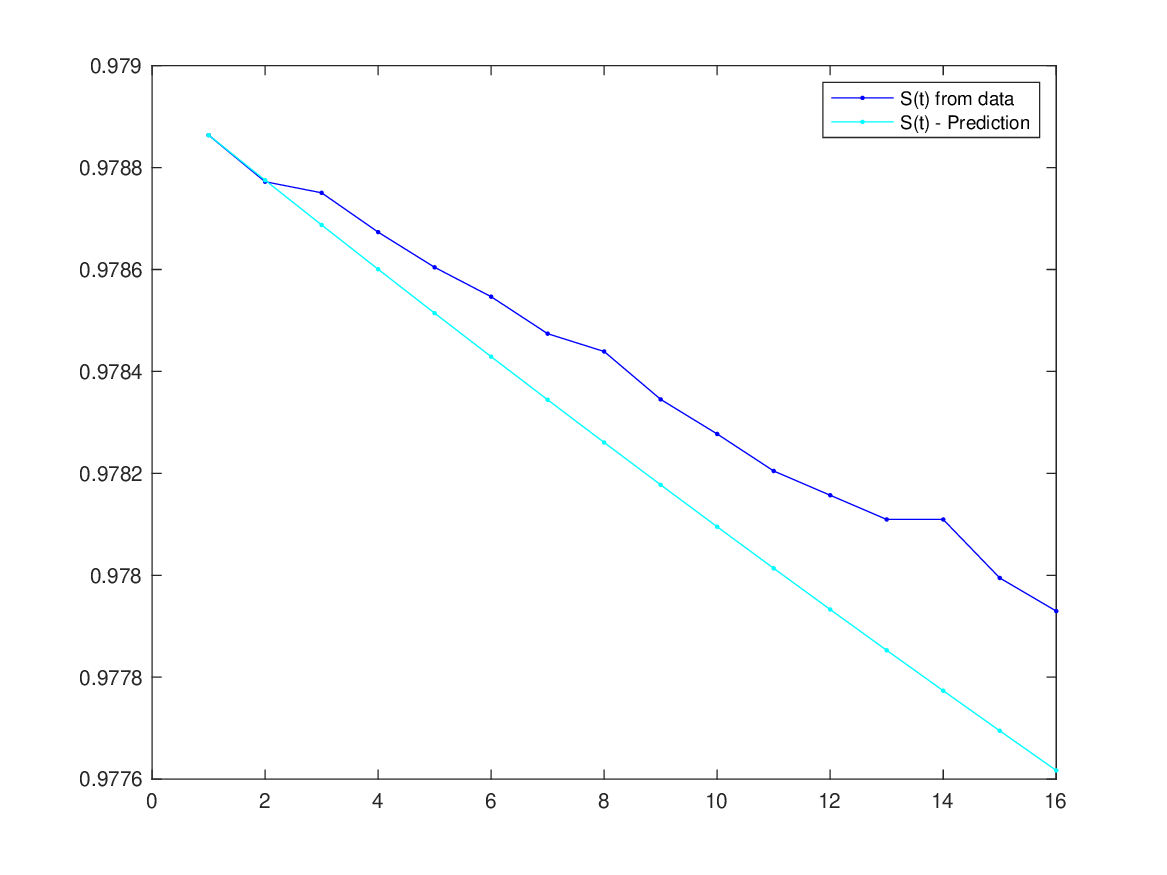}\vspace{-2mm}
\caption{t4}
\end{subfigure}
\caption{Prediction of $s(t)$ for the 4 starting dates and duration of 15 days
}\vspace{-.5cm}
\label{fig:predictionS15M}
\end{figure}

\begin{figure}[H]
\begin{subfigure} {0.5\textwidth}
\centering
\includegraphics[scale=0.32]{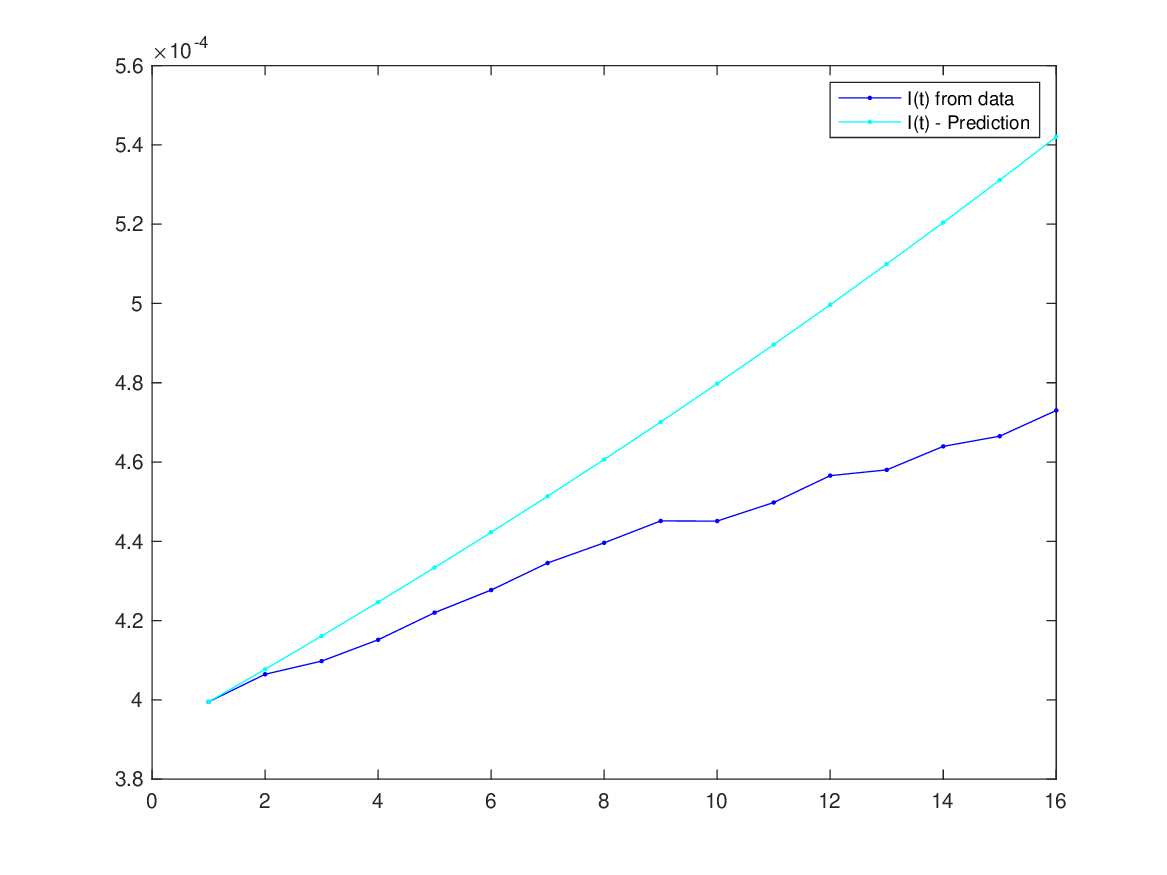}\vspace{-2mm}
\caption{t1}
\end{subfigure}%
\begin{subfigure}{.5\linewidth}
\centering
\includegraphics[scale=.32]{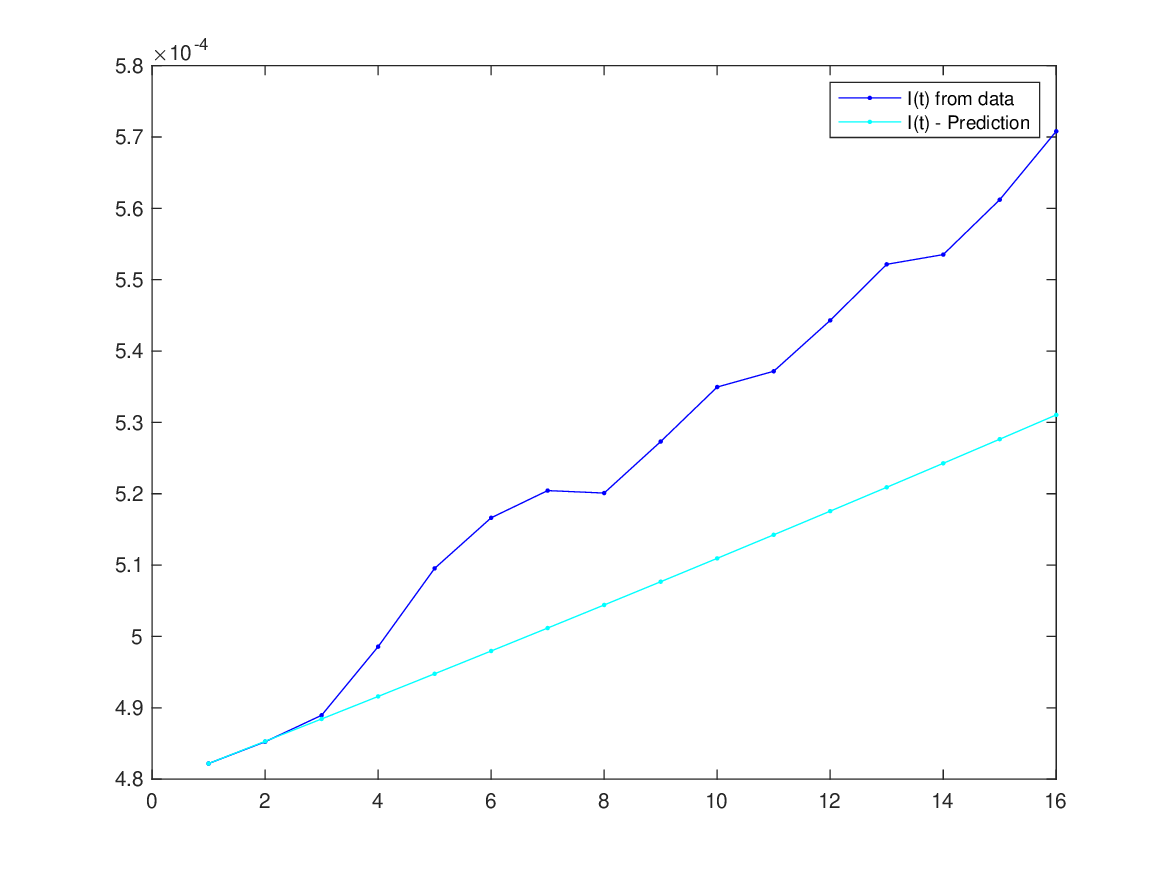}\vspace{-2mm}
\caption{t2}
\end{subfigure}
\begin{subfigure}{0.5\linewidth}
\centering
\includegraphics[scale=.32]{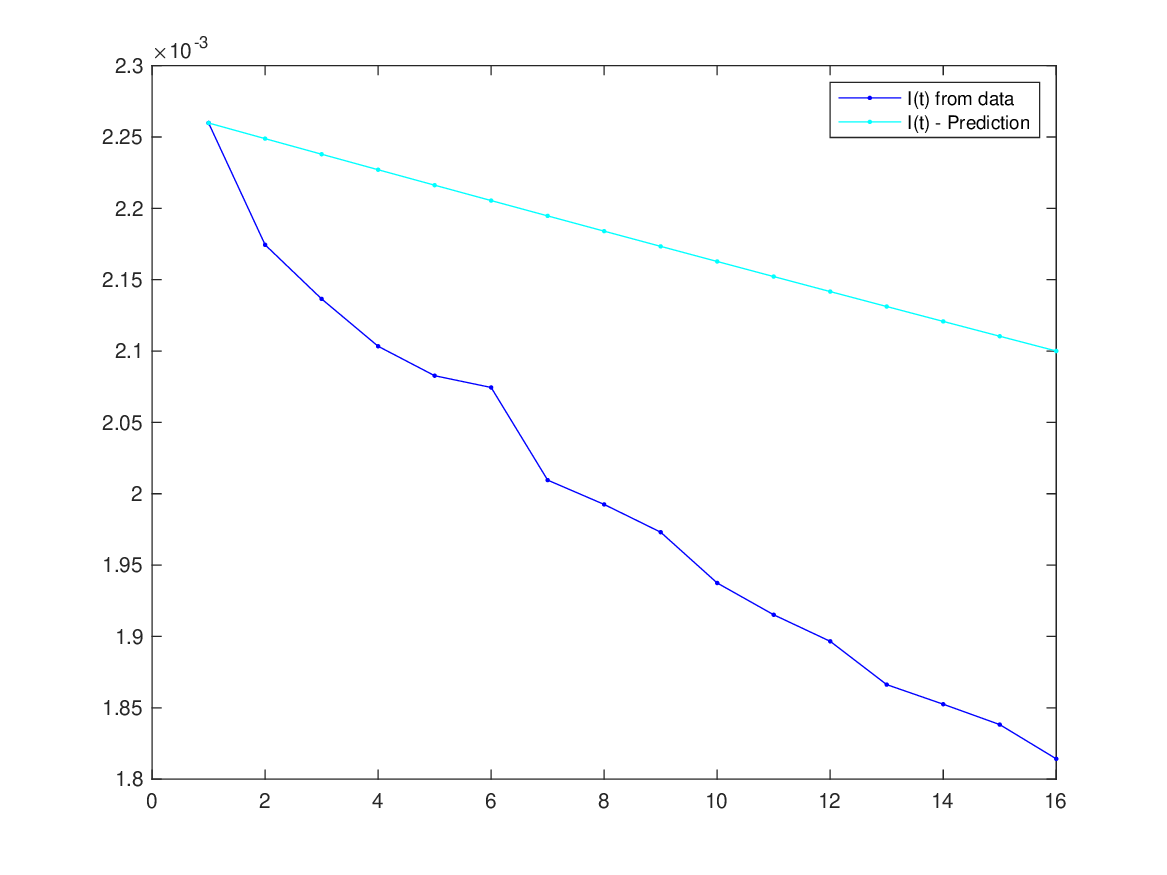}\vspace{-2mm}
\caption{t3}
\end{subfigure}
\begin{subfigure}{0.5\linewidth}
\centering
\includegraphics[scale=.32]{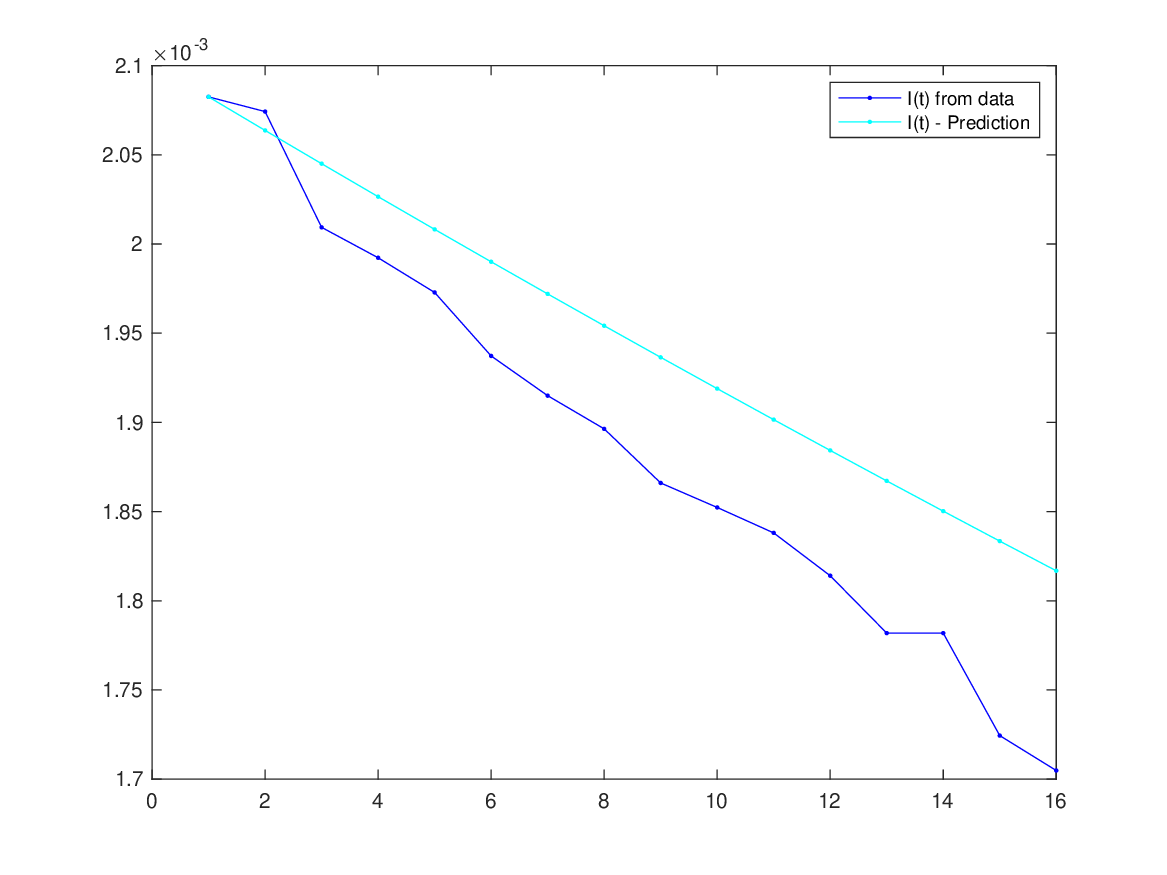}\vspace{-2mm}
\caption{t4}
\end{subfigure}
\caption{Prediction of $i(t)$ for the 4 starting dates and duration of 15 days
}\vspace{-.5cm}
\label{fig:predictionI15M}
\end{figure}

\begin{figure}[H]
\begin{subfigure} {0.5\textwidth}
\centering
\includegraphics[scale=0.32]{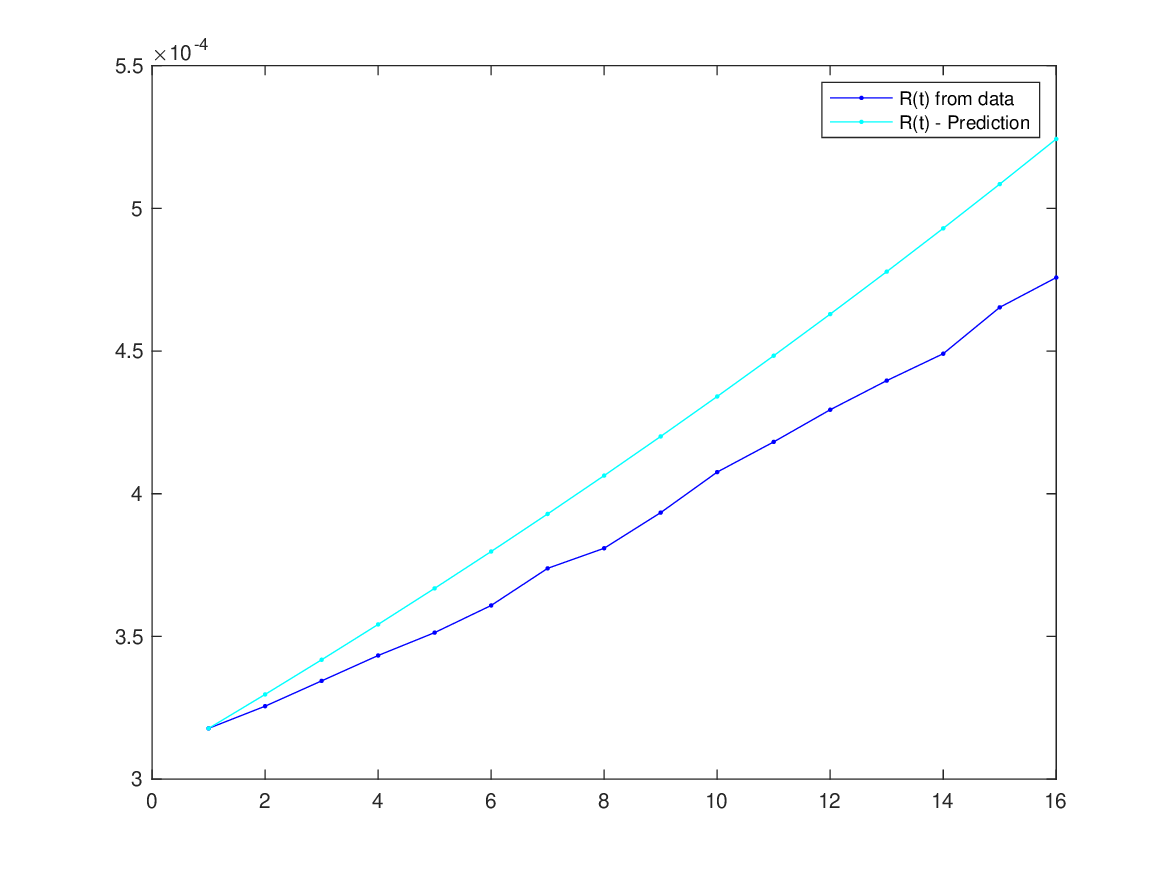}\vspace{-2mm}
\caption{t1}
\end{subfigure}%
\begin{subfigure}{.5\linewidth}
\centering
\includegraphics[scale=.32]{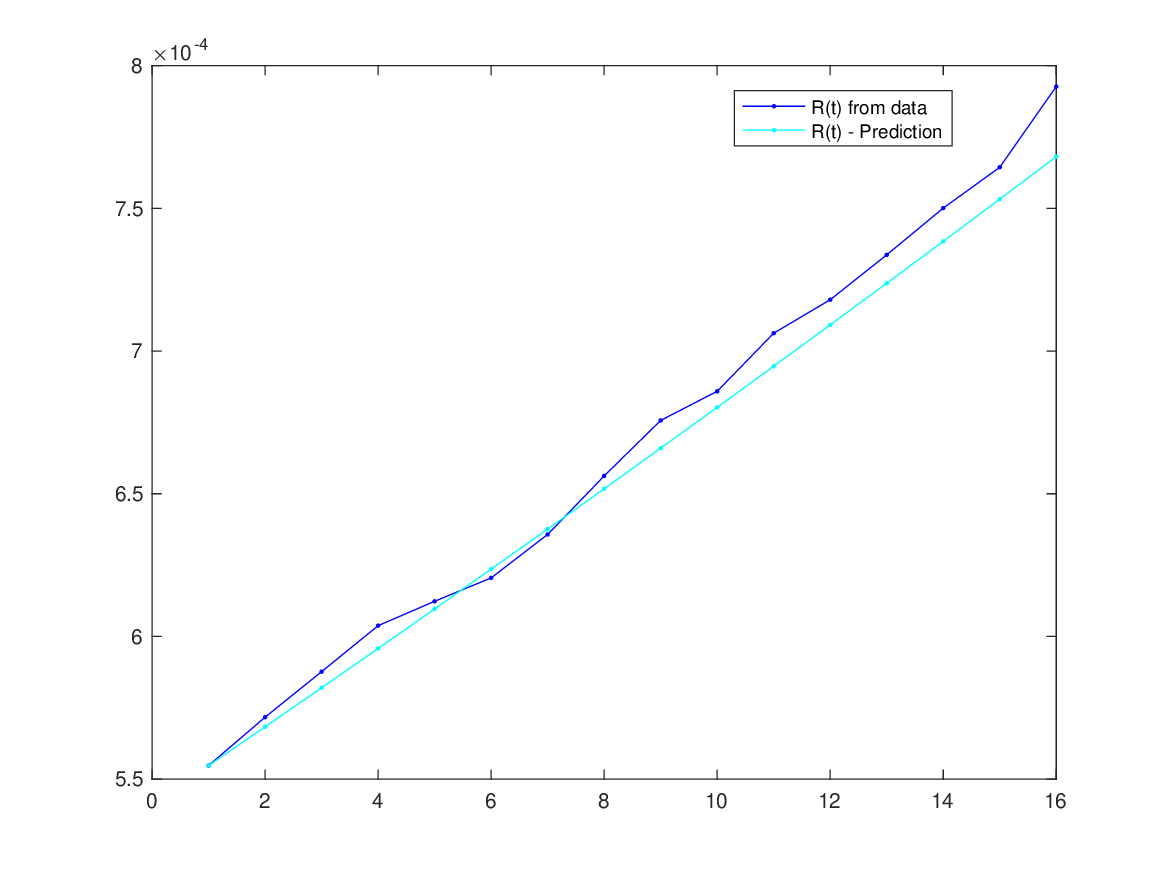}\vspace{-2mm}
\caption{t2}
\end{subfigure}
\begin{subfigure}{0.5\linewidth}
\centering
\includegraphics[scale=.32]{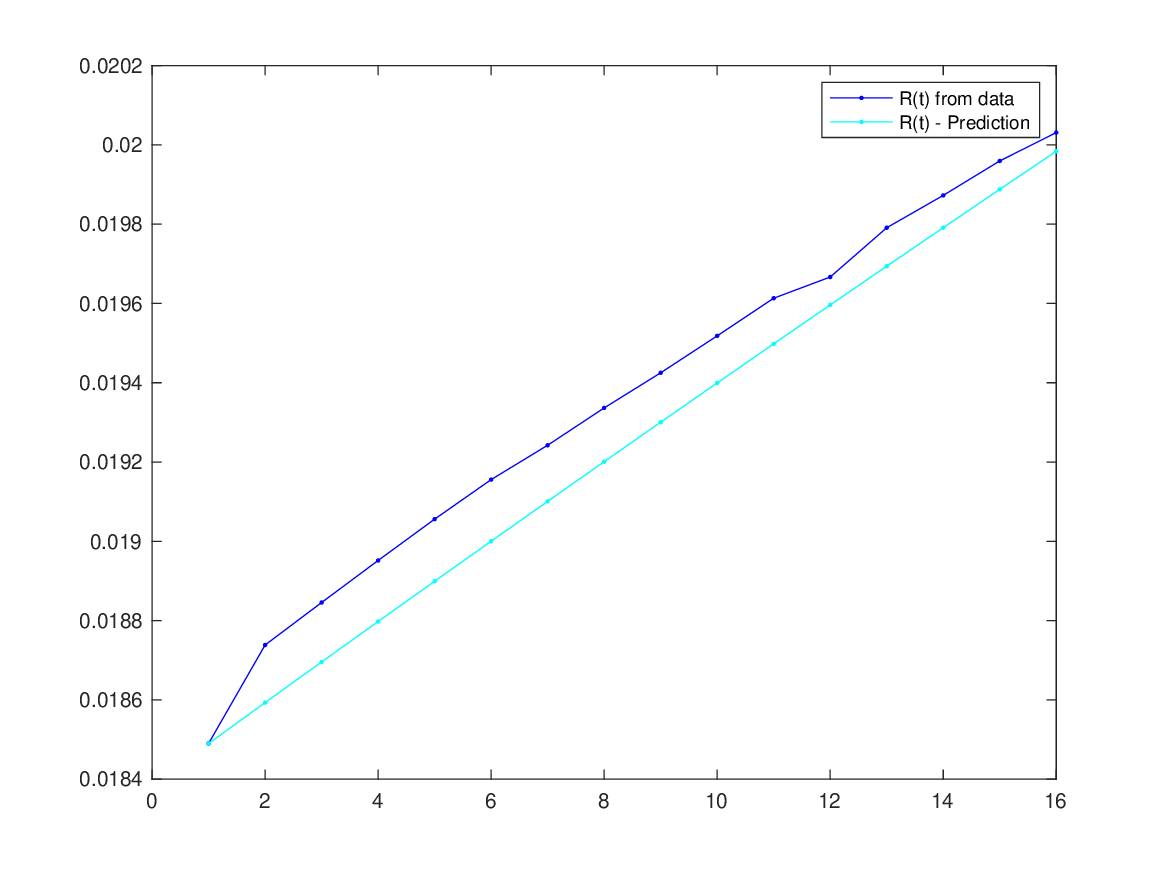}\vspace{-2mm}
\caption{t3}
\end{subfigure}
\begin{subfigure}{0.5\linewidth}
\centering
\includegraphics[scale=.32]{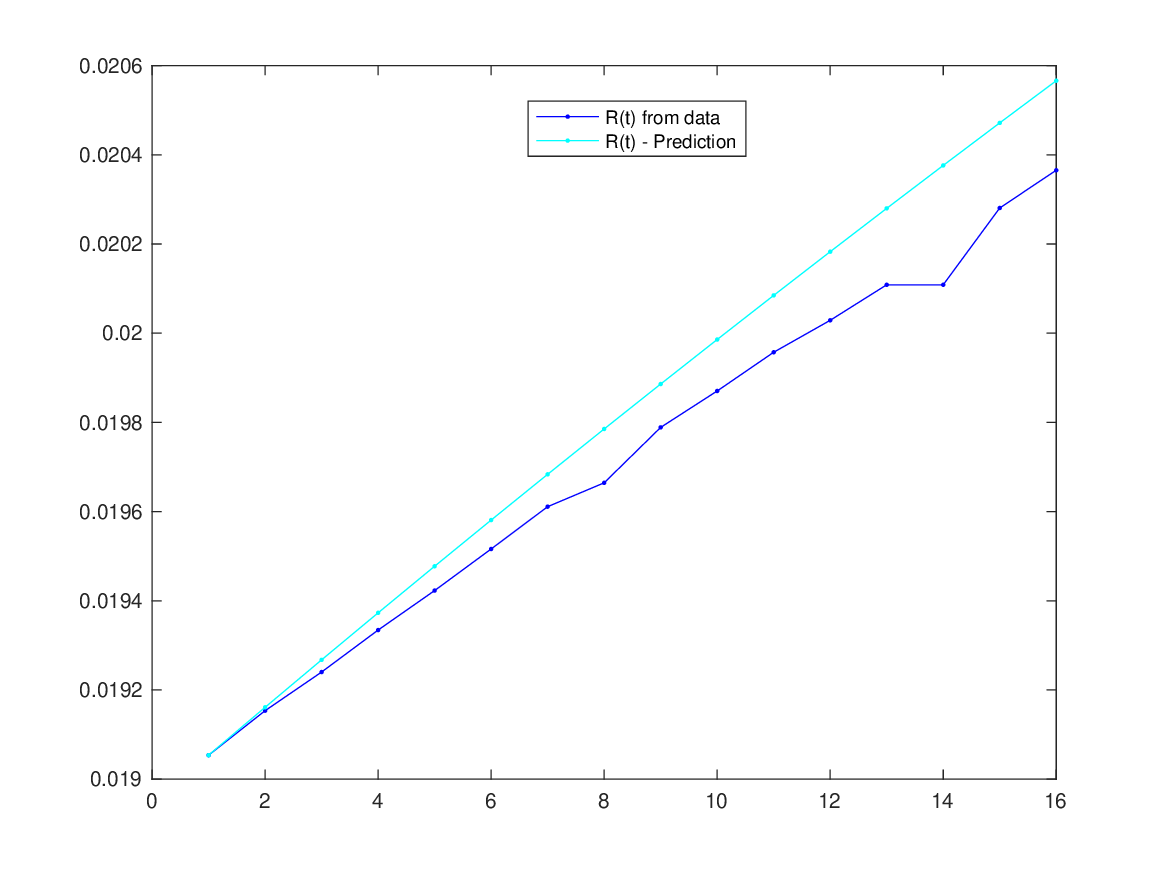}\vspace{-2mm}
\caption{t4}
\end{subfigure}
\caption{Prediction of $r(t)$ for the 4 starting dates and duration of 15 days
}\vspace{-.5cm}
\label{fig:predictionR15M}
\end{figure}

\begin{figure}[H]
\begin{subfigure} {0.5\textwidth}
\centering
\includegraphics[scale=0.32]{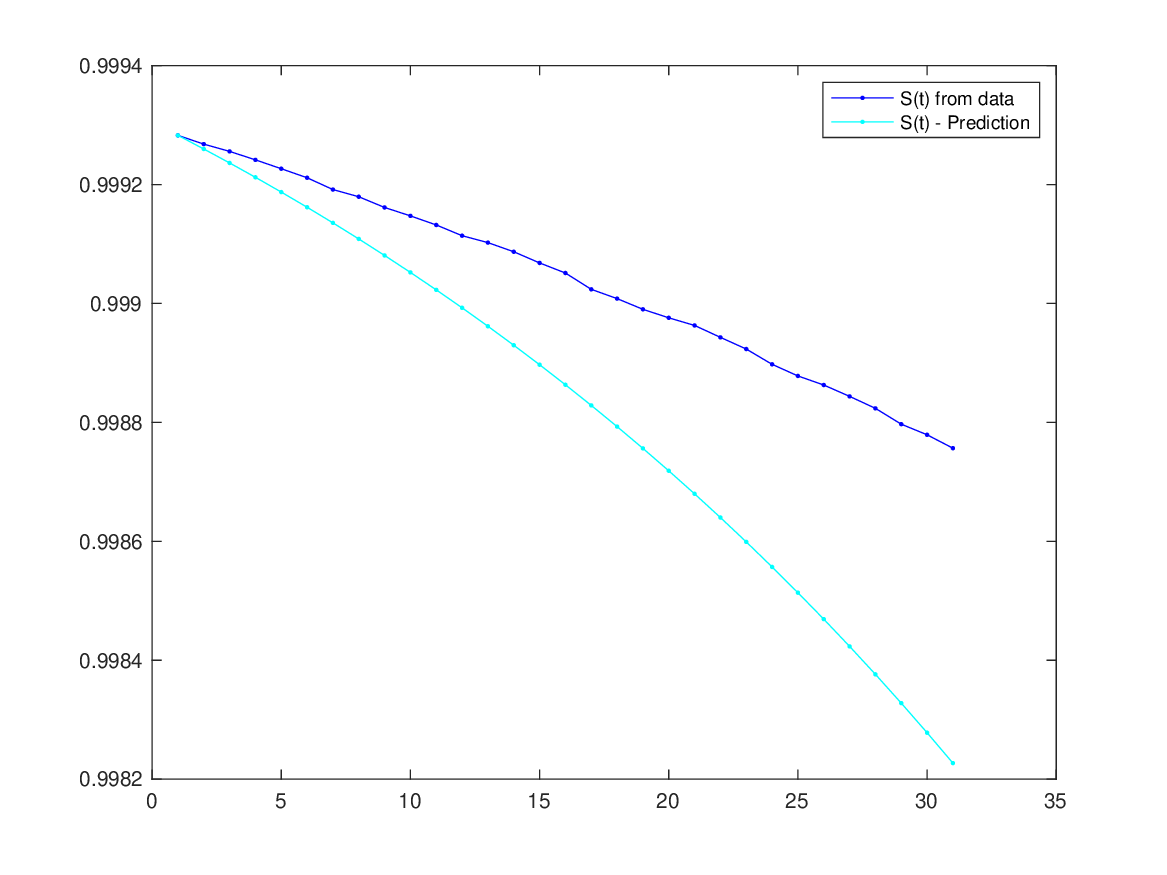}\vspace{-2mm}
\caption{t1}
\end{subfigure}%
\begin{subfigure}{.5\linewidth}
\centering
\includegraphics[scale=.32]{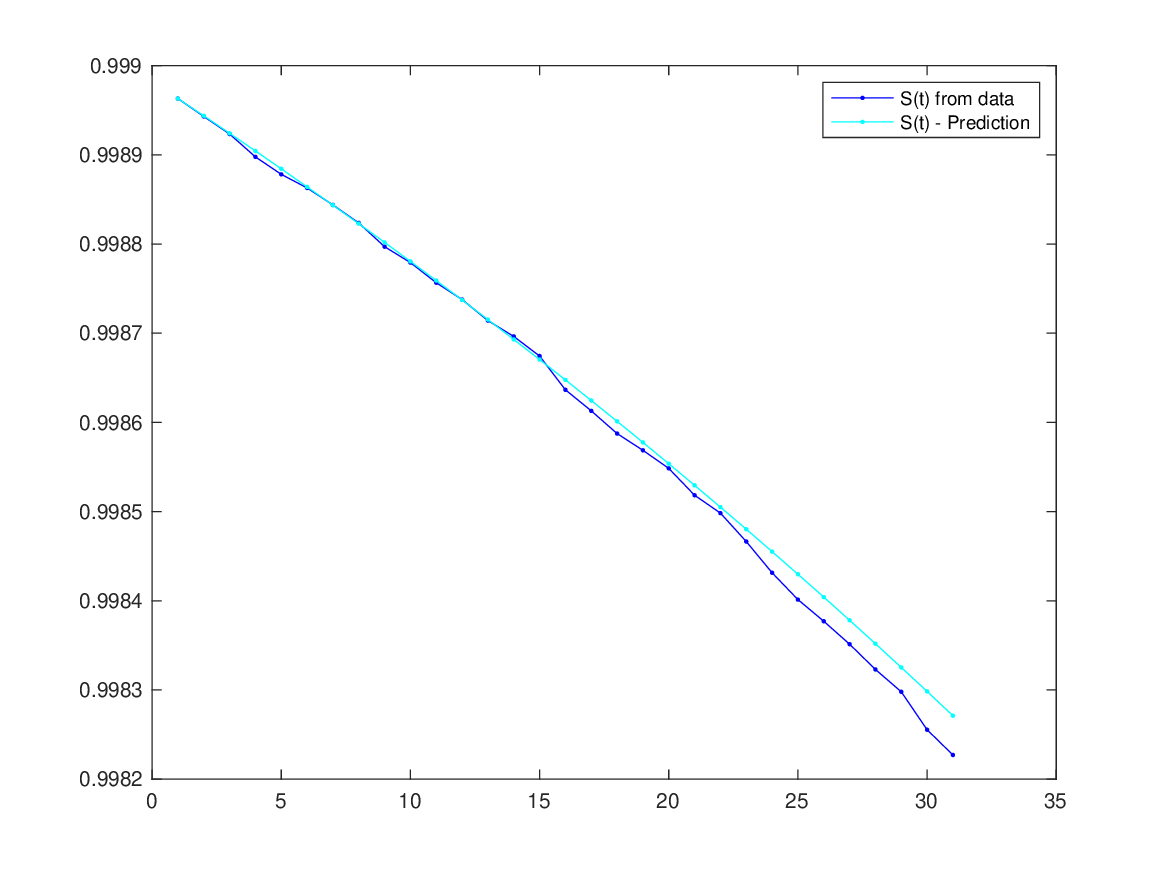}\vspace{-2mm}
\caption{t2}
\end{subfigure}
\begin{subfigure}{0.5\linewidth}
\centering
\includegraphics[scale=.32]{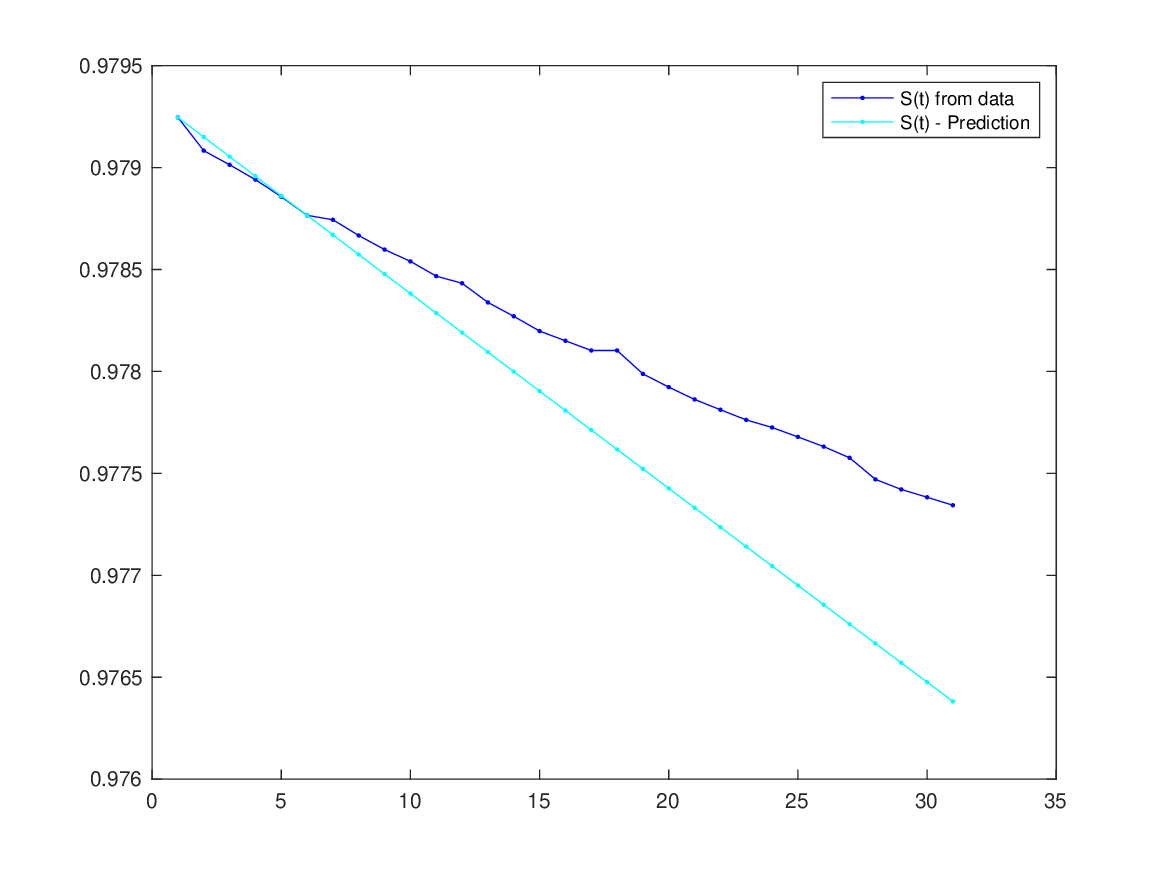}\vspace{-2mm}
\caption{t3}
\end{subfigure}
\begin{subfigure}{0.5\linewidth}
\centering
\includegraphics[scale=.32]{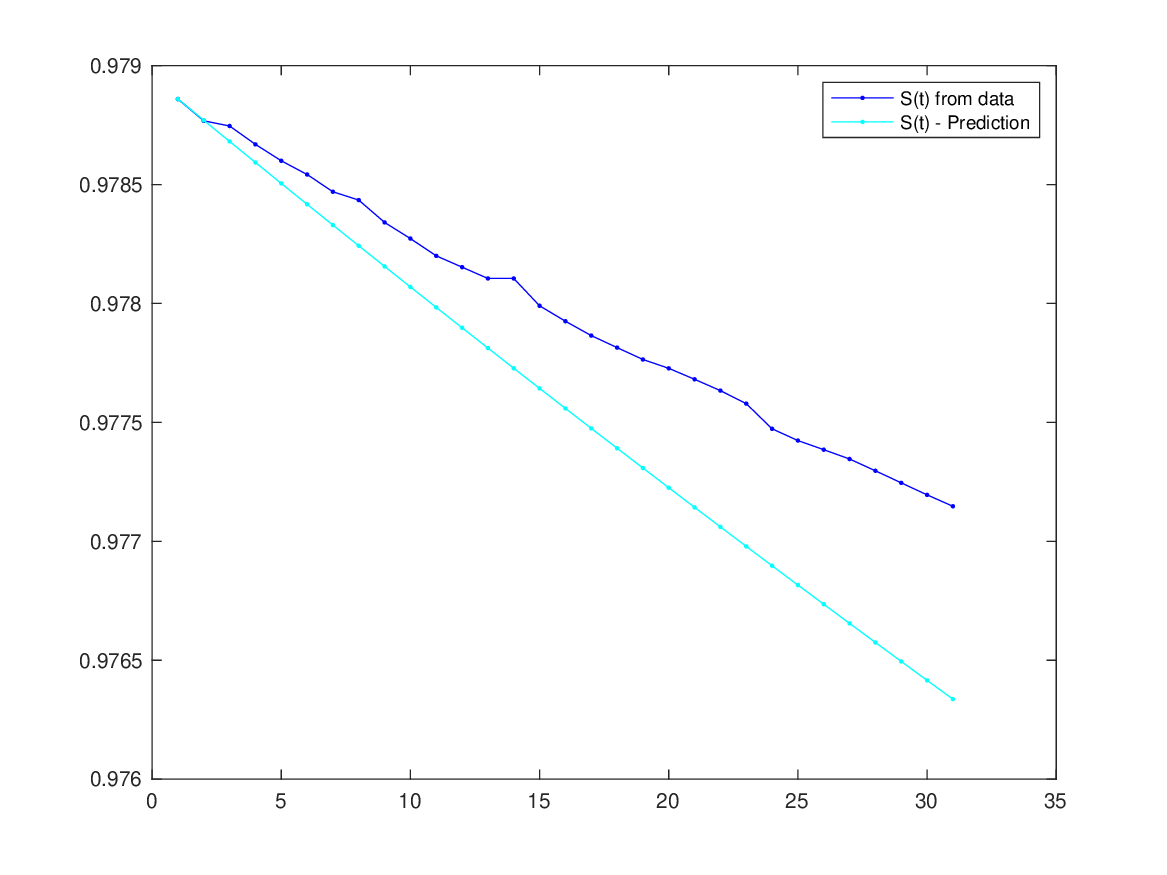}\vspace{-2mm}
\caption{t4}
\end{subfigure}
\caption{Prediction of $s(t)$ for the 4 starting dates and duration of 30 days
}\vspace{-.5cm}
\label{fig:predictionS30M}
\end{figure}

\begin{figure}[H]
\begin{subfigure} {0.5\textwidth}
\centering
\includegraphics[scale=0.32]{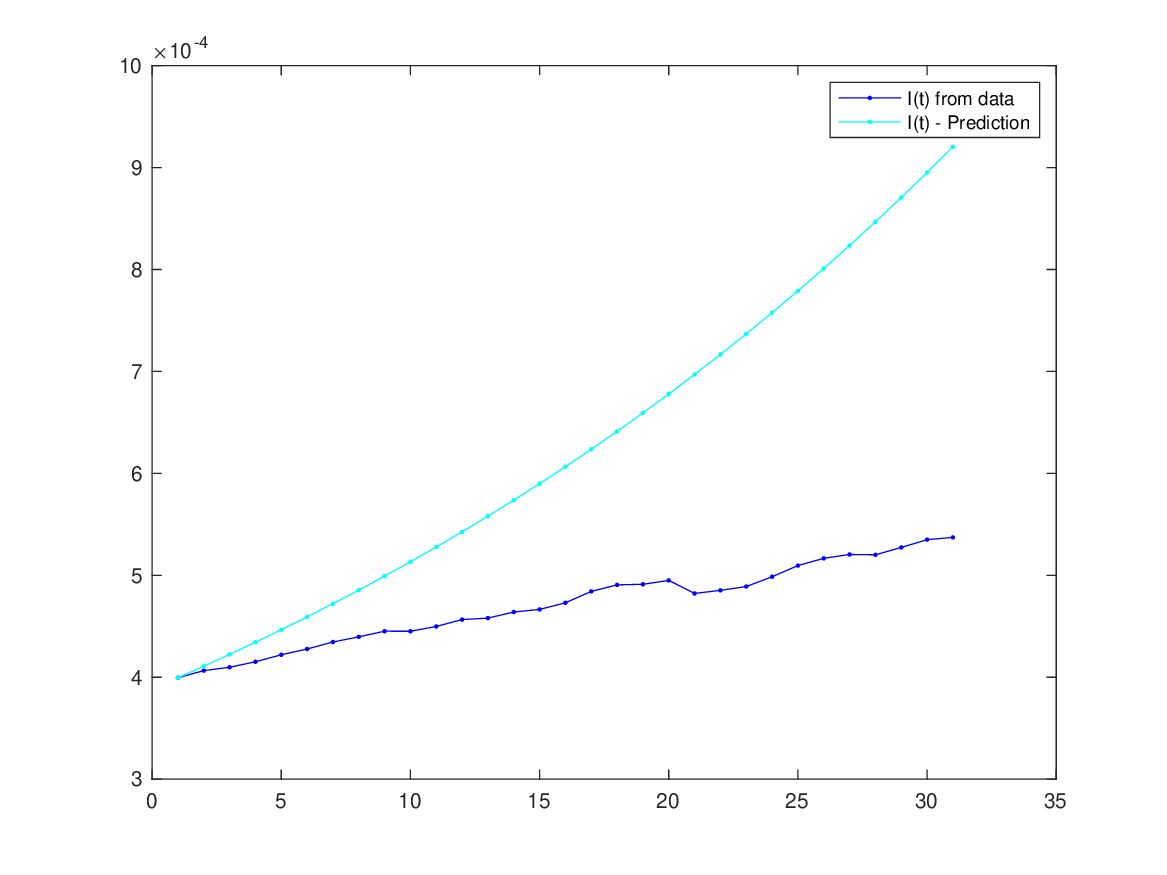}\vspace{-2mm}
\caption{t1}
\end{subfigure}%
\begin{subfigure}{.5\linewidth}
\centering
\includegraphics[scale=.32]{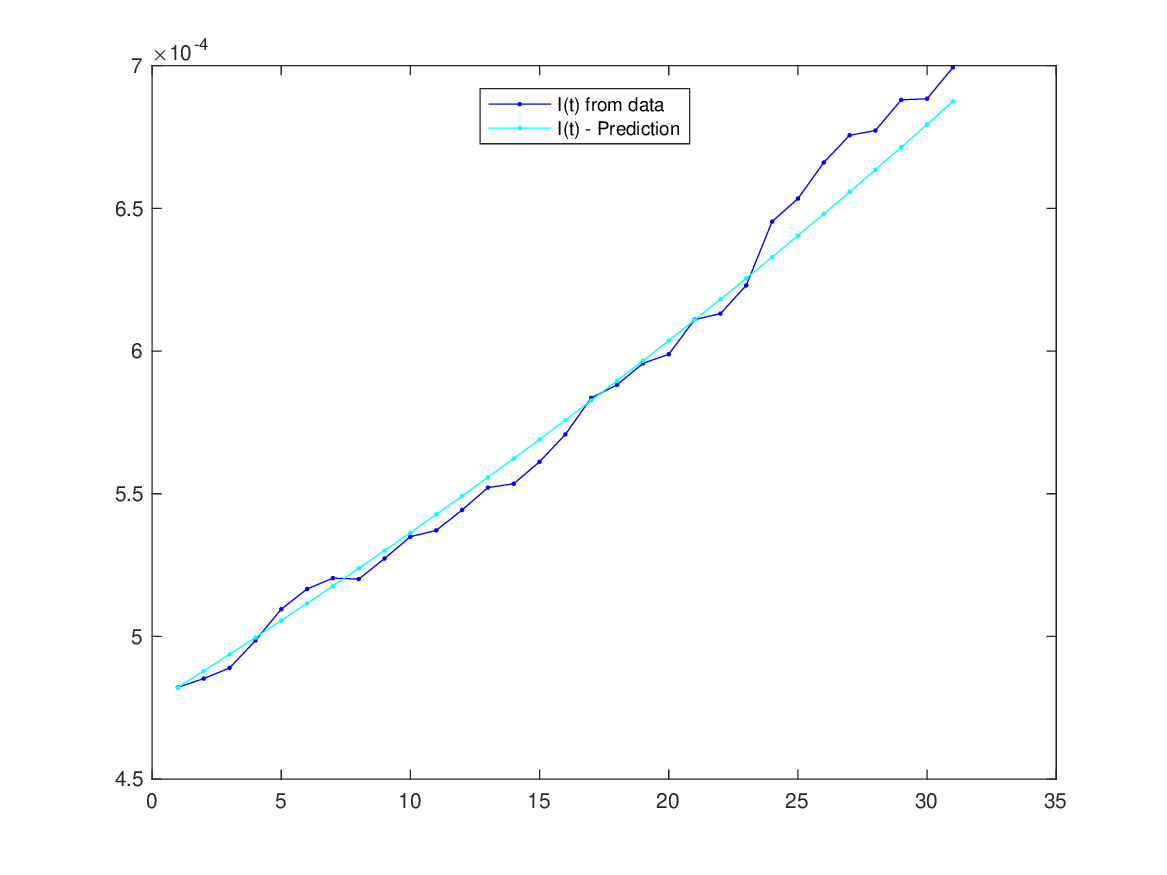}\vspace{-2mm}
\caption{t2}
\end{subfigure}
\begin{subfigure}{0.5\linewidth}
\centering
\includegraphics[scale=.32]{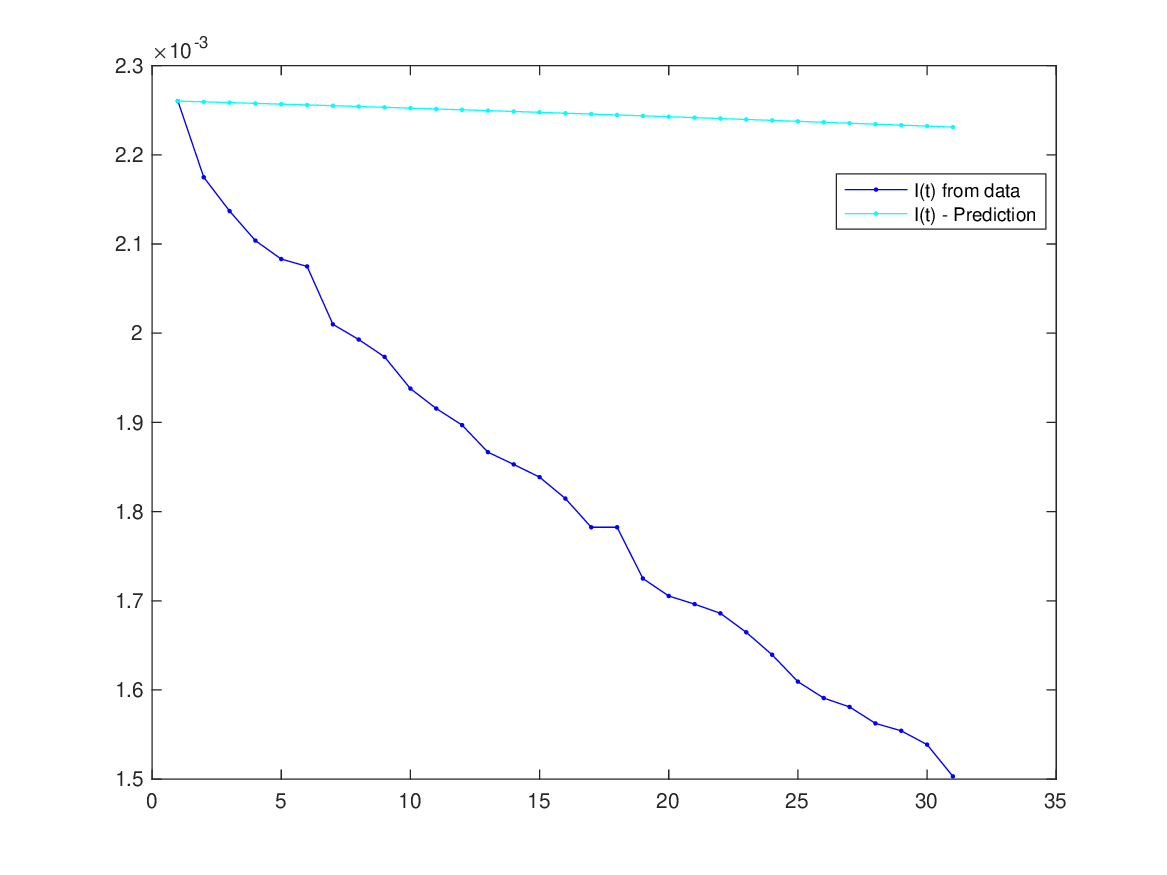}\vspace{-2mm}
\caption{t3}
\end{subfigure}
\begin{subfigure}{0.5\linewidth}
\centering
\includegraphics[scale=.32]{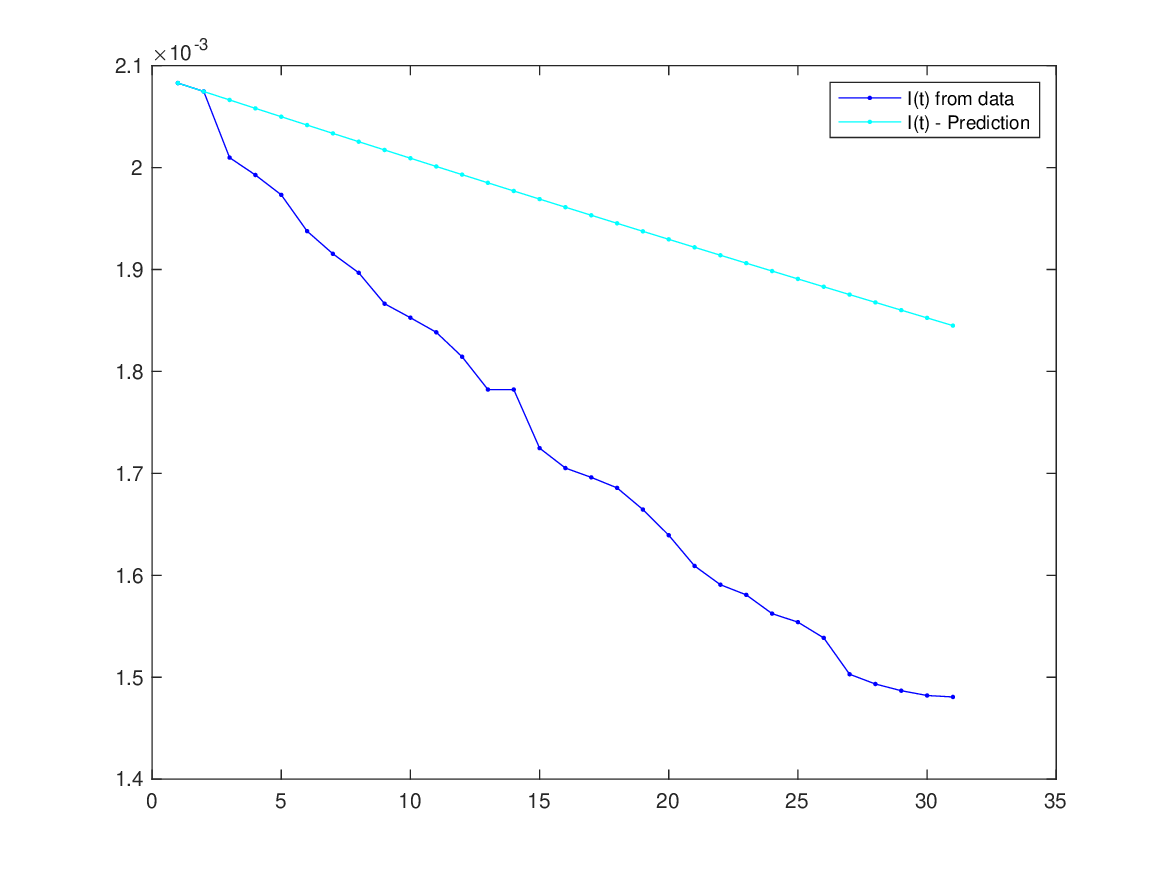}\vspace{-2mm}
\caption{t4}
\end{subfigure}
\caption{Prediction of $i(t)$ for the 4 starting dates and duration of 30 days
}\vspace{-.5cm}
\label{fig:predictionI30M}
\end{figure}

\begin{figure}[H]
\begin{subfigure} {0.5\textwidth}
\centering
\includegraphics[scale=0.32]{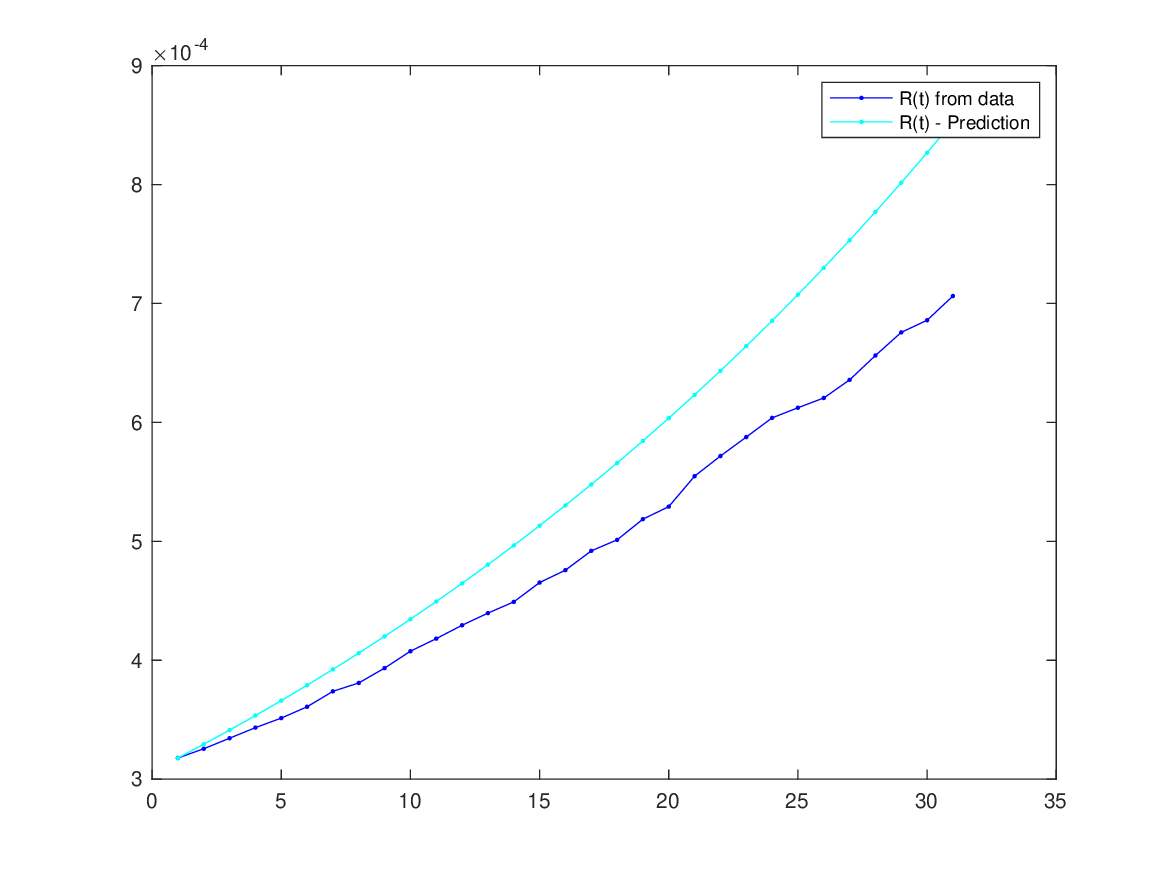}\vspace{-2mm}
\caption{t1}
\end{subfigure}%
\begin{subfigure}{.5\linewidth}
\centering
\includegraphics[scale=.32]{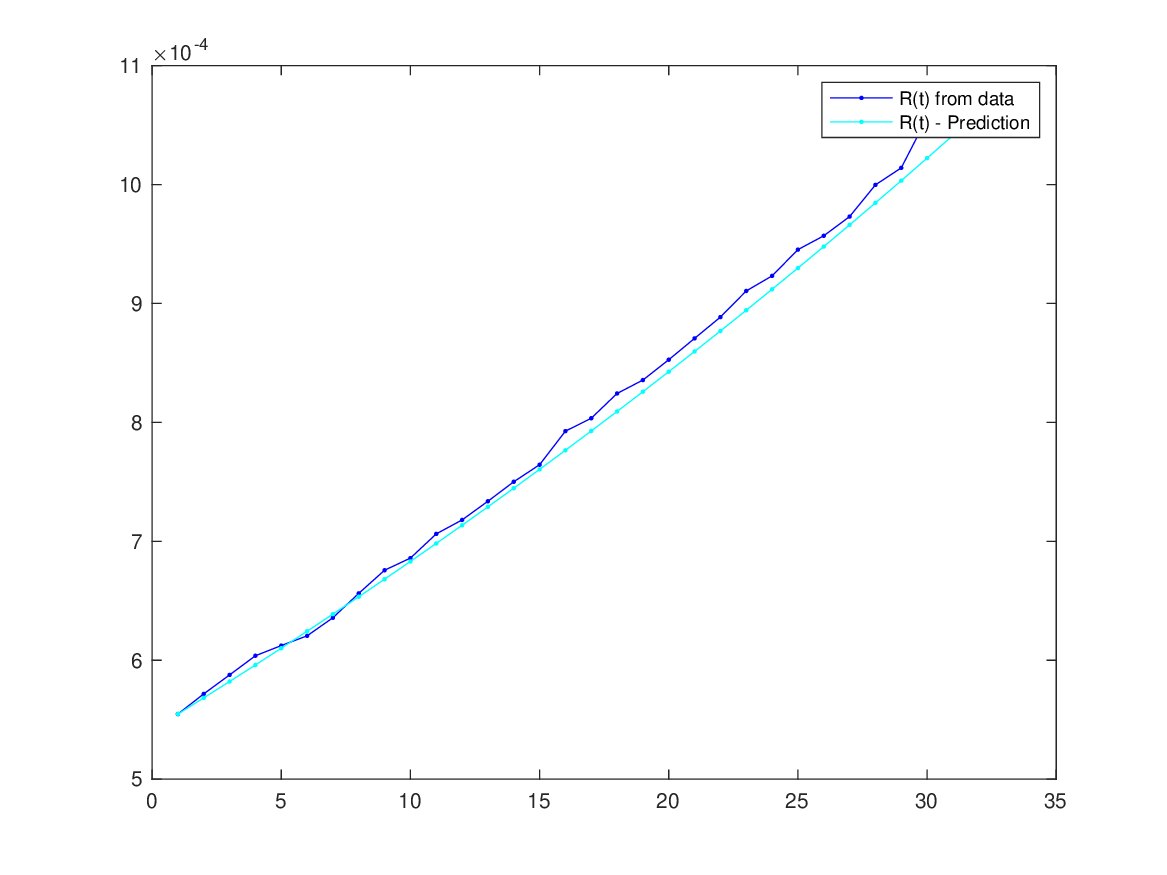}\vspace{-2mm}
\caption{t2}
\end{subfigure}
\begin{subfigure}{0.5\linewidth}
\centering
\includegraphics[scale=.32]{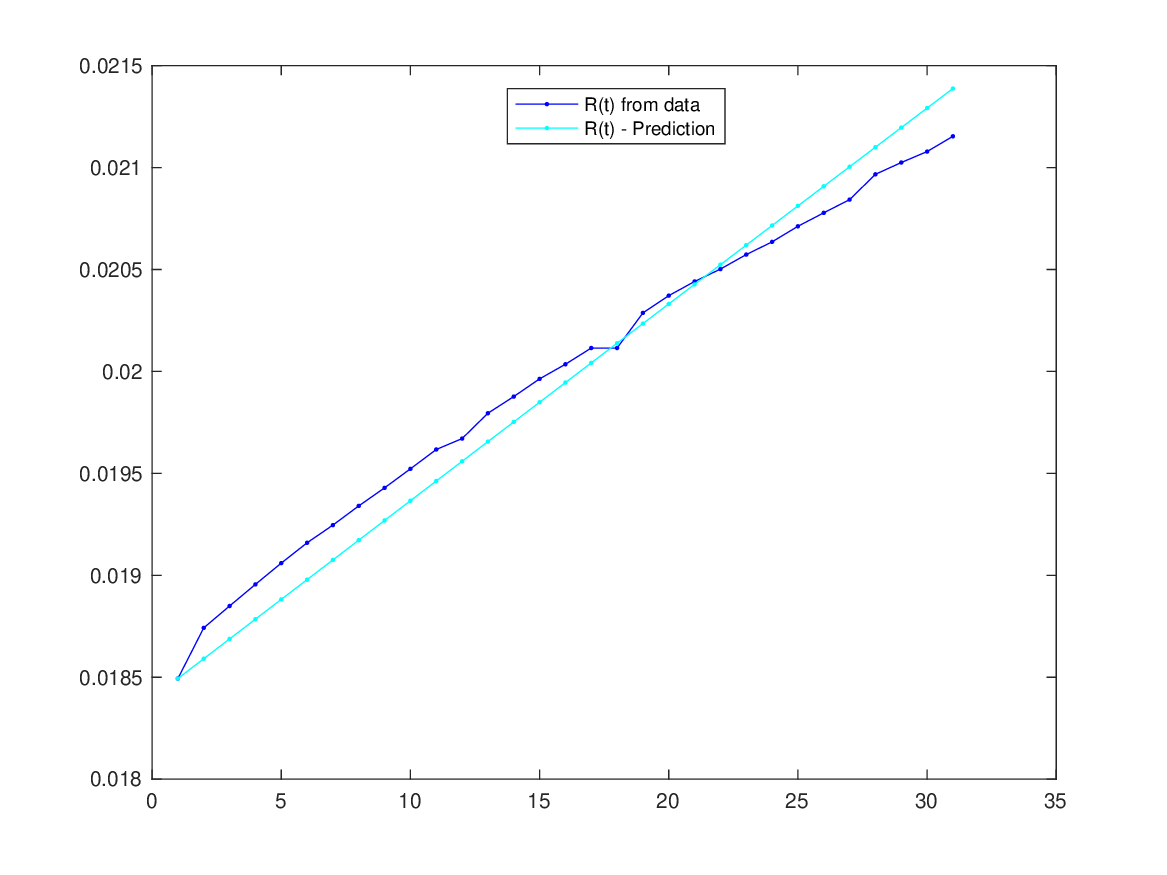}\vspace{-2mm}
\caption{t3}
\end{subfigure}
\begin{subfigure}{0.5\linewidth}
\centering
\includegraphics[scale=.32]{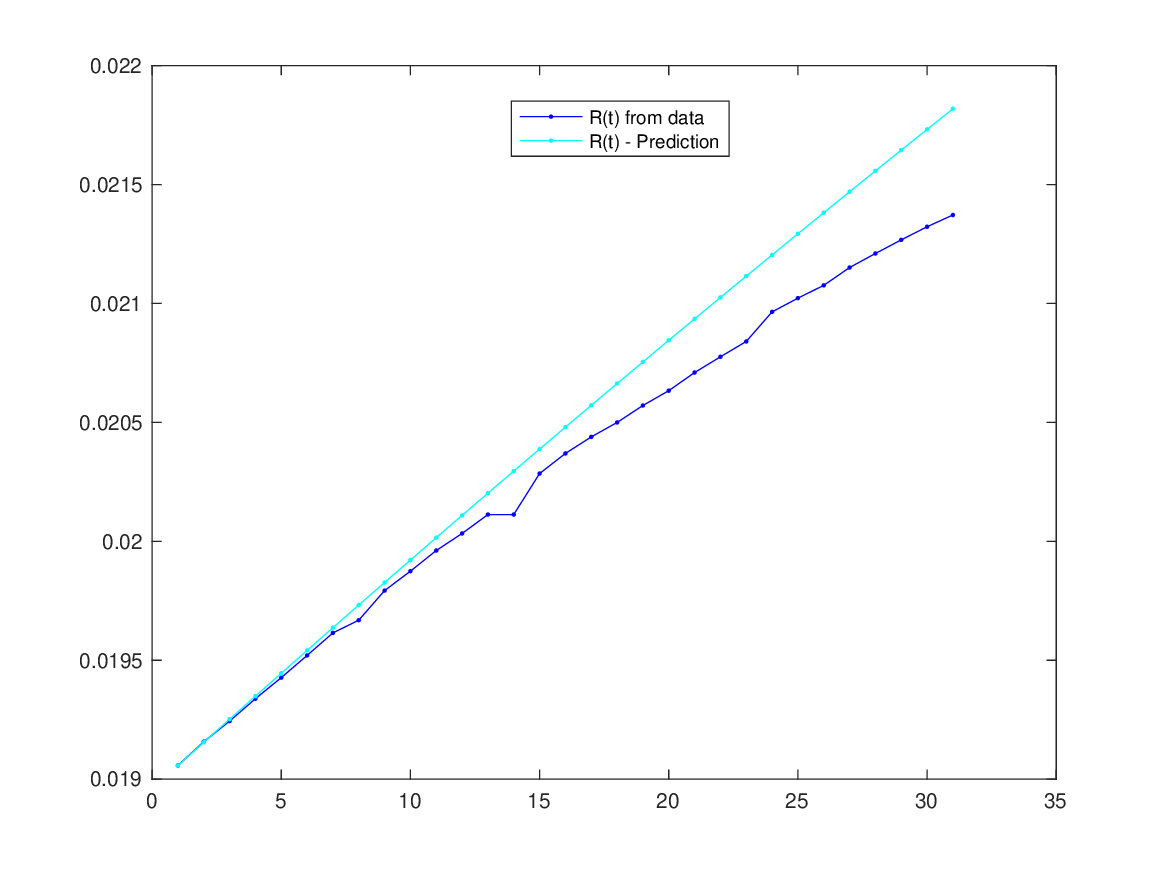}\vspace{-2mm}
\caption{t4}
\end{subfigure}
\caption{Prediction of $r(t)$ for the 4 starting dates and duration of 30 days
}\vspace{-.5cm}
\label{fig:predictionR30M}
\end{figure}

\begin{table}[H]
\centering
\setlength{\tabcolsep}{5pt}
{\renewcommand{\arraystretch}{1.4}
\begin{tabular}{||c|c||c|c|c||c|c|c||c|c|c||}
 \cline{3-11}
  \cline{3-11}
\multicolumn{2}{c||}{}  &\multicolumn{3}{c||}{k = 10 Days} & \multicolumn{3}{c||}{k = 20 Days} & \multicolumn{3}{c||}{k = 30 Days} \\
 \cline{3-11}
\multicolumn{2}{c||}{}  & S & I & R& S & I & R& S & I & R \\ 
 \hline\hline
\multirow{2}{*}{$t1$} & $L_2$ &    3.79E-5 & 3.93E-2 & 5.85E-2 & 8.70E-5 & 1.23E-1 & 7.46E-2 & 2.65E-4 & 4.09E-1 & 1.43E-1 \\
\cline{2-11}
 &$L_\infty$ &    6.69E-5 & 7.40E-2 & 8.01E-2 & 1.66E-4 & 2.44E-1 & 9.73E-2 & 5.30E-4 & 7.13E-1 & 2.07E-1 \\ 
 \hline  \hline
\multirow{2}{*}{$t2$} &$L_2$ & 5.94E-5 & 7.00E-2 & 6.68E-3 & 2.24E-4 & 1.37E-1 & 5.59E-3 & 5.10E-4 & 2.54E-1 & 6.74E-3 \\ 
\cline{2-11}
 &$L_\infty$ &   1.07E-4 & 9.26E-2 & 7.70E-3 & 4.26E-4 & 1.77E-1 & 7.89E-3 & 9.83E-4 & 3.22E-1 & 1.11E-2 \\ 
\hline  \hline
 \multirow{2}{*}{$t3$} &$L_2$ &1.12E-4 & 5.87E-3 & 5.40E-3 & 2.53E-4 & 6.01E-2 & 7.00E-3 & 4.68E-4 & 1.45E-1 & 1.03E-2 \\ 
 \cline{2-11}
 &$L_\infty$ &  1.62E-4 & 9.51E-3 & 8.39E-3 & 4.53E-4 & 8.67E-2 & 1.27E-2 & 8.28E-4 & 1.80E-1 & 2.09E-2 \\  
 \hline  \hline
  \multirow{2}{*}{$t4$} &$L_2$ & 2.18E-5 & 4.31E-2 & 6.42E-3 & 4.11E-5 & 4.16E-2 & 2.63E-2 & 1.73E-5 & 1.42E-2 & 1.53E-2 \\
 \cline{2-11}
 &$L_\infty$ &   3.77E-5 & 6.30E-2 & 1.03E-2 & 7.81E-5 & 7.26E-2 & 3.87E-2 & 4.41E-5 & 2.84E-2 & 3.16E-2 \\ 
 \hline  \hline
\end{tabular}}\caption{Relative error between between exact solution and predicted one where $\beta(t)$ and $\rho(t)$ are computed using Method 2.}
\end{table}

\end{document}